\documentclass[numbers,spanish,a4paper,singlespace,nonatbib]{ezthesis}

\usepackage[spanish]{babel}
\usepackage[latin1]{inputenc}

\usepackage{amsmath}
\usepackage{amsfonts}
\usepackage{amssymb}
\usepackage{dsfont}
\usepackage{accents}
\usepackage[mathscr]{eucal}
\usepackage{braket}

\usepackage{color}

\usepackage[style=english]{csquotes}

\spanishdecimal{.}

\newcommand{\rmd}{\ensuremath{\mathrm{d}}}
\newcommand{\rme}{\ensuremath{\mathrm{e}}}
\newcommand{\rmi}{\ensuremath{\mathrm{i}}}

\newcommand{\hfrac}[2]{\ensuremath{\left.#1\middle/#2\right.}}
\newcommand{\deriv}[3][]{\ensuremath{\frac{\rmd^{#1} #2}{{\rmd #3}^{#1}}}}
\newcommand{\pderiv}[3][]{\ensuremath{\frac{\partial^{#1} #2}{{\partial #3}^{#1}}}}
\newcommand{\fr}[1]{\ensuremath{\frac{1}{#1}}}
\newcommand{\evat}[3][]{\ensuremath{\left.#2\right|^{#1}_{#3}}}
\newcommand{\derivat}[4][]{\ensuremath{\evat{\deriv[#1]{#2}{#3}}{#4}}}
\newcommand{\derivateq}[4][]{\ensuremath{\derivat[#1]{#2}{#3}{#3=#4}}}
\newcommand{\sra}[1][]{\ensuremath{\quad \mathop{\longrightarrow}_{#1} \quad}}

\newcommand{\rs}{r^*}
\newcommand{\rc}{r_0}
\newcommand{\rst}{r_{\rm s}}
\newcommand{\ub}{\bar{u}}

\newcommand{\dt}{\Delta t_0}
\newcommand{\dtw}{\widetilde{\Delta t}_0}
\newcommand{\UH}{U_{\rm H}}
\newcommand{\uH}{u_{\rm H}}
\newcommand{\ks}{\ensuremath{\bar{\kappa}}}
\newcommand{\vl}{\ensuremath{v_l}}
\newcommand{\apr}{\ensuremath{a_{\rm p}}}
\newcommand{\rl}{\ensuremath{r_l}}
\newcommand{\tl}{\ensuremath{t_l}}
\newcommand{\xl}{\ensuremath{x_l}}

\newcommand{\ahalf}{\ensuremath{\fr{2}}}
\newcommand{\athird}{\ensuremath{\fr{3}}}

\author{Luis Cortés Barbado}
\title{Percepción de las radiaciones Hawking y Unruh por distintos observadores: aplicaciones de la función de temperatura efectiva}
\degree{doctor en física}
\supervisor{Carlos Barceló Serón\\ Luis Javier Garay Elizondo}
\institution{Instituto de Astrofísica de Andalucía\\ Consejo Superior de Investigaciones Científicas}
\faculty{Escuela de Doctorado de Ciencias, Tecnologías e Ingenierías}
\department{Programa de doctorado en física y matemáticas}
\submitdate{Granada, 2014}

\hyperlinking
\begin{document}


\begin{titlepage}
  \vspace{-4cm}
  \TitleBlock{\scshape\insertdepartment}
  \TitleBlock{\scshape\insertfaculty}
  \vspace{0.5cm}
  \TitleBlock{\includegraphics[height=4cm]{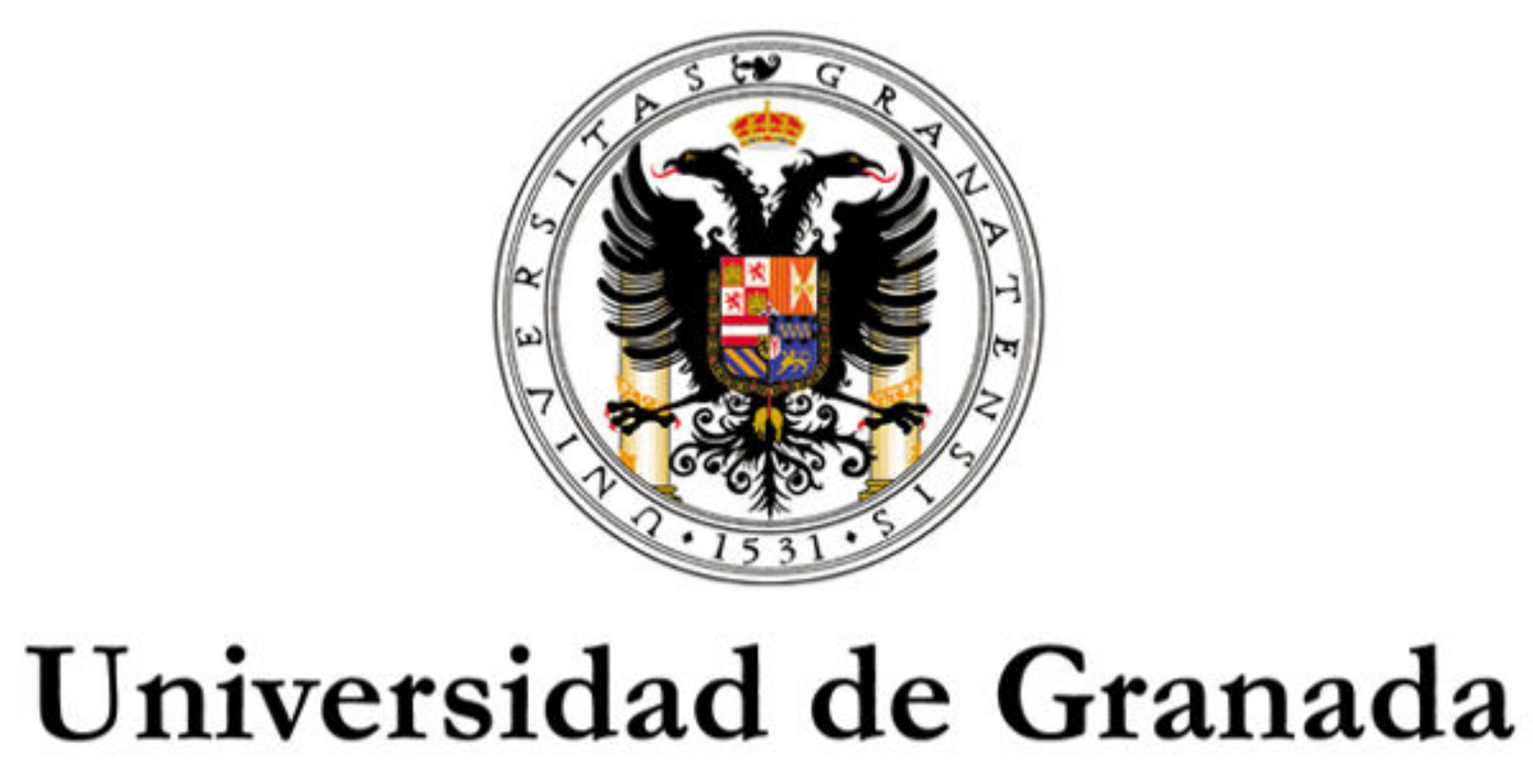}}
  \vspace{1cm}
  \TitleBlock{\Huge\textbf{\inserttitle}}
  \vspace{1cm}
  \TitleBlock{\large\scshape
    Tesis presentada por \insertauthor \\
    para optar al grado de \insertdegree}
  \vspace{1cm}
  \TitleBlock{\scshape\insertinstitution}
  \vspace{0.5cm}
  \TitleBlock{\includegraphics[height=2.5cm]{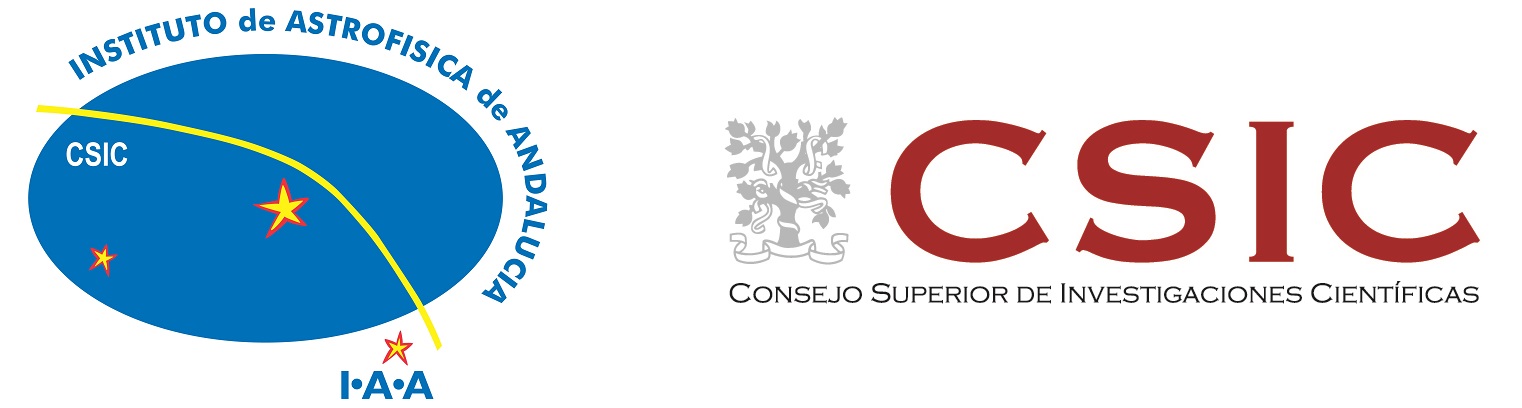}}
  \vspace{1cm}
  \TitleBlock{\scshape Directores de tesis:\\ \insertsupervisor}
  \vspace{1cm}
  \TitleBlock{\insertsubmitdate}
\end{titlepage}



\newpage
\thispagestyle{empty}
\hbox{}

\newpage
\thispagestyle{empty}
\hbox{}

\makeatletter
\def\cleardoublepage{\clearpage\if@twoside \ifodd\c@page\else
    \hbox{}
    \thispagestyle{empty}
    \newpage
    \if@twocolumn\hbox{}\newpage\fi\fi\fi}
\makeatother \clearpage{\pagestyle{empty}\cleardoublepage}

\chapter*{Resumen}
\markboth{}{}

En esta tesis se ha realizado un estudio de la percepción de los fenómenos de radiación en Teoría cuántica de Campos en sistemas de referencia no inerciales y en agujeros negros, en particular los conocidos fenómenos de la radiación de Hawking y el efecto Unruh. Se ha considerado un campo escalar de Klein-Gordon real sin masa como campo de radiación. Para llevar a cabo el estudio, se han utilizado dos herramientas: los detectores de partículas de Unruh-DeWitt macroscópicos y la función de temperatura efectiva, esta última basada en el cálculo del espectro percibido mediante transformaciones de Bogoliubov. Utilizando el modelo de detector de Unruh-DeWitt, se ha propuesto una expansión adiabática de las propiedades de detección de estos objetos cuando siguen trayectorias con aceleración lineal lentamente variable en el espacio-tiempo de Minkowski. Esta expansión adiabática permite calcular el espectro detectado a lo largo de dichas trayectorias, y en ella se encuentra el espectro térmico característico del efecto Unruh como término de orden cero en la expansión. Por otra parte, mediante la función de temperatura efectiva se ha estudiado la percepción de la radiación de Hawking por observadores siguiendo distintas trayectorias radiales en el exterior de un agujero negro de Schwarzschild. Para simular el proceso de encendido de la radiación en un escenario real de colapso, se ha introducido un estado de vacío no estacionario para el campo de radiación que interpola suavemente entre la ausencia de radiación en el pasado asintótico y la radiación de Hawking en el futuro asintótico. Uno de los resultados más importantes del análisis realizado es el hecho de que, en general, los observadores en caída libre no detectan vacío al cruzar el horizonte de sucesos, aun cuando el campo se encuentra en el vacío de Unruh, hecho que podemos explicar debido a un factor Doppler que diverge en el horizonte. Tras un estudio caso por caso de distintos observadores, procedemos a dar una expresión general para la función de temperatura efectiva, la cual tiene una clara interpretación en términos de fenómenos físicos conocidos. Discutimos especialmente qué contribución a la percepción de radiación de un observador cualquiera procede de la radiación emitida por el agujero negro, y cuál procede del efecto Unruh que experimenta el propio observador debido a su estado de movimiento. De esta discusión concluimos que, en general, el efecto Unruh no se debe únicamente a la aceleración propia del observador, ni puede definirse localmente, sino que se debe a la aceleración del observador con respecto a la región asintótica del espacio-tiempo. También aplicamos la expresión general de la función de temperatura efectiva encontrada al análisis de diversas situaciones físicas, en particular a un posible escenario de flotación en las cercanías de un agujero negro debido a la radiación de Hawking. Finalmente, proponemos un estado de vacío no estacionario, que denominamos vacío pulsante, para el campo de radiación en el exterior de un objeto celeste esféricamente simétrico y cercano a la formación de un horizonte. En este estado de vacío se tiene radiación muy similar a la radiación de Hawking emitida por el objeto, pero evitando los conocidos problemas de la paradoja de la información y el problema transplanckiano.
\newpage
\thispagestyle{empty}
\hbox{}

\makeatletter
\def\cleardoublepage{\clearpage\if@twoside \ifodd\c@page\else
    \hbox{}
    \thispagestyle{empty}
    \newpage
    \if@twocolumn\hbox{}\newpage\fi\fi\fi}
\makeatother \clearpage{\pagestyle{empty}\cleardoublepage}

\chapter*{{\large Compromiso de respeto de derechos de autor}}

\thispagestyle{empty}

El doctorando Luis Cortés Barbado y los directores de la tesis Carlos Barceló Serón y Luis Javier Garay Elizondo garantizamos, al firmar esta tesis doctoral, que el trabajo ha sido realizado por el doctorando bajo la dirección de los directores de la tesis y hasta donde nuestro conocimiento alcanza, en la realización del trabajo, se han respetado los derechos de otros autores a ser citados, cuando se han utilizado sus resultados o publicaciones.

\vspace{1.5cm}

En Granada, a 24 de febrero de 2014,

\vspace{0.5cm}

\noindent Directores de la tesis\phantom{Fdo: Carlos Barceló Serón}Doctorando

\vspace{0.5cm}

\noindent Fdo: Carlos Barceló Serón\phantom{Directores de la tesis}Fdo: Luis Cortés Barbado

\vspace{2cm}

\noindent Fdo: Luis Javier Garay Elizondo

\newpage
\thispagestyle{empty}
\hbox{}

\makeatletter
\def\cleardoublepage{\clearpage\if@twoside \ifodd\c@page\else
    \hbox{}
    \thispagestyle{empty}
    \newpage
    \if@twocolumn\hbox{}\newpage\fi\fi\fi}
\makeatother \clearpage{\pagestyle{empty}\cleardoublepage}

\chapter*{{\large Estructuración de contenidos en la memoria}}

\thispagestyle{empty}

Los contenidos exigidos en una tesis doctoral en la Universidad de Granada para los programas de doctorado regulados por el RD1393/2007 se encuentran en la presente memoria estructurados en los siguientes capítulos:

\vspace{0.5cm}

\textbf{Título}

Portada

\vspace{0.5cm}

\textbf{Compromiso de respeto de derechos de autor}

Compromiso de respeto de derechos de autor

\vspace{0.5cm}

\textbf{Resumen}

Resumen

\vspace{0.5cm}

\textbf{Introducción}

Introducción, capítulo 1

\vspace{0.5cm}

\textbf{Objetivos}

Introducción

\vspace{0.5cm}

\textbf{Metodología}

Capítulo 2

\vspace{0.5cm}

\textbf{Resultados}

Capítulos 3, 4, 5 y 6

\vspace{0.5cm}

\textbf{Conclusiones}

Conclusiones

\vspace{0.5cm}

\textbf{Bibliografía}

Bibliografía
\newpage
\thispagestyle{empty}
\hbox{}

\makeatletter
\def\cleardoublepage{\clearpage\if@twoside \ifodd\c@page\else
    \hbox{}
    \thispagestyle{empty}
    \newpage
    \if@twocolumn\hbox{}\newpage\fi\fi\fi}
\makeatother \clearpage{\pagestyle{empty}\cleardoublepage}

\prefacesection{Agradecimientos}
\markboth{Agradecimientos}{Agradecimientos}

\textit{\hfill Quiero escribir, pero me sale espuma,}

\textit{\hfill quiero decir muchísimo y me atollo;}

\vspace{0.5cm}

\hfill César Vallejo

\vspace{1cm}

Escribir los agradecimientos de una tesis es una oportunidad tentadora. Hasta tal punto, que pareciera que debe uno intentar ser comedido, no aprovecharse más de la cuenta de la oportunidad que se le brinda, y mantenerse mínimamente cercano al entorno de la tesis. Por supuesto, yo no haré tal cosa.

En primer lugar, me gustaría destacar que esta tesis ha sido el final de un largo trayecto (quizá también el comienzo de otro). Trayecto que, bien mirado, comenzó mucho, muchísimo antes... que ha pasado por muchas estaciones, pero cuyos raíles han tenido siempre un denominador común: la educación pública. Desde aquel colegio San Juan Bosco, pasando por el Instituto Albarregas, hasta las Universidades de Extremadura, la Complutense de Madrid y la de Granada, la educación pública ha sido la que me ha permitido llegar a este punto, la que se lo permite a tantos otros. No corren buenos tiempos para ella en los días que escribo esto. Sin embargo, como convencido que soy de que, algún día, la gente dejará de permitir que le arranquen lo que por derecho les pertenece, quisiera brindar esta tesis, en buena medida, a esa educación pública. A la que hubo, a la que aún queda, pero, sobre todo, a la que está por venir. Además, en este largo trayecto he tenido la oportunidad de aprender de \emph{maestros} que me han aportado mucho, a menudo no solo en lo académico. Me gustaría recordar aquí especialmente a Paqui Calle, Juana Molano, Vidal Luis Mateos, Santos Bravo, Andrés Santos y Manuel Molina.

Seguramente, la mayoría de los agradecimientos de tesis sean bastante similares. Pero quizá lo único común a todos, sin excepción, una cuestión obligada, sea la mención a los directores. Lo cual, eso sí, no tiene por qué restarle autenticidad a ninguna de ellas. Me gustaría, sin embargo, que en este caso tal mención fuera especial. Y no porque pretenda ser más que nadie. Más bien porque, en mi torpe retórica, no encuentro una forma mejor de alejarme cuanto pueda de un agradecimiento formal. Porque mi agradecimiento a los dos, a Carlos y a Luis, no puede ser más sincero. Por la oportunidad que me han brindado de llevar a cabo esta tesis, por su enseñanza, su apoyo y su comprensión durante el proceso, y por su amistad a lo largo de estos años. También a mi ``hermano de tesis'', a Raúl, por su amistad, y por su compañía como físico teórico entre tanta gente dedicada a cosas tan mundanas como planetas, estrellas, o galaxias. A los tres, por su ejemplo de genialidad, trabajo y, sobre todo, pasión por la física.

Dejando ahora a un lado bromas de teórico vanidoso, deseo expresar mi sincera gratitud al personal del Instituto de Astrofísica de Andalucía, donde he llevado a cabo esta tesis, por haberme brindado un inmejorable entorno tanto académico como humano. Entorno en el cual, además, he aprendido a disfrutar y a asombrarme con ese mundo, o universo, tan mundano como fascinante, que es la astrofísica. Gracias a muchos de vosotros (tantos que sería inútil intentar nombraros a todos) por vuestra compañía en el Instituto, en los almuerzos, y en el resto de ocasiones que hemos compartido. Un agradecimiento especial a mis compañeros de despacho en estos últimos años, a la abuela Marta y al primo Javi, por los buenos momentos, las buenas conversaciones y las muchas risas que nos hemos echado allí.

A lo largo de estos años de tesis he tenido la gran oportunidad de viajar y trabajar con varios buenos investigadores en diversos lugares del mundo. Mi agradecimiento a todos ellos. A Daniele Oriti, por su recepción y hospitalidad en el Instituto Albert Einstein. A Ivette Fuentes, Jorma Louko y Carlos Sabín, por su magnífico acogimiento en la Universidad de Nottingham, y por las muchas e interesantes charlas de física compartidas. Ojalá volvamos a trabajar juntos. Y a Matt Visser por su amistad, su acogida durante mi estancia en Nueva Zelanda y su disposición como guía turístico (y gastronómico) allí, y por las mil y una magníficas conversaciones sobre mil y un temas diferentes (entre otros, algunos de física), a menudo con mil y una jarras de buena cerveza. Gracias también al resto de personas que he conocido y me han acogido en los distintos destinos, y con las que he pasado muchos buenos momentos.

A Guillermo Mena muchos le debemos, entre otras cosas, la organización de los magníficos Jarramplas. Yo, como paisano, le debo además el haber escogido nuestra tierra, Extremadura, como enclave inmejorable para ello (ya sabes que, aunque como buen mangurrino hayas escogido el norte de Cáceres, como pacense no te guardo rencor). En esos Jarramplas he tenido la oportunidad de pasar muchos buenos momentos con Dani, Ana, Míkel, Javi, Merce, Gil, Víctor, y el resto de compañeros.

A toda la gente que, más allá del Instituto, he tenido la oportunidad de conocer en Granada, y que me han acompañado en tantos momentos estos años. A mi compañera de piso y paisana, Noelia. A Kathe, Rodrigo, y demás compañeros del máster. A las muchas personas del 15M de Granada y, en particular, de la Asamblea del barrio del Zaidín, a Lázaro, Valentín, Amparo, Varo, Ana, José Juan, Ricardo, Gabi, Sara, y tantos otros; y a los camaradas del partido, a Miguel Ángel, Javi, Adrián, Ángel, Juanjo, Paco, Álex,..., a quienes agradezco en especial su enseñanza, su compañía y su ejemplo en la lucha por una sociedad más justa.

A todos los miembros de los grupos de música de los que he formado parte: de Incorruptibles, Dsgarre, Aula 202 y, en especial, de Voodoo Soul, gracias por tantos conciertos juntos, por los discos que hemos sacado adelante y, sobre todo, por las innumerables horas de ensayo disfrutando de nuestra pasión común: la música. También, por supuesto, a los compañeros del nuevo grupo en Granada, aun sin nombre, pero ya con muchas ganas de echar a andar.

A los muchos amigos que tengo en mi tierra, en Extremadura, por tantos años de buena amistad, tanto en la cercanía como en la distancia. En Mérida, quisiera mencionar en especial a Pablo, Nano, Tutxi, Cárton, Jony, Paola y David; en Cabeza del Buey, a Mata y Toñete. También a Ana en Badajoz, y a Gema en Cáceres. Y a Beatriz, extremeña de Madrid.

Quisiera, ahora que casi estoy terminando, dar un agradecimiento especial a algunas personas en particular con las que he tenido la inmersa suerte de cruzarme en la vida. A Coral, Iván y Leo (y Mateo), por hacer de aquel piso en Carabanchel uno de los lugares que con más nostalgia guardo en mi memoria. Por vuestra amistad, por los incontables buenos momentos allí... y por las infinitas risas compartidas, por esas risas hasta la asfixia que rara vez he vuelto a repetir, y que de tantas cosas nos salvan. No sé qué hubiera hecho sin vosotros aquellos años en Madrid. A José Ferreira, Quini, Matamoros y Víctor Muñoz, por vuestra amistad y vuestra ayuda en momentos difíciles, que nunca olvidaré. A Saray, por todo lo que tuvimos oportunidad de compartir, y por los buenos momentos que pasamos juntos. A José Luis Jaramillo, a quien tuve poco tiempo de conocer, pero el suficiente para saber que es una de las mejores personas que he conocido, por su inestimable apoyo en Berlín, sin el cual probablemente esta tesis no existiría. Y a Vicky, por contagiarme su vitalidad cuando más lo necesitaba. Se diría, retóricamente, que os debo a todos una parte de esta tesis. Pero nada más lejos. La tesis, entera, es solo una pequeña parte de lo que a cada uno os debo.

Para terminar, mi más profundo agradecimiento a mi querida familia. A la familia grande y numerosa, a la que está y a la que se marchó... abuelos, tíos, primos (primeros y segundos)... con quienes tanto he compartido. En particular, en estos últimos años de tesis, agradecer a mi primo David sus tantas y tan fructíferas escapadas a Granada. También agradecer especialmente a mi abuelo Casimiro su inmenso ejemplo de vida como luchador por el conocimiento, la enseñanza, y la cultura. Y, finalmente, y muy en especial, a mis padres, que me enseñaron todo lo verdaderamente importante que sé, y a mi hermana Paula, a quienes estimo más que a nadie, por ser una fuente constante de apoyo en el camino.

\newpage
\thispagestyle{empty}
\hbox{}

\makeatletter
\def\cleardoublepage{\clearpage\if@twoside \ifodd\c@page\else
    \hbox{}
    \thispagestyle{empty}
    \newpage
    \if@twocolumn\hbox{}\newpage\fi\fi\fi}
\makeatother \clearpage{\pagestyle{empty}\cleardoublepage}

\tableofcontents

\newpage
\thispagestyle{empty}
\hbox{}

\makeatletter
\def\cleardoublepage{\clearpage\if@twoside \ifodd\c@page\else
    \hbox{}
    \thispagestyle{empty}
    \newpage
    \if@twocolumn\hbox{}\newpage\fi\fi\fi}
\makeatother \clearpage{\pagestyle{empty}\cleardoublepage}

\chapter*{\phantom{Cita}}

\thispagestyle{empty}

\textit{\hfill Un hombre pasa con un pan al hombro}

\textit{\hfill ¿Voy a escribir, después, sobre mi doble?}

\vspace{0.1cm}

\textit{\hfill ...}

\vspace{0.1cm}

\textit{\hfill Otro busca en el fango huesos, cáscaras}

\textit{\hfill ¿Cómo escribir, después, del infinito?}

\vspace{0.3cm}

\textit{\hfill Un albañil cae de un techo, muere y ya no almuerza}

\textit{\hfill ¿Innovar, luego, el tropo, la metáfora?}

\vspace{0.3cm}

\textit{\hfill Un comerciante roba un gramo en el peso a un cliente}

\textit{\hfill ¿Hablar, después, de cuarta dimensión?}

\vspace{0.1cm}

\textit{\hfill ...}

\vspace{0.1cm}

\textit{\hfill Alguien va en un entierro sollozando}

\textit{\hfill ¿Cómo luego ingresar a la Academia?}

\vspace{0.3cm}

\textit{\hfill Alguien limpia un fusil en su cocina}

\textit{\hfill ¿Con qué valor hablar del más allá?}

\vspace{0.3cm}

\textit{\hfill Alguien pasa contando con sus dedos}

\textit{\hfill ¿Cómo hablar del no-yó sin dar un grito?}

\vspace{0.5cm}

\hfill César Vallejo
\newpage
\thispagestyle{empty}
\hbox{}

\makeatletter
\def\cleardoublepage{\clearpage\if@twoside \ifodd\c@page\else
    \hbox{}
    \thispagestyle{empty}
    \newpage
    \if@twocolumn\hbox{}\newpage\fi\fi\fi}
\makeatother \clearpage{\pagestyle{empty}\cleardoublepage}

\prefacesection{Introducción}
\markboth{Introducción}{Introducción}

Desde su formulación por Albert Einstein en 1915, la relatividad general sigue siendo a fecha de hoy la teoría más satisfactoria para describir la interacción gravitatoria. Sin embargo, a pesar de su éxito al explicar los distintos fenómenos observables en gravedad, tiene una característica destacable: parece contener las semillas de su propia destrucción. El paradigma de este hecho son los conocidos \emph{teoremas de singularidad}~\cite{Penrose:1964wq,Hawking:1969sw}. Según nos indican estos teoremas, la relatividad general clásica predice de forma necesaria la formación de \emph{agujeros negros,} separados del exterior por un \emph{horizonte de sucesos} y dentro de los cuales se forma una \emph{singularidad,} en la cual la descripción clásica del espacio-tiempo deja de ser válida. Por otra parte, estas predicciones anómalas entran en el dominio de cortas distancias y altas energías, lo cual parece indicar que la relatividad general necesita de la aplicación de la mecánica cuántica para su consistencia.

Un paso que marcó un antes y un después en la comprensión del papel de la mecánica cuántica en este escenario fue la derivación por Stephen Hawking de que los agujeros negros deberían emitir radiación térmica con temperatura proporcional a su gravedad de superficie, debido a efectos cuánticos~\cite{Hawking:1974rv,Hawking:1974sw}. Se trata, sin duda, de uno de los resultados más importante hasta ahora que combina elementos de relatividad general con elementos de mecánica cuántica. Además, no solo incorpora la mecánica cuántica, sino que esta a su vez permite dotar a los agujeros negros de propiedades termodinámicas al asignar una temperatura a estos objetos. Este hallazgo permite completar la analogía entre la dinámica de los agujeros negros y las leyes de la termodinámica~\cite{Bardeen:1973gs,Wald:1995yp}, la cual requiere de la constante de Planck~$\hbar$ para poder concretarse.

Si en la evolución del agujero negro no estuvieran involucradas fuentes de materia o radiación que aumentaran el tamaño de este (como discos de acreción, o la radiación de fondo cósmico), la radiación de Hawking iría evaporándolo lentamente, hasta hacerlo desaparecer. Sin embargo, aun estando presente en todo el periodo de vida del agujero negro, se puede argumentar que la radiación de Hawking (al menos en su formulación original) tiene su origen únicamente en el proceso de colapso que da lugar a la formación del mismo. Cuando dicho proceso de colapso ha finalizado, y la geometría se aproxima a la de un agujero negro cuasi-estacionario (salvo por la propia evaporación), es cuando esta radiación es térmica, y su temperatura proporcional a la gravedad de superficie del agujero negro en el horizonte, sin que quede ningún rastro del proceso de formación~\cite{Hawking:1974sw}. El hecho de que la producción de partículas continúe hasta evaporar el agujero negro, aun después de que el proceso de colapso claramente haya finalizado, puede interpretarse como debido a que el breve lapso de tiempo que supone la formación del horizonte de sucesos, visto desde el propio horizonte, es contemplado como un tiempo indefinidamente largo desde la región asintótica, debido al factor de corrimiento en frecuencias divergente entre una zona y otra (es decir, a la ``congelación del tiempo'' en las cercanías del horizonte).

Sin embargo, el fenómeno de la radiación de Hawking también presenta, a su vez, otros problemas conceptuales. En particular, el factor divergente en el corrimiento en frecuencias mencionado es a su vez el responsable del conocido \emph{problema transplanckiano}~\cite{Unruh:1976db}, que se debe a la necesidad de involucrar frecuencias extremadamente altas en la derivación de la radiación de Hawking. Otro conocido problema del proceso de evaporación mediante la radiación de Hawking es la \emph{paradoja de la información}~\cite{Hawking:1976ra,Hawking:2005kf,Preskill:1992tc}, que se debe a la necesidad de formar un horizonte de sucesos, y que consiste en la aparente pérdida irreversible de información tras dicho horizonte, la cual no se recupera a través de la radiación emitida, debido al carácter térmico (decorrelacionado) de la misma.

Como hemos mencionado, la existencia de radiación de Hawking es probablemente uno de los resultados más importantes de la Teoría Cuántica de Campos en espacios curvos. Un resultado, como veremos relacionado, que rivaliza en importancia con el fenómeno de la radiación de Hawking es el \emph{efecto Unruh}~\cite{Unruh:1976db}. Este efecto se refiere al hecho de que un observador con aceleración propia constante en un espacio-tiempo plano, y acoplado a un campo cuántico, debería detectar radiación térmica de dicho campo cuando este se encuentra en el estado de vacío (para los observadores inerciales), siendo la temperatura de la radiación proporcional a la aceleración del observador. La relación de este fenómeno con la radiación de Hawking se observa claramente a través del \emph{principio de equivalencia.} El efecto Unruh es, sin embargo, el resultado paradigmático de un hecho más general en Teoría Cuántica de Campos en espacios curvos: el hecho de que la noción de \emph{partícula} de un campo cuántico es dependiente del observador~\cite{Unruh:1976db,Birrell:1982ix}. Esto significa que, en un estado en el que determinados observadores no perciben partículas del campo, otros observadores con distinto estado de movimiento sí pueden percibir un contenido no nulo de partículas.

Existen dos formas comunes de acercarse al problema de la percepción de radiación por distintos observadores en Teoría Cuántica de Campos en espacios curvos: las \emph{transformaciones de Bogoliubov}~\cite{Birrell:1982ix}, que permiten comparar cómo distintos observadores describen un mismo estado de vacío en términos de partículas; y los modelos de \emph{detectores cuánticos de partículas} (en especial, el \emph{modelo de Unruh-DeWitt}~\cite{Unruh:1976db,nla.cat-vn973875}), los cuales son dispositivos que experimentan interacción con el campo cuántico, excitándose ante la presencia de partículas del mismo.

Una cuestión importante en el estudio de los dos fenómenos mencionados, a saber, la radiación de Hawking y la dependencia de la percepción de partículas con el estado de movimiento del observador, es intentar comprender cómo se combinan en la percepción final del campo de radiación por un observador en el exterior de un agujero negro. Por una parte, sabemos de la presencia de radiación de Hawking emitida por el agujero negro. Sin embargo, el resultado original solo nos indica que dicha radiación es percibida como térmica por los observadores en reposo en la región asintótica y para tiempos suficientemente tardíos tras el colapso~\cite{Hawking:1974sw}. Para un observador siguiendo una trayectoria genérica en el exterior del agujero negro, la percepción de esta radiación se verá ``distorsionada'' por su propio estado de movimiento.

Aunque se ha estudiado la percepción por parte de distintos observadores de la radiación emitida por agujeros negros, los diversos estudios hasta la fecha, aunque útiles, oscilan entre el análisis de casos muy particulares (casi todos correspondientes a posiciones estáticas u órbitas circulares de los observadores), por una parte; y los desarrollos formales más genéricos, pero difícilmente aplicables a casos concretos o interpretables en términos de fenómenos físicos conocidos, por la otra. Por ejemplo, se ha estudiado la percepción de radiación mediante el uso de detectores de Unruh-DeWitt en agujeros negros de Schwarzschild~\cite{Candelas:1980zt,Sriramkumar:1994pb,Singleton:2011vh,Acquaviva:2011vq,Hodgkinson:2013tsa,Ahmadzadegan:2013iua}, agujeros negros BTZ~\cite{Hodgkinson:2013tsa,Hodgkinson:2012mr,Hodgkinson:2012ma}, o campos gravitatorios newtonianos~\cite{Louko:2007mu}. También se ha hecho un análisis de la percepción de radiación por un observador en caída libre en un agujero negro de Schwarzschild mediante el formalismo de funciones de onda de Schrödinger~\cite{Greenwood:2008zg}. Expresiones formales de la respuesta de detectores de Unruh-DeWitt en espacio-tiempos curvos generales pueden encontrarse en~\cite{Hodgkinson:2013tsa,Louko:2007mu}.

Sin embargo, una comprensión más general de la percepción de radiación en el exterior de un agujero negro para distintos observadores según sus trayectorias, pero a su vez con resultados concretos e interpretables físicamente, es el objetivo principal que nos hemos propuesto en esta tesis. En la percepción de radiación estarán involucrados tanto la radiación de Hawking como el efecto Unruh, según la trayectoria de los observadores, así como otros efectos conocidos (como el \emph{efecto Doppler} o el corrimiento en frecuencias gravitacional). Para este análisis, utilizaremos dos herramientas: los detectores de Unruh-DeWitt y, principalmente, una herramienta introducida por primera vez en~\cite{Barcelo:2010pj,Barcelo:2010xk}: la \emph{función de temperatura efectiva.} Con esta función es posible calcular, en determinados escenarios y bajo cumplimiento de ciertas condiciones, la temperatura de la radiación térmica que un determinado observador percibe en torno a un instante concreto en un estado de vacío dado. Su justificación está basada en el cálculo del espectro percibido mediante \emph{coeficientes de Bogoliubov.} Como veremos, se trata de un método especialmente sencillo, pero con mucho contenido físico, y que nos ha permitido comprender mejor el papel de los distintos fenómenos físicos involucrados en la percepción final de la radiación, así como extraer conclusiones novedosas.

Una motivación de especial relevancia para abordar el análisis de la percepción de radiación en agujeros negros es su potencial aplicación al estudio de los \emph{fenómenos de flotación} debidos a la radiación de Hawking~\cite{Unruh:1982ic,Bekenstein:1999bh}. Para tal estudio, es requisito indispensable una correcta comprensión, en primer lugar, de la \emph{percepción} de la radiación de Hawking por los distintos observadores en el exterior del agujero negro.

Comenzaremos esta memoria con una introducción a la Teoría Cuántica de Campos en espacios curvos, en particular en sistemas de referencia no inerciales y en agujeros negros (capitulo~\ref{cap_preliminares}). Prestaremos especial atención a las transformaciones de Bogoliubov, obteniendo a partir de ellas los conocidos resultados del efecto Unruh y la radiación de Hawking. En el capítulo~\ref{percepcion} describiremos las dos herramientas que utilizaremos para estudiar la percepción de partículas por distintos observadores a lo largo de la tesis: los detectores de Unruh-DeWitt y la función de temperatura efectiva. Para los primeros, introduciremos algunas consideraciones técnicas~\cite{Schlicht:2003iy,Louko:2006zv,Satz:2006kb,Barbado:2012fy} que no se encuentran en su formulación original. En la última sección de este capítulo demostraremos que, al menos en el marco en el cual trabajamos a lo largo de la tesis, el espectro obtenido mediante los coeficientes de Bogoliubov (y, por tanto, también la función de temperatura efectiva) tiene a su vez una interpretación inmediata en términos de excitación de detectores de Unruh-DeWitt. Estos dos primeros capítulos tienen un carácter introductorio. Los siguientes capítulos son ya una exposición de los resultados originales obtenidos en esta tesis.

Los fenómenos de la radiación de Hawking y el efecto Unruh comparten la característica de que la radiación involucrada presenta un espectro planckiano. Una cuestión de interés es conocer el comportamiento del espectro de radiación cuando nos alejamos de las condiciones ``ideales'' que dan lugar a tal espectro planckiano. En el capítulo~\ref{adiabatica} haremos un acercamiento a esta cuestión, en concreto para el efecto Unruh en Minkowski. Utilizaremos el modelo de detector de Unruh-DeWitt para obtener las propiedades de excitación de estos detectores para trayectorias con aceleración lineal lentamente variable en el tiempo, y por tanto fuera del régimen estrictamente constante que requiere el efecto Unruh. Para ello, propondremos una \emph{expansión adiabática} de la cantidad que denominaremos \emph{función respuesta}~\cite{Barbado:2012fy}, y estudiaremos sus propiedades. Extraeremos el efecto Unruh, con temperatura proporcional a la aceleración en cada instante, como resultado a orden cero en la expansión, siendo los órdenes superiores contribuciones debidas a las derivadas de la aceleración. Aunque en un marco distinto al del resto de la tesis (pero, por otra parte, el más habitual cuando se trata de estudiar el efecto Unruh), este análisis nos permitirá hacernos una idea de cómo se desvía el espectro percibido del espectro planckiano que caracteriza los fenómenos en los que estamos interesados, cuando nos desviamos ligeramente de las condiciones en las que tales fenómenos se producen.

El resto de la tesis se centrará exclusivamente (salvo una pequeña sección) en el estudio de la percepción de la radiación emitida por agujeros negros, haciendo uso de la función de temperatura efectiva. Al hacerlo, nos quedaremos únicamente con la ``componente térmica'' de la percepción de radiación, ignorando el resto de contribuciones que hemos encontrado en la expansión adiabática. No obstante, con esta función podemos estudiar la percepción de radiación también en espacio-tiempos curvos, que es lo que necesitamos en nuestro análisis. Además, veremos que el estudio de esta componente es sin duda el que tiene mayor interés físico. En el capítulo~\ref{hawking} utilizaremos la función de temperatura efectiva para estudiar la percepción de la radiación de Hawking por observadores siguiendo distintas trayectorias en el exterior de un agujero negro de Schwarzschild~\cite{Barbado:2011dx}. Propondremos un vacío para el campo de radiación distinto al habitual \emph{vacío de Unruh} (en el que el agujero negro emite radiación de Hawking de forma constante), que denominaremos \emph{vacío de colapso,} y el cual pretende simular el proceso de ``encendido'' de la radiación en un escenario de colapso real. Veremos que muchos de los resultados que obtengamos tienen una clara interpretación física, y en particular encontraremos un resultado especialmente interesante, al cual daremos explicación: aun estando el campo de radiación en el vacío de Unruh, los observadores en caída libre en general \emph{sí detectan radiación} al cruzar el horizonte.

El capítulo~\ref{fisica} proporciona una síntesis conceptual del desarrollo anterior. En él obtendremos una expresión general para la función de temperatura efectiva en función de las propiedades locales de la trayectoria del observador cuya percepción queremos calcular, y de las propiedades del vacío en el cual la estemos calculando~\cite{Barbado:2012pt}. Daremos una interpretación física a tal expresión, poniendo especial énfasis en discutir qué contribución a la percepción de radiación procede de la radiación propia del estado de vacío (como la radiación de Hawking), y qué contribución corresponde al efecto Unruh propio de la aceleración del observador. Obtenida y discutida la expresión general, la utilizaremos para analizar con ella algunos escenarios físicos, con especial atención a su aplicación en un posible escenario de \emph{flotación} en las cercanías de un agujero negro, debido a la radiación de Hawking.

En el capítulo~\ref{pulsante} propondremos un estado de vacío no estacionario en la región exterior de un cuerpo celeste, que denominaremos \emph{vacío pulsante}~\cite{Barbado:2011ai}, el cual presenta una radiación muy similar a la radiación de Hawking emitida a la región asintótica, sin necesidad de invocar frecuencias extremadamente altas ni la formación de un horizonte de ningún tipo en su derivación.

Finalizaremos la memoria resumiendo y concretando las conclusiones obtenidas a lo largo de la tesis.

\phantom{Espacio.}

\textbf{Notación:} La signatura de la métrica será~$(-,+,+,+)$ o~$(-,+)$, según nos encontremos en $3+1$~dimensiones o en $1+1$~dimensiones, respectivamente. Utilizaremos siempre unidades naturales~$G = c = \hbar = 1$.

\newpage
\thispagestyle{empty}
\hbox{}

\makeatletter
\def\cleardoublepage{\clearpage\if@twoside \ifodd\c@page\else
    \hbox{}
    \thispagestyle{empty}
    \newpage
    \if@twocolumn\hbox{}\newpage\fi\fi\fi}
\makeatother \clearpage{\pagestyle{empty}\cleardoublepage}

\chapter{Preliminares}
\label{cap_preliminares}

En este capítulo haremos una descripción de la Teoría Cuántica de Campos en sistemas de referencia no inerciales y en agujeros negros (en particular, en agujeros negros de Schwarzschild), lo que constituye la base teórica fundamental de todo el desarrollo de la tesis.\footnote{El material de este capítulo es elemental, y puede encontrarse en diversos manuales~\cite{Wald:1995yp,Birrell:1982ix}. Sin embargo, hemos preferido incluirlo con el fin de hacer la memoria de tesis lo más auto-contenida posible.}

El campo cuántico más sencillo es el campo de Klein-Gordon real sin masa. Aunque no existe ningún campo conocido en el mundo físico descrito correctamente por este modelo,\footnote{Tras las recientes observaciones en el LHC~\cite{Chatrchyan:2013lba}, el bosón de Higgs es el candidato más importante a ser el primer campo físico observado experimentalmente y modelizado mediante un campo escalar (aunque con una masa de aproximadamente~$125~{\rm GeV}$). Otros candidatos son el inflatón~\cite{Guth:1980zm} (que podría ser el propio bosón de Higgs) y el curvatón~\cite{Enqvist:2001zp} en cosmología.} como modelo teórico puede usarse para estudiar muchos de los fenómenos que aparecen en otros campos cuánticos físicos, especialmente en campos bosónicos y sin masa (como el campo electromagnético). En particular, el modelo resulta especialmente útil, por su sencillez, para estudiar la mayoría de los fenómenos de la Teoría Cuántica de Campos en espacios curvos, o en sistemas de referencia no inerciales, como los que estudiaremos en este capítulo.

Además, el campo de Klein-Gordon sin masa posee otra importante ventaja: en un espacio-tiempo de $1+1$~dimensiones la teoría para este campo es invariante conforme, es decir, invariante bajo cambio del factor conforme de la métrica. En los distintos escenarios físicos que estudiaremos a lo largo de la tesis, nos limitaremos a la interacción del campo cuántico con observadores que siguen trayectorias rectilíneas en el espacio-tiempo de Minkowski o trayectorias radiales en el espacio-tiempo de Schwarzschild. Por tanto, aunque los espacio-tiempos físicos sean de dimensión~$3+1$, la teoría de campos podrá reducirse de forma efectiva, con algunas aproximaciones que indicaremos, a la teoría en $1+1$~dimensiones, permitiéndonos así explotar las ventajas de la invariancia conforme.

\section{Teoría Cuántica de Campos en sistemas de referencia no inerciales}\label{sec_cuantica_no_inercial}

En esta sección haremos una descripción general de la teoría cuántica de campos en sistemas de referencia no inerciales. Es decir, aunque el espacio-tiempo en el que el campo cuántico está definido será el espacio-tiempo plano de Minkowski, nos centraremos en la descripción del campo hecha desde sistemas de referencia no inerciales, particularmente desde sistemas de referencia con aceleración constante.

\subsection{Campo de Klein-Gordon real sin masa}\label{sec_campo_klein-gordon}

Consideremos el espacio-tiempo de Minkowski en $1+1$~dimensiones, y un campo escalar real~$\phi (x)$ definido en todos los puntos~$x=(t,z)$ del espacio-tiempo, que satisface la ecuación de Klein-Gordon sin masa
\begin{equation}
\Box \phi = - \frac{\partial^2 \phi}{\partial t^2} + \frac{\partial^2 \phi}{\partial z^2} = 0.
\label{klein-gordon}
\end{equation}
En la teoría invariante conforme es especialmente conveniente utilizar las coordenadas nulas $(U, V)$, definidas por
\begin{equation}
U := t - z, \quad V := t + z.
\label{coordenadas_nulas}
\end{equation}
Utilizando estas coordenadas, la ecuación~(\ref{klein-gordon}) puede escribirse como
\begin{equation}
\frac{\partial}{\partial U} \frac{\partial}{\partial V} \phi = 0.
\label{klein-gordon_nulas}
\end{equation}
Esta ecuación tiene la solución general
\begin{equation}
\phi (U, V) = \phi_U (U) + \phi_V (V).
\label{solucion_general}
\end{equation}
Cada término de esta solución puede expandirse en modos de Fourier de la forma
\begin{align}
\phi^U_\omega (U) & := \frac{1}{\sqrt{4\pi\omega}} \rme^{-\rmi \omega U}, \label{modos-U} \\
\phi^V_\omega (V) & := \frac{1}{\sqrt{4\pi\omega}} \rme^{-\rmi \omega V}, \label{modos-V}
\end{align}
respectivamente. Los modos con~$\omega > 0$ son modos de frecuencia positiva bien definida con respecto a la coordenada temporal~$t$, viajando hacia la derecha (modos $U$) y hacia la izquierda (modos $V$). Debido a la forma de la solución~(\ref{solucion_general}), los modos viajando hacia la derecha y hacia la izquierda se encuentran completamente desacoplados en la teoría libre e invariante conforme. En adelante, explicitaremos solo el sector $\phi_U (U)$ de~(\ref{solucion_general}), siendo la teoría para los modos~$V$ completamente análoga en un espacio-tiempo plano.

Los modos~(\ref{modos-U}) son ortonormales con respecto al producto escalar
\begin{equation}
\langle \psi,\psi' \rangle := \rmi \int_{-\infty}^\infty \rmd U \left\{ \psi (U) \frac{\partial}{\partial U} \psi'(U)^* - \psi'(U)^* \frac{\partial}{\partial U} \psi (U) \right\}.
\label{producto_escalar}
\end{equation}
Esto significa que
\begin{align}
\langle \phi^U_\omega,\phi^U_{\omega'} \rangle & = \delta(\omega-\omega'), \nonumber \\
\langle \phi^{U*}_\omega,\phi^{U*}_{\omega'} \rangle & = -\delta(\omega-\omega'), \label{ortonormal} \\
\langle \phi^U_\omega,\phi^{U*}_{\omega'} \rangle & = 0. \nonumber
\end{align}

Dado que el conjunto de modos~$\{\phi^U_\omega,\phi^{U*}_{\omega'}\}$ con~$\omega > 0$ forma una base ortonormal completa, el campo puede expandirse en los modos~(\ref{modos-U}) de la forma
\begin{equation}
\phi_U (U) = \int_0^\infty \rmd \omega \left[ a^U_\omega \phi^U_\omega (U) + (a^U_\omega)^* \phi^U_\omega (U)^* \right],
\label{expansion}
\end{equation}
donde~$a^U_\omega, (a^U_\omega)^*$ son los coeficientes de la expansión (unos son los complejos conjugados de los otros debido a que estamos considerando un campo real).

A continuación, pasaremos a construir una teoría cuántica para este campo escalar. Denominaremos~$\mathcal{H}_U$ al espacio de Hilbert de los modos~$U$ del campo. Dado que los modos~$U$ y~$V$ están completamente desacoplados, el espacio de Hilbert del campo es~$\mathcal{H}_\phi = \mathcal{H}_U \otimes \mathcal{H}_V$, donde el espacio de Hilbert~$\mathcal{H}_V$ para los modos~$V$ es completamente análogo a~$\mathcal{H}_U$. El procedimiento de cuantización pasa por convertir al campo~$\phi_U$ en un operador~$\hat{\phi}_U$ sobre~$\mathcal{H}_U$. Por tanto, los coeficientes de la expansión~(\ref{expansion}) pasan también a ser operadores~$\hat{a}^U_\omega, (\hat{a}^U_\omega)^\dagger$. El operador campo en función de estos operadores se escribe de forma análoga a~(\ref{expansion}):
\begin{equation}
\hat{\phi}_U (U) = \int_0^\infty \rmd \omega \left[ \hat{a}^U_\omega \phi^U_\omega (U) + (\hat{a}^U_\omega)^\dagger \phi^U_\omega (U)^* \right].
\label{expansion_cuantica}
\end{equation}
No obstante, con el fin de simplificar la notación, en adelante suprimiremos el acento circunflejo, puesto que siempre resultará claro cuándo un determinado objeto es un operador. La \emph{cuantización canónica} es equivalente a la imposición de las siguientes relaciones de conmutación sobre los operadores~$a^U_\omega, (a^U_\omega)^\dagger$:
\begin{align}
[a^U_\omega,a^U_{\omega'}] & = 0, \nonumber \\
[(a^U_\omega)^\dagger,(a^U_{\omega'})^\dagger] & = 0, \label{conmutacion} \\
[a^U_\omega,(a^U_{\omega'})^\dagger] & = \delta (\omega-\omega'). \nonumber
\end{align}

Una posible base para~$\mathcal{H}_U$ es la dada por la denominada \emph{representación de Fock.} Esta representación se construye a partir un \emph{estado de vacío}~$\ket{0_U}$. Este estado se define como aquél que es aniquilado por todos los operadores~$a^U_\omega$, denominados \emph{operadores de aniquilación},
\begin{equation}
a^U_\omega \ket{0_U} = 0.
\label{vacio}
\end{equation}

El resto de estados de la representación de Fock se construyen usando los denominados \emph{operadores de creación}~$(a^U_\omega)^\dagger$ sobre el estado de vacío,
\begin{equation}
\ket{(n_1)_{\omega_1},\ldots,(n_m)_{\omega_m}}_U = \left(n_1! \ldots n_m! \right)^{-\frac{1}{2}} \left[(a^U_{\omega_1})^\dagger\right]^{n_1} \cdots \left[(a^U_{\omega_m})^\dagger\right]^{n_m} \ket{0_U}.
\label{estados_fock}
\end{equation}
Estos estados tienen un número de partículas bien definido $n_i$ para cada frecuencia bien definida $\omega_i$. Es decir, son autoestados de los operadores número $N^U_{\omega_i} := (a^U_{\omega_i})^\dagger a^U_{\omega_i}$, con autovalores $n_i$:
\begin{equation}
N^U_{\omega_i} \ket{(n_1)_{\omega_1},\ldots,(n_m)_{\omega_m}}_U = n_i \ket{(n_1)_{\omega_1},\ldots,(n_m)_{\omega_m}}_U.
\label{operador_numero}
\end{equation}
Cualquier otro estado de~$\mathcal{H}_U$ puede construirse como combinación lineal de estados de esta base.

\subsection{Transformaciones de Bogoliubov}\label{sec_bogoliubov}

Para llevar a cabo la cuantización del campo de Klein-Gordon, hemos considerado una base particular de modos normales, a saber, aquella cuyos modos tienen frecuencia bien definida con respecto al tiempo~$t$ y número de onda bien definido respecto a la coordenada~$z$. Podríamos usar una base de modos ortonormales distinta~$\{\phi^u_\omega (U)\}$, dados por
\begin{equation}
\phi^u_\omega (U) := \frac{1}{\sqrt{4\pi\omega}} \rme^{-\rmi \omega u(U)}.
\label{nuevos_modos}
\end{equation}
La invariancia conforme de la ecuación de Klein-Gordon sin masa en $1+1$~dimensiones~(\ref{klein-gordon_nulas}) hace que la descomposición en nuevos modos normales pueda describirse simplemente mediante la definición de una nueva coordenada nula~$u=u(U)$ (que, como función de~$U$, debe ser continua y monótona creciente), como puede verse en la definición de los modos~(\ref{nuevos_modos}).

Podemos también descomponer el campo en esta nueva base:
\begin{equation}
\phi_U (U) = \int_0^\infty \rmd \omega \left[ a^u_\omega \phi^u_\omega (U) + (a^u_\omega)^* \phi^u_\omega (U)^* \right].
\label{nueva_expansion}
\end{equation}
Por su parte, cada modo de la nueva base puede ser expandido a su vez en la base anterior~$\{\phi^U_\omega (U)\}$,
\begin{equation}
\phi^u_\omega (U) = \int_0^\infty \rmd \omega' \left[ \alpha_{\omega, \omega'} \phi^U_{\omega'} (U) + \beta_{\omega, \omega'} \phi^U_{\omega'} (U)^* \right],
\label{transformacion_bogoliubov}
\end{equation}
donde~$\alpha_{\omega, \omega'}, \beta_{\omega, \omega'}$ se denominan \emph{coeficientes de Bogoliubov.} La transformación lineal entre las dos bases de modos ortonormales dada por~(\ref{transformacion_bogoliubov}) se denomina \emph{transformación de Bogoliubov.} Los coeficientes pueden expresarse como productos escalares de los modos anteriores y los nuevos:
\begin{equation}
\alpha_{\omega, \omega'} = \langle \phi^u_\omega, \phi^U_{\omega'} \rangle, \quad \beta_{\omega, \omega'} = -\langle\phi^u_\omega, (\phi^U_{\omega'})^*\rangle.
\label{coeficientes_bogoliubov}
\end{equation}

Dada la forma de los modos de la nueva base~(\ref{nuevos_modos}), se pueden obtener expresiones cerradas para estos coeficientes. Tras hacer una integración por partes en la integral del producto escalar~(\ref{producto_escalar}), y descartando un término de frontera irrelevante, se tiene:
\begin{align}
\alpha_{\omega, \omega'} & = -\frac{1}{2\pi} \sqrt{\frac{\omega'}{\omega\phantom{'}}} \int_{-\infty}^\infty \rmd U\ \rme^{-\rmi (\omega u (U) - \omega' U)}, \label{coeficientes_alfa} \\
\beta_{\omega, \omega'} & = -\frac{1}{2\pi} \sqrt{\frac{\omega'}{\omega\phantom{'}}} \int_{-\infty}^\infty \rmd U\ \rme^{-\rmi (\omega u (U) + \omega' U)}. \label{coeficientes_beta}
\end{align}
Haciendo una integración por partes diferente, podemos expresar los coeficientes también como integrales en la coordenada nula~$u$:
\begin{align}
\alpha_{\omega, \omega'} & = \frac{1}{2\pi} \sqrt{\frac{\omega}{\omega'}} \int_{-\infty}^\infty \rmd u\ \rme^{-\rmi (\omega u - \omega' U(u))}, \label{coeficientes_alfa_en_u} \\
\beta_{\omega, \omega'} & = \frac{1}{2\pi} \sqrt{\frac{\omega}{\omega'}} \int_{-\infty}^\infty \rmd u\ \rme^{-\rmi (\omega u + \omega' U(u))}. \label{coeficientes_beta_en_u}
\end{align}
En las expresiones se observa que los coeficientes $\beta_{\omega, \omega'}$ son aquellos que mezclan modos de norma positiva y de norma negativa [según las relaciones de ortonormalidad~(\ref{ortonormal})] en el cambio de base.

Si realizamos la cuantización canónica con los nuevos modos, los coeficientes de la expansión~(\ref{nueva_expansion}) pasan a ser operadores creación y destrucción~$a^u_\omega, (a^u_\omega)^\dagger$ sobre~$\mathcal{H}_U$. Es fácil ver que la transformación de Bogoliubov~(\ref{transformacion_bogoliubov}) es equivalente a la relación entre operadores
\begin{equation}
a^u_\omega = \int_0^\infty \rmd \omega' \left[ \alpha_{\omega, \omega'}^* a^U_{\omega'} - \beta_{\omega, \omega'}^* (a^U_{\omega'})^\dagger \right].
\label{tranformacion_bogoliubov_operadores}
\end{equation}

Actuando con estos nuevos operadores de creación y destrucción obtenemos una nueva cuantización de Fock para~$\mathcal{H}_U$. Denotaremos con la etiqueta~$u$ a los estados de la representación de Fock obtenida a partir de los nuevos modos. Se puede comprobar que, en general, el vacío de una representación y de otra no son el mismo. Si calculamos el valor esperado del número de partículas con energía bien definida para los modos~$u$, pero en el vacío~$\ket{0_U}$ de la representación de Fock obtenida a partir de los modos~$U$, tenemos que
\begin{equation}
\bra{0_U} N^u_\omega \ket{0_U} = \int_0^\infty \rmd \omega' \left| \beta_{\omega, \omega'} \right|^2.
\label{particulas_vacio}
\end{equation}
Es decir, salvo que los coeficientes~$\beta_{\omega, \omega'}$ sean todos nulos, el estado de vacío~$\ket{0_U}$ contiene partículas de los modos~$u$.\footnote{En realidad, la cantidad calculada en~(\ref{particulas_vacio}) corresponde a una densidad de partículas por intervalo de frecuencia, puesto que estamos tratando con frecuencias en el continuo. Sin embargo, por abuso de lenguaje hablaremos habitualmente de ``numero de partículas''.} Es fácil comprobar que, en tal caso, esto también es cierto a la inversa: el estado de vacío~$\ket{0_u}$ asociado a los modos~$u$ contiene partículas de los modos~$U$. Por tanto, una teoría cuántica de campos en general tiene distintos vacíos, según la base de modos normales que se utilice para la cuantización. Sin información adicional, no es posible escoger uno de ellos como vacío ``preferible'' de la teoría. Para seleccionar un estado de vacío concreto necesitamos escoger un concepto de energía, de tal forma que el vacío sea el \emph{estado de mínima energía.} Cuando el espacio-tiempo en el que se describe el campo cuántico no tiene simetrías, ni tan siquiera es estático, no existe un concepto privilegiado de energía. Sin embargo, en el caso sencillo de Minkowski, el concepto de energía privilegiado es aquel que definen los observadores inerciales. El estado de mínima energía definido por estos observadores es el estado de vacío~$\ket{0_U}$, asociado a los modos normales~(\ref{modos-U}), al cual denominaremos \emph{vacío de Minkowski.}

\subsection{Coordenadas Rindler}\label{sec_rindler}

\begin{figure}[ht]
	\centering
    \includegraphics[height=7.5cm]{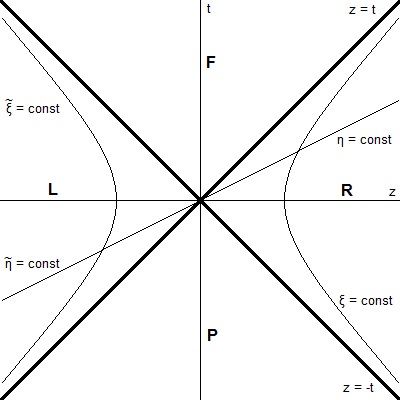}
  \caption{\footnotesize{Espacio-tiempo de Minkowski con las cuñas de Rindler~\textbf{R} y~\textbf{L}, y los universos degenerados de Kasner~\textbf{P} y~\textbf{F}.}}
  \label{fig_rindler}
\end{figure}

El Principio de Covariancia nos indica que las coordenadas carecen por sí mismas de significado físico, y que las únicas cantidades verdaderamente físicas son aquellas que permanecen invariantes ante difeomorfismos. Esto es cierto también en el espacio-tiempo de Minkowski: podemos describir este espacio-tiempo utilizando unas nuevas coordenadas, en lugar de las coordenadas cartesianas $(t, z)$. En particular, podemos usar las coordenadas~$(\eta, \xi)$, dadas implícitamente por la transformación
\begin{equation}
t = \frac{1}{a} \rme^{a \xi} \sinh (a \eta), \quad z = \frac{1}{a} \rme^{a \xi} \cosh (a \eta), \quad |t| < z,
\label{coordenadas_rindler}
\end{equation}
donde $a>0$. Estas coordenadas se denominan \emph{coordenadas Rindler.} La métrica en coordenadas Rindler tiene la forma
\begin{equation}
\rmd s^2 = \rme^{2 a \xi} (-\rmd \eta^2 + \rmd \xi^2).
\label{metrica_rindler}
\end{equation}
Se comprueba claramente que la métrica~(\ref{metrica_rindler}) es estática y \emph{conforme} a Minkowski. Debe observarse que las coordenadas~(\ref{coordenadas_rindler}) solo están definidas en una parte del espacio-tiempo: aquella para la cual $|t| < z$. Esta región del espacio-tiempo de Minkowski se denomina \emph{cuña derecha de Rindler,} y corresponde a la región denotada por~\textbf{R} en la figura~\ref{fig_rindler}. Esta región por sí misma constituye un espacio-tiempo estático y globalmente hiperbólico, sobre el que puede construirse una teoría cuántica de campos para el campo escalar sin masa de Klein-Gordon.

Podemos introducir otras coordenadas $(\tilde{\eta}, \tilde{\xi})$ para cubrir también la región $|t| < -z$, denominada \emph{cuña izquierda de Rindler,} y denotada por~\textbf{L} en la figura~\ref{fig_rindler}. Estas coordenadas están dadas por
\begin{equation}
t = -\frac{1}{a} \rme^{a \tilde{\xi}} \sinh (a \tilde{\eta}), \quad z = -\frac{1}{a} \rme^{a \tilde{\xi}} \cosh (a \tilde{\eta}), \quad |t| < -z.
\label{coordenadas_rindler_izq}
\end{equation}
Esta región es también un espacio-tiempo estático y globalmente hiperbólico.

Las dos regiones restantes~\textbf{P} y~\textbf{F} en las que se divide el espacio-tiempo en la figura~\ref{fig_rindler} se denominan universo degenerado de Kasner en contracción y en expansión, respectivamente~\cite{Crispino:2007eb}. Son también regiones globalmente hiperbólicas, pero no estáticas.

Aunque la elección de coordenadas para describir un espacio-tiempo sea físicamente irrelevante, coordenadas elegidas adecuadamente pueden guardar relación directa con magnitudes físicas medibles. Por ejemplo, las coordenadas cartesianas en Minkowski se corresponden con los tiempos propios y las distancias propias medidas en un sistema de referencia inercial concreto. En el caso de las coordenadas Rindler, su significado físico está en relación con los observadores con aceleración constante. Las trayectorias de aceleración constante en Minkowski son las curvas integrales de los campos vectoriales de Killing dados por los boosts, es decir, las transformaciones que llevan de un sistema de referencia inercial a otro con velocidad relativa respecto a este. En coordenadas Rindler, estas trayectorias están descritas sencillamente por~$\xi = \xi_0 = \text{const}$, siendo la aceleración propia de la trayectoria $a \rme^{-a \xi_0}$ (la situación es análoga para las coordenadas en la cuña izquierda). Por tanto, los observadores con aceleración constante descritos en coordenadas Rindler son observadores estáticos. El hecho de que sus trayectorias son curvas integrales de campos de Killing es lo que lleva a que ambas cuñas de Rindler sean espacio-tiempos estáticos. El tiempo propio en cada trayectoria acelerada está relacionado con la coordenada $\eta$ mediante $\tau = \rme^{a \xi_0} \eta$ (eligiendo arbitrariamente el origen $\tau = 0$). En particular, un observador con aceleración constante $a$ se describe en coordenadas Rindler simplemente como $(\xi = 0, \eta = \tau)$.

Para los observadores acelerados con aceleración constante hacia la derecha, es decir, estáticos (en coordenadas Rindler) en la cuña derecha de Rindler, la geodésica nula $t = z$ hace las veces de horizonte de sucesos futuro (pero solo para estos observadores): cualquier evento en la zona del espacio-tiempo dada por $z < t$ no puede afectar causalmente a ninguno de estos observadores. A su vez, la geodésica $t = -z$ hace las veces de horizonte de sucesos pasado para estos observadores. Estos roles se invierten en el caso de la cuña izquierda. Se concluye por tanto que ambas cuñas están desconectadas causalmente.

\subsection{Modos Rindler y modos Unruh}\label{sec_rindler_unruh}

Debido a la invariancia conforme de la teoría de Klein-Gordon sin masa en $1+1$ dimensiones, la ecuación de movimiento para el campo en coordenadas Rindler es formalmente idéntica a la ecuación escrita en coordenadas Minkowski~(\ref{klein-gordon}). Considerando el campo definido en la cuña derecha, la ecuación se escribe:
\begin{equation}
- \frac{\partial^2 \phi}{\partial \eta^2} + \frac{\partial^2 \phi}{\partial \xi^2} = 0.
\label{klein-gordon_rindler}
\end{equation}

Por tanto, la teoría cuántica de campos desarrollada en la sección~\ref{sec_campo_klein-gordon} en coordenadas Minkowski~$(t,z)$ puede desarrollarse de igual forma en coordenadas Rindler para la cuña derecha. En particular, podemos introducir las coordenadas nulas de Rindler~$u := \eta - \xi$ y~$v := \eta + \xi$, relacionadas con las coordenadas nulas de Minkowski~$(U,V)$ definidas en~(\ref{coordenadas_nulas}) mediante
\begin{equation}
u = -\frac{1}{a} \log (-a U), \quad v = \frac{1}{a} \log (a V), \quad U < 0 < V.
\label{coordenadas_nulas_rindler}
\end{equation}

Igualmente, podemos separar la solución general de la ecuación~(\ref{klein-gordon_rindler}), que en coordenadas nulas se escribe~$\partial_u \partial_v \phi = 0$, en dos términos, dependientes solamente de~$u$ y de~$v$, respectivamente. Cada uno de esos términos podemos descomponerlo en los modos normales
\begin{align}
\phi^u_\omega (u) & := \frac{1}{\sqrt{4\pi\omega}} \rme^{-\rmi \omega u}, \label{modos-u} \\
\phi^v_\omega (v) & := \frac{1}{\sqrt{4\pi\omega}} \rme^{-\rmi \omega v}. \label{modos-v}
\end{align}

Podemos hacer una construcción idéntica para la cuña izquierda. Introducimos las coordenadas nulas de Rindler para esta cuña~$\tilde{u} := \tilde{\eta} - \tilde{\xi}$ y~$\tilde{v} := \tilde{\eta} + \tilde{\xi}$, relacionadas con las coordenadas $(U, V)$ por\footnote{Debe notarse que, aunque el conjunto de cartas $\{(\eta, \xi), (\tilde{\eta}, \tilde{\xi})\}$ cubre solo las cuñas de Rindler~\textbf{R} y~\textbf{L}, el conjunto de cartas $\{(u, v), (\tilde{u}, \tilde{v}), (u, \tilde{v}), (\tilde{u}, v)\}$ cubre de nuevo todo el espacio-tiempo de Minkowski (regiones~\textbf{R},~\textbf{L},~\textbf{P} y~\textbf{F}, respectivamente).}
\begin{equation}
\tilde{u} := \frac{1}{a} \log (a U), \quad \tilde{v} := -\frac{1}{a} \log (-a V), \quad V < 0 < U.
\label{coordenadas_nulas_rindler_izq}
\end{equation}
A continuación, hacemos la misma descomposición en modos~$\tilde{u}$ y~$\tilde{v}$. El conjunto de modos~$u$, $v$, $\tilde{u}$ y $\tilde{v}$ se denominan \emph{modos Rindler,} y forman un conjunto completo de modos ortonormales. Por tanto, podemos expandir el campo en estos modos. Una vez más, nos centraremos únicamente en los modos viajando hacia la derecha (modos~$u$ y~$\tilde{u}$), siendo la teoría análoga para los modos viajando hacia la izquierda:
\begin{align}
\phi_U (U) = \int_0^\infty \rmd \omega & \left\{ \Theta (-U) \left[ a^u_\omega \phi^u_\omega (u(U)) + (a^u_\omega)^* \phi^u_\omega (u(U))^* \right] \right. \nonumber \\
& \left. + \Theta (U) \left[ a^{\tilde{u}}_\omega \phi^{\tilde{u}}_\omega (\tilde{u}(U)) + (a^{\tilde{u}}_\omega)^* \phi^{\tilde{u}}_\omega (\tilde{u}(U))^* \right] \right\},
\label{expansion_rindler}
\end{align}
donde $\Theta (U)$ es la función de Heaviside.

Utilizando las expresiones de los coeficientes de Bogoliubov~(\ref{coeficientes_alfa_en_u},~\ref{coeficientes_beta_en_u}) y de las relaciones entre coordenadas~(\ref{coordenadas_nulas_rindler},~\ref{coordenadas_nulas_rindler_izq}), podemos calcular los coeficientes de Bogoliubov entre los modos Minkowski y los modos Rindler. Por conveniencia para capítulos posteriores, haremos explícitamente el cálculo de los coeficientes~$\beta^u_{\omega, \omega'}$ (correspondientes a los modos Rindler de la cuña derecha), siendo similar el cálculo para el resto de coeficientes. Sustituyendo la relación~$U(u)$ obtenida de~(\ref{coordenadas_nulas_rindler}) en la expresión de~$\beta^u_{\omega, \omega'}$ en~(\ref{coeficientes_beta_en_u}), tenemos que
\begin{equation}
\beta^u_{\omega, \omega'} = \frac{1}{2\pi} \sqrt{\frac{\omega}{\omega'}} \int_{-\infty}^\infty \rmd u\ \exp \left[-\rmi \omega u + \rmi (\omega'/a) \rme^{-a u} \right]
\label{beta-u_integral}
\end{equation}
Si en esta integral hacemos el cambio de variable~$z = - \rmi (\omega' / a) \exp (- a u)$, tenemos finalmente
\begin{align}
\beta^u_{\omega, \omega'} & = \frac{1}{2\pi a} \sqrt{\frac{\omega}{\omega'}} \left( - \frac{\rmi \omega'}{a} \right)^{-\rmi \frac{\omega}{a}} \int_0^\infty \rmd z\ z^{\rmi \omega/a - 1} \rme^{-z} \nonumber \\
& = \frac{\rme^{-\frac{\pi \omega}{2a}}}{2\pi a} \sqrt{\frac{\omega}{\omega'}} \left( \frac{\omega'}{a} \right)^{-\rmi \frac{\omega}{a}} \Gamma (\rmi \omega/a).
\label{beta-u_resultado}
\end{align}

El resto de coeficientes se obtienen de forma similar, resultando
\begin{align}
\alpha^u_{\omega, \omega'} & = - \frac{\rme^{\frac{\pi \omega}{2a}}}{2\pi a} \sqrt{\frac{\omega}{\omega'}} \left( \frac{\omega'}{a} \right)^{-\rmi \frac{\omega}{a}} \Gamma (\rmi \omega/a), \label{alpha-u} \\
\beta^u_{\omega, \omega'} & = \frac{\rme^{-\frac{\pi \omega}{2a}}}{2\pi a} \sqrt{\frac{\omega}{\omega'}} \left( \frac{\omega'}{a} \right)^{-\rmi \frac{\omega}{a}} \Gamma (\rmi \omega/a), \label{beta-u}
\end{align}
\begin{align}
\alpha^{\tilde{u}}_{\omega, \omega'} & = - \frac{\rme^{\frac{\pi \omega}{2a}}}{2\pi a} \sqrt{\frac{\omega}{\omega'}} \left( \frac{\omega'}{a} \right)^{\rmi \frac{\omega}{a}} \Gamma (-\rmi \omega/a), \label{alpha-u_izq} \\
\beta^{\tilde{u}}_{\omega, \omega'} & = \frac{\rme^{-\frac{\pi \omega}{2a}}}{2\pi a} \sqrt{\frac{\omega}{\omega'}} \left( \frac{\omega'}{a} \right)^{\rmi \frac{\omega}{a}} \Gamma (-\rmi \omega/a). \label{beta-u_izq}
\end{align}

Dado que los coeficientes $\beta^u_{\omega, \omega'}$ y $\beta^{\tilde{u}}_{\omega, \omega'}$ no son nulos, el vacío de la base de Fock obtenida de los modos Rindler no es el mismo que el vacío de Minkowski~$\ket{0_U}$. El vacío definido por los operadores de aniquilación asociados a estos modos [definido de forma análoga al vacío de Minkowski en~(\ref{vacio})] se denomina \emph{vacío de Rindler,} y las excitaciones de tipo partícula sobre este estado se denominan \emph{partículas Rindler} [definidas de forma análoga a las partículas Minkowski en~(\ref{estados_fock})].

Los coeficientes de Bogoliubov (\ref{alpha-u}-\ref{beta-u_izq}) satisfacen las siguientes relaciones:
\begin{equation}
\beta^{\tilde{u}}_{\omega, \omega'} = - \rme^{-\pi \omega/a} (\alpha^u_{\omega, \omega'})^*, \quad \beta^u_{\omega, \omega'} = - \rme^{-\pi \omega/a} (\alpha^{\tilde{u}}_{\omega, \omega'})^*.
\label{relaciones_coeficientes_bogoliubov}
\end{equation}
Estas relaciones nos llevan a introducir unos nuevos modos, de especial utilidad para el cálculo en la siguiente sección: los \emph{modos Unruh.} Podemos definir estos modos directamente a través de una transformación de Bogoliubov desde los modos Rindler:
\begin{align}
\phi^{\rm I}_\omega (U) & := \Theta (-U) \phi^u_\omega (u (U)) + \rme^{-\frac{\pi \omega}{a}} \Theta (U) [\phi^{\tilde{u}}_\omega (\tilde{u} (U))]^*, \nonumber \\
\phi^{\rm II}_\omega (U) & := \Theta (U) \phi^{\tilde{u}}_\omega (\tilde{u} (U)) + \rme^{-\frac{\pi \omega}{a}} \Theta (-U) [\phi^u_\omega (u (U))]^*.
\label{modos_unruh}
\end{align}
A diferencia de los modos Rindler, los modos Unruh son analíticos en todo el espacio-tiempo de Minkowski. En términos de los operadores creación y destrucción, la transformación se escribe:
\begin{equation}
a^{\rm I}_\omega = a^u_\omega - \rme^{-\frac{\pi \omega}{a}} (a^{\tilde{u}}_\omega)^\dagger, \quad a^{\rm II}_\omega = a^{\tilde{u}}_\omega - \rme^{-\frac{\pi \omega}{a}} (a^u_\omega)^\dagger.
\label{bogoliubov_unruh}
\end{equation}

La principal ventaja de estos modos es que la transformación de modos Rindler a modos Unruh es diagonal. Además, los modos Unruh se componen únicamente de frecuencias positivas de modos Minkowski, como puede demostrarse a partir de las relaciones (\ref{relaciones_coeficientes_bogoliubov}, \ref{bogoliubov_unruh}). En efecto, si tenemos en cuenta la relación entre operadores~(\ref{tranformacion_bogoliubov_operadores}), se tiene que
\begin{align}
a^{\rm I}_\omega = & \int_0^\infty \rmd \omega' [ (\alpha^u_{\omega, \omega'})^* a^U_{\omega'} - (\beta^u_{\omega, \omega'})^* (a^U_{\omega'})^\dagger - \rme^{-\frac{\pi \omega}{a}} \alpha^{\tilde{u}}_{\omega, \omega'} (a^U_{\omega'})^\dagger + \rme^{-\frac{\pi \omega}{a}} \beta^{\tilde{u}}_{\omega, \omega'} a^U_{\omega'} ] \nonumber \\
= & - 2 \sinh \left( \frac{\pi \omega}{a} \right) \int_0^\infty \rmd \omega' \beta^{\tilde{u}}_{\omega, \omega'} a^U_{\omega'}, \label{rindler_minkowski_uno} \\
a^{\rm II}_\omega = & \int_0^\infty \rmd \omega' [ (\alpha^{\tilde{u}}_{\omega, \omega'})^* a^U_{\omega'} - (\beta^{\tilde{u}}_{\omega, \omega'})^* (a^U_{\omega'})^\dagger - \rme^{-\frac{\pi \omega}{a}} \alpha^u_{\omega, \omega'} (a^U_{\omega'})^\dagger + \rme^{-\frac{\pi \omega}{a}} \beta^u_{\omega, \omega'} a^U_{\omega'} ] \nonumber \\
= & - 2 \sinh \left( \frac{\pi \omega}{a} \right) \int_0^\infty \rmd \omega' \beta^u_{\omega, \omega'} a^U_{\omega'}. \label{rindler_minkowski_dos}
\end{align}
Por tanto, el vacío definido por estos modos es el vacío de Minkowski~$\ket{0_U}$. Es decir, los operadores $a^{\rm I}_\omega$ y $a^{\rm II}_\omega$ aniquilan el vacío de Minkowski:
\begin{equation}
a^{\rm I}_\omega \ket{0_U} = a^{\rm II}_\omega \ket{0_U} = 0.
\label{vacio_unruh}
\end{equation}

\subsection{El efecto Unruh}\label{sec_unruh}

El efecto Unruh, postulado por W.\ G.\ Unruh en 1976~\cite{Unruh:1976db}, se refiere al hecho de que los observadores con aceleración constante describen el estado de vacío de Minkowski de un campo cuántico como un estado térmico con temperatura proporcional a su aceleración propia, dada por 
\begin{equation}
T_{\rm U} := \frac{a}{2 \pi k_{\rm B}},
\label{temperatura_unruh}
\end{equation}
donde~$a > 0$ es la aceleración propia de los observadores. Esta temperatura se denomina \emph{temperatura de Unruh.}

Hay fundamentalmente dos enfoques complementarios del efecto Unruh: el que se refiere a la descripción del vacío de Minkowski por observadores acelerados, y el que se refiere a la probabilidad de excitación de detectores cuánticos de partículas en trayectorias con aceleración. En la sección~\ref{sec_detectores_unruh}, nos dedicaremos al segundo enfoque, en tanto que ahora pasaremos a describir el primer enfoque.\footnote{En la sección~\ref{sec_bogos_detectores} veremos que, al menos en un escenario concreto (en el cual trabajamos en esta tesis), ambos enfoques son equivalentes.}

Como vimos anteriormente, las coordenadas Rindler están ligadas a los observadores con aceleración constante: el tiempo propio de estos observadores es proporcional a la coordenada $\eta$ ($\tilde{\eta}$ en la cuña izquierda). Por tanto, la base natural del espacio de Fock para estos observadores será aquélla construida mediante los modos Rindler, ya que son estos los modos de frecuencia positiva respecto a la coordenada temporal $\eta$. Es decir, estos observadores describen de forma natural el estado del campo cuántico en términos del vacío de Rindler y de partículas Rindler.

Como un primer cálculo inmediato, a partir de los coeficientes de Bogoliubov~(\ref{beta-u}) y~(\ref{beta-u_izq}) es fácil obtener el valor esperado del número de partículas Rindler (tanto en una cuña como en otra) en el vacío de Minkowski. En el caso de la cuña derecha, tenemos
\begin{align}
\bra{0_U} (a^u_\omega)^\dagger a^u_{\omega'} \ket{0_U} & = \int_0^\infty \rmd \omega'' \beta^u_{\omega, \omega''} (\beta^u_{\omega', \omega''})^* \nonumber \\
& = \frac{\rme^{-\frac{\pi (\omega + \omega')}{2a}}}{4 \pi^2 a^2} \sqrt{\omega \omega'}\ \Gamma (\rmi \omega/a) \Gamma (-\rmi \omega' /a) \int_0^\infty \frac{\rmd \omega''}{\omega''} \left( \frac{\omega''}{a} \right)^{-\rmi \frac{\omega - \omega'}{a}} \nonumber \\
& = \frac{1}{\rme^{\frac{2 \pi \omega}{a}} - 1} \delta (\omega - \omega'),
\label{particulas_rindler}
\end{align}
donde se ha utilizado la propiedad de la función Gamma
\begin{equation}
|\Gamma (\rmi x)|^2 = \frac{\pi}{x \sinh (\pi x)}.
\label{funcion_gamma}
\end{equation}
El resultado en la cuña izquierda es idéntico. Debe notarse que, en realidad, no hemos calculado directamente el valor esperado del número de partículas con frecuencia~$\omega$, sino el valor esperado del operador~$(a^u_\omega)^\dagger a^u_{\omega'}$. De hecho, del resultado obtenido se deduce que el valor esperado del número de partículas, obtenido al tomar~$\omega = \omega'$, es divergente. Esto tiene una explicación física: los modos Rindler están asociados a observadores con aceleración constante durante toda su trayectoria. Son, por tanto, observadores irreales desde un punto de vista físico, desde el momento en que necesitan una cantidad infinita de energía para seguir su trayectoria completa. De esa energía infinita procede, en última instancia, el hecho de que el promedio de partículas descrito por observadores con aceleración constante ``eterna'' sea divergente. Otra forma alternativa de entender esta divergencia es como un efecto de \emph{volumen infinito:} en~(\ref{particulas_rindler}) se está calculando el total de partículas Rindler en toda la cuña derecha. No obstante, si prescindimos de la divergencia, reflejada en el factor~$\delta (\omega - \omega')$, vemos claramente que el valor esperado del número de partículas Rindler en el vacío de Minkowski sigue una distribución de Planck con temperatura igual a la temperatura de Unruh~(\ref{temperatura_unruh}).

Sin embargo, este resultado, aunque necesario, no es suficiente para concluir que el vacío de Minkowski es un estado térmico cuando es descrito por observadores con aceleración constante: hay infinitos estados, puros y mezcla, que dan lugar al mismo \emph{valor esperado}~(\ref{particulas_rindler}). Para entender cómo describen el vacío de Minkowski~$\ket{0_U}$ los observadores acelerados, tenemos que escribir dicho estado en la base de Fock dada por los modos Rindler.

Haciendo uso de las transformaciones de Bogoliubov entre operadores creación y destrucción de dos bases de modos, es posible encontrar igualmente la matriz de transformación entre las bases del espacio de Fock obtenidas de cada base de modos. En particular, a partir de los coeficientes de Bogoliubov~(\ref{alpha-u}-\ref{beta-u_izq}) se podría obtener directamente la expresión del vacío de Minkowski en la base dada por los modos Rindler. Sin embargo, es mucho más sencillo hacer uso de las expresiones obtenidas a partir de la definición de los modos Unruh~(\ref{bogoliubov_unruh}) y~(\ref{vacio_unruh}), introducidos precisamente a este efecto. Por otra parte, resultará necesario trabajar con paquetes de ondas, en lugar de los modos con energía bien definida que hemos utilizado hasta ahora. De forma genérica, eligiendo una resolución en frecuencias arbitraria~$\Delta \omega > 0$, considérese el conjunto de paquetes de ondas etiquetados por el par de números enteros~$(n \geq 0, \bar{n})$, y definidos por
\begin{equation}
\phi_{n,\bar{n}} (\bar{u}) := \frac{1}{\sqrt{\Delta \omega}} \int_{n \Delta \omega}^{(n+1) \Delta \omega} \rmd \omega\ \rme^{\rmi (2\pi \bar{n} / \Delta \omega) \omega} \phi_\omega (\bar{u}),
\label{paquetes_onda}
\end{equation}
siendo~$\bar{u}$ una coordenada nula arbitraria, y~$\phi_\omega(\bar{u})$ modos con energía~$\omega$ bien definida con respecto a dicha coordenada nula. Estos paquetes están centrados en la frecuencia~$(n+1/2)\Delta \omega$, con anchura~$\Delta \omega$; y en el valor~$2 \pi \bar{n}/\Delta \omega$ de la coordenada nula~$\bar{u}$, con dispersión~$1/(2\Delta \omega)$. Mediante esta expresión, podemos construir paquetes de ondas de los modos Minkowski~(\ref{modos-U}), de los modos Rindler~(\ref{modos-u}) (y de los correspondientes a la cuña izquierda), o de los modos Unruh~(\ref{modos_unruh}); según qué modos con energía bien definida~$\phi_\omega (\bar{u})$ se utilicen en~(\ref{paquetes_onda}). En el caso de los modos Minkowski y Rindler, el hecho de que los modos con energía bien definida formen una base ortonormal garantiza que el conjunto de paquetes de ondas para un~$\Delta \omega$ dado también forma una base ortonormal. Podemos, por tanto, construir bases del espacio de Fock de estados con partículas correspondientes a paquetes de ondas Minkowski o Rindler.

Es fácil ver que, si consideramos~$\Delta \omega \ll a$, las propiedades de los operadores destrucción de los modos Unruh~(\ref{bogoliubov_unruh}) y~(\ref{vacio_unruh}) tienen todas su análoga (las primeras, de forma aproximada) para los operadores creación y destrucción asociados a los paquetes de ondas. Utilizando una notación que es una extensión inmediata de la empleada hasta ahora, tenemos que
\begin{equation}
a^{\rm I}_{n,\bar{n}} \approx a^u_{n,\bar{n}} - \rme^{-\frac{\pi n \Delta \omega}{a}} (a^{\tilde{u}}_{n,\bar{n}})^\dagger, \quad a^{\rm II}_{n,\bar{n}} \approx a^{\tilde{u}}_{n,\bar{n}} - \rme^{-\frac{\pi n \Delta \omega}{a}} (a^u_{n,\bar{n}})^\dagger,
\label{bogoliubov_unruh_paquetes}
\end{equation}
\begin{equation}
a^{\rm I}_{n,\bar{n}} \ket{0_U} = a^{\rm II}_{n,\bar{n}} \ket{0_U} = 0.
\label{vacio_unruh_paquetes}
\end{equation}
De estas ecuaciones (consideradas en adelante como igualdades estrictas) es inmediato deducir~\cite{Crispino:2007eb} que
\begin{equation}
[(a^u_{n,\bar{n}})^\dagger a^u_{n,\bar{n}} - (a^{\tilde{u}}_{n,\bar{n}})^\dagger a^{\tilde{u}}_{n,\bar{n}}] \ket{0_U} = 0,
\label{igual_numero}
\end{equation}
de lo cual se concluye que existen el mismo número de partículas correspondientes a paquetes de ondas Rindler en una cuña que en otra. Esto limita la expresión del vacío de Minkowski escrito en la base de paquetes de ondas Rindler a la forma genérica
\begin{equation}
\ket{0_U} = \prod_{n,\bar{n}} \left( \sum_{k=0}^\infty K_{n, m} \ket{m_{n,\bar{n}}}_u \otimes \ket{m_{n,\bar{n}}}_{\tilde{u}} \right).
\label{expresion_generica}
\end{equation}
donde los~$K_{n, m}$ son coeficientes por determinar (dada la invariancia bajo traslación en la coordenada nula, estos no pueden depender de la etiqueta~$\bar{n}$). Utilizando de nuevo cualquiera de las ecuaciones~(\ref{bogoliubov_unruh_paquetes}) se puede deducir fácilmente una relación de recurrencia para estos coeficientes, dada por
\begin{equation}
K_{n, m+1} = \rme^{-\frac{\pi n \Delta \omega}{a}} K_{n, m}.
\label{recurrencia}
\end{equation}
De esta forma, salvo un factor de proporcionalidad global, tenemos
\begin{equation}
\ket{0_U} \propto \prod_{n,\bar{n}} \left( \sum_{m=0}^\infty \rme^{-\frac{\pi m n \Delta \omega}{a}} \ket{m_{n,\bar{n}}}_u \otimes \ket{m_{n,\bar{n}}}_{\tilde{u}} \right).
\label{vacio_minkowski_modos_rindler}
\end{equation}
El estado~(\ref{vacio_minkowski_modos_rindler}) es no normalizable en la base del espacio de Fock construida a partir de los paquetes de ondas Rindler. Esto se debe a que, estrictamente hablando, la cuantización en modos Minkowski y la cuantización en modos Rindler no son equivalentes. El origen físico de este problema es el mismo que el de la divergencia que encontramos en el valor esperado~(\ref{particulas_rindler}), a saber, el carácter no físico de los observadores Rindler, acelerados por un tiempo infinito. De hecho, es fácil ver que la divergencia en la norma aparece al sumar sobre el índice~$\bar{n}$, es decir, al sumar sobre ese tiempo infinito. Si cancelamos esta divergencia dividiendo la cantidad por un tiempo total~$T$ al que se limite la suma, tomando a continuación el límite~$T \to \infty$, y calculamos el valor esperado del número de partículas Rindler para un paquete de ondas concreto, obtenemos
\begin{equation}
\bra{0_U} (a^u_{n,\bar{n}})^\dagger a^u_{n,\bar{n}} \ket{0_U} = \bra{0_U} (a^{\tilde{u}}_{n,\bar{n}})^\dagger a^{\tilde{u}}_{n,\bar{n}} \ket{0_U} \propto \frac{1}{\rme^{\frac{2 \pi n \Delta \omega}{a}} - 1},
\label{espectro_termico_paquetes}
\end{equation}
es decir, un espectro térmico con la temperatura de Unruh~(\ref{temperatura_unruh}).

Finalmente, debemos tener en cuenta que los observadores cuya trayectoria completa se encuentra dentro de la cuña derecha están causalmente desconectados de la cuña izquierda. Por tanto, en su descripción del estado de campo, trazarán los modos Rindler definidos en la cuña izquierda, ``invisibles'' para ellos, obteniendo una matriz densidad como estado del campo. De esta forma, trazando los modos en la cuña izquierda (equivalentemente los de la derecha),
\begin{equation}
\rho \propto \prod_{n,\bar{n}} \left( \sum_{m=0}^\infty \rme^{-2 \pi m n \Delta \omega/a} \ket{m_{n,\bar{n}}}_u \bra{m_{n,\bar{n}}}_u \right).
\label{matriz_densidad}
\end{equation}
Esta es la matriz densidad de un estado térmico con temperatura dada por la temperatura de Unruh~(\ref{temperatura_unruh}), con lo que queda formalmente descrito el efecto Unruh.

Conviene notar que, de los observadores estáticos en coordenadas Rindler, solo aquel situado en $\xi = 0$ lleva aceleración $a$. Por tanto, solo para este observador la temperatura del baño térmico está relacionada con su aceleración propia mediante~(\ref{temperatura_unruh}). Sin embargo, se debe tener en cuenta que la geometría de la cuña de Rindler no es trivial. La temperatura percibida por cada observador debe ir además multiplicada por el factor de Tolman $(g_{\eta \eta})^{-1/2}$, donde~$g_{\eta \eta}$ es la correspondiente componente de la métrica de Rindler~(\ref{metrica_rindler}), ya que su tiempo propio es, en general, solo \emph{proporcional} a~$\eta$. Para la métrica~(\ref{metrica_rindler}), este factor es $\rme^{-a \xi}$. Teniendo en cuenta que la aceleración propia de cada observador es precisamente~$a \rme^{-a \xi}$, esto hace que la temperatura \emph{percibida} venga dada por la temperatura de Unruh~(\ref{temperatura_unruh}), donde $a$ se reemplaza por la aceleración propia de cada observador.

Por último, es importante enfatizar la diferencia entre los términos \emph{planckiano} y \emph{térmico.} El primero es aplicable a los espectros obtenidos de valores esperados~(\ref{particulas_rindler}) y~(\ref{espectro_termico_paquetes}), en tanto que el segundo lo es a la matriz densidad~(\ref{matriz_densidad}). Aunque, como decimos, el efecto Unruh se refiere estrictamente hablando a este último resultado (que implica los anteriores), en los sucesivos capítulos haremos uso de resultados que involucran únicamente el cálculo del valor esperado del número de partículas, como~(\ref{particulas_rindler}) y~(\ref{espectro_termico_paquetes}), por lo que en esta tesis el término correcto a usar es \emph{planckiano.} No obstante, por abuso del lenguaje se hará uso también del la denominación \emph{espectro térmico} con el mismo significado, sin que ello pueda dar lugar a confusión, pues siempre nos referiremos al mismo concepto: un espectro planckiano.

\section{Teoría Cuántica de Campos en agujeros negros}\label{sec_cuantica_agujeros_negros}

En la anterior sección hemos descrito una teoría cuántica de campos en el espacio-tiempo de Minkowski desde distintos sistemas de referencia, centrándonos en particular en sistemas de referencia con aceleración constante. En este capítulo describiremos la teoría cuántica de campos en espacio-tiempos curvos, centrándonos en dos casos en particular: el espacio-tiempo estático de Schwarzschild, que describe un~\emph{agujero negro de Schwarzschild} eterno; y un espacio-tiempo dinámico que describe un proceso de colapso, cuyo resultado final es un agujero negro de Schwarzschild. Aunque estos serán los escenarios en los que trabajaremos a lo largo de la tesis (además de Minkowski), los resultados que encontraremos son cualitativamente válidos para cualquier configuración de agujero negro con simetría esférica.

El resultado más importante de la teoría cuántica de campos en sistemas de referencia no inerciales es el efecto Unruh. Veremos que, en la teoría cuántica de campos en agujeros negros el resultado central, estrechamente relacionado con el efecto Unruh, será la \emph{radiación de Hawking}~\cite{Hawking:1974rv,Hawking:1974sw}.

\subsection{Agujero negro de Schwarzschild}\label{sec_schwarzschild}

La primera solución exacta de las ecuaciones de Einstein fue la denominada \emph{solución de Schwarzschild,} obtenida por Karl Schwarzschild en 1916, poco después de la publicación de las \emph{ecuaciones de Einstein} en noviembre de 1915. Se trata de la solución de las ecuaciones de Einstein en el exterior de un cuerpo esféricamente simétrico. Por tanto, describe con un excelente grado de aproximación el campo gravitatorio en el exterior de numerosos objetos astrofísicos.

La métrica de Schwarzschild en $3+1$~dimensiones en el exterior de un objeto de masa~$M$ se escribe
\begin{equation}
\rmd s^2 = -\left( 1 - \frac{2M}{r} \right) \rmd t^2 + \left(1 - \frac{2M}{r}\right)^{-1} \rmd r^2+ r^2 \rmd \Omega_2^2,
\label{schwarzschild}
\end{equation}
donde $r$ es la denominada \emph{coordenada radial,} $t$ la \emph{coordenada temporal de Schwarzschild,} y $\rmd \Omega_2^2$ el elemento de línea de la 2-esfera. Físicamente, la coordenada radial es igual a la longitud de una circunferencia máxima de la superficie dada por $r = {\rm const}$, dividida por $2 \pi$. Coincide con la distancia propia radial para un observador en la denominada \emph{región asintótica} $r~\to~\infty$. La coordenada temporal de Schwarzschild se corresponde con el tiempo propio de un observador estático en la región asintótica.

Como se ve explícitamente, la métrica~(\ref{schwarzschild}) es estática. El \emph{teorema de Birkhoff} demuestra además que se trata de la única solución esféricamente simétrica de las ecuaciones de Einstein en vacío~\cite{0226870332}.\footnote{El objeto que genera el campo gravitatorio no necesariamente tiene que ser estático, ni tan siquiera estacionario. Basta con que sea esféricamente simétrico. Esto incluye por tanto a un objeto en proceso de colapso, siempre que dicho colapso tenga simetría esférica.}

\begin{figure}[ht]
	\centering
    \includegraphics[height=7.5cm]{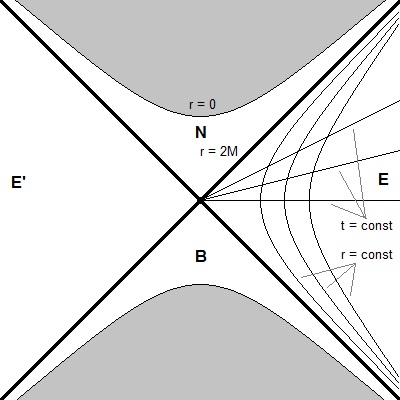}
  \caption{\footnotesize{Sector radial de la geometría de Schwarzschild. La región exterior al agujero negro está denotada por~\textbf{E}, y la región interior por~\textbf{N}. Están separadas por el horizonte de sucesos~$r=2M$. Se observa la singularidad en~$r=0$ (la zona sombreada no es parte del espacio-tiempo). La región~\textbf{B} corresponde a un~\emph{agujero blanco,} y la región \textbf{E'} a una región exterior de la geometría de Schwarzschild paralela.~\textbf{E} y \textbf{E'} están totalmente incomunicadas entre sí.}}
  \label{fig_schwarzschild}
\end{figure}

La solución de Schwarzschild~(\ref{schwarzschild}) deja de ser válida en el interior del objeto material. En el caso en el que la región exterior abarca hasta~$r = 2M$, nos encontramos ante un \emph{agujero negro de Schwarzschild,} el cual es fruto del colapso gravitatorio esféricamente simétrico de un objeto suficientemente masivo. Este se produce cuando la presión que pueden ejercer distintos procesos físicos en el interior del objeto no es suficiente para contrarrestar la atracción gravitatoria. El agujero negro de Schwarzschild es el caso más sencillo de agujero negro. No posee momento angular ni carga, quedando caracterizado únicamente por su masa~$M$. Escrita en las coordenadas $(t, r)$, la métrica~(\ref{schwarzschild}) presenta una singularidad en $r = 2M$. Sin embargo, no se trata de una singularidad física, sino de una \emph{singularidad coordenada,} fruto de la elección de unas coordenadas concretas. La superficie $r = 2M$ está generada por geodésicas nulas, y se corresponde con el \emph{horizonte de sucesos} del agujero negro. El horizonte de sucesos separa la región del espacio-tiempo correspondiente al agujero negro ($r < 2M$) de la región exterior a este ($r > 2M$). Cualquier evento en el interior del agujero negro no puede afectar causalmente al exterior. Finalmente, en~$r=0$ se encuentra una singularidad. En la figura~\ref{fig_schwarzschild} se muestra el sector radial de esta geometría.

Debido a la simetría esférica de la métrica de Schwarzschild en $3+1$~dimensiones, podemos centrarnos en el sector radial de la misma, y considerar una métrica efectiva en $1+1$~dimensiones. Esto supondrá aceptar algunas aproximaciones en la teoría cuántica de campos para este espacio-tiempo, que indicaremos más adelante. La métrica de Schwarzschild~(\ref{schwarzschild}) reducida a $1+1$~dimensiones se escribe
\begin{equation}
\rmd s^2 = -\left( 1 - \frac{2M}{r} \right) \rmd t^2 + \left(1 - \frac{2M}{r}\right)^{-1} \rmd r^2.
\label{schwarzschild_una_dimension}
\end{equation}

Esta métrica puede escribirse en unas coordenadas que la hagan conforme a Minkowski. Para ello, basta definir la denominada \emph{coordenada ``tortuga''} como
\begin{equation}
r^* := r + 2 M \ln \left( \frac{r}{2 M} - 1 \right).
\label{tortuga}
\end{equation}
Esta nueva coordenada cubre únicamente la región exterior al agujero negro. En coordenadas $(t,r^*)$, el sector radial de la métrica de Schwarzschild se escribe
\begin{equation}
\rmd s^2 = \left( 1 - \frac{2M}{r(r^*)} \right) \left[ - \rmd t^2 + (\rmd r^*)^2 \right],
\label{schwarzschild_conforme}
\end{equation}
donde $r(r^*)$ es la función implícita dada por la definición de~$r^*$~(\ref{tortuga}). La métrica~(\ref{schwarzschild_conforme}) es claramente conforme a Minkowski.

\subsection{Teoría cuántica de campos en el espacio-tiempo de Schwarzschild}\label{sec_cuantica_schwarzschild}

Nos disponemos ahora a construir una teoría cuántica para el campo real de Klein-Gordon sin masa sobre el espacio-tiempo de Schwarzschild. Como ya hemos indicado, nos centraremos en el sector radial de dicho espacio-tiempo, por lo que solo estudiaremos modos con simetría esférica (ondas~\emph{s}). Es decir, el campo, que denotaremos~$\psi$, dependerá únicamente de las coordenadas~$(t, r)$ ($\psi = \psi (t, r)$). En una situación general, esto constituye solo una aproximación al comportamiento real del campo. Sin embargo, en los casos que consideraremos en este trabajo, la única fuente de excitaciones para el campo será la propia geometría, y esta siempre tendrá simetría esférica. Por tanto, las únicas excitaciones presentes en el campo serán ondas~\emph{s.}

En el espacio-tiempo de Schwarzschild en $3+1$~dimensiones dado por la métrica~(\ref{schwarzschild}), la ecuación de ondas $\Box \psi = 0$ se escribe:
\begin{equation}
- \frac{\partial^2 \psi}{\partial t^2} + \frac{\partial^2 \psi}{\partial (r^*)^2} + \frac{2}{r(r^*)} \left( 1 - \frac{2 M}{r(r^*)} \right) \frac{\partial \psi}{\partial (r^*)} = 0,
\label{klein-gordon_schwarzschild_tres}
\end{equation}
donde se ha empleado de nuevo la definición~(\ref{tortuga}) de la coordenada~$r^*$ de forma implícita. Vemos por tanto que la teoría en $3+1$~dimensiones no es invariante conforme, y por ello la ecuación~(\ref{klein-gordon_schwarzschild_tres}) no es análoga a la ecuación de ondas en Minkowski~(\ref{klein-gordon}). Esta ecuación se comprende mejor si sustituimos~$\psi = r^{-1} \phi$, quedando
\begin{equation}
- \frac{\partial^2 \phi}{\partial t^2} + \frac{\partial^2 \phi}{\partial (r^*)^2} - \frac{2M}{r(r^*)^3} \left( 1 - \frac{2 M}{r(r^*)} \right) \phi = 0.
\label{klein-gordon_potencial}
\end{equation}
Por tanto, con respecto a la teoría libre en $1+1$~dimensiones en Minkowski dada por~(\ref{klein-gordon}), esta ecuación presenta además un \emph{potencial efectivo} $V(r^*) := (2 M/r^3)(1-2M/r)$. Si tomamos la aproximación de suprimir el término correspondiente al potencial, obtenemos la ecuación:
\begin{equation}
- \frac{\partial^2 \phi}{\partial t^2} + \frac{\partial^2 \phi}{\partial (r^*)^2} = 0.
\label{klein-gordon_schwarzschild}
\end{equation}
En adelante consideraremos la teoría invariante conforme dada por la ecuación~(\ref{klein-gordon_schwarzschild}) en la métrica de Schwarzschild reducida a $1+1$~dimensiones~(\ref{schwarzschild_conforme}). Por tanto, con respecto a la descripción del campo en un agujero negro real en $3+1$~dimensiones dada por la ecuación~(\ref{klein-gordon_potencial}), únicamente hemos despreciado el término de potencial efectivo indicado anteriormente. La consecuencia física de la presencia de este término es el denominado \emph{backscattering} de los modos en la métrica. El potencial impide separar completamente los sectores de modos salientes y entrantes, puesto que parte de los modos salientes ``rebotan'' en la métrica curva y se vuelven entrantes, y viceversa. Además, este efecto no es independiente de la frecuencia, lo que provoca una distorsión del espectro real de emisión de un agujero negro con respecto al espectro que nosotros encontraremos, que será exactamente el de un cuerpo negro. Los factores que determinan esta distorsión se denominan \emph{factores de cuerpo gris.} Puede comprobarse que la influencia de estos factores es pequeña, y por tanto la teoría reducida a $1+1$~dimensiones invariante conforme constituye una buena aproximación~\cite{Page:1976df,Unruh:1976fm,Andersson:2000tf}.

En la ecuación de ondas~(\ref{klein-gordon_schwarzschild}), análogamente a las coordenadas nulas de Minkowski introducimos las \emph{coordenadas nulas de Eddington-Finkelstein}~$\bar{u} := t - r^*$ y~$\bar{v} := t+ r^*$, de tal manera que la ecuación queda
\begin{equation}
\frac{\partial}{\partial \bar{u}} \frac{\partial}{\partial \bar{v}} \phi = 0,
\label{klein-gordon_schwarzschild_nulas}
\end{equation}
y podemos separar la solución general en la parte dependiente de~$\bar{u}$ (modos \emph{salientes} del agujero negro) y la parte dependiente de~$\bar{v}$ (modos \emph{entrantes} al agujero negro). Cada parte, respectivamente, podemos descomponerla en los modos normales
\begin{align}
\phi^{\bar{u}}_\omega (\bar{u}) & := \frac{1}{\sqrt{4 \pi \omega}} \rme^{-\rmi \omega \bar{u}},
\label{modos_boulware_u} \\
\phi^{\bar{v}}_\omega (\bar{v}) & := \frac{1}{\sqrt{4 \pi \omega}} \rme^{-\rmi \omega \bar{v}},
\label{modos_boulware_v}
\end{align}
denominados \emph{modos Boulware.} Una vez hecha la cuantización, de forma análoga a~(\ref{expansion_cuantica}) en Minkowski, tenemos la expansión de cada parte siguiente:
\begin{align}
\phi_{\bar{u}} (\bar{u}) & = \int_0^\infty \rmd \omega \left[ a^{\bar{u}}_\omega \phi^{\bar{u}}_\omega (\bar{u}) + (a^{\bar{u}}_\omega)^\dagger \phi^{\bar{u}}_\omega (\bar{u})^* \right], \label{campo_schwarzschild_u} \\
\phi_{\bar{v}} (\bar{v}) & = \int_0^\infty \rmd \omega \left[ a^{\bar{v}}_\omega \phi^{\bar{v}}_\omega (\bar{v}) + (a^{\bar{v}}_\omega)^\dagger \phi^{\bar{v}}_\omega (\bar{v})^* \right]. \label{campo_schwarzschild_v}
\end{align}

Podemos también construir una base de Fock para el espacio de Hilbert del campo cuántico utilizando los operadores creación y destrucción asociados a estos modos. En particular, tendremos un estado de vacío~$\ket{0_{\bar{u}}} \otimes \ket{0_{\bar{v}}}$, asociado a los modos~$\bar{u}$ y~$\bar{v}$, denominado \emph{vacío de Boulware}~\cite{Boulware:1974dm}. Si recordamos que las coordenadas $t$ y $r$ coinciden con el tiempo propio y la distancia radial propia para los observadores estáticos en la región asintótica $r \to \infty$, y que en esa zona $\rmd r^* \to \rmd r$, concluimos que los modos~(\ref{modos_boulware_u}, \ref{modos_boulware_v}) son los modos naturales para los observadores inerciales en la región asintótica. Por tanto, el vacío de Boulware es el vacío asociado a estos observadores, los cuales detectan de forma natural partículas asociadas a los modos Boulware.

Restringiremos en adelante nuestro análisis (salvo en algunas ocasiones) a los modos~$\bar{u}$, ya que estamos interesados únicamente en la radiación saliente del agujero negro. Como sucedía en Minkowski, podemos usar una base de modos ortonormales distinta~$\{ \phi^U_\omega (\bar{u}) \}$ para descomponer el campo. Esta base queda descrita unívocamente con la elección de una nueva coordenada nula radial~$U = U (\bar{u})$:
\begin{equation}
\phi^U_\omega (\bar{u}) := \frac{1}{\sqrt{4 \pi \omega}} \rme^{-\rmi \omega U (\bar{u})}.
\label{modos_U}
\end{equation}

Análogamente a los coeficientes de Bogoliubov~(\ref{coeficientes_alfa_en_u}, \ref{coeficientes_beta_en_u}), los coeficientes de Bogoliubov de la transformación de los modos Boulware~(\ref{modos_boulware_u}) a los nuevos modos~(\ref{modos_U}) se escriben
\begin{align}
\alpha_{\omega, \omega'} & = -\frac{1}{2\pi} \sqrt{\frac{\omega'}{\omega\phantom{'}}} \int_{-\infty}^\infty \rmd \bar{u}\ \rme^{-\rmi (\omega U (\bar{u}) - \omega' \bar{u})}, \label{alfa_schwarzschild} \\
\beta_{\omega, \omega'} & = -\frac{1}{2\pi} \sqrt{\frac{\omega'}{\omega\phantom{'}}} \int_{-\infty}^\infty \rmd \bar{u}\ \rme^{-\rmi (\omega U (\bar{u}) + \omega' \bar{u})}. \label{beta_schwarzschild}
\end{align}

\subsection{Colapso gravitacional y radiación de Hawking}\label{sec_colapso_hawking}

Tal y como las hemos descrito hasta ahora, las transformaciones de Bogoliubov son solo cambios entre dos descripciones de un campo cuántico igualmente válidas. Es decir, solo nos indican cómo ``traducir'' la descripción del estado del campo entre una base de modos y otra. En particular, hemos comprobado que el estado de vacío construido con un conjunto de modos normales puede ser descrito como un estado con partículas en la base del espacio de Fock construida con otro conjunto de modos normales. Sin embargo, para tener fenómenos físicos de creación o detección de partículas, se necesita un proceso dinámico que involucre energía. Por ejemplo, en la sección~\ref{sec_detectores_unruh} veremos cómo detectores cuánticos de partículas acelerados en el vacío de Minkowski detectan un baño término de partículas del campo, lo cual da un sentido físico a la descripción del efecto Unruh dada en la sección~\ref{sec_unruh}. La energía involucrada en este caso es la que se necesita para acelerar cualquier detector físico.

\begin{figure}[ht]
	\centering
    \includegraphics[height=8.5cm]{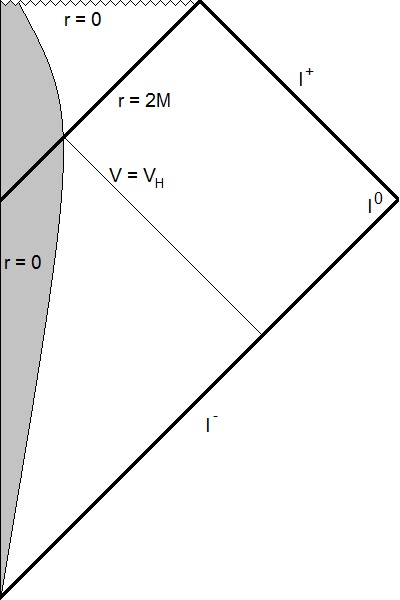}
  \caption{\footnotesize{Diagrama de Penrose de un proceso de colapso y formación de un agujero negro. La materia que colapsa corresponde a la zona sombreada. Se observa cómo se forma el horizonte de sucesos en~$r=2M$ y la singularidad en~$r=0$. $I^-$ e~$I^+$ corresponden al \emph{infinito pasado nulo} y \emph{al infinito futuro nulo,} respectivamente, en tanto que~$I^0$ corresponde al \emph{infinito espacial.} La geodésica nula~$V=V_{\rm H}$ se corresponde con la última que entra al agujero negro sin salir (por formarse el horizonte de sucesos).}}
  \label{fig_colapso}
\end{figure}

En el caso del espacio-tiempo de Schwarzschild, el proceso dinámico que da lugar a la emisión de partículas que denominaremos radiación de Hawking es el \emph{colapso gravitatorio.} Por tanto, para comprender la radiación de agujeros negros hay que estudiar su proceso de colapso. Como ya se ha dicho, los agujeros negros son fruto del colapso gravitatorio de un objeto astronómico suficientemente masivo. En la figura~\ref{fig_colapso} se representa el diagrama de Penrose de un proceso de colapso. La métrica de Schwarzschild es una descripción correcta del campo gravitatorio en el exterior del objeto esféricamente simétrico que colapsa; región que, para tiempos suficientemente largos tras el comienzo del colapso, cuando el objeto ya es efectivamente un agujero negro, abarca hasta el horizonte de sucesos.

Por simplificar la descripción, se considera que en el pasado asintótico el objeto en colapso ocupa un volumen arbitrariamente grande. Por tanto, en la zona exterior al objeto la métrica de Schwarzschild~(\ref{schwarzschild_conforme}) se aproxima a la de Minkowski. En tanto no exista otra fuente externa de radiación, el estado en el que se encuentra el campo de radiación será el estado de vacío asociado a los observadores inerciales en el exterior del objeto en el pasado asintótico y en la zona~$r \to \infty$. Es decir, definiendo las coordenadas nulas~$U := t - r$ y~$V := t + r$ el estado del campo es el vacío asociado a los modos
\begin{align}
\phi^U_\omega (U) & := \frac{1}{\sqrt{4 \pi \omega}} \rme^{-\rmi \omega U},
\label{modos_U_pasado} \\
\phi^V_\omega (V) & := \frac{1}{\sqrt{4 \pi \omega}} \rme^{-\rmi \omega V},
\label{modos_V_pasado}
\end{align}
definidos en el pasado asintótico. Este vacío lo denotaremos~$\ket{0_{\rm in}}$. Trabajaremos en la imagen de Heisenberg, por lo que este será el estado del campo en todo momento, mientras que los operadores creación y destrucción asociados a los distintos observadores serán los que evolucionen. Conviene destacar que el vacío~$\ket{0_{\rm in}}$ es el asociado a los modos que adquieren la forma (\ref{modos_U_pasado}, \ref{modos_V_pasado}) \emph{en el pasado asintótico.} En el futuro asintótico, cuando el proceso de colapso ya ha formado un agujero negro y todo el espacio es indistinguible de Schwarzschild, los modos naturales para los observadores inerciales en $r \to \infty$ son los modos que se comportan como (\ref{modos_boulware_u}, \ref{modos_boulware_v}), y que definen el vacío que denominaremos~$\ket{0_{\rm out}}$. Para comparar los modos en una y otra región, es necesario \emph{propagarlos} a través de la geometría de colapso. Debido a que la geometría de colapso es dinámica, la evolución de los modos no es trivial, y de hecho veremos que mezcla frecuencias positivas y negativas entre unos modos y otros. Esto es lo que hace que los vacíos~$\ket{0_{\rm in}}$ y~$\ket{0_{\rm out}}$ no sean el mismo estado.

Veamos entonces cómo se describe el vacío~$\ket{0_{\rm in}}$ utilizando los modos~(\ref{modos_boulware_u}, \ref{modos_boulware_v}) en el futuro asintótico. En particular, nos interesa saber el contenido de partículas correspondientes a modos salientes~(\ref{modos_boulware_u}), es decir, a la radiación emitida. Para ello, tenemos que propagar estos modos hacia atrás en el tiempo, hasta el pasado asintótico, donde podremos compararlos con los modos que definen el vacío~$\ket{0_{\rm in}}$. Esta propagación tiene varias partes. En primer lugar, la propagación desde el futuro asintótico hasta la superficie del objeto en colapso. En segundo lugar, la propagación a través del objeto en colapso (pasando por el origen $r=0$). Y, en último lugar, la propagación desde este al pasado asintótico.

Por una parte, tenemos que la ecuación de ondas~(\ref{klein-gordon_schwarzschild}) corresponde a una propagación libre.\footnote{Aquí es donde se realiza la aproximación de ignorar los factores de cuerpo gris que introduciría el \emph{backscattering} dado por el potencial de la ecuación~(\ref{klein-gordon_potencial}).} Por tanto, la propagación será libre hasta alcanzar la superficie del objeto en colapso. Consideremos modos salientes generados en el futuro asintótico para tiempos suficientemente tardíos, es decir, muy cercanos al horizonte de sucesos. En las cercanías del horizonte, se tiene que~$r \gtrsim 2M$, y pueden hacerse una serie de aproximaciones. En primer lugar, de la definición de la coordenada~$r^*$ en~(\ref{tortuga}) tenemos
\begin{equation}
r^* \approx 2m \log \left( \frac{r}{2M} - 1 \right) \quad \Rightarrow \quad \frac{r}{2M} - 1 \approx \rme^{r^*/2M} = \rme^{(\bar{v} - \bar{u})/(4M)},
\label{aproximacion_estrella}
\end{equation}
y
\begin{equation}
1 - \frac{2M}{r} = \frac{\frac{r}{2M} - 1}{1 + \left(\frac{r}{2M} - 1\right)} \approx \frac{r}{2M} - 1 \approx \rme^{(\bar{v} - \bar{u})/(4M)}.
\label{aproximacion_conforme}
\end{equation}
Si a continuación introducimos las coordenadas nulas de \emph{Kruskal-Szeckeres}, dadas por
\begin{equation}
\hat{u} := - 4M \rme^{-\bar{u}/(4M)}, \quad \hat{v} := 4M \rme^{\bar{v}/(4M)},
\label{kruskal}
\end{equation}
puede comprobarse que la métrica de Schwarzschild dimensionalmente reducida~(\ref{schwarzschild_conforme}) queda
\begin{equation}
ds^2 = - \rmd \hat{u} \rmd \hat{v}.
\label{schwarzschild_kruskal}
\end{equation}
Por tanto, la dependencia de los modos Boulware salientes~(\ref{modos_boulware_u}) con la distancia nula afín~$\hat{u}$ al horizonte, teniendo en cuenta la definición~(\ref{kruskal}), está dada por~$\phi^{\bar{u}}_\omega \propto \exp \{ \rmi \omega (4 M) \log [\hat{u} / (4 M)] \}$. Según nos acercamos al horizonte de sucesos ($\hat{u} \to 0$) estos modos tienden a oscilar infinitamente rápido en el parámetro afín. Por tanto, los modos procedentes del futuro asintótico alcanzan frecuencias arbitrariamente altas\footnote{La necesidad de invocar frecuencias arbitrariamente altas (mayores que la frecuencia de Planck) para obtener la radiación de agujeros negros se conoce como el \emph{problema transplanckiano,} problema que discutiremos brevemente en el capítulo~\ref{pulsante}.} en las cercanías del objeto que colapsa, y al propagarse a través de este. Es por ello que la probabilidad de interaccionar con el material de dicho objeto durante su propagación a través del mismo es despreciable. Tenemos por tanto que todo modo saliente en el futuro asintótico se propaga \emph{completamente} hacia el pasado asintótico. Además, las altas frecuencias que alcanzan los modos permiten hacer uso de la aproximación de la óptica geométrica. De esta forma, el modo propagado hacia el pasado asintótico, pasando a través del objeto que colapsa, acabará a la misma distancia afín~$\hat{u}$ de la geodésica~$V=V_{\rm H}$ que da lugar al horizonte de sucesos (ver figura~\ref{fig_colapso}). Teniendo en cuenta que en el pasado asintótico la coordenada~$V$ es ella misma un parámetro afín, y que la propagación desde las cercanías del horizonte hasta allí ha sido libre, el modo toma finalmente la forma~$\phi^{\bar{u}}_\omega (\bar{u}) \propto \exp \{ \rmi \omega (4 M) \log [A (V_{\rm H} - V)] \}$, siendo~$A$ una constante. El cambio~$1/(4 M) \to A$ respecto a la expresión anterior se debe al cambio en el campo gravitatorio, debido a la dinámica de colapso, durante el tiempo que tarda el rayo en cruzar el objeto que colapsa. Por tanto, un modo saliente de frecuencia~$\omega$, tras propagarlo hacia el pasado asintótico, se escribe
\begin{equation}
\phi^{\bar{u}}_\omega (\bar{u}) = \frac{1}{\sqrt{4 \pi \omega}} \rme^{\rmi \omega (4 M) \log[A(V_{\rm H}-V)]}.
\label{modo_boulware_en_V}
\end{equation}
Dicho de otra forma, la relación entre las coordenadas nulas $\bar{u}$ y $V$ en el futuro y pasado asintóticos, respectivamente, que establece un rayo de luz que conecta ambas regiones a través de la geometría de colapso es, para tiempos suficientemente tardíos,
\begin{equation}
\bar{u} = -4 M \log[A(V_{\rm H}-V)].
\label{relacion_u_V}
\end{equation}

Esta relación es formalmente idéntica a la relación~$u(U)$ entre las coordenadas nulas de Rindler y de Minkowski dada en~(\ref{coordenadas_nulas_rindler}) sustituyendo~$a \to 1/(4M)$, salvo por elección arbitraria de los orígenes de coordenadas de~$V$ y~$\bar{u}$, fijados por las constantes $V_{\rm H}$ y $A$, respectivamente. De la relación~$u(U)$ en~(\ref{coordenadas_nulas_rindler}) obteníamos el espectro planckiano~(\ref{particulas_rindler}). Por tanto, de~(\ref{relacion_u_V}) parece inmediato concluir el resultado
\begin{equation}
\bra{0_{\rm in}} (a^{\bar{u}}_\omega)^\dagger a^{\bar{u}}_{\omega'} \ket{0_{\rm in}} = \frac{1}{\rme^{8 \pi M \omega} - 1} \delta (\omega - \omega').
\label{particulas_boulware}
\end{equation}
De esta forma, obtenemos que en este estado del campo cuántico tras el proceso de colapso los observadores en la región asintótica~$r \to \infty$ detectan una radiación cuyo espectro sigue una distribución de Planck con temperatura
\begin{equation}
T_{\rm H} := \frac{1}{8 \pi M k_{\rm B}}.
\label{temperatura_hawking}
\end{equation}
Esta temperatura se denomina \emph{temperatura de Hawking,} y es proporcional a la gravedad de superficie del agujero negro, dada por~$\kappa := 1/(4M)$. El resultado~(\ref{particulas_boulware}) describe el espectro de la denominada \emph{radiación de Hawking.} Es la radiación saliente que detectan observadores inerciales en la región asintótica para tiempos suficientemente largos tras el proceso de colapso que da lugar a la formación de un agujero negro.

No obstante, hay que resaltar un punto delicado a la hora de obtener el resultado~(\ref{particulas_boulware}). Para obtener el resultado análogo~(\ref{particulas_rindler}), correspondiente al efecto Unruh, se hizo uso de los coeficientes de Bogoliubov~(\ref{beta-u}), calculados en el caso de que la relación logarítmica en~(\ref{coordenadas_nulas_rindler}) entre coordenadas nulas se cumplía \emph{para todo tiempo,} lo cual es cierto entre las coordenadas nulas Minkowski y Rindler. Por el contrario, la relación entre coordenadas nulas en el escenario de colapso~(\ref{relacion_u_V}) se cumple únicamente de forma asintótica \emph{para tiempos suficientemente tardíos} tras dicho colapso. De esta forma, considerar que los coeficientes de Bogoliubov que encontraríamos en este caso son análogos a los que encontramos al deducir el efecto Unruh~(\ref{beta-u}), lo cual es un paso necesario para obtener el espectro~(\ref{particulas_boulware}), constituye una aproximación. Sin embargo, como discutiremos en profundidad (y generalizaremos) en la sección~\ref{sec_kappa_paquetes}, se trata de una aproximación excelente, que tiende a ser exacta en el futuro asintótico, si se consideran los coeficientes de Bogoliubov obtenidos como coeficientes \emph{efectivos} para paquetes de ondas con soporte en tiempos suficientemente largos tras el colapso. La idea fundamental tras esto es que dichos paquetes de ondas no pueden ``explorar'' otra relación entre coordenadas nulas que no sea la logarítmica~(\ref{relacion_u_V}).

En este caso nos hemos limitado a obtener la distribución de Planck, sin demostrar que el estado~$\ket{0_{\rm in}}$ es un estado térmico descrito por los observadores inerciales en la región asintótica. Esto último puede hacerse si definimos también modos en el interior del agujero negro. Al describir el estado por observadores en el exterior, debemos trazar por dichos modos, ya que están separados del exterior por el horizonte de sucesos. El resultado es un estado térmico con la temperatura de Hawking~(\ref{temperatura_hawking}) para los modos salientes. El cálculo es, una vez más, muy similar al realizado en la sección anterior.

Para tiempos suficientemente largos tras el colapso, la percepción del estado de vacío~$\ket{0_{\rm in}}$ no depende de las particularidades de este, por lo que podemos prescribir un estado de vacío en un agujero negro estático eterno con el cual obtengamos el mismo resultado de radiación saliente para los observadores inerciales en la región asintótica. Como ya hemos visto, en la teoría invariante conforme reducida a $1+1$~dimensiones, una nueva base de modos normales, y por tanto una nueva base del espacio de Fock y un (posiblemente distinto) estado de vacío se consiguen simplemente definiendo una nueva coordenada nula~$U(\bar{u})$. En este caso, es obvio por la relación logarítmica~(\ref{relacion_u_V}) que la nueva coordenada nula debe venir dada por
\begin{equation}
U = U_{\rm H} - C \rme^{-\bar{u}/(4 M)},
\label{relacion_U_u}
\end{equation}
donde de nuevo las constantes~$U_{\rm H}$ y~$C$ fijan arbitrariamente los orígenes de coordenadas. El vacío asociado a los modos normales definidos por esta coordenada nula~$\phi^U_\omega (U) = (4 \pi \omega)^{-1/2} \exp (- \rmi \omega U)$ lo denotaremos~$\ket{0_U}$. Dado que la relación exponencial~(\ref{relacion_U_u}) es equivalente a la logarítmica~(\ref{relacion_u_V}), es inmediato concluir el resultado $\bra{0_U} (a^{\bar{u}}_\omega)^\dagger a^{\bar{u}}_{\omega'} \ket{0_U} = [\exp(8 \pi M \omega) - 1]^{-1} \delta (\omega - \omega')$, es decir, que en el vacío~$\ket{0_U}$ los observadores inerciales en la región asintótica perciben radiación saliente térmica con la temperatura de Hawking.

La construcción del vacío~$\ket{0_U}$ puede parecer una trivialidad, correspondiente a una simple definición~$U(u) := V(u)$ dada por~(\ref{relacion_u_V}). Sin embargo, la diferencia física entre un vacío y otro se encuentra en que el vacío~$\ket{0_{\rm in}}$ está definido en una geometría de colapso, en tanto que el vacío~$\ket{0_U}$ está definido en un agujero negro de Schwarzschild eterno. Además, dado que en este caso la relación~$U(u)$ en~(\ref{relacion_U_u}) es por definición válida para todo tiempo (y no solo de forma asintótica), para los modos~$\phi^U_\omega (U)$ el cálculo de los coeficientes de Bogoliubov es exacto.

Si incluimos el sector de radiación entrante en la descripción, debemos diferenciar también entre los distintos vacíos posibles para este sector. Además del vacío de Boulware ya definido, existen otros dos vacíos que suelen aparecer en la literatura~\cite{Fabbri:2005mw}. El vacío dado por~$\ket{0_U} \otimes \ket{0_{\bar{v}}}$, donde el vacío del sector entrante es el correspondiente a los modos Boulware~(\ref{modos_boulware_v}), es el denominado \emph{vacío de Unruh}~\cite{Unruh:1976db}. En él, los observadores en la región asintótica detectan radiación saliente pero no entrante. Es evidente que este estado representa, en una geometría estática, el estado final del campo tras un proceso de colapso real: en dicho proceso, los modos entrantes desde la región asintótica vienen de ``más allá'' de la región asintótica, por lo que únicamente se han propagado por Minkowski. El vacío dado por~$\ket{0_U} \otimes \ket{0_V}$, donde el vacío del sector entrante es el correspondiente a los modos normales asociados a la coordenada nula $V = V_{\rm H} - C' \exp[-\bar{v}/(4 M)]$ [definida de forma análoga a $U(u)$ en~(\ref{relacion_U_u})], es el denominado \emph{vacío de Hartle-Hawking}~\cite{Hartle:1976tp}. En él, los observadores en la región asintótica detectan radiación entrante idéntica a la saliente. En este caso, se trata de un vacío en el cual la emisión de radiación por parte del agujero negro se encuentra en equilibrio con otra radiación procedente de la región asintótica. Lo que se observa en el exterior del agujero es un \emph{baño térmico.}

Volviendo a centrarnos únicamente en la radiación saliente (para la cual el vacío de Unruh y el de Hartle-Hawking son equivalentes), como mostraremos, el vacío~$\ket{0_U}$, que en adelante denominaremos por sencillez vacío de Unruh, es el vacío definido por los observadores inerciales (en caída libre) arbitrariamente cercanos al horizonte de sucesos y con velocidad nula respecto al agujero negro. Para estos observadores, el tiempo propio es igual a la coordenada nula~$U$ ($\tau = U$), y por tanto los modos normales que detectan de forma natural son los que definen este vacío. Uno de los resultados centrales de esta tesis, que trataremos en el capítulo~\ref{hawking}, consiste en demostrar que, si bien el vacío de Unruh es aquel estado que definen como vacío los observadores en caída libre arbitrariamente cercanos al horizonte y con velocidad nula, los observadores que \emph{cruzan} el horizonte en caída libre con velocidad no nula sí detectan partículas cuando el estado es el vacío de Unruh. El origen de esta detección está en un factor Doppler que diverge en el horizonte de sucesos.

\newpage
\thispagestyle{empty}
\hbox{}

\makeatletter
\def\cleardoublepage{\clearpage\if@twoside \ifodd\c@page\else
    \hbox{}
    \thispagestyle{empty}
    \newpage
    \if@twocolumn\hbox{}\newpage\fi\fi\fi}
\makeatother \clearpage{\pagestyle{empty}\cleardoublepage}

\chapter{Percepción de partículas} 
\label{percepcion}

\section{Detectores cuánticos de partículas}\label{sec_detectores}

En el capítulo anterior, hemos descrito el contenido de partículas de un campo cuántico utilizando distintas bases de Fock para el espacio de Hilbert del campo cuántico. Una aproximación más \emph{operacional} al concepto de partícula puede hacerse a través de la construcción de modelos teóricos de \emph{detectores de partículas.} Estos son sistemas físicos localizados con grados de libertad acoplados al campo cuántico mediante algún tipo de interacción. Estos grados de libertad son típicamente internos, si bien podrían utilizarse también grados de libertad externos (como la posición relativa de un elemento anexo al detector, o la propia posición del detector). La interacción con el campo hace que los grados de libertad del detector puedan ser excitados absorbiendo energía del campo de forma discreta. La absorción de esa cantidad de energía se considera una \emph{detección} de una o varias partículas del campo. El hecho de que el concepto de partícula es relativo al observador se corresponde con el hecho de que, como veremos, la detección de partículas por estos dispositivos dependerá de la trayectoria que sigan. En este capítulo trataremos cómo se describen los detectores cuánticos de partículas y sus cantidades asociadas, centrándonos en el modelo más sencillo: \emph{el detector de Unruh-DeWitt}~\cite{Unruh:1976db,nla.cat-vn973875}. Veremos cómo de su comportamiento puede obtenerse como resultado el efecto Unruh.

\subsection{El modelo de Unruh-DeWitt}\label{sec_unruh-dewitt}

El sistema físico completo que queremos estudiar consta de tres partes: un espacio-tiempo físico de fondo, el cual permanecerá inalterado; un campo cuántico cuyo contenido de partículas queremos medir, definido en ese espacio-tiempo; y un detector cuántico de partículas acoplado a dicho campo, que se mueve en el espacio-tiempo siguiendo una trayectoria concreta. A lo largo de este capítulo, consideraremos únicamente el espacio-tiempo de Minkowski como espacio-tiempo de fondo.

El espacio de Hilbert del campo cuántico será~$\mathcal{H}_{\rm F}$, y el espacio de Hilbert de los grados de libertad internos del detector será~$\mathcal{H}_{\rm D}$. Una base para~$\mathcal{H}_{\rm D}$ vendrá dada por un conjunto discreto de estados con energía bien definida~$\ket{\omega_i},\ i \in \mathds{N}$. El espacio de Hilbert completo es~$\mathcal{H} := \mathcal{H}_{\rm F} \otimes \mathcal{H}_{\rm D}$. En adelante, consideraremos que solo los grados de libertad internos del detector interaccionan con el campo. La trayectoria del detector se considerará, junto con el espacio-tiempo en el que se mueve, elementos fijos de la descripción que no forman parte de la dinámica del problema.\footnote{Considerar la trayectoria del detector como parte de la dinámica es necesario para estudiar la \emph{reacción} que los cambios en el campo cuántico tienen sobre la propia trayectoria del detector~\cite{Parentani:1995iw,Parentani:1995ts}.}

Trabajaremos en la denominada \emph{imagen de interacción,} en la que la evolución libre tanto del campo como del grado de libertad interno del detector queda absorbida en los operadores, en tanto que el estado del sistema (perteneciente al espacio de Hilbert completo~$\mathcal{H}$) evoluciona con un Hamiltoniano de interacción entre el campo y el detector. Si denotamos el campo genéricamente por~$\phi$, sin necesidad de considerar (de momento) su naturaleza, el Hamiltoniano de interacción viene dado por:
\begin{equation}
H_{\rm I} (\tau) := c\ \xi (\tau) m_{\mu \nu \ldots} (\tau) F[\phi]^{\mu \nu \ldots}(\tau).
\label{interaccion_general}
\end{equation}
Aquí~$c$ es una constante de acoplo, que regula la intensidad de la interacción; $\tau$ es el tiempo propio en algún punto fijo del detector; $\xi (\tau)$  es la denominada \emph{función de encendido} del detector, que permite un acoplo variable en el tiempo; $m_{\mu \nu \ldots} (\tau)$ es el momento del detector (monopolar, dipolar, etc., según el número de índices); y~$F[\phi]^{\mu \nu \ldots}$ es un funcional del campo. La forma de estos factores nos indica el modelo de detector. En particular, el funcional~$F[\phi]^{\mu \nu \ldots}$ nos indica como este se acopla al campo. Salvo la constante de acoplo, todas las cantidades tienen (en general) dependencia temporal con~$\tau$.

El modelo más sencillo de detector es el \emph{modelo de Unruh-DeWitt.} Se trata de un detector \emph{puntual} con acoplo constante monopolar a un campo escalar real~$\phi$. Para este modelo, el Hamiltoniano de interacción~(\ref{interaccion_general}) se escribe simplemente
\begin{equation}
H_{\rm I} (\tau) := c\ m (\tau) \phi (x(\tau)).
\label{interaccion}
\end{equation}
donde~$x(\tau)$ es la trayectoria seguida por el detector puntual. Este será el modelo que utilizaremos en adelante, si bien para algunas consideraciones introduciremos funciones de encendido~$\xi(\tau)$ no triviales. Además, consideraremos que el campo~$\phi$ es un campo escalar \emph{sin masa.} Otros modelos que pueden encontrarse en la literatura son detectores con extensión espacial (no puntuales), en los que el funcional~$F[\phi]$ hace un promediado del campo siguiendo un perfil rígido alrededor de la trayectoria~$x(\tau)$~\cite{Schlicht:2003iy,Louko:2006zv,Takagi:1986kn}; detectores multipolares, bien porque se acoplan a determinadas componentes de un campo no escalar (como el electromagnético), o porque se acoplan a derivadas direccionales de un campo escalar~\cite{Hinton:1983dq,Hinton:1983uk,0264-9381-2-3-013}; o detectores direccionales, acoplados solo a radiación procedente de ciertos ángulos~\cite{0264-9381-2-3-013,Israel1983329}.

El operador de evolución entre dos instantes~$\tau_0$ y~$\tau$ para los estados del espacio de Hilbert~$\mathcal{H}$ viene dado por
\begin{equation}
U (\tau_0, \tau) = \mathcal{T} \exp \left( \rmi \int_{\tau_0}^\tau \rmd \tau' H_{\rm I} (\tau') \right),
\label{evolucion}
\end{equation}
donde~$\mathcal{T}$ es el operador de ordenación temporal. Consideremos que, en el instante inicial~$\tau_0$, el campo se encuentra en el estado~$\ket{\psi_0} \in \mathcal{H}_{\rm F}$, en tanto que el detector se encuentra en un estado con energía bien definida~$\ket{\omega_0} \in \mathcal{H}_{\rm D}$. La amplitud de probabilidad de que en el instante~$\tau$ el detector haya ``saltado'' a otro estado con energía bien definida~$\ket{\omega_{\rm F}}$, y el campo a un estado cualquiera~$\ket{\psi_{\rm F}}$ es
\begin{equation}
\mathcal{A}_{\psi_{\rm F}, \omega_{\rm F}} (\tau_0, \tau) = \rmi \bra{\psi_{\rm F}, \omega_{\rm F}} U (\tau_0, \tau) \ket{\psi_0, \omega_0}.
\label{amplitud_prob}
\end{equation}

Para calcular la probabilidad de que el detector se encuentre con una determinada energía, sin importar lo que le suceda al campo, debemos trazar sobre una base de~$\mathcal{H}_{\rm F}$, quedando
\begin{equation}
\mathcal{P}_{\omega_{\rm F}} (\tau_0, \tau) = \sum_i \left| \mathcal{A}_{\psi_i, \omega_{\rm F}} (\tau_0, \tau) \right|^2.
\label{prob_transicion}
\end{equation}

Esta probabilidad es, por definición, una cantidad no negativa menor o igual a la unidad. Durante su evolución, no puede sino oscilar. Sin embargo, debemos buscar una cantidad que capture la intuición física de que, en los periodos en los que se dan las circunstancias para que exista detección, el detector encontrará continuamente nuevas partículas, produciendo un efecto acumulativo. Esto es lo que veremos a continuación.

\subsection{Detector macroscópico y función respuesta}\label{sec_respuesta}

Para modelizar este comportamiento, consideremos un conjunto de~$N$ detectores de Unruh-DeWitt idénticos, todos ellos preparados en el mismo estado inicial, y siguiendo trayectorias idénticas. Asumamos ahora que cada detector tiene una constante de acoplo~$c = C / \sqrt{N}$ en su Hamiltoniano de interacción~(\ref{interaccion}), con~$C > 0$ constante. Dado un periodo de evolución específico~$(\tau_0, \tau)$, siempre puede tomarse~$N$ suficientemente grande, de forma que, al ser la constante de acoplo pequeña, el operador de evolución~(\ref{evolucion}) puede aproximarse a primer orden en~$c$. En este caso, tenemos que la evolución para cada detector individual es\footnote{En este modelo, los detectores interaccionan entre sí solo \emph{a través del campo,} por lo que tal interacción es de segundo orden en~$c$.}
\begin{equation}
\ket{\Psi (\tau)} = U (\tau_0, \tau) \ket{\psi_0, \omega_0} \approx \ket{\psi_0, \omega_0} + \rmi \int_{\tau_0}^\tau \rmd \tau' H_{\rm I} (\tau') \ket{\psi_0, \omega_0}.
\label{estado_tiempo}
\end{equation}
De esta forma, la amplitud de probabilidad de transición a un estado concreto~$\ket{\psi_{\rm F}, \omega_{\rm F}}$ con energía bien definida para el detector, y con~$\omega_{\rm F} \neq \omega_0$, se escribe (salvo un factor de fase irrelevante)
\begin{equation}
\mathcal{A}_{\psi_{\rm F}, \omega_{\rm F}} (\tau_0, \tau) = c \bra{\omega_{\rm F}} m(\tau_0) \ket{\omega_0} \int_{\tau_0}^\tau \rmd \tau' \rme^{\rmi (\omega_{\rm F}-\omega_0) \tau'} \bra{\psi_{\rm F}} \phi(x(\tau')) \ket{\psi_0},
\label{amplitud_prob_colec}
\end{equation}
donde se ha usado el Hamiltoniano de interacción~(\ref{interaccion}), y se ha tenido en cuenta que la evolución del operador~$m(\tau')$ (en imagen de interacción) es tal que
\begin{equation}
\bra{\omega_{\rm F}} m(\tau') \ket{\omega_0} = \rme^{\rmi \omega_{\rm F} (\tau'-\tau_0)} \rme^{-\rmi \omega_0 (\tau'-\tau_0)} \bra{\omega_{\rm F}} m(\tau_0) \ket{\omega_0}.
\label{evolucion_detector}
\end{equation}
Teniendo en cuenta, además, la relación de cierre
\begin{equation}
\sum_i \ket{\psi_i}\bra{\psi_i} = {\rm I},
\label{relacion_cierre}
\end{equation}
donde la suma se hace sobre una base de~$\mathcal{H}_{\rm F}$, e~${\rm I}$ es el operador identidad, la probabilidad de transición del detector a un estado de energía bien definida~$\omega_{\rm F}$ distinto del inicial [trazando sobre los grados de libertad del campo, como en~(\ref{prob_transicion})] es
\begin{equation}
\mathcal{P}_{\omega_{\rm F}} (\tau_0, \tau) = c^2 \left| \bra{\omega_0} m (\tau_0) \ket{\omega_{\rm F}} \right|^2 \int_{\tau_0}^\tau \rmd \tau'' \int_{\tau_0}^\tau \rmd \tau' \rme^{-\rmi (\omega_{\rm F} - \omega_0) (\tau'' - \tau')} \mathcal{W} (\tau'', \tau').
\label{prob_excitacion}
\end{equation}
Aquí~$\mathcal{W} (\tau'', \tau')$ es la \emph{distribución de Wightman} escrita en función de los tiempos propios de la trayectoria, esto es,
\begin{equation}
\mathcal{W} (\tau'', \tau') := \bra{\psi_0} \phi(x(\tau'')) \phi(x(\tau')) \ket{\psi_0}.
\label{wightman_def}
\end{equation}
Para $3+1$~dimensiones en Minkowski, esta función se puede expresar explícitamente en coordenadas cartesianas~$(t, \mathbf{x})$ como
\begin{equation}
\mathcal{W}_3 (\tau'', \tau') = - \frac{1}{4 \pi^2 \left\{ \left[t(\tau'') - t(\tau') - \rmi \varepsilon \right]^2 - \left[\mathbf{x} (\tau'') - \mathbf{x} (\tau') \right]^2 \right\} },
\label{wightman_tres}
\end{equation}
mientras que en $1+1$~dimensiones (también en Minkowski) se escribe como
\begin{equation}
\mathcal{W}_1 (\tau'', \tau') = - \frac{1}{4 \pi} \log \left\{ \left[t(\tau'') - t(\tau') - \rmi \varepsilon \right]^2 - \left[z (\tau'') - z (\tau') \right]^2 \right\},
\label{wightman_una}
\end{equation}
donde, en ambos casos, se considera implícitamente el límite~$\varepsilon \to 0^+$ \emph{después} de tomar las integrales en la probabilidad~(\ref{prob_excitacion}). En adelante en esta sección consideraremos el espacio-tiempo de Minkowski en $3+1$~dimensiones. El caso de $1+1$~dimensiones también se considerará posteriormente.

Es fácil concluir que el número medio de detectores (de los~$N$ totales) con energía~$\omega_{\rm F}$ en el intervalo~$(\tau_0, \tau)$ está dado por~$\mathcal{N}_{\omega_{\rm F}} := N \mathcal{P}_{\omega_{\rm F}}$, de manera que
\begin{equation}
\mathcal{N}_{\omega_{\rm F}} (\tau_0, \tau) = C^2 \left| \bra{\omega_0} m (\tau_0) \ket{\omega_{\rm F}} \right|^2 \int_{\tau_0}^\tau \rmd \tau'' \int_{\tau_0}^\tau \rmd \tau' \rme^{-\rmi (\omega_{\rm F} - \omega_0) (\tau'' - \tau')} \mathcal{W} (\tau'', \tau').
\label{numero_excitados}
\end{equation}
Esta cantidad ya no está restringida a tener un valor menor que uno, y de hecho presenta (en general) efectos acumulativos. El conjunto de~$N$ detectores individuales funciona como un \emph{detector macroscópico.}

El número medio de detectores dado por~(\ref{numero_excitados}) consta de dos factores de naturaleza completamente distinta. El primero de ellos depende de cómo el momento monopolar~$m$ acopla los distintos estados de energía del detector. Se trata por tanto de un factor que depende por completo de la naturaleza del propio detector. En cambio el segundo factor, dado por
\begin{equation}
\mathcal{F} (\omega; \tau_0, \tau) := \int_{\tau_0}^\tau \rmd \tau'' \int_{\tau_0}^\tau \rmd \tau' \rme^{-\rmi \omega (\tau'' - \tau')} \mathcal{W} (\tau'', \tau'),
\label{funcion_respuesta_acum}
\end{equation}
con~$\omega := \omega_{\rm F} - \omega_0$, no depende de las características particulares del detector, siendo función únicamente del salto de energía~$\omega$ entre los estados inicial y final, y de la trayectoria del detector entre los tiempos inicial y final (a través de la dependencia concreta de la distribución de Wightman con los tiempos propios). Además, retiene la posibilidad acumulativa del número de excitaciones~(\ref{numero_excitados}).

Finalmente, para comprender mejor el comportamiento del detector en un determinado instante, vamos a considerar el límite~$\tau_0 \to -\infty$ (lo que significa considerar que la interacción entre el campo y el detector ha estado presente desde el pasado remoto), y a tomar la derivada temporal de~(\ref{funcion_respuesta_acum}), dada por
\begin{equation}
\mathcal{R} (\omega, \tau) := \frac{\partial \mathcal{F}}{\partial \tau} (\omega; -\infty, \tau),
\label{funcion_respuesta_def}
\end{equation}
cantidad que ya no conserva la propiedad acumulativa de la anterior. Explícitamente, esta cantidad se escribe
\begin{equation}
\mathcal{R} (\omega, \tau) := 2\ \Re \left[ \int_{-\infty}^0 \rmd s\ \rme^{-\rmi \omega s} \mathcal{W} (\tau+s, \tau) \right].
\label{funcion_respuesta}
\end{equation}

En el caso de trayectorias estacionarias (inerciales, con aceleración constante, o circulares) esta expresión se puede simplificar, pues en este caso tenemos que la distribución de Wightman es invariante bajo traslaciones temporales a lo largo de la trayectoria, de manera que podemos definirla en función de un único argumento:
\begin{equation}
\mathcal{W} (s) := \mathcal{W} (\tau + s, \tau) = \mathcal{W} (\tau+ \Delta \tau + s, \tau + \Delta \tau).
\label{wightman_estacionaria}
\end{equation}
Además, tenemos la propiedad de que
\begin{equation}
\mathcal{W}^* (s) = \mathcal{W}^* (s,0) = \mathcal{W} (0,s) = \mathcal{W} (-s,0) = \mathcal{W} (-s).
\label{wightman_propiedad}
\end{equation}
Esto permite simplificar la expresión de~$\mathcal{R} (\omega, \tau)$ en~(\ref{funcion_respuesta}), puesto que en tal caso no depende de~$\tau$:
\begin{equation}
\mathcal{R} (\omega) = \int_{-\infty}^\infty \rmd s\ \rme^{-\rmi \omega s} \mathcal{W} (s).
\label{funcion_respuesta_est}
\end{equation}
Esta expresión coincide con la habitual para la frecuencia de detección del detector de Unruh-DeWitt en situaciones estacionarias que encontramos en la literatura~\cite{Birrell:1982ix}. Por esta razón, podría pensarse en denominar también ``frecuencia de detección'' a la cantidad más general~(\ref{funcion_respuesta}), entendiendo que caracteriza el número de excitaciones producidas por unidad de tiempo. Sin embargo, esta interpretación no es siempre correcta, debido a que puede comprobarse que dicha cantidad puede tomar valores negativos, lo que indicaría la presencia de desexcitaciones del detector (lo cual se da, por ejemplo, en algunas situaciones con aceleración \emph{decreciente}). En realidad, la función~(\ref{funcion_respuesta_acum}) es proporcional al número de detectores que encontraremos excitados si dejamos evolucionar \emph{unitariamente} al conjunto y medimos el estado de los detectores a un tiempo~$\tau$, y es perfectamente posible encontrar menos detectores excitados en tiempos posteriores~\cite{Louko:2006zv}. La función~(\ref{funcion_respuesta}) es la derivada de dicha cantidad, por lo que puede tomar valores negativos. Es por ello que la denominaremos simplemente \emph{función respuesta.} Podrían elaborarse modelos más realistas de detectores macroscópicos, con nociones de irreversibilidad (decoherencia) o latencia, pero estas cuestiones no resultan necesarias para el análisis que haremos en este trabajo.

\subsection{Tiempos de acoplo finitos y función de Wightman regularizada}\label{regularizacion}

Para llegar a la expresión de~$\mathcal{F} (\omega; -\infty, \tau)$ en~(\ref{funcion_respuesta_acum}) hemos considerado un conjunto de~$N$ detectores que permanecen siempre ``encendidos'', es decir, siempre acoplados al campo (al tomar el límite~$\tau_0 \to -\infty$), y hemos medido la cantidad media de detectores excitados a un tiempo~$\tau$. En general, puede considerarse que los detectores tienen un acoplo variable en el tiempo mediante un Hamiltoniano de interacción de la forma
\begin{equation}
H_{\rm I} (\tau') := c\ \xi(\tau') m (\tau') \phi (x(\tau')),
\label{interaccion_encendido}
\end{equation}
donde, como dijimos, $\xi(\tau')$ es la función de encendido, la cual toma valores positivos durante los periodos de interacción con el campo, anulándose en los periodos en los que no hay interacción. A la hora de calcular la función respuesta, la función de encendido aparecerá en las integrales involucradas.

Como se explica detalladamente en~\cite{Satz:2006kb}, el resultado de integrar la distribución de Wightman solo está bien definido cuando se integra multiplicada con una función suave ($\mathrm{C}^\infty$) de soporte compacto (denominada en inglés ``bump function''). Este criterio, además de matemáticamente necesario, es físicamente razonable, pues ningún detector real permanece en interacción con el campo durante un tiempo infinito, ni tampoco es encendido o apagado de forma (estrictamente) brusca.

Puede verse fácilmente que calcular la función respuesta utilizando la definición~(\ref{funcion_respuesta_def}) con~$\mathcal{F} (\omega; -\infty, \tau)$ dado por~(\ref{funcion_respuesta_acum}) corresponde a escoger como función de encendido la función de Heaviside:
\begin{equation}
\xi (\tau') = \Theta (\tau - \tau').
\label{heaviside}
\end{equation}
Esta función no cumple ninguno de los dos requisitos para que la integral de la función de Wightman esté bien definida. Por tanto, estrictamente hablando, los resultados que obtengamos utilizando la función respuesta en~(\ref{funcion_respuesta}) en general no serán correctos. Por ejemplo, en~\cite{Schlicht:2003iy} se demuestra que con esta expresión se obtienen resultados que no son invariantes Lorentz, resultando en frecuencias de detección no nulas incluso para detectores en trayectorias inerciales en el vacío de Minkowski.

\begin{figure}[ht]
	\centering
    \includegraphics{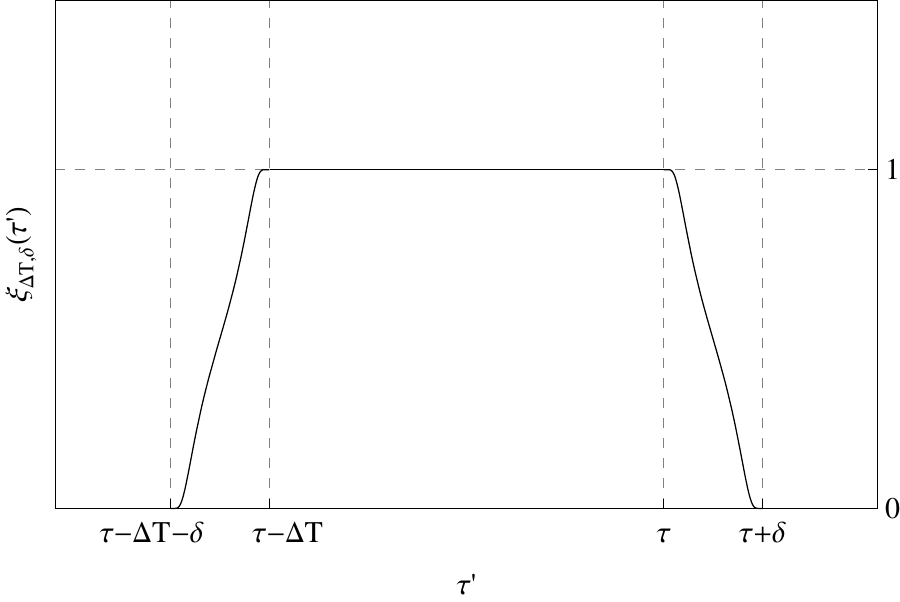}
  \caption{\footnotesize{Ejemplo de una función~$\xi_{\Delta T, \delta} (\tau')$.}}
  \label{fig_bump}
\end{figure}

Sin embargo, en la referencia~\cite{Satz:2006kb} se comprueba que podemos aproximarnos cuanto queramos al régimen con tiempo de interacción infinito y apagado brusco de los detectores [función de encendido~(\ref{heaviside})] con funciones pertenecientes a una familia~$\{\xi_{\Delta T, \delta} (\tau')\}$, etiquetadas con los parámetros reales positivos~$\Delta T$ y~$\delta$, todas las cuales son funciones suaves de soporte compacto. Dada una cualquiera de las funciones, esta se anula desde el pasado remoto hasta un tiempo~$\tau_0 = \tau - \Delta T - \delta$. A continuación, la función crece suave y monótonamente hasta el valor~$\xi = 1$ durante un tiempo~$\delta$, permanece constantemente igual a~$1$ durante un tiempo~$\Delta T$ hasta el instante~$\tau$, y finalmente decrece suave y monótonamente a cero durante un tiempo~$\delta$, permaneciendo nula en adelante. En la figura~\ref{fig_bump} se muestra un ejemplo de una de estas funciones. Físicamente, estas funciones describe un detector que permanece apagado hasta un tiempo~$\tau_0 = \tau - \Delta T - \delta$, momento en el que se enciende suavemente, permaneciendo encendido durante un tiempo~$\Delta T$, con periodos de encendido y apagado suaves de duración~$\delta$. Es inmediato ver que el límite~$\Delta T \to \infty,\ \delta \to 0$ tomado en esta familia de funciones es precisamente la función de Heaviside~(\ref{heaviside}).

Si calculamos la función respuesta para los detectores empleando como función de encendido una función de esta familia, podemos obtener la aproximación~\cite{Satz:2006kb}
\begin{multline}
\mathcal{R}_{\Delta T, \delta} (\omega, \tau) = \\
-\frac{\omega}{4 \pi} - \frac{1}{2 \pi^2} \int_{-\Delta T}^0 \rmd s \left\{ \frac{\cos (\omega s)}{\left[t(\tau + s) - t(\tau) \right]^2 - \left[\mathbf{x} (\tau + s) - \mathbf{x} (\tau) \right]^2} - \frac{1}{s^2} \right\} \\
+ \frac{1}{2 \pi^2 \Delta T} + O \left( \frac{\delta}{\Delta T^2} \right).
\label{funcion_respuesta_aprox}
\end{multline}
En el límite~$\Delta T \to \infty,\ \delta \to 0$ tenemos
\begin{equation}
\mathcal{R} (\omega, \tau) = 2 \int_{-\infty}^0 \rmd s \cos (\omega s) W (\tau + s, \tau),
\label{funcion_respuesta_reg}
\end{equation}
donde
\begin{equation}
W (\tau'', \tau') := -\frac{1}{4 \pi^2} \left\{ \frac{1}{\left[t(\tau'') - t(\tau') \right]^2 - \left[\mathbf{x} (\tau'') - \mathbf{x} (\tau') \right]^2} - \frac{1}{(\tau'' - \tau')^2} \right\}.
\label{wightman_reg}
\end{equation}

Esta expresión es idéntica a la obtenida anteriormente~(\ref{funcion_respuesta}), pero reemplazando la distribución de Wightman~(\ref{wightman_tres}) por la que denominaremos \emph{función de Wightman regularizada,} dada por~(\ref{wightman_reg}). Esta cantidad regularizada está definida en el límite~$\tau'' \to \tau'$ en el que los dos argumentos coinciden (solo es necesario que la trayectoria~$(t(\tau), \mathbf{x} (\tau))$ sea~${\rm C}^2$). Se trata por tanto no solo de una distribución, sino de una función bien definida. El polo que presentaba la distribución~(\ref{wightman_tres}), que entonces se ``esquivaba'' al introducir el término~$-\rmi \varepsilon$, ahora se elimina al restar la cantidad~$1/(\tau'' - \tau')^2$. Debe notarse, además, que se trata de una cantidad manifiestamente invariante Lorentz, a diferencia de la distribución de Wightman, la cual en su expresión~(\ref{wightman_tres}) no lo es.

Si comparamos~$\mathcal{R}_{\Delta T, \delta} (\omega, \tau)$ en~(\ref{funcion_respuesta_aprox}) con~$\mathcal{R} (\omega, \tau)$ en~(\ref{funcion_respuesta_reg}), tenemos que
\begin{multline}
\left| \mathcal{R}_{\Delta T, \delta} (\omega, \tau) - \mathcal{R} (\omega, \tau) \right| = \\
\left| \frac{1}{2 \pi^2} \int_{-\infty}^{-\Delta T} \rmd s \left\{ \frac{\cos (\omega s)}{\left[t(\tau + s) - t(\tau) \right]^2 - \left[\mathbf{x} (\tau + s) - \mathbf{x} (\tau) \right]^2} - \frac{1}{s^2} \right\} \right. \\
\left. + \frac{1}{2 \pi^2 \Delta T} + O \left( \frac{\delta}{\Delta T^2} \right) \right|,
\label{comparativa_funciones_respuesta}
\end{multline}
de donde se deduce fácilmente que
\begin{multline}
\left| \mathcal{R}_{\Delta T, \delta} (\omega, \tau) - \mathcal{R} (\omega, \tau) \right| \\
\leq \frac{1}{2 \pi^2} \int_{-\infty}^{-\Delta T} \rmd s \left\{ \left| \frac{1}{\left[t(\tau + s) - t(\tau) \right]^2 - \left[\mathbf{x} (\tau + s) - \mathbf{x} (\tau) \right]^2} \right| + \left| \frac{1}{s^2} \right| \right\} \\
+ \frac{1}{2 \pi^2 \Delta T} + O \left( \frac{\delta}{\Delta T^2} \right) \leq \frac{3}{2 \pi^2 \Delta T} + O \left( \frac{\delta}{\Delta T^2} \right),
\label{comparativa_funciones_respuesta2}
\end{multline}
(donde se ha empleado la \emph{desigualdad triangular} de la métrica en la última línea), lo que demuestra que $\mathcal{R}_{\Delta T, \delta} \to \mathcal{R}$ cuando~$\Delta T \to \infty$ (como puede comprobarse, ni tan siquiera es necesario minimizar la duración de los periodos de encendido y apagado~$\delta$ para obtener este resultado). Esto nos indica que, siempre que el conjunto de detectores permanezca activado durante un periodo de tiempo suficientemente largo, puede alcanzarse una precisión arbitraria mediante la función respuesta en~(\ref{funcion_respuesta_reg}).

\subsection{El efecto Unruh en detectores}\label{sec_detectores_unruh}

Como se adelantó en el capítulo anterior, podemos aproximarnos al fenómeno del efecto Unruh mediante la utilización de detectores cuánticos de partículas. En concreto, el sencillo modelo de Unruh-DeWitt descrito en este capítulo es suficiente para encontrar este efecto.

Consideremos un observador que sigue una trayectoria de aceleración propia constante~$a$ en la dirección del eje~$z$ (en coordenadas cartesianas). Su trayectoria en función del tiempo propio~$\tau$ es
\begin{equation}
t = \frac{1}{a} \cosh (a \tau), \quad z = \frac{1}{a} \sinh (a \tau), \quad x = 0, \quad y = 0,
\label{trayectoria_acelerada}
\end{equation}
donde se han fijado por conveniencia las constantes arbitrarias que determinan el origen de coordenadas, la velocidad inicial y el origen de tiempo propio. La distribución de Wightman~(\ref{wightman_tres}) para esta trayectoria está dada por
\begin{equation}
\mathcal{W} (\tau'',\tau') = - \frac{1}{4 \pi^2}\ \frac{a^2}{4 \sinh^2 \left[ (a/2) (\tau'' - \tau') - \rmi \varepsilon \right]}.
\label{wightman_acelerada}
\end{equation}
La distribución de Wightman depende únicamente de la diferencia de tiempos propios, debido a que se trata de una trayectoria estacionaria.

En la derivación más conocida del efecto Unruh mediante detectores~\cite{Birrell:1982ix} se utiliza la expresión~(\ref{wightman_acelerada}) para la función respuesta~(\ref{funcion_respuesta_est}) (puesto que se trata de una trayectoria estacionaria), con $s = \tau'' - \tau'$:
\begin{equation}
\mathcal{R} (\omega) = - \frac{1}{4 \pi^2} \int_{-\infty}^\infty \rmd s\ \frac{a^2 \rme^{-\rmi \omega s}}{4\sinh^2 \left[ (a/2) s - \rmi \varepsilon \right]}.
\label{funcion_respuesta_unruh}
\end{equation}
La integral que aparece en esta expresión puede calcularse utilizando el teorema de los residuos, cerrando el contorno en el semiplano inferior del plano complejo. En esta región, el integrando presenta polos en los puntos del eje imaginario~$s=-(2 \pi \rmi / a) n$, con~$n \in \mathds{N}^*$. El polo que encontraríamos en el origen, y por tanto en el camino de integración, sale fuera del contorno de integración gracias al término~$-\rmi \varepsilon$. El valor de los residuos en los polos dentro del contorno de integración es
\begin{equation}
\frac{\rmi \omega\ \rme^{-(2\pi \omega/a) n}}{4 \pi^2},
\label{residuos}
\end{equation}
y el resultado de la integración
\begin{equation}
\mathcal{R} (\omega) = \frac{\omega}{2 \pi} \frac{1}{\rme^{\frac{2 \pi \omega}{a}} - 1}.
\label{funcion_respuesta_acelerada}
\end{equation}
Es decir, el detector percibe partículas con una distribución planckiana con la temperatura de Unruh dada por~(\ref{temperatura_unruh}).

No obstante, en~\cite{Schlicht:2003iy} se demuestra que esta derivación, aunque aparentemente obtiene el resultado esperado, contiene un error que procede de la cuestión comentada en la sección~\ref{regularizacion}, a saber, que la distribución de Wightman solo debe integrarse con funciones suaves de soporte compacto para obtener resultados válidos. Si, por ejemplo, hiciésemos desde el principio la misma derivación, pero en un sistema de referencia inercial distinto, aunque considerando exactamente la misma trayectoria, los resultados en general no coincidirían, y no obtendríamos el espectro térmico.

Sin embargo, el espectro térmico~(\ref{funcion_respuesta_acelerada}) puede obtenerse, en este caso de forma correcta, haciendo uso de la expresión~(\ref{funcion_respuesta_reg}) para la función respuesta, que involucra a la función de Wightman regularizada~(\ref{wightman_reg}). La derivación es muy similar a la que se sigue para obtener~(\ref{funcion_respuesta_acelerada}). En este caso, tenemos que la integral a calcular, tras sustituir la trayectoria~(\ref{trayectoria_acelerada}) en~(\ref{wightman_reg}), y la función resultante en~(\ref{funcion_respuesta_reg}), y teniendo en cuenta las propiedades de simetría, es
\begin{equation}
\mathcal{R} (\omega) = - \frac{1}{4 \pi^2} \int_{-\infty}^\infty \rmd s\ \rme^{-\rmi \omega s} \left[ \frac{a^2}{4 \sinh^2 \left( \frac{a}{2} s \right)} - \frac{1}{s^2} \right].
\label{funcion_respuesta_unruh_reg}
\end{equation}
Esta integral se resuelve igualmente utilizando el teorema de los residuos, cerrando el contorno en el semiplano inferior del plano complejo. En este caso, el polo en el origen sencillamente no existe. El resultado es, una vez más, el espectro planckiano~(\ref{funcion_respuesta_acelerada}). Sin embargo, en este caso la función de Wightman regularizada es manifiestamente invariante Poincaré, por lo que el cálculo en cualquier sistema de referencia inercial reproduce el mismo resultado.

Este acercamiento al efecto Unruh, a partir de detectores y probabilidades de excitación, es un acercamiento más físico al fenómeno que el desarrollado en la sección~\ref{sec_unruh}: no se trata solamente de cómo se describe un estado de vacío empleando diferentes bases de Fock, sino de proponer un proceso físico dinámico que dé lugar a una detección real de partículas siguiendo un espectro térmico, debida únicamente a la aceleración propia del sistema físico que las detecta.

\section{La función de temperatura efectiva}\label{sec_kappa}

En el primer capítulo hemos descrito el efecto Unruh y la radiación de Hawking, respectivamente, mediante el cálculo de coeficientes de Bogoliubov. En la expresión de los coeficientes de Bogoliubov~$\beta_{\omega, \omega'}$ en~(\ref{coeficientes_beta_en_u}) hemos comprobado que, mientras exista invariancia conforme (como es el caso de Minkowski en $1+1$~dimensiones, y de la métrica de Schwarzschild reducida a $1+1$~dimensiones), para el cálculo de estos coeficientes solo es necesario conocer la relación~$U=U(u)$ entre las coordenadas nulas asociadas a los modos de una y otra base.

En particular, al estudiar los fenómenos que nos ocupan estaremos siempre interesados en la percepción de un determinado estado de vacío por un determinado observador o conjunto de observadores. Por tanto, necesitamos la relación entre la coordenada nula~$U$ asociada a los modos que definen el estado de vacío en el que se encuentra el campo, y la coordenada nula~$u$ asociada a los modos naturales para los observadores cuya percepción queremos conocer.\footnote{En adelante, las coordenadas nulas~$U$ y~$u$ no se corresponderán, en general, con ninguna definición anterior (coordenadas nulas de Minkowski, de Rindler, etc.). Serán coordenadas nulas genéricas, asociadas únicamente al vacío y al observador, respectivamente, salvo cuando se les asigne un valor concreto en función de coordenadas geométricas, lo cual se hará al concretar un vacío o un observador particulares.}

Como se comprobó claramente en el primer capítulo, el resultado fundamental para obtener el espectro térmico de la radiación de Hawking y del efecto Unruh consiste en que la relación~$U(u)$ sea exponencial. Es decir, genéricamente,
\begin{equation}
U(u) = U^*_{\rm H} - A_* \rme^{- \kappa_* u}.
\label{exponencial_u}
\end{equation}
Esta es la forma exacta (salvo los orígenes de coordenadas arbitrarios), en todo momento, de la relación entre ambas coordenadas nulas para observadores con aceleración constante en el vacío de Minkowski [ecuación~(\ref{coordenadas_nulas_rindler}), con~$\kappa_* = a$] y para los observadores inerciales en la región asintótica de un agujero negro de Schwarzschild con el campo en el estado de vacío de Unruh [ecuación~(\ref{relacion_U_u}), con~$\kappa_* = 1/(4M)$]. En el caso de los observadores inerciales en la región asintótica de una geometría de colapso (esféricamente simétrica), la relación entre coordenadas nulas únicamente se aproxima asintóticamente a la forma~(\ref{exponencial_u}) [con~$\kappa_* = 1/(4M)$] en el límite de tiempos suficientemente tardíos tras el proceso de colapso.

En esta sección demostraremos que también podemos encontrar como resultado un espectro térmico de radiación cuando la relación~(\ref{exponencial_u}) se cumple solo de forma aproximada durante un intervalo de tiempo suficientemente largo. Dicho espectro térmico solo aparecerá durante ese intervalo, y su temperatura vendrá determinada por el objeto central de esta sección, y de gran parte de la tesis: la \emph{función de temperatura efectiva.} Este concepto fue introducido en~\cite{Barcelo:2010pj,Barcelo:2010xk}, y en esta memoria se hace un uso extensivo del mismo, aplicándolo a situaciones físicas concretas.

\subsection{Función de temperatura efectiva y relación exponencial}

Dada una relación entre las coordenadas nulas~$U(u)$, la función de temperatura efectiva~$\kappa (u)$ se define como
\begin{equation}
\kappa (u) := - \frac{\left. \rmd^2 U \middle/ \rmd u^2 \right.}{\left. \rmd U \middle/ \rmd u \right.}.
\label{funcion_temperatura}
\end{equation}
De esta forma, $\kappa(u)$ es una función que caracteriza la relación entre ambas coordenadas nulas. Tomemos una geodésica nula saliente arbitraria, caracterizada por $u=u_* = {\rm const}$. En torno a este valor arbitrario, podemos escribir la relación~$U(u)$ en términos de~$\kappa(u)$:
\begin{equation}
U (u) = U_* + C_* \int_{u_*}^u \rmd u' \exp \left[ -\int_{u_*}^{u'} \rmd u'' \kappa(u'') \right],
\label{relacion_kappa}
\end{equation}
donde~$U_* := U(u_*)$ y~$C_* := \rmd U / \rmd u |_{u=u_*}$ son constantes de integración. Si definimos~$\kappa_* := \kappa(u_*)$ y
\begin{equation}
\delta \kappa (u) := \kappa(u) - \kappa_*,
\label{delta_kappa}
\end{equation}
podemos escribir de nuevo~(\ref{relacion_kappa}), de forma exacta, como
\begin{equation}
U (u) = U_* + C_* \int_{u_*}^u \rmd u' \left\{ \exp \left[ -\kappa_* (u' - u_*) \right] \exp \left[ -\int_{u_*}^{u'} \rmd u'' \delta \kappa(u'') \right] \right\}.
\label{relacion_kappa_delta}
\end{equation}

Consideremos ahora el intervalo abierto de valores de~$u$ en torno a~$u_*$, que denotaremos~$\mathcal{S}_*$, para el cual se cumple la siguiente condición:
\begin{equation}
\left| \int_{u_*}^u \rmd u' \delta \kappa(u') \right| < \epsilon^2 \ll 1, \quad u \in \mathcal{S}_*.
\label{condicion_integral}
\end{equation}
Debe observarse que el intervalo~$\mathcal{S}_*$ siempre existe mientras la relación~$U(u)$ sea~${\rm C}^1$ (lo que se traduce en que la integral de $\delta \kappa (u)$ sea continua). Solo hay que solucionar~(\ref{condicion_integral}) como igualdad para un~$\epsilon$ dado y obtener los dos límites de~$\mathcal{S}_*$ (los dos valores de~$u$ que satisfacen la igualdad más cercanos a~$u_*$ por cada lado). Dentro de este intervalo podemos escribir, hasta orden~$\epsilon^2$,
\begin{equation}
U(u) = U_* + C_* \int_{u_*}^u \rmd u' \left\{ \exp \left[ -\kappa_* (u' - u_*) \right] \left[ 1 + O(\epsilon)^2 \right] \right\}.
\label{relacion_kappa_aproximada}
\end{equation}
Ignorando el término~$O(\epsilon)^2$, se obtiene como aproximación la relación exponencial~(\ref{exponencial_u})
\begin{equation}
U(u) \approx U^*_{\rm H} - A_* \rme^{- \kappa_* u},
\label{exponencial_aproximada}
\end{equation}
con los valores
\begin{equation}
U^*_{\rm H} = U_* + \frac{C_*}{\kappa_*}, \quad A_* = \frac{C_*}{\kappa_*} \rme^{\kappa_* u_*}.
\label{constantes_exponencial}
\end{equation}

La relación~(\ref{exponencial_aproximada}) se cumple de forma aproximada en el intervalo~$\mathcal{S}_*$. Veremos que, si este intervalo es suficientemente largo, podremos obtener como resultado que la percepción de los observadores (cuyos modos naturales están asociados a la coordenada~$u$) durante el intervalo~$\mathcal{S}_*$, en el vacío que estamos considerando (aquel que corresponde a los modos normales asociados a la coordenada~$U$), es la de un espectro térmico con temperatura dada por
\begin{equation}
T = \frac{|\kappa_*|}{2 \pi k_{\rm B}}.
\label{temperatura_kappa_especifica}
\end{equation}

En este punto, cabe señalar que~$U^*_{\rm H}$ en~(\ref{exponencial_aproximada}) no corresponde necesariamente a la localización de un horizonte de sucesos, como ocurría con~$U_{\rm H}$ en la relación exponencial~(\ref{relacion_U_u}) del capítulo anterior. Lo hará si la relación exponencial se mantiene de forma indefinida en el tiempo (la \emph{misma} relación exponencial, con parámetros constantes). Sin embargo, ya hemos indicado que para obtener el resultado que buscamos solo necesitamos que la relación se cumpla en un intervalo~$\mathcal{S}_*$ suficientemente largo, por lo que en general no tiene por qué alcanzarse el límite~$U \to U_{\rm H}^*$ cuando~$u \to \infty$.

Finalmente, dado que~$\kappa_*$ no es más que el valor de~$\kappa(u)$ en el instante \emph{arbitrario}~$u = u_*$, tenemos que en todos los valores de~$u$ para los que el intervalo~$\mathcal{S}(u)$ (construido de la misma forma en cada punto) sea suficientemente largo, la función~$|\kappa(u)|$ es proporcional a la temperatura del espectro percibido en el entorno de~$u$. Es decir, podemos definir la temperatura efectiva
\begin{equation}
T (u) := \frac{|\kappa (u)|}{2 \pi k_{\rm B}}.
\label{temperatura_kappa}
\end{equation}

\subsection{Condición adiabática}\label{sec_adiabatica}

La siguiente cuestión es cuál debe ser la anchura mínima del intervalo~$\mathcal{S}_*$ [caracterizado por~(\ref{condicion_integral})] para que, de la relación exponencial aproximada, pueda obtenerse el espectro térmico con temperatura dada por~(\ref{temperatura_kappa_especifica}). Físicamente, resulta razonable imponer que durante el intervalo~$\mathcal{S}_*$ sea posible detectar frecuencias suficientemente bajas como para caracterizar adecuadamente el espectro planckiano; es decir, frecuencias del orden de~$|\kappa_*|$ e inferiores. Para ello, si definimos~$u_-$ y~$u_+$ como los límites inicial y final, respectivamente, del intervalo~$\mathcal{S}_*$, debemos tener que
\begin{equation}
|\kappa_*| | u_+ - u_- | \gtrsim \frac{1}{\epsilon} \gg 1,
\label{condicion_resolucion}
\end{equation}
donde empleamos la misma cantidad~$\epsilon \ll 1$ para determinar aproximaciones en todos los cálculos por consistencia. Más adelante, al calcular los coeficientes de Bogoliubov y el espectro percibido, veremos que esta condición sobre el intervalo~$\mathcal{S}_*$ es suficiente para obtener el resultado matemático buscado. Denominaremos \emph{condición adiabática} a la condición de que exista \emph{algún} intervalo~$\mathcal{S}_*$ que verifique~(\ref{condicion_integral}) y~(\ref{condicion_resolucion}). Esta será (como demostraremos) la condición que permita entender~$|\kappa_*|$ como (proporcional a) la temperatura del espectro percibido en tal intervalo.

Dado que~(\ref{condicion_integral}) y~(\ref{condicion_resolucion}) son condiciones a cumplir en \emph{algún} entorno~$\mathcal{S}_*$ de~$u_*$, puede determinarse la existencia de un intervalo de~$u_*$ que cumpla ambos criterios mediante una condición sobre propiedades locales de la función~$\kappa(u)$ en~$u = u_*$; es decir, mediante una condición que involucre~$\kappa_*$ y las derivadas~$\kappa^{(n)}_* := \rmd^n \kappa / \rmd u^n |_{u_*}$. Consideremos que la función~$\kappa (u)$ es analítica en~$u=u_*$, y que puede definirse la siguiente constante:
\begin{equation}
D_* := \sup_{n \geq 1} \left\{ \left[ \frac{1}{(n+1)!} \frac{| \kappa^{(n)}_* |}{|\kappa_*|^{n+1}} \right]^{1/(n+1)} \right\}.
\label{definicion_d}
\end{equation}
Definimos a continuación el intervalo abierto~$\mathcal{S}_*$ para un~$\epsilon \ll 1$, dado por
\begin{equation}
u \in (u_-, u_+), \quad \text{con} \quad u_\pm := u_* \pm \frac{\epsilon}{\sqrt{2} D_* |\kappa_*|}.
\label{definicion_s}
\end{equation}

Comprobemos que dicho intervalo cumple la condición~(\ref{condicion_integral}). Por una parte, todos los puntos dentro del intervalo~$\mathcal{S}_*$ verifican, por definición,
\begin{equation}
2 D^2_* \kappa^2_* (u - u_*)^2 < \epsilon^2 \ll 1, \quad u \in \mathcal{S}_*.
\label{condicion_intervalo}
\end{equation}
Por otra parte, escogiendo~$\epsilon$ suficientemente pequeño, siempre es posible hacer que~$\mathcal{S}_*$ se encuentre incluido en el entorno de~$u_*$ en el que el desarrollo en serie de potencias de~$\kappa (u)$ en~$u_*$ es convergente. Para un valor~$u \in \mathcal{S}_*$, se tiene entonces que
\begin{align}
\left| \int_{u_*}^u \rmd u' \delta \kappa(u') \right| & = \left| \sum_{n=1}^\infty \frac{1}{(n+1)!} \kappa^{(n)}_* (u - u_*)^{n+1} \right| \nonumber \\
& \leq \sum_{n=1}^\infty \frac{1}{(n+1)!} | \kappa^{(n)}_* | |u - u_*|^{n+1} \leq \sum_{n=1}^\infty D^{n+1}_* | \kappa_* |^{n+1} |u - u_*|^{n+1} \nonumber \\
& = \frac{D^2_* \kappa^2_* |u - u_*|^2}{1 - D_* |\kappa_*| |u - u_*|} \leq 2 D^2_* \kappa^2_* (u - u_*)^2 < \epsilon^2 \ll 1, \quad u \in \mathcal{S}_*,
\label{condicion_serie}
\end{align}
donde las dos últimas desigualdades se obtienen de~(\ref{condicion_intervalo}). Por tanto, se cumple la condición~(\ref{condicion_integral}).

Una vez hemos comprobado que se cumple esta condición, imponemos la condición~(\ref{condicion_resolucion}) sobre~$\mathcal{S}_*$, lo cual, teniendo en cuenta la definición de los límites~$u_\pm$ en~(\ref{definicion_s}), nos lleva a la siguiente condición sobre~$D_*$ (y, por tanto, sobre~$\kappa (u)$ en~$u_*$):
\begin{equation}
D_* \lesssim \frac{\epsilon^2}{\sqrt{2}} \ll 1.
\label{condicion_adiabatica}
\end{equation}

La desigualdad~(\ref{condicion_adiabatica}) es un criterio únicamente sobre los valores locales de la función de temperatura efectiva y sus derivadas. Conviene dejar claro su significado: \emph{bajo las condiciones impuestas a la función~$\kappa (u)$, si se verifica~(\ref{condicion_adiabatica}) en un punto~$u_*$, entonces se verifica la condición adiabática en ese punto.} No obstante, es la condición adiabática la que en última instancia se necesitará para obtener como resultado un espectro térmico. Dado que~(\ref{condicion_intervalo}) es condición suficiente, \emph{pero no necesaria,} de~(\ref{condicion_integral}), la condición~(\ref{condicion_adiabatica}) es, en cierta medida, innecesariamente restrictiva. Sin embargo, tiene la gran ventaja de ser un criterio únicamente sobre los valores locales de~$\kappa (u)$ y sus derivadas.

La condición~(\ref{condicion_adiabatica}) tiene por sí misma un significado físico claro. Sabemos que el intervalo de tiempo necesario para detectar una partícula con frecuencia característica del espectro planckiano es~$\Delta u \sim 1/ |\kappa_*|$. Por tanto, para poder caracterizar el espectro durante ese intervalo, es necesario que la variación en la temperatura~$\Delta \kappa_*$ en ese tiempo sea \emph{relativamente} pequeña. Es decir, $|\Delta \kappa_* / \kappa_*| \ll 1$. Y en efecto, si se cumple~(\ref{condicion_adiabatica}), se tiene que
\begin{align}
\left| \frac{\Delta \kappa_*}{\kappa_*} \right| = & \ \left| \frac{1}{\kappa_*} \sum_{n=1}^\infty \frac{1}{n!} \kappa^{(n)}_* \Delta u^{n+1} \right| \lesssim \sum_{n=1}^\infty \frac{1}{n!} \frac{| \kappa^{(n)}_* |}{| \kappa^{n+1}_* |} \leq \sum_{n=1}^\infty (n+1) D^{n+1}_* \nonumber \\
= & \ \frac{D^2_* (D_* - 2)}{(D_* - 1)^2} \leq \sqrt{2} D_* \lesssim \epsilon^2 \ll 1.
\label{justificacion_adiabatica}
\end{align}
Se observa, además, claramente que la deducción~(\ref{justificacion_adiabatica}) es esencialmente la derivada de~(\ref{condicion_serie}).

En la mayoría de los casos, el término que domina la sucesión que aparece en la definición de~$D_*$ en~(\ref{definicion_d}) es~$n=1$, por lo que la condición~(\ref{condicion_adiabatica}) sobre~$D_*$ se escribe simplemente
\begin{equation}
\frac{| \dot \kappa_* |}{ \kappa^2_* } \lesssim \epsilon^4 \ll 1,
\label{condicion_adiabatica_derivada}
\end{equation}
lo cual puede, en tal caso, considerarse suficiente para el cumplimiento de la condición adiabática. Este será el criterio que empleemos a lo largo de la tesis cuando hagamos uso de la función de temperatura efectiva.

\subsection{Coeficientes de Bogoliubov y espectro térmico}\label{sec_kappa_paquetes}

A continuación, vamos a demostrar el significado que hemos atribuido a la función de temperatura efectiva. Es decir, vamos a demostrar que, bajo cumplimiento de la condición adiabática en un instante~$u_*$, el valor absoluto de dicha función~$|\kappa_*|=|\kappa(u_*)|$ es proporcional a la temperatura del espectro térmico percibido en torno a ese instante. Conviene notar que, para llevar a cabo tal demostración, no basta con calcular los coeficientes de Bogoliubov~$\beta_{\omega, \omega'}$ entre los modos~$U$ y los modos~$u$, puesto que dichos coeficientes dependen de \emph{toda} la relación~$U(u)$, en tanto que el resultado que aquí se busca es un resultado local en torno a~$u=u_*$. Para obtener este resultado local, introduciremos paquetes de ondas de anchura en frecuencias~$\Delta \omega$, centrados  espectralmente en~$\omega_n := (n+1/2) \Delta \omega$ y espacio-temporalmente en~$u_*$, dados por
\begin{equation}
\phi^{u_*}_{\omega_n} (u) := \frac{1}{\sqrt{\Delta \omega}} \int_{n \Delta \omega}^{(n + 1) \Delta \omega} \rmd \bar{\omega}\ \rme^{\rmi \bar{\omega} u_*} \phi^u_{\bar{\omega}} (u).
\label{paquetes_onda_dos}
\end{equation}
Estos paquetes de ondas pertenecen al conjunto de paquetes de ondas~(\ref{paquetes_onda}) definido anteriormente, los cuales forman una base ortonormal siempre que los modos normales~$\phi^u_{\bar{\omega}} (u)$ que se utilizan en su definición también formen una base ortonormal, lo cual es cierto en este caso. Por tanto, podemos considerar transformaciones de Bogoliubov entre estos paquetes de ondas y otras bases ortonormales de modos.

Los coeficientes de Bogoliubov en los que estamos interesados son los que relacionan los modos~$U$ con los paquetes de ondas~(\ref{paquetes_onda_dos}), a los cuales denominaremos~$\beta^P_{u_*;\omega_n, \omega'}$. Teniendo en cuenta que la integral en la expresión del producto escalar~(\ref{producto_escalar}) puede escribirse igualmente en la coordenada~$u$, estos coeficientes son
\begin{align}
\beta^P_{u_*;\omega_n, \omega'} = & \ -\langle\phi^{u_*}_{\omega_n}, (\phi^U_{\omega'})^*\rangle \nonumber \\
= & \ \rmi \int_{-\infty}^\infty \rmd u \left( \phi^{u_*}_{\omega_n} (u) \frac{\partial}{\partial u} \phi^U_{\omega'} (U(u)) - \phi^U_{\omega'} (U(u)) \frac{\partial}{\partial u} \phi^{u_*}_{\omega_n} (u) \right).
\label{coeficientes_beta_paquetes}
\end{align}

Supongamos que para el valor~$u=u_*$ se verifica la condición adiabática, y por tanto existe un entorno~$\mathcal{S}_*$ que verifica~(\ref{condicion_integral}). Imponemos entonces que la región temporal con peso significativo de los paquetes de ondas~(\ref{paquetes_onda_dos}) esté incluida en este entorno~$\mathcal{S}_*$. Si denominamos~$\Delta u$ a la dispersión en el tiempo de los paquetes de ondas, y suponiendo razonablemente que~$u_*$ se encuentra aproximadamente en el centro de~$\mathcal{S}_*$, esto implica que
\begin{equation}
\Delta u \lesssim |u_+ - u_-|.
\label{imposicion_delta_u}
\end{equation}
En tal caso, la región con peso significativo en la integral del producto escalar~(\ref{coeficientes_beta_paquetes}) estará también contenida en~$\mathcal{S}_*$, y por tanto sustituir la relación exacta~$U(u)$ por la relación exponencial~(\ref{exponencial_aproximada}) válida en~$\mathcal{S}_*$ será una buena aproximación (hasta orden~$\epsilon$). Dado que la relación~$U(u)$ únicamente aparece en el argumento de los modos normales~$\phi^U_{\omega'} (U(u))$, definimos los modos~$\phi^{\rm exp}_{u_*;\omega'} (u)$ como los modos normales~$\phi^U_{\omega'} (U(u))$ en los que se ha sustituido la relación exponencial~(\ref{exponencial_aproximada}). Es decir,
\begin{equation}
\phi^{\rm exp}_{u_*;\omega'} (u) := \phi^U_{\omega'} (U^*_{\rm H} - A_* \rme^{- \kappa_* u}) = \frac{1}{\sqrt{4 \pi \omega'}} \exp \left[- \rmi \omega' \left( U^*_{\rm H} - A_* \rme^{- \kappa_* u} \right) \right].
\label{modos_exp}
\end{equation}

De esta forma, los coeficientes de Bogoliubov~(\ref{coeficientes_beta_paquetes}) pueden aproximarse por
\begin{align}
\beta^P_{u_*;\omega_n, \omega'} \approx -\langle\phi^{u_*}_{\omega_n}, (\phi^{\rm exp}_{u_*;\omega'})^*\rangle & = -\left\langle \frac{1}{\sqrt{\Delta \omega}} \int_{n \Delta \omega}^{(n +1) \Delta \omega} \rmd \bar{\omega}\ \rme^{\rmi \bar{\omega} u_*} \phi^u_{\bar{\omega}} (u) , (\phi^{\rm exp}_{u_*;\omega'})^* \right\rangle \nonumber \\
& = \frac{1}{\sqrt{\Delta \omega}} \int_{n \Delta \omega}^{(n +1) \Delta \omega} \rmd \bar{\omega}\ \rme^{\rmi \bar{\omega} u_*} \beta^{\rm exp}_{u_*;\bar{\omega}, \omega'},
\label{coeficientes_beta_paquetes_dos}
\end{align}
donde se ha usado la definición de los paquetes de ondas~(\ref{paquetes_onda_dos}), y los coeficientes
\begin{align}
\beta^{\rm exp}_{u_*; \bar{\omega}, \omega'} & := -\langle\phi^u_{\bar{\omega}}, (\phi^{\rm exp}_{u_*;\omega'})^*\rangle \nonumber \\
& = \frac{\rme^{-\rmi \omega' U^*_{\rm H}}}{2\pi} \sqrt{\frac{\bar{\omega}\phantom{'}}{\omega'}} \int_{-\infty}^\infty \rmd u\ \exp[-\rmi ( \bar{\omega} u + \omega' A_* \rme^{- \kappa_* u} ) ]
\label{coeficientes_beta_extension}
\end{align}
deben entenderse como los coeficientes de Bogoliubov entre los modos~$U$ y los modos~$u$ calculados \emph{extendiendo la relación exponencial~(\ref{exponencial_aproximada}) a toda la relación~$U(u)$.} Estos coeficientes se pueden obtener siguiendo un cálculo idéntico al realizado en la sección~\ref{sec_rindler_unruh} [ecuación~(\ref{beta-u_resultado})], resultando\footnote{Debe tenerse en cuenta que, dada la definición de~$A_*$ en~(\ref{constantes_exponencial}), y sabiendo que~$C_*>0$ [puesto que, según~(\ref{relacion_kappa}), $C_* := \rmd U / \rmd u |_{u=u_*}$], se tiene que $A_*$ tiene el mismo signo que~$\kappa_*$. El procedimiento es ligeramente diferente según~$\kappa_*$ sea positiva o negativa.}
\begin{equation}
\beta^{\rm exp}_{u_*; \bar{\omega}, \omega'} = \rme^{-\rmi \omega' U^*_{\rm H}} \frac{\rme^{-\frac{\pi \bar{\omega}}{2|\kappa_*|}}}{2\pi \kappa_*} \sqrt{\frac{\bar{\omega}}{\omega'}} ( |A_*| \omega')^{-\rmi \frac{\bar{\omega}}{\kappa_*}} \Gamma (\rmi \bar{\omega}/\kappa_*).
\label{coeficientes_beta_resultado}
\end{equation}
Teniendo en cuenta esta expresión, es fácil ver que la integral~(\ref{coeficientes_beta_paquetes_dos}) puede aproximarse, para~$\omega_n \gg \Delta \omega$, y salvo una fase global irrelevante, por
\begin{equation}
|\beta^P_{u_*;\omega_n, \omega'}| \approx \sqrt{\Delta \omega}\ \frac{\rme^{-\frac{\pi \omega_n}{2|\kappa_*|}}}{2\pi \kappa_*} \sqrt{\frac{\omega_n}{\omega'}} |\Gamma (\rmi \omega_n/\kappa_*)| \frac{\sin \left[ \Delta \omega \log(C_* \omega'/|\kappa_*|) / (2 \kappa_*) \right]}{\Delta \omega \log(C_* \omega'/|\kappa_*|) / (2 \kappa_*)},
\label{bogolibov_paquetes_approx}
\end{equation}
donde se ha tenido en cuenta la expresión de~$A_*$ en~(\ref{constantes_exponencial}). A diferencia de lo que sucedía con los resultados obtenidos anteriormente para el efecto Unruh~(\ref{particulas_rindler}) y la radiación de Hawking~(\ref{particulas_boulware}), en este caso sí que podemos calcular directamente el espectro en frecuencias observadas~$\omega_n$ de los paquetes de ondas, puesto que no es divergente. Teniendo en cuenta la propiedad de la función~$\Gamma(\rmi x)$ dada en~(\ref{funcion_gamma}), e introduciendo el cambio de variable~$z = \Delta \omega \log(C_* \omega'/|\kappa_*|) / (2 \kappa_*)$ al integrar en~$\omega'$, se obtiene fácilmente el resultado
\begin{equation}
\int_0^\infty \rmd \omega' |\beta^P_{u_*;\omega_n, \omega'}|^2 \approx \frac{1}{\rme^{2 \pi \omega_n / |\kappa_*|} - 1},
\label{espectro_exacto}
\end{equation}
el cual resulta ser el espectro térmico buscado.

Hasta el momento, de los dos criterios que debe verificar~$\mathcal{S}_*$ para que se cumpla la condición adiabática, solo hemos necesitado que se verifique~(\ref{condicion_integral}), a fin de poder utilizar la relación exponencial~(\ref{exponencial_aproximada}). Sin embargo, la condición~(\ref{condicion_resolucion}) (es decir, que~$\mathcal{S}_*$ sea suficientemente largo) es necesaria para poder caracterizar adecuadamente el espectro térmico, es decir, para que resultado~(\ref{espectro_exacto}) sea válido en frecuencias~$\omega_n \sim |\kappa_*|$. Ignorando, de momento, la aproximación~$\omega_n \gg \Delta \omega$ realizada para obtener dicho resultado, solamente el uso de paquetes de ondas de dispersión en frecuencias~$\Delta \omega$ ya impone una cota inferior~$\omega_{\rm min} \sim \Delta \omega$ a las frecuencias que pueden resolverse con dichos paquetes. Teniendo en cuenta que en un paquete de ondas~$\Delta \omega \sim 1/\Delta u $, y que anteriormente hemos necesitado la condición~(\ref{imposicion_delta_u}) sobre~$\Delta u$, esto implica que
\begin{equation}
\omega_{\rm min} \sim \Delta \omega \sim \frac{1}{\Delta u} \gtrsim \frac{1}{|u_+ - u_-|}
\label{condicion_omega_min}
\end{equation}
Por tanto, para que puedan explorarse frecuencias del orden de~$|\kappa_*|$, se requiere~$1/|u_+ - u_-| \lesssim |\kappa_*|$. Como hemos partido de que se verifica la condición adiabática, esta desigualdad se cumple gracias a que se cumple la condición~(\ref{condicion_resolucion}), por lo que efectivamente es posible caracterizar suficientemente el espectro térmico. Es más, la condición~(\ref{condicion_resolucion}) garantiza que~$1/|u_+ - u_-| \ll |\kappa_*|$, por lo que la aproximación~$\omega_n \gg \Delta \omega$ realizada para obtener dicho espectro térmico es legítima, puesto que el límite inferior a la resolución en frecuencias dado por~$\omega_{\rm min} \sim \Delta \omega$ está muy por debajo de~$|\kappa_*|$.

Con esto queda clarificada la interpretación física de la función de temperatura efectiva: bajo las condiciones apuntadas, nos indica la temperatura aproximada a la que un observador concreto, en un vacío concreto, percibe los paquetes de ondas que llegan en torno a un rayo de luz concreto. Uno de los resultados centrales de esta tesis será demostrar que la función de temperatura efectiva no es un mero ``artefacto numérico'' con el que estimar la temperatura de un espectro que no es estrictamente térmico. Se trata de una cantidad con mucho contenido físico, incluso en las situaciones en las que no cumple las condiciones que permiten interpretarla estrictamente como una temperatura.

\section{Coeficientes de Bogoliubov y detectores}\label{sec_bogos_detectores}

En la sección~\ref{sec_detectores}, nos hemos acercado al fenómeno de la percepción de radiación por distintos observadores mediante el uso de detectores de partículas, en concreto del modelo de detector de Unruh-DeWitt. Como dijimos, este acercamiento responde a la descripción de un proceso físico concreto (excitaciones de detectores). Además, obtuvimos como resultado el efecto Unruh en Minkowski en $3+1$~dimensiones, debido a que es el escenario más simple y habitual en el que se trata tal fenómeno.

Sin embargo, para analizar la percepción de radiación en los siguientes capítulos (salvo en el capítulo~\ref{adiabatica}) utilizaremos la función de temperatura efectiva introducida en la sección anterior. El significado de esta función se basa en el uso de coeficientes de Bogoliubov, y en cómo distintos conjuntos de observadores describen un estado de vacío dado, en función de qué modos normales son los naturales para ellos. Esta descripción no hace alusión a ningún proceso físico concreto, sino a cómo distintos observadores construyen su Teoría Cuántica de Campos para el campo de radiación. Para darle una justificación física a los resultados que encontraremos en adelante, comprobaremos a continuación que, al menos en la teoría invariante conforme en $1+1$~dimensiones (tanto en la métrica de Minkowski como en la métrica de Schwarzschild reducida), las aproximaciones al problema en términos de coeficientes de Bogoliubov o de detectores de Unruh-DeWitt son esencialmente equivalentes.

Por una parte, en la aproximación al problema mediante detectores, el resultado fundamental (no por ser el más importante, sino por ser el resultado ``raíz'' del que obtuvimos la función respuesta) es el número total de detectores excitados de la colectividad que constituye lo que denominamos detector macroscópico [ecuación~(\ref{numero_excitados})]. En lugar de considerar el número de detectores excitados entre dos tiempos~$(\tau_0, \tau)$ (lo que equivale, como dijimos, a un encendido y un apagado ``bruscos'' de los detectores), introduciremos desde el principio una función de encendido~$\xi(\tau)$ suave y de soporte compacto. En tal caso, el número de detectores excitados desde el estado inicial~$\ket{\omega_0}$ (del detector) hasta un estado final~$\ket{\omega_{\rm F}}$ tras el apagado resulta proporcional a
\begin{equation}
\mathcal{N}_\xi (\omega_0,\omega_{\rm F}) := C^2 \left| \bra{\omega_0} m (\tau_0) \ket{\omega_{\rm F}} \right|^2 \int_{-\infty}^\infty \rmd \tau\ \xi(\tau) \int_{-\infty}^\infty \rmd \tau' \xi (\tau') \rme^{-\rmi \omega (\tau - \tau')} \mathcal{W} (\tau, \tau'),
\label{numero_excitados_xi}
\end{equation}
con~$\omega := \omega_{\rm F} - \omega_0$.

Por otra parte, en la aproximación al problema mediante coeficientes de Bogoliubov, el resultado fundamental es el valor esperado del número de partículas [ecuación~(\ref{particulas_vacio})]. Para compararlo con el resultado anterior, consideremos el valor esperado del número de partículas de los siguientes paquetes de ondas:
\begin{equation}
\phi^\xi_\omega (u) \propto \int_{-\infty}^\infty \rmd \bar{\omega} \left( \int_{-\infty}^\infty \rmd \bar{u}\ \rme^{\rmi (\bar{\omega}-\omega) \bar{u}} \xi (\bar{u}) \right) \phi^u_{\bar{\omega}} (u),
\label{paquetes_ondas_xi}
\end{equation}
donde se omite el factor de normalización [pues este resultado se comparará con el anterior~(\ref{numero_excitados_xi}), que incluye una constante de proporcionalidad arbitraria]. En~(\ref{paquetes_ondas_xi}), los modos~$u$ son los modos naturales para los observadores cuya percepción queremos conocer. Es decir, $\phi^\xi_\omega (u)$ son paquetes de ondas centrados en la frecuencia~$\omega$ y con distribución temporal dada por~$\xi(u)$ para tales observadores. Siguiendo con la notación utilizada en anteriores capítulos, los modos que definen el estado de vacío en el que calcularemos nuestro valor esperado serán los modos~$U$. De manera análoga al cálculo hecho en la sección~\ref{sec_kappa_paquetes}, los coeficientes de Bogoliubov entre los modos~$U$ y los paquetes de ondas~(\ref{paquetes_ondas_xi}) se pueden escribir como\footnote{Siempre puede construirse una base ortonormal de paquetes de ondas tal que los paquetes de ondas~(\ref{paquetes_ondas_xi}) formen parte de ella~\cite{Dragan:2012hy}, de manera que pueda hablarse de transformaciones de Bogoliubov entre estos paquetes de ondas y otra base ortonormal de modos.}
\begin{equation}
\beta^\xi_{\omega, \omega'} \propto \int_{-\infty}^\infty \rmd \bar{\omega} \left( \int_{-\infty}^\infty \rmd \bar{u}\ \rme^{\rmi (\bar{\omega}-\omega) \bar{u}} \xi (\bar{u}) \right) \beta_{\bar{\omega}, \omega'},
\label{bogoliubov_paquetes_xi}
\end{equation}
donde los coeficientes de Bogoliubov~$\beta_{\bar{\omega}, \omega'}$ están dados por la expresión~(\ref{coeficientes_beta_en_u})
\begin{equation}
\beta_{\bar{\omega}, \omega'} = -\frac{1}{2\pi} \sqrt{\frac{\bar{\omega}}{\omega'}} \int_{-\infty}^\infty \rmd u\ \rme^{-\rmi (\bar{\omega} u + \omega' U(u))}.
\label{coeficientes_beta_generico_dos}
\end{equation}
El valor esperado del número de excitaciones de estos paquetes de ondas, en el vacío definido por los modos~$U$, resulta
\begin{align}
\int_0^\infty \rmd \tilde{\omega} | \beta^\xi_{\omega, \tilde{\omega}} |^2 \propto & \int_{-\infty}^\infty \rmd \bar{\omega}\ \sqrt{\bar{\omega}} \left( \int_{-\infty}^\infty \rmd \bar{u}\ \rme^{\rmi (\bar{\omega}-\omega) \bar{u}} \xi (\bar{u}) \right) \int_{-\infty}^\infty \rmd u\ \rme^{-\rmi \bar{\omega} u} \nonumber \\
\times & \int_{-\infty}^\infty \rmd \bar{\omega}' \sqrt{\bar{\omega}'} \left( \int_{-\infty}^\infty \rmd \bar{u}' \rme^{-\rmi (\bar{\omega}'-\omega) \bar{u}'} \xi (\bar{u}') \right) \int_{-\infty}^\infty \rmd u' \rme^{\rmi \bar{\omega}' u'} \nonumber \\
\times & \int_0^\infty \frac{\rmd \tilde{\omega}}{\tilde{\omega}}\ \rme^{-\rmi \tilde{\omega} (U(u) - U(u'))}.
\label{valor_esperado_xi}
\end{align}
Es fácil ver que el factor en la última línea de~(\ref{valor_esperado_xi}) es proporcional a la distribución de Wightman~$\mathcal{W} (u,u')$ en $1+1$~dimensiones \emph{calculada en el vacío definido por los modos~$U$.} Por otra parte, consideremos que la función~$\xi(u)$ tiene una dispersión~$\Delta u$ suficientemente grande, de manera que su transformada [las expresiones entre paréntesis de las dos primeras líneas de~(\ref{valor_esperado_xi})] está muy picada en torno a la frecuencia~$\omega$ (es decir, $\omega \gg \Delta \omega \sim 1/\Delta u$).\footnote{Debe notarse que esta suposición es compatible con la deducción del espectro térmico a partir de la función de temperatura efectiva que llevamos a cabo la sección~\ref{sec_kappa_paquetes}, puesto que en tal deducción se hizo la misma suposición.} Entonces, una buena aproximación se obtiene sacando los factores~$\sqrt{\bar{\omega}}$ y~$\sqrt{\bar{\omega}'}$ de las integrales como~$\sqrt{\omega}$. De esta forma, las integrales en~$\bar{\omega}$ y~$\bar{\omega}'$ resultan en deltas de Dirac que eliminan otras dos integrales, quedando el resultado final
\begin{equation}
\int_0^\infty \rmd \tilde{\omega} | \beta^\xi_{\omega, \tilde{\omega}} |^2 \propto \omega \int_{-\infty}^\infty \rmd u\ \xi(u) \int_{-\infty}^\infty \rmd u' \xi (u') \rme^{-\rmi \omega (u - u')} \mathcal{W} (u, u').
\label{valor_esperado_xi_dos}
\end{equation}

Si comparamos este resultado con el valor esperado del número de detectores excitados~(\ref{numero_excitados_xi}), vemos que
\begin{equation}
\mathcal{N}_\xi (\omega_0,\omega_{\rm F}) \propto \frac{1}{\omega} \left| \bra{\omega_0} m (\tau_0) \ket{\omega_{\rm F}} \right|^2 \int_0^\infty \rmd \tilde{\omega} | \beta^\xi_{\omega, \tilde{\omega}} |^2.
\label{comparacion}
\end{equation}
La conclusión que se obtiene es que en una colectividad de detectores de Unruh-DeWitt, con un tiempo de encendido mucho mayor que el inverso de la menor frecuencia a detectar, se excitan detectores en una cantidad proporcional al valor esperado del número de partículas con la frecuencia~$\omega = \omega_{\rm F} - \omega_0$ correspondientes a los modos naturales para los observadores que siguen la trayectoria de los detectores, modulada por las características particulares de los propios detectores.

Esto nos lleva, a su vez, a dos conclusiones físicas. Primero, que el detector de Unruh-DeWitt no es solo un modelo ``especialmente sencillo'' de detector, sino aquel que mide precisamente el valor esperado del número de partículas, en el sentido apuntado (otros modelos más complejos, en general, medirán cantidades distintas del campo). Y segundo, que todo el formalismo en términos de coeficientes de Bogoliubov, y en particular la función de temperatura efectiva, tiene una interpretación física inmediata en términos de excitación de detectores.

\newpage
\thispagestyle{empty}
\hbox{}

\makeatletter
\def\cleardoublepage{\clearpage\if@twoside \ifodd\c@page\else
    \hbox{}
    \thispagestyle{empty}
    \newpage
    \if@twocolumn\hbox{}\newpage\fi\fi\fi}
\makeatother \clearpage{\pagestyle{empty}\cleardoublepage}

\chapter{Expansión adiabática en espacio-tiempo plano} 
\label{adiabatica}

En la sección~\ref{sec_detectores} describimos el modelo de detector de Unruh-DeWitt macroscópico. Cuando este detector sigue una trayectoria de aceleración constante en el espacio-tiempo de Minkowski, con el campo cuántico en el estado de vacío de Minkowski, se obtiene como resultado el efecto Unruh. Sin embargo, los detectores cuánticos de partículas, y en concreto el modelo de Unruh-DeWitt, permiten estudiar escenarios mucho más generales. A primer orden en teoría de perturbaciones, y considerando el espacio-tiempo de Minkowski en $3+1$~dimensiones, el resultado más general que encontramos para el detector de Unruh-DeWitt en el vacío de Minkowski es la función respuesta~(\ref{funcion_respuesta_reg}) con la función de Wightman regularizada~(\ref{wightman_reg}). Teniendo en cuenta la interpretación física de la función respuesta descrita en la sección~\ref{sec_detectores}, con este resultado se pueden estudiar trayectorias~$(t(\tau), \mathbf{x}(\tau))$ generales.

Un régimen especialmente interesante es aquel en el cual la trayectoria seguida por el detector presenta una aceleración lineal propia~$g(\tau)$ (es decir, la aceleración lineal medida en un sistema de referencia inercial y co-móvil con el detector en el instante~$\tau$) que varía lentamente con el tiempo propio~$\tau$ del detector. Es esperable que, en este régimen, el detector perciba radiación térmica con temperatura variable, proporcional a la aceleración en cada instante de la trayectoria según la fórmula de Unruh~(\ref{temperatura_unruh}); es decir, $T(\tau) = |g(\tau)|/(2 \pi k_{\rm B})$. A este régimen podemos llamarle ``régimen adiabático''. Si ignoramos derivadas de la aceleración más allá de la primera, este régimen podría definirse como aquel en el cual $|g'(\tau)|/g(\tau)^2 \ll 1$ (esto es una ligadura sobre la derivada de la aceleración en términos de la aceleración instantánea). Se trata de un requisito físico razonable: Si queremos que el detector perciba un espectro térmico con temperatura proporcional a~$|g(\tau)|$, el cambio relativo en la aceleración $|g'(\tau)/g(\tau)|$ durante el tiempo necesario para detectar una partícula de la energía característica de ese espectro, $1/|g(\tau)|$, debe ser pequeño. En este capítulo encontraremos el régimen adiabático precisamente como la aproximación de orden cero de una expansión asintótica general, que denominaremos ``expansión adiabática''. Se trata de una expansión en potencias de las derivadas de~$g(\tau)$ para un tipo particular de trayectorias: las trayectorias rectilíneas con aceleración variable. Como hemos dicho, encontraremos el espectro térmico con temperatura proporcional a la aceleración en el orden más bajo. Es fácil ver que este resultado, aunque en un contexto diferente, está íntimamente relacionado con el resultado del espectro térmico que encontramos a través de la función de temperatura efectiva en la sección~\ref{sec_kappa}. El resto de términos serán correcciones debidas a la no adiabaticidad, y serán de mayor importancia cuanto más nos alejemos del régimen adiabático.

\section{Trayectorias rectilíneas con aceleración variable}

Las trayectorias rectilíneas con aceleración variable tienen una ventaja importante: las ecuaciones de las coordenadas como funciones del tiempo propio se pueden escribir explícitamente, de forma sencilla, en términos de la aceleración lineal propia~$g(\tau)$ en cada instante (ver, por ejemplo, las referencias~\cite{Padmanabhan:2003gd,moller1972theory}). Considerando, por simplificar, una trayectoria rectilínea paralela al eje~$x$, tenemos las ecuaciones
\begin{align}
t(\tau) &= t_0 + \int_{\tau_0}^\tau \rmd \tau' \cosh \chi (\tau'), \nonumber\\
x(\tau) &= x_0 + \int_{\tau_0}^\tau \rmd \tau' \sinh \chi (\tau'), \label{trayectoria_acelerada_dos}\\
y &= y_0, \quad z = z_0, \nonumber
\end{align}
donde
\begin{equation}
\chi (\tau) := \chi_0 +  \int_{\tau_0}^\tau \rmd \tau' g(\tau').
\label{definicion_xi}
\end{equation}
Los valores~$(t_0,x_0,y_0,z_0)$ son las coordenadas iniciales en el instante~$\tau = \tau_0$. La velocidad inicial en ese instante es~$(\cosh \chi_0, \sinh \chi_0,0,0)$. Al haber escogido una trayectoria paralela al eje~$x$, hemos fijado cinco de los diez parámetros que caracterizan una transformación de Poincaré general entre los posibles sistemas de referencia inerciales arbitrarios en los que describir la trayectoria. Los otros cinco parámetros son~$(t_0,x_0,y_0,z_0,\chi_0)$. Como la función respuesta es, por supuesto, invariante Poincaré, dicha función no depende ni de los parámetros fijados ni, como veremos, de los libres.

Trayectorias descritas de esta forma también se utilizan en~\cite{Kothawala:2009aj}, donde también se estudia la respuesta de un detector de Unruh-DeWitt con aceleración variable. Sin embargo, la elección de la ``variable temporal'' respecto a la cual se define la función respuesta en estos trabajos difiere de la nuestra y, de hecho, da lugar a una dependencia de dicha función que no respeta la causalidad: El detector, tras haber permanecido en una trayectoria inercial durante todo el tiempo pasado, podría eventualmente detectar una partícula solo por ``estar preparado'' para acelerar un tiempo \emph{después} de la detección. Debido a esta no causalidad en el cálculo, sus resultados no son directamente comparables con los que aquí presentamos.

Insertando la trayectoria dada por~(\ref{trayectoria_acelerada_dos}) y~(\ref{definicion_xi}) en la función de Wightman regularizada~(\ref{wightman_reg}), tenemos la expresión
\begin{align}
W(\tau+s,\tau;g] = \frac{1}{4 \pi^2 s^2} - \frac{1}{4 \pi^2} & \left\{ \left[ \int_0^s \rmd s' \cosh \left( \int_0^{s'} \rmd s'' g(\tau + s'') \right) \right]^2 \right. \nonumber \\
- & \left. \left[ \int_0^s \rmd s' \sinh \left( \int_0^{s'} \rmd s'' g(\tau + s'') \right) \right]^2 \right\}^{-1}.
\label{wightman_reg_g}
\end{align}
Por conveniencia, hemos escrito explícitamente la dependencia funcional con la aceleración~$g$ entre los argumentos de la función de Wightman regularizada. Debe notarse que, como adelantamos, ninguna de las condiciones iniciales aparece en esta última expresión.

\section{Expansión adiabática de la función respuesta}

La función respuesta~(\ref{funcion_respuesta_reg}) y la función de Wightman regularizada~(\ref{wightman_reg_g}), formalmente, son suficientes para calcular completamente el valor de la función respuesta en el caso de una aceleración lineal arbitraria~$g(\tau)$. Sin embargo, excepto para aceleraciones con una dependencia temporal muy simple, debido a las integrales involucradas en las expresiones no es posible calcular explícitamente ni tan siquiera la función de Wightman regularizada~(\ref{wightman_reg_g}). En lugar de intentar obtener resultados explícitos, lo que haremos será, como ya dijimos, utilizar las expresiones~(\ref{funcion_respuesta_reg}) y~(\ref{wightman_reg_g}) para calcular una expansión adiabática de la función respuesta. Esta expansión dará una aproximación tan buena como se quiera al valor exacto de la función respuesta en el límite en el que la aceleración varía de forma suficientemente lenta.

Debe notarse que, para calcular el valor de~$\mathcal{R} (\omega, \tau)$, debemos conocer toda la historia pasada de la trayectoria del detector, a través de la función~$g(\tau'), \tau' \leq \tau$. Como veremos, en el marco de la expansión adiabática podemos calcular adecuadamente la función respuesta usando únicamente propiedades locales de la trayectoria en el instante~$\tau$. Para ello, asumiremos que la función~$g(\tau')$ es una función real analítica, al menos para~$\tau' \leq \tau$. Para obtener una expansión en serie asintótica de la función respuesta, introduciremos un parámetro auxiliar~$\alpha$, y definiremos una función auxiliar~$\bar{W}$ de la siguiente forma: En primer lugar, hacemos la sustitución
\begin{equation}
g(\tau') \to g(\alpha (\tau' - \tau) + \tau)
\label{reemplazar_g_alpha}
\end{equation}
en la expresión formal de la función de Wightman regularizada~(\ref{wightman_reg_g}), de manera que
\begin{align}
\bar{W}(\tau+s,\tau,\alpha;g] := \frac{1}{4 \pi^2 s^2} - \frac{1}{4 \pi^2} & \left\{ \left[ \int_0^s \rmd s' \cosh \left( \int_0^{s'} \rmd s'' g(\tau + \alpha s'') \right) \right]^2 \right. \nonumber \\
- & \left. \left[ \int_0^s \rmd s' \sinh \left( \int_0^{s'} \rmd s'' g(\tau + \alpha s'') \right) \right]^2 \right\}^{-1}.
\label{wightman_aux}
\end{align}
Es evidente que la función regularizada de Wightman original se obtiene de esta función auxiliar fijando~$\alpha=1$. A continuación, podemos encontrar un desarrollo en serie \emph{formal} para la función~(\ref{wightman_aux}) en potencias de~$\alpha$. Los coeficientes de esta expansión dependerán de~$s$ y de derivadas de orden cada vez mayor de~$g(\tau')$, evaluadas en~$\tau' = \tau$. Explícitamente,
\begin{align}
\bar{W}(\tau+s,\tau,\alpha;g] = & \ W_0 (s,g(\tau)) + W_1 (s,g(\tau),g'(\tau)) \alpha \nonumber \\
& + W_2 (s,g(\tau),g'(\tau),g''(\tau)) \alpha^2 + \ldots,
\label{wightman_expansion}
\end{align}
donde
\begin{align}
W_0 (s,g(\tau)) & := \lim_{\alpha \to 0^+} \bar{W} (\tau+s,\tau,\alpha;g], \nonumber \\
W_n (s,g(\tau),g'(\tau),\ldots,g^{(n)}(\tau)) & := \frac{1}{n!} \left. \frac{\partial^n \bar{W} (\tau+s,\tau,\alpha;g]}{\partial \alpha^n} \right|_{\alpha \to 0^+}, \ n \geq 1.
\label{wightman_coefs}
\end{align}

Una vez hemos hecho esta expansión en potencias de~$\alpha$, evaluamos \emph{formalmente} la expresión en~$\alpha=1$ para recobrar el valor original de la función de Wightman regularizada en~(\ref{wightman_reg_g}) [lo cual es inmediato de~(\ref{wightman_aux})], pero escrito en forma de serie asintótica
\begin{align}
W(\tau+s,\tau;g] \sim & \ W_0 (s,g(\tau)) + W_1 (s,g(\tau),g'(\tau)) \nonumber \\
& + W_2 (s,g(\tau),g'(\tau),g''(\tau)) + \ldots.
\label{wightman_expansion_asintotica}
\end{align}
Finalmente, calcularemos (de nuevo, \emph{formalmente}) la integral que aparece en la función respuesta~(\ref{funcion_respuesta_reg}) término a término usando~(\ref{wightman_expansion_asintotica}). Es claro que cada término en la serie es integrable. Primero, para cada valor de~$s$, el objeto~$\bar{W}(\tau+s,\tau,\alpha;g]$ es una función real analítica de~$\alpha$, y por tanto los coeficientes~$W_n$ son finitos para cada~$s$. Y segundo, todos los coeficientes tienden a cero para $s \to -\infty$ más rápido que~$1/s^2$ ya que, para~$\alpha$ suficientemente pequeña en la expansión~(\ref{wightman_expansion}), tomar un número finito de términos en la serie en~$\alpha$ ofrece una aproximación arbitrariamente buena a~$\bar{W}(\tau+s,\tau,\alpha;g]$, que de hecho se anula en dicho límite más rápido que~$1/s^2$, como es fácil de comprobar. Tenemos entonces que
\begin{align}
\mathcal{R}(\omega, \tau) \sim & \ \mathcal{R}_0 (\omega,g(\tau)) + \mathcal{R}_1 (\omega,g(\tau),g'(\tau)) \nonumber \\
& + \mathcal{R}_2 (\omega,g(\tau),g'(\tau),g''(\tau)) + \ldots,
\label{funcion_respuesta_expansion}
\end{align}
donde
\begin{equation}
\mathcal{R}_n (\omega,g(\tau),g'(\tau),\ldots,g^{(n)}(\tau)) := 2 \int_{-\infty}^0 \rmd s \cos (\omega s) W_n (s,g(\tau),g'(\tau),\ldots,g^{(n)}(\tau)).
\label{funcion_respuesta_coefs}
\end{equation}
La expresión~(\ref{funcion_respuesta_expansion}) es lo que denominaremos \emph{expansión adiabática de la función respuesta.} Debemos remarcar que no se trata de una expansión a tiempos cortos en torno a~$\tau$, ni una expansión a altas energías, aunque con esta última guarde alguna relación (como veremos a continuación).

En particular, los primeros cuatro coeficientes de la expansión adiabática se pueden escribir (con la notación~$\hat{\omega} := \omega/ |g(\tau)|$) como
\begin{alignat}{2}
\mathcal{R}_0 \left(\omega,g(\tau)\right) = & |g(\tau)| && \ \frac{\hat{\omega}}{2 \pi} \; \frac{1}{\rme^{2 \pi \hat{\omega}} - 1}, \nonumber \\
\mathcal{R}_1 \left(\omega,g(\tau),g'(\tau)\right) = & \ g(\tau) && \left[ \frac{g'(\tau)}{g(\tau)^2} \; f_{1,1} (\hat{\omega}) \right], \nonumber \\
\mathcal{R}_2 \left(\omega,g(\tau),g'(\tau),g''(\tau)\right) = & |g(\tau)| && \left[ \frac{g''(\tau)}{g(\tau)^3} \; f_{2,1} (\hat{\omega}) + \frac{g'(\tau)^2}{g(\tau)^4} \; f_{2,2} (\hat{\omega}) \right], \nonumber \\
\mathcal{R}_3 \left(\omega,g(\tau),g'(\tau),g''(\tau),g'''(\tau)\right) = & \ g(\tau) && \left[ \frac{g'''(\tau)}{g(\tau)^4}\; f_{3,1} (\hat{\omega}) + \frac{g'(\tau) g''(\tau)}{g(\tau)^5} \; f_{3,2} (\hat{\omega}) \right. \nonumber \\
& && \left. + \frac{g'(\tau)^3}{g(\tau)^6} \; f_{3,3} (\hat{\omega}) \right],
\label{respuesta_terminos_serie}
\end{alignat}
donde~$f_{i,j} (\hat{\omega})$ son funciones adimensionales de la variable adimensional~$\hat{\omega}$.\footnote{El número de funciones~$f_{i,n} (\hat{\omega})$ que aparecen en el término~$n$ de la serie está dado por la \emph{función de partición}~$p(n)$, es decir, el número de formas posibles de sumar~$n$ mediante suma de números enteros positivos.} El primer término es simplemente el espectro térmico con temperatura proporcional a la aceleración~$T = |g(\tau)|/(2 \pi k_{\rm B})$. Es decir, como ya adelantamos, recobramos el espectro térmico característico del efecto Unruh como aproximación de orden cero en la expansión adiabática. Debe notarse que~$\mathcal{R}_0$ solo depende de la aceleración, en tanto que~$\mathcal{R}_1$ depende también de la primera derivada de esta, $\mathcal{R}_2$ también de la segunda derivada, etc. Más aún, $\mathcal{R}_0$ únicamente depende del \emph{valor absoluto} de la aceleración. Esto es esperable, ya que el signo de la aceleración~$g(\tau)$ \emph{por sí mismo} no tiene ningún significado físico en el espacio-tiempo de Minkowski. Sin embargo, el término~$\mathcal{R}_1$ depende del signo del cociente $g'(\tau)/g(\tau)$. Este signo sí tiene un significado físico: indica aceleración creciente o decreciente. Este es el motivo por el que en el término~$\mathcal{R}_0$ aparece multiplicando~$|g(\tau)|$, mientras que en el término~$\mathcal{R}_1$ aparece~$g(\tau)$. Esta dependencia con el signo se alterna entre los términos pares y los impares. Aparte de este factor, las cantidades entre corchetes en~(\ref{respuesta_terminos_serie}) son sumas de funciones adimensionales multiplicadas por potencias de las cantidades adimensionales $\{ g^{(l)} (\tau) / g(\tau)^{l+1} \}$.

Las expresiones explícitas para las tres primeras funciones~$f_{i,j} (\hat{\omega})$ son
\begin{align}
f_{1,1} (\hat{\omega}) = & -\frac{1}{4 \pi^2} \left[1 + 2 \hat{\omega} F (\hat{\omega}) + \hat{\omega}^2 G(\hat{\omega}) \right], \nonumber \\
f_{2,1} (\hat{\omega}) = & -\frac{\hat{\omega}}{24 \sinh^4 (\pi \hat{\omega})} \left\{ \hat{\omega} \left[ \left(3+4\pi^2\right) + \left(-3+2\pi^2\right) \cosh(2 \pi \hat{\omega}) \right] \right. \nonumber \\
& \phantom{-\frac{\hat{\omega}}{24 \sinh^4 (\pi \hat{\omega})}} \ \ \left. - 3 \pi \sinh (2 \pi \hat{\omega}) \right\}, \nonumber \\
f_{2,2} (\hat{\omega}) = & \ \frac{1}{12 \sinh^2(\pi \hat{\omega})} \left\{ \pi \hat{\omega} \left[\left(-8+\hat{\omega}^2\left(1-3\pi^2\right)\right)\coth(\pi \hat{\omega}) \right. \right. \nonumber \\
& \phantom{\frac{1}{12 \sinh^2(\pi \hat{\omega})}} \ \ \left. \left. - 9 \pi \hat{\omega} \left(-2 + \pi \hat{\omega} \coth(\pi \hat{\omega}) \right) \sinh^{-2}(\pi \hat{\omega}) \right] \right. \nonumber \\
& \phantom{\frac{1}{12 \sinh^2(\pi \hat{\omega})}} \ \ \left. 3 \hat{\omega}^2\left(-3+4 \pi^2\right) - 1 \right\},
\label{f_ij}
\end{align}
donde hemos definido las funciones
\begin{equation}
F(\hat{\omega}) := \int_{-\infty}^0 \rmd u \frac{u \sin (\hat{\omega} u)}{1-\rme^{-u}}, \quad G(\hat{\omega}) := \int_{-\infty}^0 \rmd u \frac{u^2 \cos (\hat{\omega} u)}{1-\rme^{-u}}.
\label{funciones_F_G}
\end{equation}
La evaluación explícita de estas integrales da lugar a fórmulas en términos de derivadas de la función \emph{digamma}~$\psi(x) := \Gamma'(x) / \Gamma (x)$:
\begin{align}
F(\hat{\omega}) = & \ \frac{\rmi}{2} \left[ \psi'(1-\rmi \hat{\omega}) - \psi'(1+\rmi \hat{\omega}) \right],  \nonumber\\\
G(\hat{\omega}) = & \ \frac{1}{2} \left[ \psi''(1-\rmi \hat{\omega}) + \psi''(1+\rmi \hat{\omega}) \right].
\label{digamma}
\end{align}

\begin{figure}[ht]
	\centering
    \includegraphics{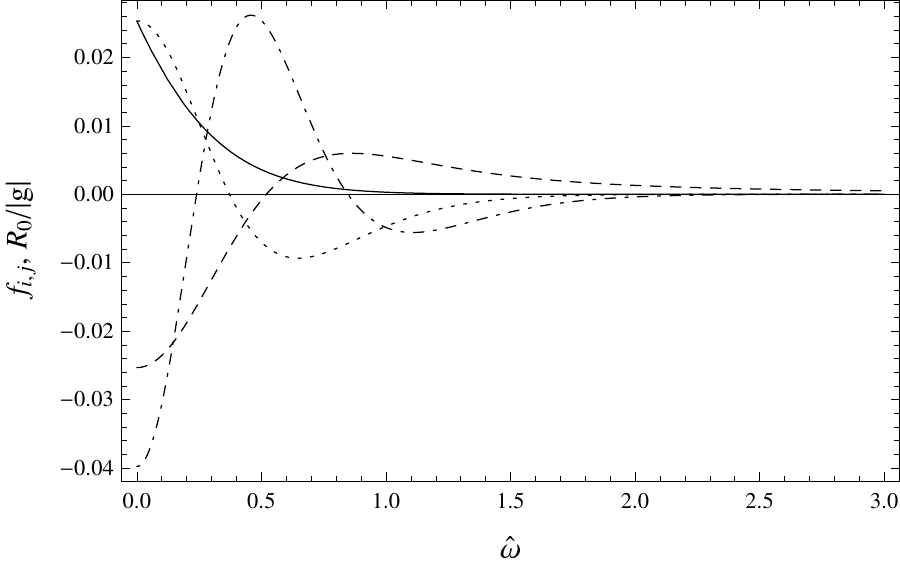}
  \caption{\footnotesize{Valores de~$\mathcal{R}_0 (\omega, g(\tau))/ |g(\tau)|$ (el espectro térmico normalizado, línea continua), $f_{1,1} (\hat{\omega})$ (línea discontinua), $f_{2,1} (\hat{\omega})$ (línea punteada) y $f_{2,2} (\hat{\omega})$ (línea punto-raya), como funciones de~$\hat{\omega}$.}}
  \label{fig_espectros}
\end{figure}

En la figura~\ref{fig_espectros}, se muestran los valores de las tres primeras funciones~$f_{i,j} (\hat{\omega})$, junto al espectro térmico normalizado~$\mathcal{R}_0 (\omega, g(\tau))/ |g(\tau)|$. Debe notarse que todos los términos que contribuyen a la serie adiabática tienen una forma bien definida como funciones de la cantidad~$\hat{\omega}$. Cambiar las derivadas de~$g(\tau)$ únicamente aumenta o disminuye la contribución de cada término al resultado final, pero no su forma. Es decir, podemos considerar la función respuesta total como una superposición de diferentes espectros. El primero es el espectro térmico. Los siguientes tienen distintas formas, y su contribución es proporcional a diferentes potencias de los cocientes $\{ g^{(l)} (\tau) / g(\tau)^{l+1} \}$. Esto puede parecer trivial, ya que lo que hemos hecho desde el principio no es más que una expansión en potencias de las derivadas de~$g(\tau)$. Sin embargo, debe notarse que cada coeficiente~$f_{i,j} (\hat{\omega})$ es un espectro \emph{bien definido,} sin polos y que tiende a anularse en el límite~$\hat{\omega} \to \infty$ (sin catástrofe ultravioleta).

Finalmente, es necesario decir que, en general, no se puede garantizar la convergencia matemática de la serie adiabática~(\ref{funcion_respuesta_expansion}), ya que hemos integrado formalmente la expansión de la función de Wightman~(\ref{wightman_expansion_asintotica}) sin ningún conocimiento acerca de su convergencia. Lo único que puede afirmarse es que se trata de una serie asintótica en las cantidades adimensionales $\{ g^{(l)} (\tau) / g(\tau)^{l+1} \}$.

\section{Comportamiento en altas energías}

Todos los términos~$\mathcal{R}_n (\omega,g(\tau),g'(\tau),\ldots,g^{(n)}(\tau))$, y por tanto todas las funciones~$f_{i,j} (\hat{\omega})$, tienden a cero para~$\omega \to \infty$, como puede verse fácilmente de su expresión integral~(\ref{funcion_respuesta_coefs}). A continuación, estudiaremos en detalle el comportamiento de esta caída a altas energías. Las integrales en~(\ref{funcion_respuesta_coefs}), que definen la expansión adiabática~(\ref{funcion_respuesta_expansion}), tienen el siguiente comportamiento asintótico en serie de potencias de~$1/\omega$, para~$\omega \to \infty$~\cite{Satz:2006kb,r2001asymptotic}:
\begin{multline}
\mathcal{R}_n \left(\omega, g(\tau),g'(\tau),\ldots,g^{(n)}(\tau) \right) \\
\sim 2 \sum_{m=1}^\infty \left(-\frac{1}{\omega^2}\right)^m W_n^{(2m-1)}\left(0, g(\tau),g'(\tau),\ldots,g^{(n)}(\tau) \right),
\label{altas_energias}
\end{multline}
donde, utilizando las expresiones~(\ref{wightman_coefs}) para las funciones~$W_n$, los coeficientes~$W_n^{(l)}$ de esta expansión pueden escribirse como
\begin{equation}
W_n^{(l)}\left(0, g(\tau),g'(\tau),\ldots,g^{(n)}(\tau) \right) = \frac{1}{n!} \left. \frac{\partial^l}{\partial s^l} \left. \frac{\partial^n}{\partial \alpha^n} \bar{W} \left(\tau+s,\tau,\alpha; g\right] \right|_{\alpha \to 0^+} \right|_{s=0},
\label{coeficientes_altas_energias}
\end{equation}
donde~$l = 2m -1$ es siempre impar. A continuación, utilizando la expresión para la función auxiliar~$\bar{W}(\tau+s,\tau,\alpha;g]$ en~(\ref{wightman_aux}), es fácil ver que esta función tiene la siguiente propiedad de escala:
\begin{equation}
Q^2 \bar{W}(\tau+ Q s,\tau,\alpha;g] = \bar{W}(\tau+ s,\tau,Q \alpha;Q g], \quad Q \in \mathds{R}.
\label{wightman_escala}
\end{equation}
Utilizando esta propiedad para el valor concreto~$Q=-1$, y teniendo en cuenta la invariancia del funcional~$\bar{W}$ ante el cambio $g \to -g$, encontramos el siguiente resultado:
\begin{equation}
W_n^{(l)}\left(0; g(\tau),g'(\tau),\ldots,g^{(n)}(\tau) \right) = 0, \quad {\rm para} \quad n + l {\rm \ impar}.
\label{derivadas_impares}
\end{equation}
Es decir, cualquier ``combinación impar'' de derivadas respecto a~$s$ y a~$\alpha$ se anula al evaluarla en~$s=0$ y~$\alpha \to 0^+$. Teniendo en cuenta que en la expansión~(\ref{altas_energias}) todas las derivadas respecto a~$s$ son impares ($l$~impar), se concluye que todos los términos en la expansión se cancelan para~$n$ par. Es decir, los términos pares en la expansión adiabática~(\ref{funcion_respuesta_expansion}) caen a cero más rápido que cualquier potencia inversa de~$\omega$ en el límite de altas energías. Por otra parte, el comportamiento en este límite de los términos impares en~(\ref{funcion_respuesta_expansion}) solo contendrá potencias inversas pares de~$\omega$.

También gracias a la propiedad de escala~(\ref{wightman_escala}), podemos mostrar que los coeficientes~$W_n^{(l)}$ se anulan para~$l<n$. Con el fin de obtener este resultado, se debe notar primeramente que el orden en el que se toman las derivadas con respecto a~$s$ y~$\alpha$, y por tanto las evaluaciones en~$s=0$ y~$\alpha \to 0^+$, en la ecuación~(\ref{coeficientes_altas_energias}), pueden intercambiarse entre sí, ya que estamos considerando funciones analíticas reales~$g(\tau)$, y por tanto~$\bar{W}(\tau+s,\tau,\alpha;g]$ también es analítica. A continuación, comprobamos que la cantidad~$\partial^l \bar{W} / \partial s^l |_{s=0}$ es un polinomio en~$\alpha$ de grado~$l$ o menor. Consideremos una constante arbitraria no nula~$T$ con dimensiones de tiempo. En efecto, tenemos que
\begin{align}
\left. \frac{\partial^l \bar{W}(\tau+s,\tau,\alpha;g]}{\partial s^l} \right|_{s=0} & = \left. \left. \frac{\partial^l \bar{W}(\tau+ (q / T) s,\tau,\alpha;g]}{\partial s^l} \right|_{s=0} \right|_{q = T} \nonumber \\
& = \left. \left. \frac{\partial^l \bar{W}(\tau+ (q / T) s,\tau,\alpha;g]}{\partial q^l} \right|_{q=0} \right|_{s = T} \nonumber \\
& = \left. \frac{\partial^l}{\partial q^l} \left[ \left( \frac{T}{q} \right)^2 \bar{W}(\tau+s,\tau,(q / T)\alpha;(q / T)g] \right]_{q=0} \right|_{s = T}.
\label{demostracion_polinomio}
\end{align}
La cantidad final es claramente un polinomio en~$\alpha$ de grado~$l$ o inferior. Teniendo en cuenta que, según la expresión de~$W_n^{(l)}$ en~(\ref{coeficientes_altas_energias}), cada coeficiente es proporcional a la derivada $n$-ésima de~(\ref{demostracion_polinomio}) respecto de~$\alpha$, necesariamente debe anularse para~$l < n$. Esto significa que la expansión del término~$n$ de la expansión adiabática, $\mathcal{R}_n \left(\omega, g(\tau),g'(\tau),\ldots,g^{(n)}(\tau) \right)$, en potencias inversas de~$\omega$ comenzará en el término~$\omega^{-(n+1)}$. Por tanto, la expansión adiabática hasta orden~$n$ nos da automáticamente la expansión a altas energías hasta orden~$n+1$. Esta relación es natural, puesto que el comportamiento en energías más altas dependerá de forma más crucial solo de lo que suceda en intervalos más cortos de tiempo. Y un intervalo corto de tiempo ``representa'' mejor lo que sucede a intervalos largos de tiempo cuanto más lenta sea la evolución de la aceleración, es decir, cuanto mejor se ajuste la expansión adiabática. Esto conecta conceptualmente la expansión adiabática con la expansión a altas energías.

En particular, hasta cuarto orden, el comportamiento a altas energías es
\begin{align}
& \mathcal{R}_1 \left(\omega, g(\tau), g'(\tau) \right) \sim \frac{g(\tau) g'(\tau)}{24 \pi^2}\omega^{-2} + \frac{g(\tau)^3 g'(\tau)}{40 \pi^2}\omega^{-4} + O(\omega)^{-6}, \nonumber \\
& \mathcal{R}_3 \left(\omega, g(\tau), g'(\tau), g''(\tau), g'''(\tau) \right) \nonumber \\
& \hspace{3.2cm} \sim - \frac{1}{12 \pi^2}\left(\frac{g(\tau)g'''(\tau)}{5} + \frac{g'(\tau)g''(\tau)}{2} \right)\omega^{-4} + O(\omega)^{-6}.
\label{altas_energias_cuarto}
\end{align}
Sumando, obtenemos
\begin{align}
\mathcal{R} \left(\omega, \tau \right) \sim & \ \frac{g(\tau) g'(\tau)}{24 \pi^2}\omega^{-2} \nonumber \\
& + \frac{1}{4 \pi^2}\left(\frac{g(\tau)^3 g'(\tau)}{10} - \frac{g(\tau)g'''(\tau)}{15} - \frac{g'(\tau)g''(\tau)}{6} \right)\omega^{-4} + O(\omega)^{-6}.
\label{altas_energias_cuarto_total}
\end{align}
El primer término en la expansión coincide con el que se encuentra en las referencias~\cite{Louko:2006zv,Satz:2006kb}, mientras que en la referencia~\cite{Obadia:2007qf} se encuentran términos hasta sexto orden en la expansión, los cuales igualmente coinciden con los aquí encontrados.

Es conveniente hacer un último comentario respecto a la expansión en altas energías. En la referencia~\cite{Satz:2006kb}, se demuestra que, para un detector acoplado mediante una función de encendido suave y de soporte compacto (como las que utilizamos en la sección~\ref{regularizacion}) la función respuesta decae más rápido que cualquier potencia de~$\omega$ en el límite~$\omega \to \infty$. Por tanto, puede parecer que la expansión a altas energías que hemos encontrado aquí es un artefacto que se debe a un tiempo de encendido infinito, y a un apagado brusco. Esto es cierto si se considera la expansión en potencias como la tendencia de la función respuesta para energías arbitrariamente altas. Como se menciona en~\cite{Satz:2006kb}, el comportamiento como potencia inversa siempre será suplantado por un régimen de caída mucho más rápido en este límite, cuando el detector se enciende y apaga ``adecuadamente'' (véase también la referencia~\cite{Sriramkumar:1994pb}). No obstante, siempre se puede emplear la expansión en altas energías como herramienta para calcular la función respuesta de una frecuencia dada~$\omega$ con precisión arbitraria. Si se recuerdan las consideraciones hechas en la sección~\ref{regularizacion}, siempre que se tenga la seguridad de que los efectos debido al proceso de encendido y apagado son despreciables respecto a la precisión requerida, el uso de la serie de potencias está perfectamente justificado y, como veremos a continuación en un ejemplo, da una aproximación válida a la función respuesta del detector.

\section{Resultados numéricos}

Consideraremos a continuación un caso particular de aceleración propia en función del tiempo propio~$g(\tau)$, dada por
\begin{equation}
g(\tau) = \frac{1}{2} \left[ 1 + \tanh \left( \frac{\tau}{\Delta \tau} \right) \right].
\label{g_ejemplo}
\end{equation}
El detector comienza sin aceleración en el pasado asintótico, a continuación comienza a incrementar su aceleración suavemente y, tras un periodo transitorio de duración~$\Delta \tau$, aproximadamente, en tiempo propio, tiende asintóticamente a alcanzar una aceleración final que normalizamos al valor~$1$ (lo cual termina de fijar un sistema completo de unidades naturales para el problema). Debe notarse que el parámetro~$\Delta \tau$ está muy relacionado con el (inverso del) parámetro~$\alpha$ utilizado para obtener la expansión adiabática.

Para esta dependencia particular de la aceleración, puede encontrarse analíticamente la función de Wightman regularizada~(\ref{wightman_reg}) con la ayuda de software de manipulación simbólica (en este caso, se utilizó \emph{Mathematica}). Sin embargo, la función respuesta debe ser calculada numéricamente, lo cual se consigue con una precisión muy alta (para los valores de~$\Delta \tau$ considerados), por lo que podemos comparar los resultados obtenidos directamente de la integración numérica de la función respuesta~(\ref{funcion_respuesta_reg}) con los resultados obtenidos mediante la expansión adiabática. En este sentido, describiremos dos escenarios: aceleración de evolución lenta y de evolución rápida.

\subsection{Aceleración de evolución lenta}

\begin{figure}[htbp!]
\begin{minipage}[b]{0.47\linewidth}
\centering
\includegraphics[width=6.3cm]{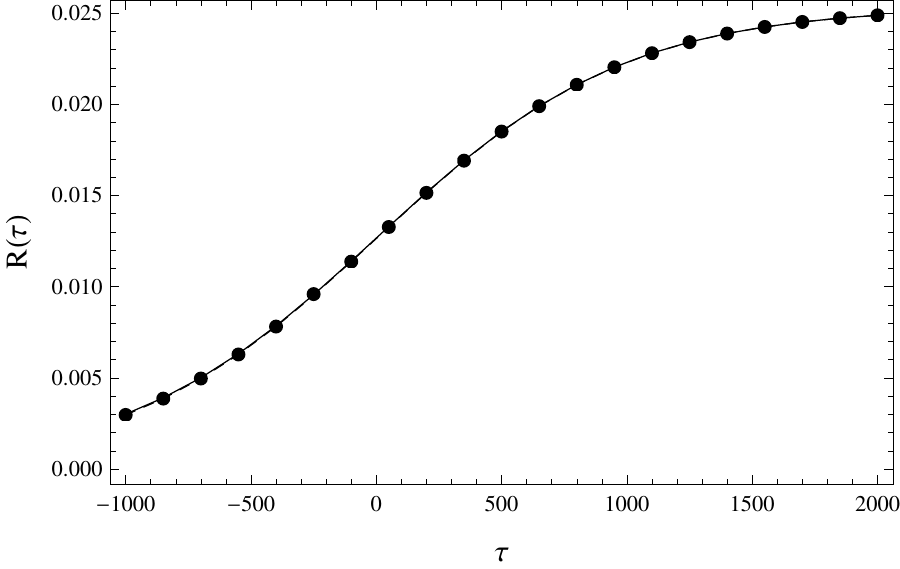}
\caption{\footnotesize{Función respuesta como función de $\tau$ para $\omega \to 0$ y $\Delta \tau = 1000$. Los puntos son la solución numérica, la línea continua es la aproximación de espectro térmico, y la línea discontinua (totalmente cubierta por la continua) la aproximación adiabática hasta tercer orden.}}
\label{dt-1000-En-0}
\end{minipage}
\hspace{0.04\linewidth}
\begin{minipage}[b]{0.47\linewidth}
\centering
\includegraphics[width=6.5cm]{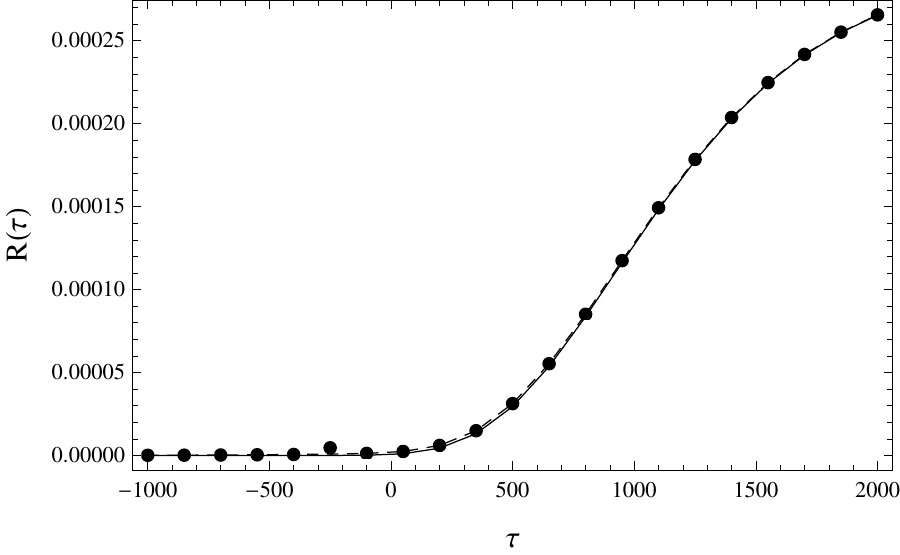}
\caption{\footnotesize{Función respuesta como función de $\tau$ para $\omega = 1$ y $\Delta \tau = 1000$. Los puntos son la solución numérica, la línea continua es la aproximación de espectro térmico, y la línea discontinua (casi totalmente cubierta por la continua) la aproximación adiabática hasta tercer orden.}}
\label{dt-1000-En-1}
\end{minipage}
\end{figure}

\begin{figure}[htbp!]
\begin{minipage}[b]{0.47\linewidth}
\centering
\includegraphics[width=6.4cm]{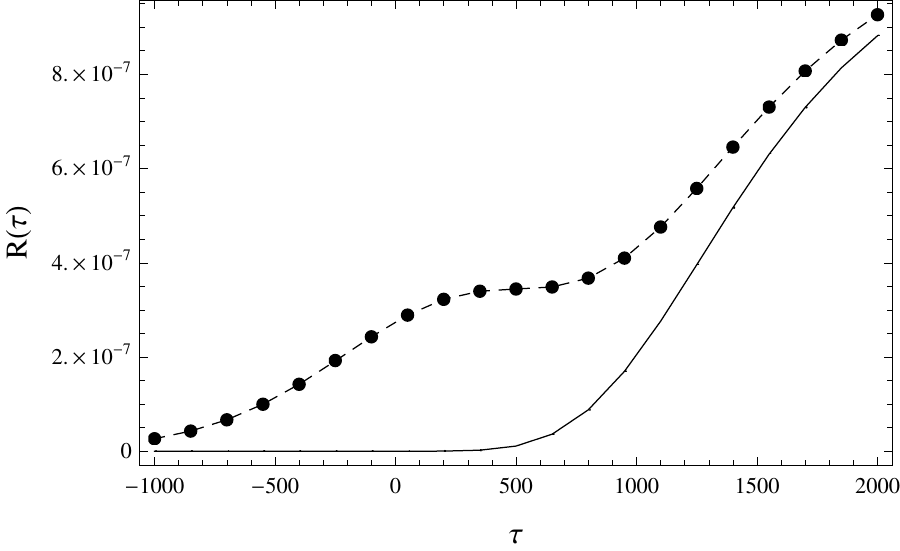}
\caption{\footnotesize{Función respuesta como función de $\tau$ para $\omega = 2$ y $\Delta \tau = 1000$. Los puntos son la solución numérica, la línea continua es la aproximación de espectro térmico, y la línea discontinua la aproximación adiabática hasta tercer orden.}}
\label{dt-1000-En-2}
\end{minipage}
\hspace{0.04\linewidth}
\begin{minipage}[b]{0.47\linewidth}
\centering
\includegraphics[width=6.5cm]{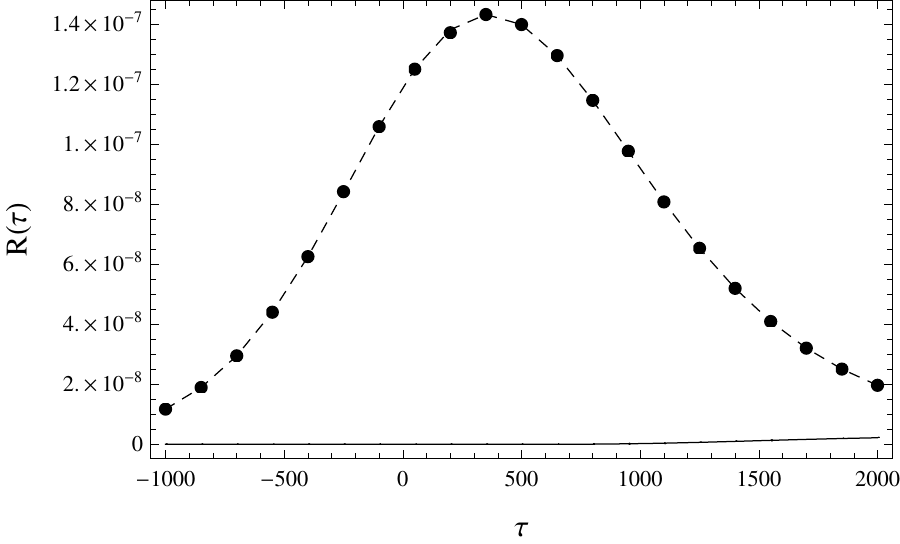}
\caption{\footnotesize{Función respuesta como función de $\tau$ para $\omega = 3$ y $\Delta \tau = 1000$. Los puntos son la solución numérica, la línea continua es la aproximación de espectro térmico, y la línea discontinua la aproximación adiabática hasta tercer orden.}}
\label{dt-1000-En-3}
\end{minipage}
\end{figure}

Para este caso, consideraremos el valor numérico~$\Delta \tau = 1000$, de manera que la aceleración varía de forma muy lenta (en comparación con la escala de tiempo fijada por su valor final, igual a~$1$). En las figuras~\ref{dt-1000-En-0}--\ref{dt-1000-En-3}, mostramos la dependencia de la función respuesta~$\mathcal{R} (\omega, \tau)$ con el tiempo propio para distintos valores de~$\omega$. Dibujamos tres cantidades: el resultado numérico, la aproximación de orden cero en la expansión adiabática (es decir, lo que se obtendría si en todo instante se detectase únicamente un baño térmico con temperatura proporcional a la aceleración), y la aproximación de tercer orden en la expansión adiabática.
%
\noindent
\noindent

En primer lugar, vemos que, para todas las energías consideradas, el ajuste de la expansión adiabática hasta el tercer orden es perfecto. Por otra parte, para energías bajas (al menos hasta la energía característica del espectro térmico final~$\omega \approx 1$), la coincidencia con el espectro térmico es prácticamente perfecta, en tanto que para altas energías ambos resultados comienzan a separarse, hasta resultar completamente diferentes. Esto puede resultar sorprendente, teniendo en cuenta lo dicho en la sección anterior: La expansión adiabática hasta orden~$n$ coincide con la expansión a altas energías hasta orden~$n+1$, por lo que la aproximación de espectro térmico (de orden cero) debería resultar \emph{mejor} a altas energías. En particular, la aproximación adiabática a orden cero debería ser \emph{exacta} para energías arbitrariamente altas. Tal afirmación es cierta, pero sencillamente porque tanto la aproximación como el resultado exacto se anulan en ese límite. Pero es cierta para la diferencia \emph{absoluta} entre ambas cantidades. Y lo que nos muestran las figuras~\ref{dt-1000-En-0}--\ref{dt-1000-En-3} es la diferencia \emph{relativa}, en comparación con la magnitud del resultado exacto (ténganse en cuenta los valores numéricos en el eje vertical). Y, si bien la diferencia absoluta entre el resultado numérico y la aproximación de espectro térmico es \emph{mayor} para energías arbitrariamente bajas (figura~\ref{dt-1000-En-0}) que para~$\omega = 3$ (figura~\ref{dt-1000-En-3}), en un sentido relativo la diferencia es despreciable para bajas energías, en tanto que para altas energías es mucho más importante. Más aún, para energías del orden de~$\omega \approx 3$, es la contribución del espectro térmico la que es despreciable respecto a contribuciones de orden superior. En particular, la forma ``de campana'' que aparece en la figura correspondiente, se debe, prácticamente en su totalidad, al primer término en la expansión a altas energías~(\ref{altas_energias_cuarto_total}), proporcional a~$g(\tau)g'(\tau)$.

Finalmente, debe notarse que en los límites~$\tau \to \pm \infty$ las tres curvas convergen en todas las figuras. Esto es esperable, ya que en esas regiones la aceleración se acerca a valores constantes, por lo que la aproximación de espectro térmico tiende a ser exacta.

\subsection{Aceleración de evolución rápida}

\begin{figure}[htbp!]
\begin{minipage}[b]{0.47\linewidth}
\centering
\includegraphics[width=6.4cm]{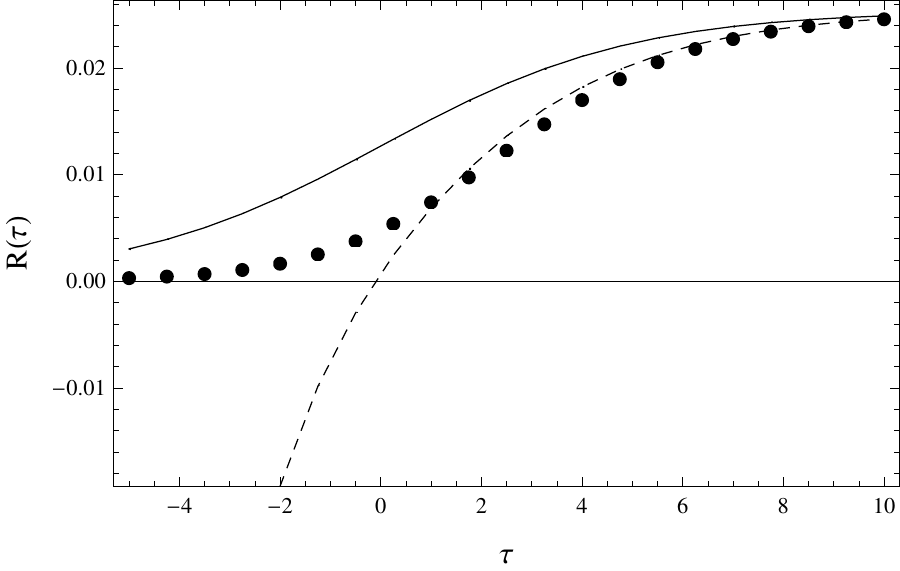}
\caption{\footnotesize{Función respuesta como función de $\tau$ para $\omega \to 0$ y $\Delta \tau = 5$. Los puntos son la solución numérica, la línea continua es la aproximación de espectro térmico, y la línea discontinua la aproximación adiabática hasta tercer orden.}}
\label{dt-5-En-0}
\end{minipage}
\hspace{0.04\linewidth}
\begin{minipage}[b]{0.47\linewidth}
\centering
\includegraphics[width=6.5cm]{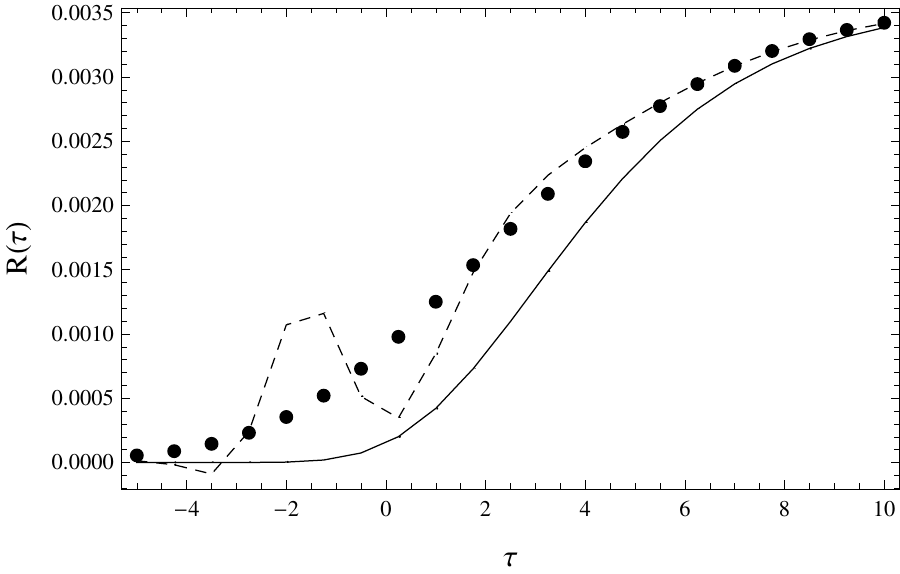}
\caption{\footnotesize{Función respuesta como función de $\tau$ para $\omega = 0.5$ y $\Delta \tau = 5$. Los puntos son la solución numérica, la línea continua es la aproximación de espectro térmico, y la línea discontinua la aproximación adiabática hasta tercer orden.}}
\label{dt-5-En-05}
\end{minipage}
\end{figure}

\begin{figure}[htbp!]
\begin{minipage}[b]{0.47\linewidth}
\centering
\includegraphics[width=6.4cm]{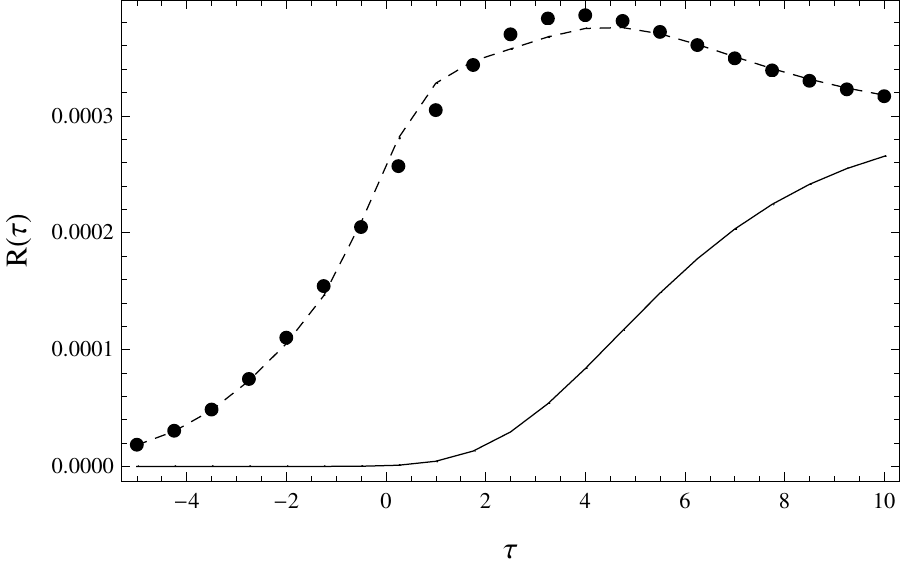}
\caption{\footnotesize{Función respuesta como función de $\tau$ para $\omega = 1$ y $\Delta \tau = 5$. Los puntos son la solución numérica, la línea continua es la aproximación de espectro térmico, y la línea discontinua la aproximación adiabática hasta tercer orden.}}
\label{dt-5-En-1}
\end{minipage}
\hspace{0.04\linewidth}
\begin{minipage}[b]{0.47\linewidth}
\centering
\includegraphics[width=6.5cm]{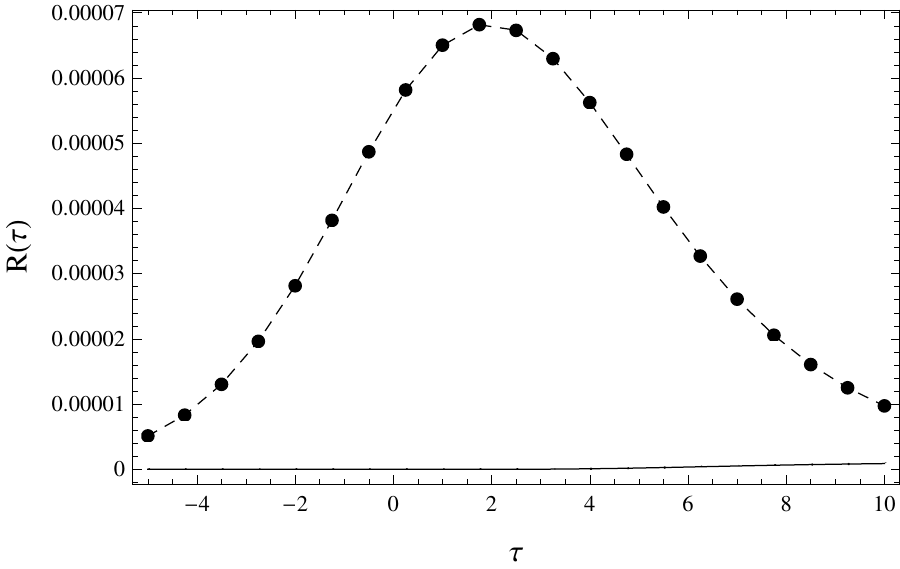}
\caption{\footnotesize{Función respuesta como función de $\tau$ para $\omega = 2$ y $\Delta \tau = 5$. Los puntos son la solución numérica, la línea continua es la aproximación de espectro térmico, y la línea discontinua la aproximación adiabática hasta tercer orden.}}
\label{dt-5-En-2}
\end{minipage}
\end{figure}

A continuación, consideraremos el valor numérico~$\Delta \tau = 5$ (figuras~\ref{dt-5-En-0}--\ref{dt-5-En-2}). En este caso, nos encontramos mucho más lejos del régimen adiabático. En contraste con el caso anterior, incluso para energías arbitrariamente bajas, se observan claramente las desviaciones respecto a la respuesta térmica. La aproximación adiabática hasta tercer orden falla para bajas energías, pero sigue siendo prácticamente exacta para energías altas. Aquí, las diferencias con la aproximación a tercer orden son mayores para bajas energías también en un sentido \emph{relativo.} Esto concuerda, una vez más, con el hecho de que la expansión adiabática nos da, ``por el mismo precio'', la expansión a altas energías. Es conveniente también notar cómo en la figura~\ref{dt-5-En-0} la aproximación a tercer orden resulta una aproximación peor al resultado numérico que la aproximación a orden cero. Esto es algo esperable al alejarnos del régimen adiabático, puesto que estamos tratando con series asintóticas, sin ninguna garantía de convergencia. En tal caso, debería aplicarse la regla del truncamiento óptimo para obtener el resultado más preciso posible al sumar la serie. Aquí, sin embargo, hemos representado las mismas cantidades que en el caso anterior (el caso de evolución lenta), puesto que no estamos tan interesados en conocer el espectro concreto para esta historia de aceleración (que ya conocemos numéricamente) como en conocer el comportamiento de la propia expansión adiabática.

\noindent
\noindent

\newpage
\thispagestyle{empty}
\hbox{}

\makeatletter
\def\cleardoublepage{\clearpage\if@twoside \ifodd\c@page\else
    \hbox{}
    \thispagestyle{empty}
    \newpage
    \if@twocolumn\hbox{}\newpage\fi\fi\fi}
\makeatother \clearpage{\pagestyle{empty}\cleardoublepage}

\chapter{Percepción de la radiación de Hawking} 
\label{hawking}

En este capítulo estudiaremos en profundidad cómo perciben la radiación de Hawking diferentes observadores en el exterior de un agujero negro de Schwarzschild, según las distintas trayectorias seguidas por los mismos. Como ya hicimos en la sección~\ref{sec_cuantica_agujeros_negros}, nos centraremos en el sector radial de la geometría, de forma que trabajaremos en $1+1$~dimensiones. Utilizaremos para este estudio la herramienta introducida en la sección~\ref{sec_kappa}: la función de temperatura efectiva~$\kappa(u)$ definida en~(\ref{funcion_temperatura}), aplicando tal herramienta al estudio de la percepción de la radiación por distintos observadores, y analizando además su rango de validez, mediante la comprobación del cumplimiento de la que denominamos condición adiabática. Los distintos tipos de observadores estudiados serán: observadores estáticos en un radio, observadores en caída libre desde el infinito, y observadores en caída libre desde un radio finito.

Como novedad, introduciremos un estado de vacío que no es ninguno de los estados estacionarios habituales (Boulware, Unruh o Hartle-Hawking), sino un estado no estacionario que denominaremos \emph{vacío de colapso.} Este vacío interpola de forma suave entre el vacío de Boulware en el pasado asintótico y el vacío de Unruh en el futuro asintótico. Con este vacío, se busca imitar el proceso de encendido de la radiación de Hawking en un proceso de colapso real. Entre otros resultados físicos, encontramos uno que ya adelantamos anteriormente: los observadores en caída libre al agujero negro, en general, no detectan vacío al cruzar el horizonte de sucesos. Muy al contrario, detectan una radiación que, si bien no es térmica, es en cualquier caso no nula, siendo su temperatura \emph{efectiva} cuatro veces mayor que la temperatura asintótica de la radiación de Hawking (\ref{temperatura_hawking}).

\section{Elección del estado de vacío}\label{fijado_vacio}

Estamos interesados en analizar cómo es percibida la radiación de Hawking por distintos observadores en el espacio-tiempo de Schwarzschild. Como sabemos, para tener radiación de Hawking necesitamos en primer lugar seleccionar un estado de vacío adecuado. En lugar de hacer la elección obvia del vacío de Unruh, elegiremos como estado de vacío un estado no estacionario que interpola entre el vacío de Boulware (sin radiación) en el pasado asintótico, y el vacío de Unruh en el futuro asintótico. Este estado dinámico (o no estacionario) en la métrica de Schwarzschild permite analizar cualitativamente lo que sucedería en un proceso de colapso, en el cual se genera un agujero negro a partir de un espacio-tiempo que inicialmente es (casi) Minkowski, sin tener que elegir un modelo clásico de colapso particular. Al contrario de lo que sucede en el vacío de Unruh, en esta situación la radiación de Hawking no estará presente inicialmente, sino que aparecerá tras un determinado transitorio de encendido (que correspondería al momento de formación del agujero negro).

Para imponer este estado de vacío al campo cuántico, consideraremos un observador que sigue una geodésica radial de género tiempo correspondiente a una caída libre al agujero negro desde el infinito espacial. Calcularemos entonces cómo etiqueta este observador de forma natural los distintos rayos (trayectorias nulas) salientes que se encuentra en su camino al horizonte de sucesos. Es decir, cuál es la coordenada nula, que denotaremos~$U$, asociada a los modos normales naturales para este observador. Por conveniencia, para definir unívocamente~$U$ la expresaremos como función de la coordenada nula de Eddington-Finkelstein~$\bar{u} := t - r^*$.\footnote{En el capítulo~\ref{fisica} se comprenderá que esta es una elección especialmente acertada.} Los modos~$U$, naturales para este observador, serán entonces los que definan el vacío en el que trabajaremos, que denotaremos~$\ket{0_{\rm col}}$.

Calculemos primeramente las trayectorias geodésicas radiales de género tiempo resolviendo la ecuación de las geodésicas
\begin{equation}
\deriv[2]{r}{\tau} = - \Gamma^r_{r r} \left( \deriv{r}{\tau} \right)^2 - 
\Gamma^r_{t t} \left( \deriv{t}{\tau} \right)^2.
\label{geodesica_radial}
\end{equation}
Para ello, tendremos también en cuenta la relación
\begin{equation}
\left(\deriv{t}{\tau}\right)^2 = \left( 1-\frac{2M}{r} \right)^{-1} \left[1 + \left( 1-\frac{2M}{r} \right)^{-1} \left(\deriv{r}{\tau}\right)^2 \right],
\label{metric.deriv}
\end{equation}
que se sigue del elemento de línea en la métrica reducida a $1+1$~dimensiones~(\ref{schwarzschild_una_dimension}) (de forma equivalente, podríamos haber usado directamente un principio variacional sobre el intervalo espacio-temporal). Aquí, $\tau$~es el tiempo propio de la trayectoria, y~$\Gamma^r_{r r}$,~$\Gamma^r_{t t}$ son los únicos símbolos de Christoffel no nulos relevantes para el cálculo de geodésicas radiales:
\begin{equation}
\Gamma^r_{r r} = -\frac{M}{r^2} \left( 1-\frac{2M}{r} \right)^{-1}, \quad \Gamma^r_{t t} = \frac{M}{r^2} \left( 1-\frac{2M}{r} \right).
\label{christoffel}
\end{equation}
Reemplazando su valor en la ecuación de las geodésicas se tiene
\begin{equation}
\deriv[2]{r}{\tau} = -\frac{M}{r^2},
\label{Newton}
\end{equation}
ecuación que recuerda a la expresión de la fuerza gravitacional de Newton. Integrando un vez, obtenemos
\begin{align}
\deriv{r}{\tau} = & -\sqrt{\frac{2M}{r}}\left(1  -\frac{r}{\rc} \right)^{1/2},
\label{r.tau.first.integral}
\\
\deriv{t}{\tau} = & \left(1- \frac{2M}{r}\right)^{-1} \left(1  -\frac{2M}{\rc} \right)^{1/2},
\label{t.tau.first.integral}
\end{align}
donde utilizamos el radio~$r_0$ para el cual se anula la velocidad como constante de integración.

De momento, estamos interesados en describir una trayectoria en caída libre que comienza en el infinito espacial, lo cual significa que~$r_0 \to \infty$. Integrando entonces la ecuación radial~(\ref{r.tau.first.integral}) obtenemos
\begin{equation}
r = \left[ \frac{3 \sqrt{2M}}{2} (\tau_0-\tau) \right]^{2/3},
\label{r.of.tau}
\end{equation}
siendo~$\tau_0$ una constante de integración. Debe notarse que~$\tau$ recorre los valores desde~$-\infty$ hasta~$\tau_0$, momento en el que la trayectoria alcanza la singularidad en~$r=0$. Antes, en~$\tau = \tau_0 - 4M/3$, la trayectoria cruza el horizonte de sucesos. También podemos encontrar el comportamiento de la coordenada temporal~$t$ con respecto a~$\tau$. De hecho, dividiendo las ecuaciones~(\ref{r.tau.first.integral}) y~(\ref{t.tau.first.integral}) (con~$\rc \to \infty$), obtenemos
\begin{equation}
\deriv{r}{t} = - \sqrt{\frac{2M}{r}} \left( 1-\frac{2M}{r} \right),
\label{dr.of.t}
\end{equation}
la cual se puede integrar directamente, resultando
\begin{equation}
t= t_0 -4M \left[ \sqrt{\frac{r}{2M}} + \athird \left( \frac{r}{2M} \right)^{3/2} + \ahalf \log \left( \frac{\sqrt{r/(2M)}-1}{\sqrt{r/(2M)}+1} \right) \right],
\label{t.of.r}
\end{equation}
donde~$t_0$ es una constante de integración. Para encontrar la relación~$t(\tau)$, sustituimos la relación~(\ref{r.of.tau}) obtenida para~$r(\tau)$ en esta expresión:
\begin{equation}
t= t_0 - 4M \left\{ \left[\frac{3 (\tau_0-\tau)}{4 M} \right]^{1/3} + \frac{\tau_0-\tau}{4M} + \ahalf \log \left[ \frac{(3 (\tau_0-\tau) / 4M)^{1/3} - 1}{(3 (\tau_0-\tau) / 4M)^{1/3} + 1} \right] \right\}.
\label{t.of.tau}
\end{equation}

Teniendo el par~$(t,r)$ como función del tiempo propio~$\tau$ en~(\ref{r.of.tau}) y~(\ref{t.of.tau}), podemos finalmente escribir la coordenada nula de Eddington-Finkelstein~$\bar{u} = t - r^*$ también como función de~$\tau$:
\begin{align}
\ub = & \ t_0- 4M \left\{ \left[ \frac{3 (\tau_0-\tau)}{4 M} \right]^{1/3} + \ahalf \left[ \frac{3 (\tau_0-\tau)}{4 M} \right]^{2/3} + \frac{\tau_0-\tau}{4M} \right. \nonumber \\
& \left. + \log \left[ \left(\frac{3 (\tau_0-\tau)}{4M}\right)^{1/3}  - 1 \right] \right\}.
\label{ub.of.tau}
\end{align}

El observador que sigue esta trayectoria utiliza de forma natural su tiempo propio~$\tau$ para etiquetar los distintos rayos salientes que encuentra; es decir, puede hacer la asignación~$U(\bar{u}) = \tau(\bar{u})$ dada por la inversa de~(\ref{ub.of.tau}) a cada rayo saliente~$\bar{u} = {\rm const}$ con el que se cruza en el tiempo propio~$\tau$. Definiendo~$\UH := \tau_0-4M/3$ y escogiendo las constantes~$t_0 = \tau_0$ de forma que las coordenadas nulas~$U$ y~$\bar{u}$ coincidan en~$\bar{u} \to -\infty$, la relación final entre ambas coordenadas es
\begin{align}
\ub (U) = & \ \UH + \frac{4M}{3} - 4M \left\{ \left[ \frac{3}{4M}(\UH-U) +1 \right]^{1/3} \right. \nonumber \\ 
& \left. + \ahalf \left[ \frac{3}{4M}(\UH-U) +1 \right]^{2/3} + \athird \left[ \frac{3}{4M}(\UH-U) +1 \right] \right. \nonumber \\ 
& \left. + \log \left[ \left(\frac{3}{4M}(\UH-U) +1\right)^{1/3}  - 1 \right] \right\}.
\label{ub.of.U}
\end{align}
Debe notarse que, mientras el rango de validez de la coordenada~$\bar{u}$ es toda la recta real~$(-\infty,\infty)$, el de la nueva coordenada~$U$ es únicamente~$(-\infty,U_{\rm H})$, donde~$U_{\rm H}$ marca el instante en el que el observador en caída libre cruza el horizonte de sucesos. Como ya hemos dicho, asociado a esta coordenada nula tenemos el estado de vacío~$\ket{0_{\rm col}}$, el cual será el objeto de nuestro análisis mediante la función de temperatura efectiva.\footnote{En principio, se podría haber elegido otra función~$U(\bar{u})$ que interpolase entre un régimen lineal (vacío de Boulware) y uno exponencial (vacío de Unruh). Sin embargo, en general los resultados serían incluso cualitativamente muy diferentes en el transitorio, pudiendo presentar comportamientos muy alejados de lo esperable. Como veremos, la elección de una trayectoria de caída libre como forma de interpolación produce un transitorio suave y monótono.}

\section{Observadores estáticos}\label{observadores_estaticos}

Comenzaremos analizando cómo perciben el estado de vacío~$\ket{0_{\rm col}}$ observadores estáticos situados en un radio fijo~$r_{\rm s}$. Para ello, debemos encontrar la nueva coordenada nula~$u$ asociada a los modos naturales para estos observadores. El procedimiento para ello es análogo al empleado para definir la coordenada nula~$U$: resolver la trayectoria para los observadores (en este caso estáticos), escribir la coordenada~$\bar{u}$ en función del tiempo propio~$\tau$, e identificar~$u=\tau$. De esta forma, obtenemos~$\bar{u}(u)$, que junto a la relación conocida~$\ub (U)$ dada por~(\ref{ub.of.U}), nos permite construir (aunque sea de forma implícita) la relación~$U(u)$. Y esta es la relación que necesitamos para calcular la función de temperatura efectiva.

La trayectoria de un observador que mantiene constante su posición radial se describe mediante las ecuaciones
\begin{align}
r = & \ r_{\rm s} = {\rm const}, \\
t= & \left(1-\frac{2M}{\rst}\right)^{-1/2} (\tau-\tau_0),
\label{t.of.r.static}
\end{align}
de manera que
\begin{equation}
\ub = \left(1-\frac{2M}{\rst}\right)^{-1/2} (\tau-\tau_0) - \rs (\rst),
\label{ub.of.u.static.t0}
\end{equation}
donde~$r^* (r_{\rm s})$ es la coordenada tortuga [ecuación~(\ref{tortuga})] asociada al radio constante~$r_{\rm s}$. Como en la sección anterior, hacemos ahora la identificación~$u = \tau$. Salvo una constante aditiva irrelevante, obtenemos la relación~$\bar{u}(u)$ entre las dos coordenadas nulas
\begin{equation}
\ub (u) = \left(1-\frac{2M}{\rst}\right)^{-1/2} u.
\label{ub.of.u.static}
\end{equation}
Debe notarse que en este caso no podemos sincronizar ambas coordenadas nulas en el pasado remoto, como hicimos con~$\bar{u}(U)$. Lo que impide tal cosa es el factor de dilatación temporal gravitacional.

En este punto, ya podemos construir la relación~$U(u)$ que buscamos. La complejidad de~(\ref{ub.of.U}) no nos permite escribir la función explícitamente, pero sí podemos escribir su inversa, utilizando las relaciones~$\bar{u}(U)$ en~(\ref{ub.of.U}) y~$\bar{u}(u)$ en~(\ref{ub.of.u.static}):
\begin{align}
u (U) = & \left(1-\frac{2M}{\rst}\right)^{1/2} \left\{ \UH + \frac{4M}{3} - 4M \left[ \left( \frac{3}{4M}(\UH-U) +1 \right)^{1/3}
\right. \right.
\nonumber \\ 
 & + \left. \left.\ahalf \left( \frac{3}{4M}(\UH-U) +1 \right)^{2/3} + \athird \left( \frac{3}{4M}(\UH-U) +1 \right)
\right. \right.
\nonumber \\ 
 & + \left. \left.\log \left( \left(\frac{3}{4M}(\UH-U) +1\right)^{1/3}  - 1 \right) \right] \right\}.
\label{u.of.U.static}
\end{align}

Podemos estudiar el comportamiento de esta relación en el pasado y el futuro asintóticos, $u \to -\infty$ y $u \to \infty$, respectivamente. En el pasado asintótico, el término lineal en el miembro de la derecha de~(\ref{u.of.U.static}) es el más importante, y la relación queda
\begin{equation}
U \approx \left(1-\frac{2M}{\rst}\right)^{-1/2} u, \quad u \to -\infty.
\label{u.of.U.static.past}
\end{equation}
Teniendo en cuenta la definición de~$\kappa(u)$ en~(\ref{funcion_temperatura}), es fácil ver que en este régimen $\kappa (u) = 0$, por lo que el observador estático no percibe radiación alguna.

Por el contrario, en el infinito futuro el comportamiento de~$U(u)$ está dominado por el término logarítmico en~(\ref{u.of.U.static}), por lo que la expresión aproximada resultante es
\begin{align}
U \approx & \ \UH -\frac{4M}{3} \left\{ \left[ \exp \left(\frac{\UH}{4M} - \frac{3}{2} -\left(1-\frac{2M}{\rst}\right)^{-1/2} \frac{u}{4M} \right) +1 \right]^3 - 1 \right\} 
\nonumber\\
\approx & \ \UH - 4M \exp \left[\frac{\UH}{4M} - \frac{3}{2} -\left(1-\frac{2M}{\rst}\right)^{-1/2} \frac{u}{4M} \right], \quad  u \to \infty.
\label{u.of.U.static.future}
\end{align}
Esta expresión tiene directamente la ya conocida forma exponencial
\begin{equation}
U(u) = U_{\rm H} - A \rme^{- \kappa_{r_{\rm s}} u},
\label{exponencial_rs}
\end{equation}
donde~$U_{\rm H}$ y~$A$ son constantes sin relevancia, y~$\kappa_{r_{\rm s}}$ es una constante dada por
\begin{equation}
\kappa_{r_{\rm s}} := \left(1-\frac{2M}{\rst}\right)^{-1/2} \frac{1}{4M}.
\label{kappa.static.observer}
\end{equation}
En estas condiciones, sabemos que~$\kappa(u) = \kappa_{r_{\rm s}}$, y que el observador percibe un espectro térmico con esta temperatura.\footnote{Con el fin de agilizar el lenguaje, a menudo nos referiremos a~$\kappa(u)$ directamente como ``temperatura'', si bien, como se ha repetido en numerosas ocasiones, solo puede hablarse de temperatura bajo cumplimiento de la condición adiabática (como en este caso), y en tal caso la verdadera temperatura del espectro está dada por la expresión~$T(u) = |\kappa (u)|/(2 \pi k_{\rm B})$ en~(\ref{temperatura_kappa}).} Por tanto, para tiempos tardíos los observadores estáticos detectan radiación con la temperatura de Hawking~$1/(4M)$ multiplicada por un factor de corrimiento al azul gravitacional. Este factor es igual a la unidad para los observadores en el infinito espacial, los cuales (como ya es conocido) detectan la radiación con la temperatura de Hawking; y alcanza valores arbitrariamente altos según el observador se encuentra en posiciones más cercanas al horizonte de sucesos. Este resultado coincide con el obtenido en~\cite{Hodgkinson:2013tsa} utilizando detectores de Unruh-DeWitt.

\subsection{Resultados numéricos}

\begin{figure}[ht]
	\centering
    \includegraphics{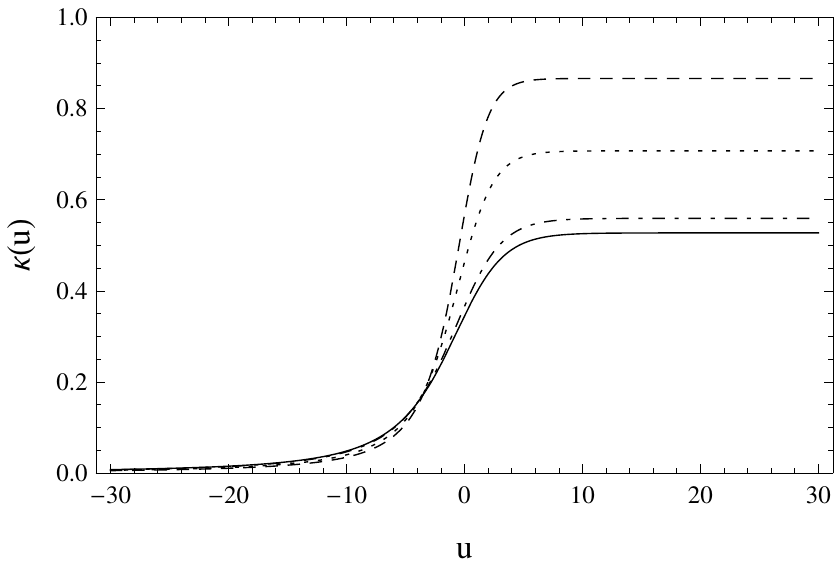}
  \caption{\footnotesize{Temperatura efectiva como función de~$u$ para diferentes observadores estáticos en~$r_{\rm s} = (3M,\ 4M,\ 10M,\ 20M)$ (línea discontinua, punteada, punto-raya y continua, respectivamente). Utilizamos unidades~$2M = 1$.}}
  \label{fig_1}
\end{figure}

A continuación describiremos la percepción de partículas de uno de estos observadores a lo largo de toda su trayectoria estática en el exterior del agujero negro. De la definición de~$\kappa(u)$ dada en~(\ref{funcion_temperatura}), es fácil comprobar que también puede escribirse tal función como
\begin{equation}
\kappa (u) = \evat{\deriv[2]{u}{U} \left( \deriv{u}{U} \right)^{-2}}{U(u)}.
\label{alt.kappa}
\end{equation}
De esta forma, evaluando las derivadas podemos obtener una forma cerrada de~$\kappa(u)$ en función de la relación implícita~$U(u)$:
\begin{equation}
\kappa (u) = \fr{4M} \left(1-\frac{2M}{\rst}\right)^{-1/2} \left\{ \frac{3}{4M}[\UH-U(u)] +1 \right\}^{-4/3}.
\label{kappa.static}
\end{equation}
Resolviendo numéricamente la relación~$U(u)$ a partir de~(\ref{u.of.U.static}), podemos mostrar~$\kappa(u)$ para diferentes observadores estáticos (figura~\ref{fig_1}). En las curvas resultantes, se observa claramente cómo el estado de vacío es tal que en el pasado no contiene partículas, pero en un determinado periodo comienza a ``calentarse'', alcanzando una temperatura final asintótica. Como dijimos al comienzo, el estado de vacío que hemos escogido en un espacio-tiempo estático reproduce cualitativamente los resultados esperables en un escenario real de formación de un agujero negro por colapso gravitacional. Solo cuando el agujero negro está muy cerca de su formación se entra en el régimen asintótico exponencial~(\ref{u.of.U.static.future}), y la radiación de Hawking aparece tal y como se describe habitualmente. Podemos estimar cuándo sucede esto, pero para ello primero estudiaremos el comportamiento de la condición adiabática, a fin de comprobar la validez del uso de la función de temperatura efectiva.

\subsection{Verificación de la condición adiabática}

En la sección~\ref{sec_adiabatica}, donde analizamos la condición adiabática para la función de temperatura efectiva, encontramos que el cumplimiento de esta condición puede expresarse localmente mediante la condición suficiente~$D_* \ll 1$ [ecuación~(\ref{condicion_adiabatica})], siendo~$D_*$ función de los valores de~$\kappa(u)$ y sus derivadas en un punto dado~$u_*$ [ecuación~(\ref{definicion_d})]. Sin embargo, en la mayoría de los casos es igualmente suficiente con tener en cuenta únicamente la primera derivada de~$\kappa(u)$, reduciéndose la condición al criterio más sencillo~$|\dot{\kappa}_*|/ \kappa_*^2 \ll 1$ [ecuación~(\ref{condicion_adiabatica_derivada})]. Para estudiar la validez de la condición adiabática a lo largo de las distintas trayectorias de los observadores, en este capítulo introducimos una función que denominaremos \emph{función de control adiabática,} dada por
\begin{equation}
\epsilon (u) := \fr{\kappa(u)^2} \left| \deriv{\kappa}{u} \right|.
\label{adiabatic.condition}
\end{equation}
En los intervalos en los que esta función verifique~$\epsilon(u) \ll 1$, la función~$\kappa(u)$ puede interpretarse inequívocamente como la temperatura del espectro de radiación percibido por el observador en cuestión en cada instante. En los intervalos en los que no se verifique este criterio, la función no tiene estrictamente tal interpretación. Sin embargo, no dejará de ser un valor orientativo de la cantidad de partículas percibidas por el observador.

En el caso que nos ocupa (observadores estáticos), de las expresiones~$u(U)$ en~(\ref{u.of.U.static}) y~$\kappa(u)$ en~(\ref{kappa.static}) obtenemos
\begin{equation}
\epsilon (u) = 4 \left\{\left[ \frac{3}{4M}(\UH-U(u)) +1 \right]^{1/3} - 1 \right\}.
\label{epsilon.static}
\end{equation}
Por tanto, $\epsilon(u)$ decae desde infinito en el pasado asintótico a cero cuando el observador que fija el vacío alcanza el horizonte ($U = U_{\rm H}$), lo cual sucede en el futuro asintótico para los observadores estáticos. Este comportamiento a tiempos tardíos es consistente con el hecho de que la radiación de Hawking está finalmente encendida entonces, de manera que la radiación que los observadores estáticos detectan es perfectamente térmica. Por otra parte, en el pasado remoto, $\epsilon(u)$ diverge. Sin embargo, esto es simplemente un artefacto procedente del hecho de que~$\kappa(u)$ tiende a anularse en esa región, por lo que la condición adiabática pierde su sentido: sencillamente, no hay radiación.

\begin{figure}[ht]
	\centering
    \includegraphics{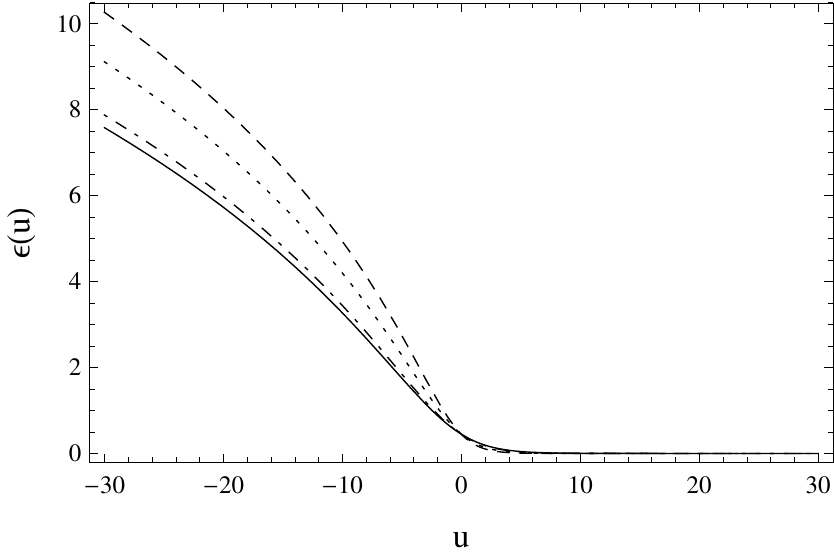}
  \caption{\footnotesize{Función de control adiabática como función de~$u$ para diferentes observadores estáticos en~$r_{\rm s} = (3M,\ 4M,\ 10M,\ 20M)$ (línea discontinua, punteada, punto-raya y continua, respectivamente). Utilizamos unidades~$2M = 1$.}}
  \label{fig_2}
\end{figure}

Podemos trazar numéricamente también~$\epsilon(u)$ para las diferentes posiciones estáticas (figura~\ref{fig_2}). Vemos que su valor decae hasta prácticamente anularse cuando la temperatura alcanza el régimen asintótico.

Introducimos ahora dos tiempos de referencia. Por una parte, definimos el \emph{instante de encendido}~$U_{\rm ig}$ del agujero negro como aquel para el cual~$\ddot{\kappa} (u) = 0$ (este criterio se basa en obtener el punto de inflexión de las curvas de la figura~\ref{fig_1}). Si calculamos este valor en términos de~$U$, es fácil ver que resulta~$U_{\rm ig} = U_{\rm H} - (676/1029) M$. Esto se corresponde con el instante en el que la trayectoria en caída libre que define el vacío alcanza el radio~$r_{\rm ig} \simeq 2 M + 0.61 M$. Vemos que, escrito en función de~$U$, este instante no depende del valor de~$r_{\rm s}$ de la trayectoria estática que se utilice para calcularlo, lo cual es perfectamente razonable. Por otra parte, definimos el \emph{instante de termalización}~$U_{\rm th}$ como aquel a partir del cual puede decirse que la radiación saliente tiene un espectro térmico. Este instante puede calcularse en términos de la coordenada~$U$ mediante la condición~$\epsilon (U_{\rm th}) = \bar{\epsilon}$ [con~$\epsilon(U) := \epsilon(u(U))$ dada por~(\ref{epsilon.static})], donde~$\bar{\epsilon}$ es un valor que fija la precisión a partir de la cual se considera que se ha alcanzado \emph{de facto} el espectro térmico. Fijando~$\bar{\epsilon} = 0.01$, esta condición se cumple en~$U_{\rm th} \simeq U_{\rm H} - 0.01M$, lo que corresponde (en un sentido análogo a~$r_{\rm ig}$) a $r_{\rm th} \simeq 2 M + 0.01 M$.

Pongamos algunos números. Si una estrella de neutrones con $1.5$~veces la masa del Sol (radio de Schwarzschild $2M = 4.43$~km) y un radio de $12$~km iniciara un colapso en caída libre, alcanzaría~$r_{\rm ig} \simeq 2 M + 1.36$~km en un tiempo Schwarzschild $t_{\rm ig} \simeq 0.13 \times 10^{-3}$~s, y~$r_{\rm th} \simeq 2 M + 0.02$~km en un tiempo $t_{\rm th} \simeq 0.20 \times 10^{-3}$~s. De esta forma, puede verse claramente que la detección de la radiación de Hawking en el infinito no supone un gran retraso asociado con la ``congelación'' de la estructura que colapsa según se aproxima a la formación del horizonte. También es importante destacar que el espectro nítidamente térmico aparece cuando el objeto que colapsa está aún a~$0.02$~km de la formación del horizonte. Se trata pues de un ejemplo concreto de lo que adelantamos en la sección~\ref{sec_kappa}: no es necesaria la formación estricta de un horizonte para obtener radiación térmica.

\section{Observadores en caída libre desde el infinito}\label{caida_libre_infinito}

En esta sección, analizaremos la percepción de partículas por observadores que siguen una trayectoria en caída libre desde el infinito espacial. Para encontrar la nueva coordenada nula~$u$ asociada a estos observadores, debemos utilizar las geodésicas de género tiempo radiales que comienzan en el infinito espacial con velocidad nula. Sin embargo, debe recordarse que el cálculo realizado para fijar el estado de vacío en la sección~\ref{fijado_vacio} hacía uso precisamente de esas trayectorias, obtenidas finalmente en~(\ref{ub.of.tau}). En aquel cálculo, teníamos dos constantes de integración a fijar, a saber, $t_0$ y~$\tau_0$. Ahora consideraremos trayectorias que difieren de la que define el vacío en un retraso temporal~$\Delta t_0$ en la constante~$t_0$. Es decir, en lugar de fijar~$t_0 = \tau_0 = U_{\rm H} + 4 M/3$, utilizaremos los valores~$t_0 = \tau_0 = U_{\rm H} + 4 M/3 + \Delta t_0 = u_{\rm H} + 4 M/3$, donde~$u_{\rm H} := U_{\rm H} + \Delta t_0$. Definimos la nueva relación~$\bar{u} (u)$ mediante
\begin{align}
\ub (u) = & \ \uH + \frac{4M}{3} - 4M \left\{ \left[ \frac{3}{4M}(\uH-u) +1 \right]^{1/3} + \ahalf \left[ \frac{3}{4M}(\uH-u) +1 \right]^{2/3}
\right. 
\nonumber \\ 
 & \left. 
+ \athird \left[ \frac{3}{4M}(\uH-u) +1 \right] + \log \left[ \left(\frac{3}{4M}(\uH-u) +1\right)^{1/3}  - 1 \right] \right\}.
\label{ub.of.u.falling}
\end{align}
Ahora, como hicimos en el caso anterior, comparamos las coordenadas nulas~$U$ en~(\ref{ub.of.U}) y~$u$ en~(\ref{ub.of.u.falling}), obteniendo una relación implícita dada por
\begin{multline}
- 4M \left\{ \left[ \frac{3}{4M}(\UH-U) +1 \right]^{1/3} + \ahalf \left[ \frac{3}{4M}(\UH-U) +1 \right]^{2/3}
\right. 
\\ 
\left. 
+ \athird \left[ \frac{3}{4M}(\UH-U) +1 \right] + \log \left[ \left(\frac{3}{4M}(\UH-U) +1\right)^{1/3}  - 1 \right] \right\}
\\ 
= \dt - 4M \left\{ \left[ \frac{3}{4M}(\uH-u) +1 \right]^{1/3} + \ahalf \left[ \frac{3}{4M}(\uH-u) +1 \right]^{2/3}
\right. 
\\ 
\left. 
+ \athird \left[ \frac{3}{4M}(\uH-u) +1 \right] + \log \left[ \left(\frac{3}{4M}(\uH-u) +1\right)^{1/3}  - 1 \right] \right\}.
\label{U.of.u.falling}
\end{multline}

Las soluciones a esta ecuación no algebraica deben encontrarse numéricamente. No obstante, podemos examinar los límites del pasado asintótico y del horizonte de sucesos de forma analítica. En el pasado asintótico, obtenemos~$U \approx u$, lo que refleja la condición de sincronización $U \approx \bar{u} \approx u$ que hemos impuesto al hacer~$t_0 = \tau_0$. Cerca del horizonte, las funciones logarítmicas en~(\ref{U.of.u.falling}) son las que tienen la contribución dominante en la ecuación y, por tanto,
\begin{equation}
U (u) \approx \UH -\frac{4M}{3} \left\{\left[ \rme^{-\dt/(4M)}\left( \left( \frac{3}{4M}(\uH-u) +1 \right)^{1/3} - 1 \right) + 1 \right]^3 - 1 \right\}.
\label{U.of.u.falling.EH}
\end{equation}

Aunque no se tenga la relación explícita~$U(u)$, ni su inversa, la función~$\kappa (u)$ puede encontrarse de forma explícita como función de~$u$ y de~$U(u)$. Si denominamos~$p(U)$ a la \emph{forma funcional} del segundo miembro de~(\ref{ub.of.U}) con respecto a~$U$, es fácil ver que la ecuación~(\ref{U.of.u.falling}) que relaciona implícitamente~$U$ y~$u$ se puede escribir como
\begin{equation}
p(U) = p(u- \Delta t_0) + \Delta t_0.
\label{relacion_implicita_U_u_caida}
\end{equation}
Derivando con respecto a~$u$, obtenemos
\begin{equation}
\deriv{U}{u}  = \deriv{p(u-\dt)}{u} \left(\derivat{p(U)}{U}{U(u)}\right)^{-1},
\label{deriv.eq.in.f.falling}
\end{equation}
y, derivando una vez más
\begin{equation}
\deriv[2]{U}{u} = \left[ \deriv[2]{p(u-\dt)}{u}  - 
\derivat[2]{p(U)}{U}{U(u)} \left( \deriv {U}{u}  \right)^2 \right] \left(\derivat{p(U)}{U}{U(u)}\right)^{-1},
\label{second.deriv.eq.in.f.falling}
\end{equation}
de forma que $\kappa(u)$ puede expresarse como
\begin{equation}
\kappa (u) = \left[ \derivat[2]{p(U)}{U}{U(u)} \left( \deriv{U}{u} \right)^2 - \deriv[2]{p(u-\dt)}{u} \right] \left(\deriv{p(u-\dt)}{u}\right)^{-1}.
\label{kappa.falling.derivs}
\end{equation}
El resultado explícito en términos de~$u$ y~$U(u)$ es
\begin{align}
\kappa (u) = & \ \fr{4M} \left[ \frac{3}{4M}(\uH-u) +1 \right]^{-1} \left\{ \left[\frac{3}{4M}(\uH-u) +1 \right]^{1/3} - 1 \right\}^{-1} 
\nonumber\\
& \times \left\{ \left[\frac{(3 / 4M)(\uH-u) + 1}{(3 / 4M)(\UH-U(u)) + 1}\right]^{4/3} - 1 \right\}.
\label{kappa.falling}
\end{align}
Debe tenerse en cuenta que, aunque se trata de una expresión explícita si se escribe en función de~$u$ y~$U$, ello no significa que sea una función de dos variables, sino tan solo de~$u$. La dependencia con~$\Delta t_0$ está oculta en la relación implícita~$U(u)$ entre las dos coordenadas, dada por~(\ref{U.of.u.falling}).

Podemos utilizar la expresión~$U(u)$ válida en las cercanías del horizonte de sucesos~(\ref{U.of.u.falling.EH}) (la cual tiende a ser exacta en ese límite) para encontrar qué temperatura efectiva detectará el observador caracterizado por el retraso~$\Delta t_0$ en el instante preciso en que cruza el horizonte. Esta temperatura efectiva resulta ser
\begin{equation}
\kappa_{\rm hor}(\dt) = \fr{M} \left( 1- \rme^{-\dt/(4M)} \right).
\label{kappa.HC.falling}
\end{equation}
Por tanto, la temperatura efectiva percibida pasa de ser inicialmente nula a alcanzar asintóticamente el valor~$1/M$, lo que supone cuatro veces la temperatura de la radiación de Hawking~$1/(4M)$, para retardos~$\Delta t_0$ mucho mayores que~$4M$. Se trata de un resultado interesante y, a primera vista, enigmático. Con posterioridad, el mismo resultado ha sido encontrado nuevamente en~\cite{Smerlak:2013sga,Smerlak:2013cha}, con un procedimiento análogo. En la sección~\ref{sec_incremento_final} daremos una explicación clara de este resultado que, adelantamos, se debe a un factor Doppler que diverge en el horizonte de sucesos.

\subsection{Resultados numéricos}

\begin{figure}[ht]
	\centering
    \includegraphics{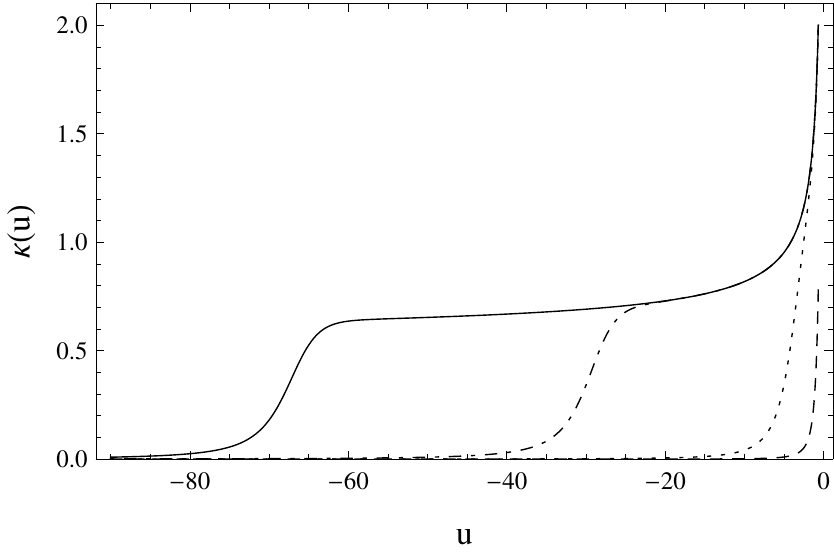}
  \caption{\footnotesize{Temperatura efectiva como función de~$u$ para diferentes observadores en caída libre con retrasos~$\Delta t_0 = (2M,\ 20M,\ 100M,\ 200M)$ (línea discontinua, punteada, punto-raya y continua, respectivamente). Utilizamos unidades~$2M = 1$ y~$u_{\rm H} = 0$.}}
  \label{fig_3}
\end{figure}

Mediante cálculo numérico podemos obtener una gráfica completa de la función~$\kappa(u)$. Los resultados se muestran en la figura~\ref{fig_3}. Para valores pequeños de~$\Delta t_0$ (menos de~$20 M$, aproximadamente) la temperatura percibida por el observador en caída libre es prácticamente nula a lo largo de casi toda su trayectoria. Solo cuando se acerca al horizonte, la temperatura crece hasta alcanzar un valor máximo en el propio horizonte, dado por~(\ref{kappa.HC.falling}). Para retrasos~$\Delta t_0$ suficientemente largos, antes de este incremento final en la temperatura aparece una ``meseta'' intermedia en la cual la temperatura es casi constante. Es fácil comprender este comportamiento: si el observador en caída libre se encuentra aún lejos del horizonte cuando aparece la radiación de Hawking, en el momento en que esto sucede el observador comenzará a percibir esa radiación, la cual permanecerá aproximadamente constante hasta que el observador se encuentre cerca del horizonte, donde tiene lugar el (hasta ahora inesperado) incremento final de la temperatura. La temperatura a lo largo de la meseta es algo mayor que la temperatura de Hawking, y ligeramente creciente. Como veremos, esto se debe principalmente al factor Doppler asociado a la velocidad radial del observador en caída libre (el mismo factor responsable de que aparezca el pico final antes de cruzar el horizonte).

\subsection{Verificación de la condición adiabática}

Habiendo encontrado una expresión para~$\kappa(u)$ en términos de~$u$ y~$U(u)$, podemos utilizarla para obtener una expresión en los mismos términos de la función de control adiabática~(\ref{adiabatic.condition}). La expresión resultante es, no obstante, excesivamente complicada, por lo que no resulta útil incluirla aquí. Sin embargo, al igual que hicimos con~$\kappa(u)$, podemos también en este caso sustituir la relación~$U(u)$ válida en las cercanías del horizonte~(\ref{U.of.u.falling.EH}) para encontrar una expresión explícita~$\epsilon(u)$ válida en esa región; y, a continuación, tomar el límite~$u \to u_{\rm H}$ para encontrar el valor de~$\epsilon(u)$ que se tiene al cruzar el horizonte. El resultado es
\begin{equation}
\epsilon_{\rm hor}(\dt) = \frac{3}{8} + \frac{7}{4 \left[ \rme^{\dt/(4M)} - 1 \right]}.
\label{epsilon.falling.EH}
\end{equation}
Este resultado diverge para~$\Delta t_0 \to 0$, decreciendo monótonamente hasta alcanzar el valor asintótico~$\epsilon = 3/8$ cuando~$\Delta t_0$ es varias veces mayor que~$4M$. La divergencia para~$\Delta t_0 \to 0$ se debe, una vez más, a que es la propia temperatura~$\kappa_{\rm hor} (\Delta t_0)$ en~(\ref{kappa.HC.falling}) la que se anula cuando el observador en caída libre se aproxima a aquel que define el vacío. Cabe destacar que la condición~$\epsilon \ll 1$ nunca se satisface al cruzar el horizonte. Sin embargo, es llamativo que, para~$\Delta t_0$ suficientemente largos, $\epsilon_{\rm hor}(\dt)$ siempre se mantiene del orden de la unidad, por lo que en tales casos no nos alejamos significativamente del régimen adiabático. Cabe conjeturar que el espectro percibido al cruzar el horizonte en tiempos suficientemente tardíos no diferirá de forma dramática del espectro de Planck con temperatura cuatro veces superior a la temperatura de Hawking.

\begin{figure}[ht]
	\centering
    \includegraphics{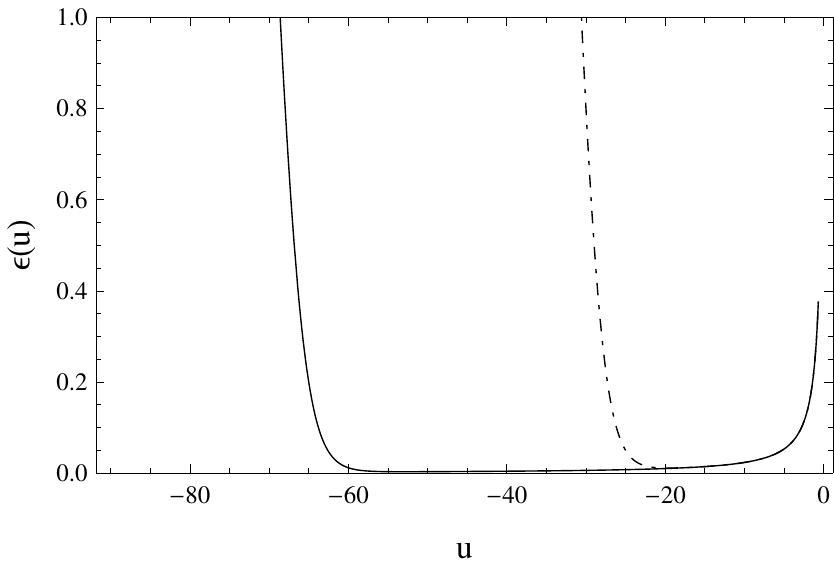}
  \caption{\footnotesize{Función de control adiabática como función de~$u$ para diferentes observadores en caída libre con retrasos~$\Delta t_0 = (100M,\ 200M)$ (línea punto-raya y continua, respectivamente). Utilizamos unidades~$2M = 1$ y~$u_{\rm H} = 0$.}}
  \label{fig_6}
\end{figure}

Por otra parte, podemos estudiar numéricamente el valor de~$\epsilon(u)$ a lo largo de toda la trayectoria (figura~\ref{fig_6}). Podemos ver que, para tiempos suficientemente tardíos hay una parte de la trayectoria para la cual la condición adiabática es sin duda válida, pues el valor de~$\epsilon(u)$ es prácticamente nulo en ella. Tal región, por supuesto, coincide con la meseta que se obtuvo para~$\kappa(u)$ (figura~\ref{fig_3}, en los casos en los que esta aparecía), el la cual~$\kappa(u)$ tomaba un valor casi constante. Una vez que el incremento final en~$\kappa(u)$ aparece, la condición adiabática comienza a violarse, aunque, como hemos visto, no de forma dramática.

\section{Observadores en caída libre desde un radio finito}

El último caso de observadores que consideraremos en este capítulo serán aquéllos que son ``soltados'' en caída libre desde un radio finito~$r_0$ con velocidad inicial nula, tras un tiempo~$\dtw$ (en tiempo de Schwarzschild) desde que la trayectoria usada para definir el vacío cruzara la posición radial de salida~$r_0$. De nuevo, en primer lugar calcularemos dicha trayectoria. Para ello, debemos integrar la expresión~$\rmd r / \rmd \tau$ en~(\ref{r.tau.first.integral}), obtenida como primera cuadratura de la ecuación de las geodésicas~(\ref{Newton}), pero manteniendo en este caso el radio inicial~$r_0$ como una constante finita. El resultado es
\begin{equation}
{\tilde\tau}_0 -\tau = - r_0 \sqrt{\frac{r_0}{2M}} \left\{ \left( \frac{r_0}{r} - 1 \right)^{1/2} \frac{r}{r_0} + \arctan \left[\left( \frac{r_0}{r} - 1 \right)^{1/2} \right] \right\},
\label{tau.of.r.radius}
\end{equation}
donde~${\tilde\tau}_0$ representa el tiempo propio en el que el observador comienza a caer (debe notarse que, en toda esta sección, el rango de valores de~$r$ es~$2M < r \leq r_0$). Esta expresión nos define la función implícita~$r(\tau;r_0,{\tilde\tau}_0)$, donde se ha escrito la dependencia con las constantes de integración por conveniencia. El instante en el que se cruza el horizonte siguiendo esta trayectoria es
\begin{equation}
{\tilde u}_{\rm H} := {\tilde\tau}_0 + 2M \sqrt{\frac{r_0}{2M}} \left\{ \left( \frac{r_0}{2M} - 1 \right)^{1/2} + \frac{r_0}{2M} \arctan \left[\left( \frac{r_0}{2M} - 1 \right)^{1/2} \right] \right\}.
\label{u_H.radius}
\end{equation}
Por otra parte, mediante las derivadas~$\rmd r / \rmd \tau$ en~(\ref{r.tau.first.integral}) y~$\rmd t / \rmd \tau$ en~(\ref{t.tau.first.integral}) obtenemos
\begin{equation}
\deriv{r}{t} = -\sqrt{\frac{2M}{r}}\left(1- {\frac{2M}{r}}\right) \left(1  -\frac{r}{\rc} \right)^{1/2}
\left(1  -\frac{2M}{\rc} \right)^{-1/2},
\label{dr.of.t.radius}
\end{equation}
expresión que puede integrarse analíticamente, resultando
\begin{multline}
t (r) = {\tilde t}_0 + \log \left\{ \left( \frac{r}{2M} - 1\right)^{-1} \right. \\
\left. \times \left[ \frac{r}{2M} + 2 \left(1- \frac{2M}{r_0}\right)^{1/2} \left(1- \frac{r}{r_0}\right)^{1/2} \sqrt{\frac{r}{2M}} - \frac{2 r}{r_0} + 1 \right] \right\} \\
+ r_0 \left( \frac{r_0}{2M} - 1 \right)^{1/2} \left\{ \left( \frac{r_0}{r} - 1 \right)^{1/2} \frac{r}{r_0} + \left( 1 + \frac{4M}{r_0} \right) \arctan \left[\left( \frac{r_0}{r} - 1 \right)^{1/2} \right] \right\},
\label{t.of.r.radius}
\end{multline}
donde~${\tilde t}_0$ es de nuevo una constante de integración, que en este caso representa el tiempo de Schwarzschild en el que el observador comienza a caer desde~$r_0$. Si denominamos por conveniencia~$t = \bar{t}(r)$ a la dependencia entre~$t$ y~$r$ para la trayectoria que define el vacío, expresada en~(\ref{t.of.r}), podemos escribir~${\tilde t}_0$ en función de las constantes~$r_0$, $t_0$ (que es una constante de integración \emph{de la trayectoria que define el vacío}) y~$\dtw$ (que es la constante que manipularemos libremente). En efecto, tal y como hemos definido~$\dtw$, lo que debemos imponer es que
\begin{align}
{\tilde t}_0 & = \bar{t} (r_0) + \dtw \nonumber \\
& = t_0 -4M \left[ \sqrt{\frac{r_0}{2M}} + \athird \left( \frac{r_0}{2M} \right)^{3/2} + \ahalf \log \left( \frac{\sqrt{r_0/(2M)}-1}{\sqrt{r_0/(2M)}+1} \right) \right] + \dtw.
\label{delay.fixing.radius}
\end{align}
Sustituyendo este resultado, y la función implícita~$r(\tau;r_0,{\tilde\tau}_0)$ dada por~(\ref{tau.of.r.radius}), en~(\ref{t.of.r.radius}), se obtiene la función~$t(\tau;r_0,{\tilde\tau}_0,t_0,\dtw)$.

Construyamos ahora la trayectoria completa de un observador que permanece en una posición fija~$r=r_0$ hasta un tiempo propio~$\tau = {\tilde\tau}_0$, instante en el cual se deja caer libremente hacia el agujero negro. Esta trayectoria~$(t(\tau),r(\tau))$ está compuesta por dos partes que conectan en~$\tau = {\tilde\tau}_0$: la primera parte, para~$\tau < {\tilde\tau}_0$, es tal que el radio permanece fijo~$r(\tau) = r_0$ y el tiempo de Schwarzschild satisface
\begin{equation}
t(\tau) = {\tilde t}_0 + \left(1-{\frac{2M}{r_0}}\right)^{-1/2}(\tau - {\tilde\tau}_0),
\label{t.first.part}
\end{equation}
con~${\tilde t}_0$ dado por~(\ref{delay.fixing.radius}); y la segunda parte, para~$\tau \geq {\tilde\tau}_0$, viene dada por las expresiones implícitas para la trayectoria en caída libre encontradas previamente, $r(\tau;r_0,{\tilde\tau}_0)$ en~(\ref{tau.of.r.radius}) y~$t(\tau;r_0,{\tilde\tau}_0,t_0,\dtw)$ en~(\ref{t.of.r.radius}) [donde se han sustituido, a su vez, (\ref{tau.of.r.radius}) y~(\ref{delay.fixing.radius})].

Como hemos hecho en las dos secciones anteriores, reemplazamos ahora~$\tau = u$. También re-nombramos la constante arbitraria~${\tilde\tau}_0$ como~$u_0$, y escogemos una sincronización entre~$t_0$ y~$u_0$ de tal forma que, en la parte estática de la trayectoria, $U$ y~$u$ tiendan a coincidir para radios iniciales lejanos. De esta forma, desaparece la constante~$t_0$, quedando únicamente tres constantes en juego: dos con significado físico, el radio inicial~$r_0$ y el ``tiempo de espera''~$\dtw$; y una arbitraria, el instante \emph{en tiempo propio}~$u_0$ en el que se comienza la caída libre.

El cálculo de la función~$\kappa(u)$ para la primera parte de la trayectoria resulta trivial (idéntico al de los observadores estáticos, ya realizado en la sección~\ref{observadores_estaticos}). En cambio, la parte en caída libre requiere más pasos. Primero, escribimos la expresión para la coordenada nula~$\bar{u}$
\begin{equation}
\ub (r) = t(r;r_0,u_0,\dtw)-r^*(r),
\label{ubar_r}
\end{equation}
donde es importante enfatizar que, por conveniencia, de momento se mantiene~$r$, y no~$u$, como variable. Si de nuevo llamamos~$p(U)$ a la \emph{forma funcional} del segundo miembro de~(\ref{ub.of.U}) con respecto a~$U$, la relación~$U(u)$ que buscamos para obtener~$\kappa(u)$ queda implícita en la ecuación
\begin{equation}
p(U) = \ub (r(u;r_0,u_0)),
\label{eq.in.f.radius}
\end{equation}
con~$p(U)$ dado por~(\ref{ub.of.U}), $\bar{u} (r)$ dado por~(\ref{ubar_r}), y~$r(u;r_0,u_0)$ dado implícitamente por~(\ref{tau.of.r.radius}) (con las sustituciones mencionadas en el párrafo anterior). Tomando derivadas con respecto a~$u$, encontramos
\begin{equation}
\deriv{U}{u} = \left( \deriv{p}{U} \right)^{-1} \deriv{\bar{u}}{r} \deriv{r}{u},
\label{deriv.eq.in.f.radius}
\end{equation}
donde~$\rmd r/ \rmd u$ no es más que~$\rmd r/ \rmd \tau$ en~(\ref{r.tau.first.integral}) con~$u=\tau$; y, derivando de nuevo,
\begin{equation}
\deriv[2]{U}{u} = \left[ \deriv[2]{\bar{u}}{r} \left( \deriv{r}{u} \right)^2 + \deriv{\bar{u}}{r} \deriv[2]{r}{u} - \deriv[2]{p}{U} \left( \deriv{U}{u} \right)^2 \right] \left( \deriv{p}{U} \right)^{-1}.
\label{second.deriv.eq.in.f.radius}
\end{equation}
De esta forma, $\kappa(u)$ puede expresarse como
\begin{equation}
\kappa(u) = \left( \deriv{p}{U} \right)^{-1} \deriv[2]{p}{U} \deriv{U}{u} - \left( \deriv{\bar{u}}{r} \right)^{-1} \deriv[2]{\bar{u}}{r} \deriv{r}{u} - \left( \deriv{r}{u} \right)^{-1} \deriv[2]{r}{u}.
\label{kappa.radius.derivs}
\end{equation}

Reemplazando las expresiones correspondientes, y simplificando, obtenemos finalmente la siguiente expresión para la función de temperatura efectiva
\begin{align}
\kappa(u) = & \ \fr{4M} \left(1-\frac{2M}{r_0}\right)^{-1/2} \left[ \frac{3}{4M}(\UH-U) +1 \right]^{-4/3}, \quad {\rm para} \quad u < u_0,
\label{kappa.static.radius}
\\
\kappa(u) = & \ \fr{4M} \left\{\frac{r}{2M}+1 +\frac{r}{M} \left[ \left(1-\frac{2M}{r_0}\right)^{1/2} \sqrt{\frac{2M}{r}} \left(1-\frac{r}{r_0}\right)^{1/2}  - \frac{2M}{r_0} \right]\right\}
\nonumber \\
& \times \left\{\left[ \left(1-\frac{2M}{r_0}\right)^{1/2} + \sqrt{\frac{2M}{r}}\left(1-\frac{r}{r_0}\right)^{1/2} \right] \left(\frac{r}{2M}-1\right)\right\}^{-1}
\nonumber \\
& \times \left\{\left[\frac{3}{4M}(\UH-U)+1\right]^{-4/3} - \left(\frac{2M}{r}\right)^2 \right\},
\quad {\rm para} \quad u \geq u_0,
\label{kappa.radius}
\end{align}
donde~$r=r(u)$ y~$U=U(u)$ deben entenderse como funciones de~$u$ [las cuales, una vez más, solo pueden resolverse numéricamente a partir de~(\ref{tau.of.r.radius}) y~(\ref{eq.in.f.radius}), respectivamente]. Como veremos, en general esta función presenta un salto finito en el instante~$u_0$ en el que el observador inicia la caída desde la posición previamente fija.

Durante el tiempo previo a la caída libre, $\kappa(u)$ sigue el patrón descrito en la sección~\ref{observadores_estaticos} para los observadores estáticos, dado por~(\ref{kappa.static}): crece desde cero e intenta alcanzar el valor asintótico asociado al observador en reposo en el radio~$r_0$. Sin embargo, en el instante~$u_0$ el observador comienza a caer libremente. En este instante, se produce un salto finito brusco en~$\kappa(u)$. El valor de~$\kappa(u)$ justo \emph{tras} este salto, que denotaremos~$\kappa_{\text{caída}}$, se encuentra evaluando~(\ref{kappa.radius}) en~$r=r_0$, lo cual resulta
\begin{align}
\kappa_{\text{caída}}(r_0,\dtw) = & \ \fr{4M} \left(1 - \frac{2M}{r_0} \right)^{-1/2} \nonumber \\
& \times \left\{ \left[ \frac{3}{4M} (\UH - U_0(r_0,\dtw)) + 1 \right]^{-4/3} - \left(\frac{2M}{r_0}\right)^{-2} \right\}.
\label{kappa.init.radius}
\end{align}
Aquí, $U_0 := U(u_0)$ es el valor de la coordenada nula~$U$ asociado con~$u_0$, y depende del parámetro~$\dtw$. Un caso particular es~$\dtw=0$, lo que significa que el observador comienza a caer justo cuando el observador que fija el vacío \emph{pasa a su lado.} Esto implica, por supuesto, $U=U(r_0)$ según las expresiones de la sección~\ref{fijado_vacio} para el observador que define el estado de vacío [puede obtenerse invirtiendo~(\ref{r.of.tau}), teniendo en cuenta que~$U$ hace las veces de~$\tau$]. Haciendo esta sustitución, es fácil comprobar que se obtiene~$\kappa_{\text{caída}} = 0$, como resulta esperable. Esto se debe a que, en este caso, en el instante inicial la única diferencia entre nuestro observador y aquel que define el vacío es su velocidad relativa. Si este último no percibe radiación, nuestro observador tampoco lo hará, puesto que entre una percepción y otra solo puede mediar un efecto Doppler multiplicativo.

En el caso opuesto en el que el observador permanece un tiempo largo en reposo antes de soltarse, la radiación de Hawking estará prácticamente encendida entonces, es decir, el estado de vacío será prácticamente indistinguible del estado de Unruh. Este límite es, evidentemente, $\dtw \to \infty$, lo cual equivale a sustituir~$U = U_{\rm H}$ en~$\kappa_{\text{caída}}$ en~(\ref{kappa.init.radius}). El resultado, que denotaremos~$\kappa_{\rm Unruh}$, es
\begin{equation}
\kappa_{\rm Unruh} (r_0):=\evat{\kappa_{\text{caída}}(r_0)}{\dtw \to \infty} = \fr{4M} \left(1 +\frac{2M}{r_0}\right) \left(1 -\frac{2M}{r_0}\right)^{1/2}.
\label{kappa.Unruh.radius}
\end{equation}
A su vez, en este resultado el límite del infinito espacial~$r \to \infty$ reproduce la temperatura de Hawking~$\kappa = 1/(4M)$, en tanto que el límite del horizonte de sucesos~$r \to 2M$ devuelve~$\kappa = 0$. Estos resultados coinciden con los conocidos para el vacío de Unruh: los observadores inerciales y en reposo en la zona asintótica detectan radiación de Hawking saliente, en tanto que los observadores en caída libre arbitrariamente cercanos al horizonte (y obsérvese que, también, \emph{instantáneamente en reposo}) no detectan radiación. Por tanto, estos resultados concuerdan con el hecho de que el vacío de colapso~$\ket{0_{\rm col}}$ reproduce el vacío de Unruh para tiempos tardíos.

\begin{figure}[ht]
	\centering
    \includegraphics{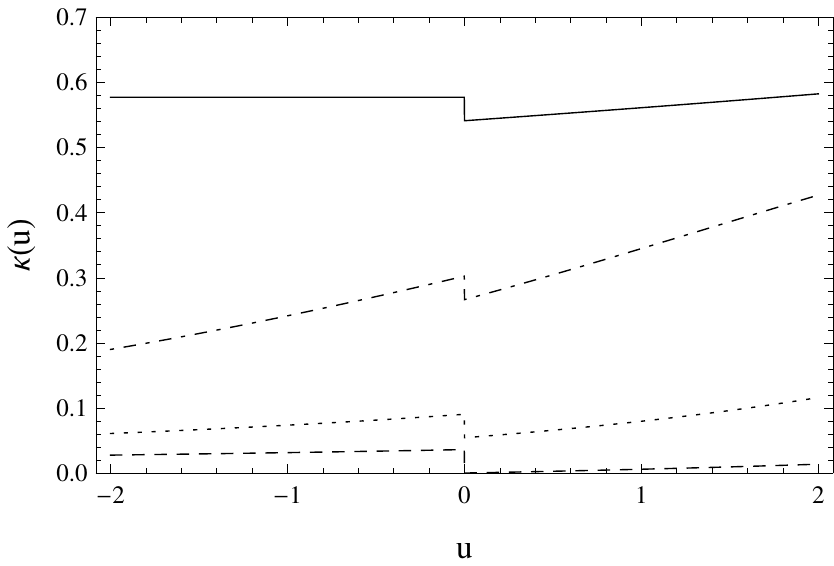}
  \caption{\footnotesize{Temperatura efectiva alrededor del instante~$u_0$ de inicio de la caída como función de~$u$ para diferentes observadores con radio inicial~$r_0 = 8M$ y retrasos~$\dtw = (0,\ 12M,\ 24M,\ \infty)$ (línea discontinua, punteada, punto-raya y continua, respectivamente). Utilizamos unidades~$2M = 1$ y~$u_0 = 0$.}}
  \label{fig_7}
\end{figure}

Otra cantidad de interés a calcular es el salto brusco en el valor de~$\kappa(u)$, que denotaremos~$\Delta \kappa (r_0)$, en el instante de iniciar la caída desde~$r_0$. Puede encontrarse directamente sustrayendo~$\kappa_{\text{caída}}$ en~(\ref{kappa.init.radius}) de~$\kappa(u \to u_0)$ previa a la caída en~(\ref{kappa.static.radius}):
\begin{equation}
\Delta \kappa (r_0) = \fr{4M} \left(1-\frac{2M}{r_0}\right)^{-1/2}\left(\frac{2M}{r_0}\right)^2=
\left(1-\frac{2M}{r_0}\right)^{-1/2} \frac{M}{r_0^2}.
\label{jump.kappa}
\end{equation}
Lo primero que llama la atención es que esta cantidad no depende de~$U$, o en otras palabras, no depende de~$\dtw$. En la figura~\ref{fig_7} se muestra el valor numérico de~$\kappa(u)$ alrededor del instante~$u_0$ para distintos tiempos de espera~$\dtw$ y un radio inicial fijo~$r_0 = 8M$. En ella se observa gráficamente cómo el salto discreto en el valor de~$\kappa(u)$ no depende del tiempo de espera.

Una explicación a este hecho que parece razonable (y la que se dio en~\cite{Barbado:2011dx}) es la siguiente: Para comenzar la caída libre, el observador estático debe \emph{apagar los motores} que lo sostienen fijo en un determinado radio, ``perdiendo'' así la aceleración correspondiente, la cual solo depende del radio~$r_0$ y viene dada, precisamente, por~$\Delta \kappa (r_0)$ en~(\ref{jump.kappa}). De hecho, vemos que~(\ref{jump.kappa}) no es más que la aceleración gravitatoria en~$r_0$ multiplicada por el factor de corrimiento gravitacional correspondiente. Por tanto, cabría concluir que~$\Delta \kappa (r_0)$ no es más que el efecto Unruh procedente de la aceleración necesaria para mantenerse fijo en un radio. Sin embargo, aunque esta explicación resulta satisfactoria a primera vista, en el capítulo siguiente discutiremos en detalle qué parte de la función de temperatura efectiva cabe asignar a la radiación de Hawking y qué parte al efecto Unruh, y veremos que la interpretación hecha aquí no es del todo adecuada.

A continuación, estudiaremos la percepción de radiación en el instante de cruzar el horizonte. El procedimiento es análogo al que llevamos a cabo en la sección~\ref{caida_libre_infinito}: en la relación implícita~$U(u)$ en~(\ref{eq.in.f.radius}), identificamos los términos divergentes en el horizonte de sucesos y, de la ecuación resultante de quedarnos solo con tales términos, despejamos~$U$ y obtenemos una expresión~$U(r)$ válida cerca del horizonte. Tal expresión es muy compleja, y no resulta interesante por sí misma, pero una vez más, introduciéndola en la expresión de~$\kappa(u)$ en~(\ref{kappa.radius}) obtenemos una expresión explícita~$\kappa(r)$ válida cerca del horizonte de sucesos. Finalmente, tomamos el límite~$r \to 2M$ en tal expresión y obtenemos el valor de~$\kappa(u)$ en el instante de cruzar el horizonte, que en este caso resulta
\begin{align}
\kappa_{\rm hor}(r_0,\dtw) = & \ \fr{M} \left\{ \left(1-\frac{2M}{r_0}\right)^{1/2} - \left( \frac{\sqrt{r_0/(2M)} - 1}{\sqrt{r_0/(2M)} + 1} \right)^{1/2}
\right.
\nonumber\\
& \times \left.\exp \left[ -\frac{5}{6} - \frac{r_0+\dtw}{4M} + \sqrt{\frac{r_0}{2M}} + \athird \left( \frac{r_0}{2M} \right)^{3/2}
\right. \right.
\nonumber\\
& \left. \left. -\left(\frac{r_0}{4M} + 1 \right) \left( \frac{r_0}{2M} - 1 \right)^{1/2} \arctan \left( \left( \frac{r_0}{2M} - 1 \right)^{1/2} \right) \right]\right\}.
\label{kappa.HC.radius}
\end{align}
Cuando el observador espera un tiempo suficientemente largo antes de soltarse de su posición inicial (matemáticamente, cuando~$\dtw \to \infty$), se simplifica a
\begin{equation}
\evat{\kappa_{\rm hor}}{\dtw \to \infty} = \fr{M} \left(1-\frac{2M}{r_0}\right)^{1/2},
\label{kappa.HC.Unruh.radius}
\end{equation}
lo cual reproduce el resultado~$\kappa = 1/M$ para un observador en caída libre desde el infinito que obtuvimos en la sección~\ref{caida_libre_infinito} [es decir, el límite~$\Delta t_0 \to \infty$ de~$\kappa_{\rm hor}(\Delta t_0)$ en~(\ref{kappa.HC.falling})]. También resulta interesante analizar el límite de~$\kappa_{\rm hor}(r_0,\dtw)$ cuando~$r_0 \to \infty$. Este límite es
\begin{equation}
\evat{\kappa_{\rm hor}}{r_0 \to \infty} = \fr{M},
\label{kappa.HC.infinity.radius}
\end{equation}
el cual, inesperadamente a primera vista, no depende de~$\dtw$. Es decir, no podemos reproducir la fórmula completa de~$\kappa_{\rm hor}(\Delta t_0)$ en~(\ref{kappa.HC.falling}) tomando el límite~$r_0 \to \infty$. El motivo es que la trayectoria utilizada para fijar el estado de vacío \emph{nunca} coincide, ni se acerca arbitrariamente, a la trayectoria de un observador soltado desde un radio finito, aun cuando el punto de partida~$r_0$ se tome cada vez más lejos. Aunque la diferencia de velocidades entre estas dos trayectorias en~$r_0$ tiende a anularse en ese límite, ambas llegan a posiciones cercanas al horizonte en tiempos de Schwarzschild muy diferentes. Los observadores en caída libre desde un radio finito soltados desde un punto suficientemente lejano \emph{siempre} alcanzan el horizonte en tiempos tardíos respecto a la trayectoria que define el vacío, de manera que siempre ven la radiación de Hawking casi completamente encendida, es decir, casi en el estado de Unruh final.

\begin{figure}[ht]
	\centering
    \includegraphics{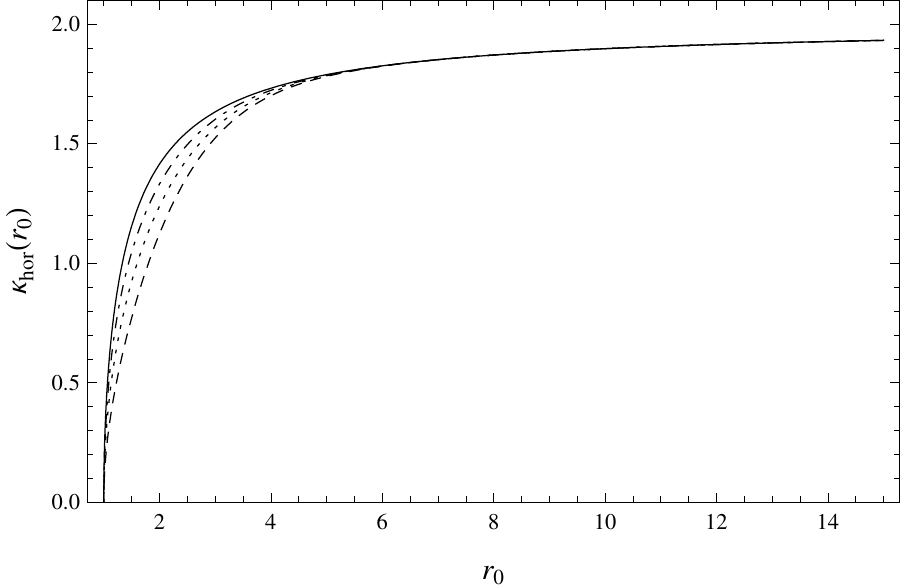}
  \caption{\footnotesize{Temperatura efectiva al cruzar el horizonte como función del radio inicial~$r_0$ para diferentes observadores con retrasos~$\dtw = (0,\ 2M,\ 5M,\ \infty)$ (línea discontinua, punteada, punto-raya y continua, respectivamente). Utilizamos unidades~$2M = 1$.}}
  \label{fig_8}
\end{figure}

Dibujando la expresión de~$\kappa_{\rm hor}(r_0, \dtw)$ en~(\ref{kappa.HC.radius}) como función de~$r_0$ para distintos valores de~$\dtw$ (figura~\ref{fig_8}), vemos que el tiempo de espera apenas tiene influencia sobre la temperatura de la radiación en el instante de cruzar el horizonte.

\subsection{Resultados numéricos}

A continuación mostraremos algunas figuras con la resolución numérica de la función~$\kappa(u)$ para distintas trayectorias. La intención es simplemente obtener una comprensión visual y cualitativa de la evolución global de la temperatura, así como comprobar que los límites calculados de forma analítica se reproducen numéricamente.

\begin{figure}[ht]
	\centering
    \includegraphics{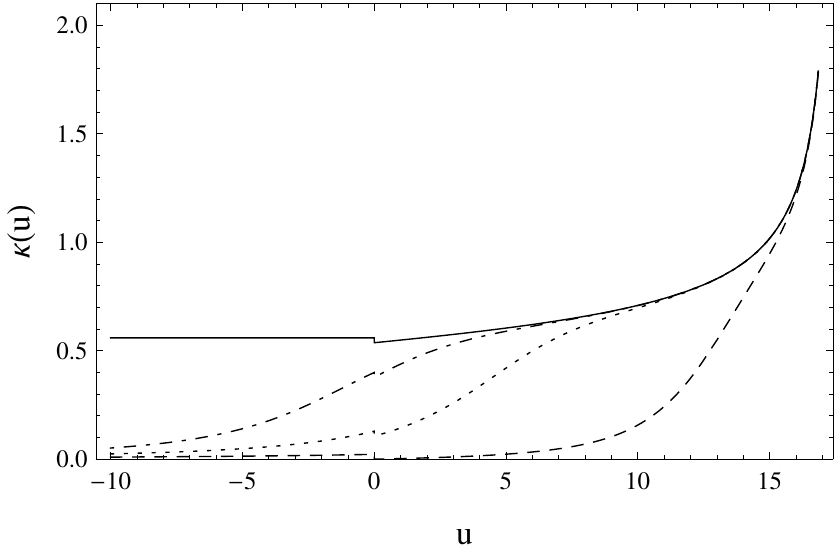}
  \caption{\footnotesize{Temperatura efectiva como función de~$u$ para diferentes observadores con radio inicial~$r_0 = 10M$ y retrasos~$\dtw = (0,\ 24M,\ 36M,\ \infty)$ (línea discontinua, punteada, punto-raya y continua, respectivamente). Utilizamos unidades~$2M = 1$ y~$u_0 = 0$.}}
  \label{fig_9}
\end{figure}

En primer lugar, mostramos el valor de~$\kappa(u)$ para observadores cuyo radio inicial es~$r_0=10M$ con diferentes tiempos de espera (figura~\ref{fig_9}). Las figuras muestran como~$\kappa(u)$ parte de valores muy diferentes cuando el observador comienza la caída, pero converge siempre a un valor casi idéntico al cruzar el horizonte, sin importar el tiempo de espera, puesto que el radio de partida es suficientemente lejano.

\begin{figure}[ht]
	\centering
    \includegraphics{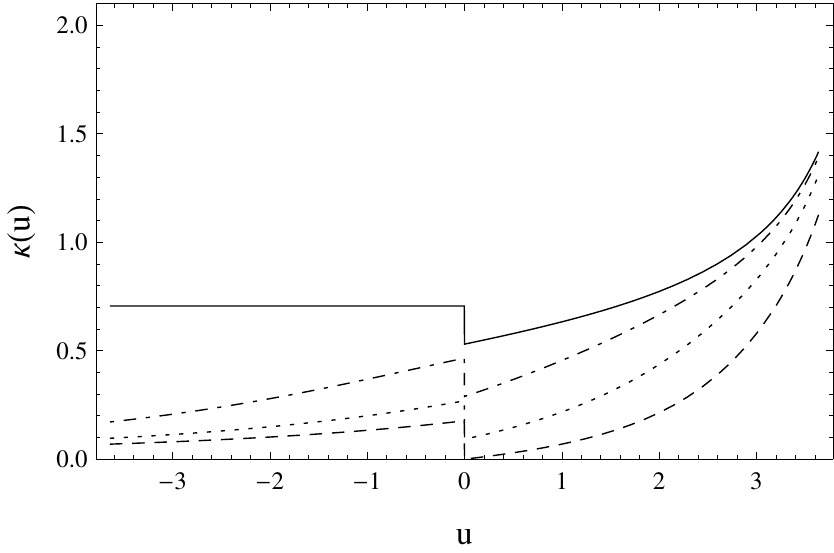}
  \caption{\footnotesize{Temperatura efectiva como función de~$u$ para diferentes observadores con radio inicial~$r_0 = 4M$ y retrasos~$\dtw = (0,\ 4M,\ 10M,\ \infty)$ (línea discontinua, punteada, punto-raya y continua, respectivamente). Utilizamos unidades~$2M = 1$ y~$u_0 = 0$.}}
  \label{fig_10}
\end{figure}

\begin{figure}[ht]
	\centering
    \includegraphics{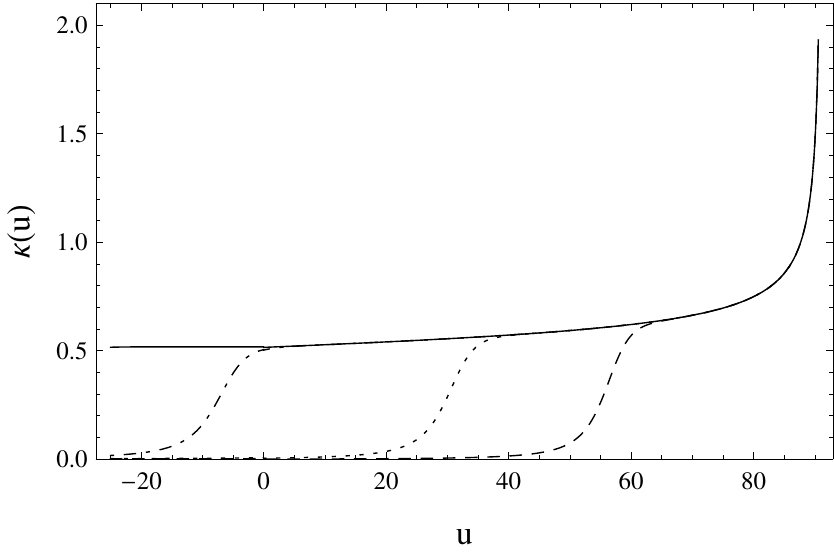}
  \caption{\footnotesize{Temperatura efectiva como función de~$u$ para diferentes observadores con radio inicial~$r_0 = 30M$ y retrasos~$\dtw = (0,\ 60M,\ 140M,\ \infty)$ (línea discontinua, punteada, punto-raya y continua, respectivamente). Utilizamos unidades~$2M = 1$ y~$u_0 = 0$.}}
  \label{fig_11}
\end{figure}

Si consideramos un radio inicial más cercano ($r_0=4M$, figura~\ref{fig_10}), se encuentra una diferencia pequeña pero apreciable en el valor de~$\kappa(u)$ al alcanzarse el horizonte de sucesos, dependiendo del tiempo de espera. También comprobamos cómo el salto en el valor de~$\kappa(u)$ al comenzar la caída es mucho más importante en este caso, puesto que la aceleración necesaria para mantenerse fijo en radios más cercanos es mucho mayor [y, al margen de interpretaciones, la ecuación~(\ref{jump.kappa}) para~$\Delta \kappa (r_0)$ nos muestra que ambas cantidades son numéricamente iguales].

Finalmente, si el radio inicial se encuentra muy alejado del horizonte ($r_0=30M$, figura~\ref{fig_11}), el salto en~$\kappa(u)$ es completamente despreciable. Los diferentes tiempos de espera solo determinan cuándo se comienza a percibir la radiación, lo cual puede suceder antes o después del comienzo de la caída libre. La meseta encontrada en la sección~\ref{caida_libre_infinito}, y por supuesto el pico final de radiación, se reproducen de nuevo, como debe suceder para radios iniciales alejados. Pero debe notarse que, como ya discutimos, no pueden reproducirse los casos de la sección~\ref{caida_libre_infinito} para los cuales~$\Delta t_0$ es pequeño. El límite~$r_0 \to \infty$ aquí da lugar al mismo resultado que el límite~$\Delta t_0 \to \infty$ en la sección~\ref{caida_libre_infinito}, sin importar el valor de~$\dtw$ aquí.

\subsection{Verificación de la condición adiabática}

\begin{figure}[ht]
	\centering
    \includegraphics{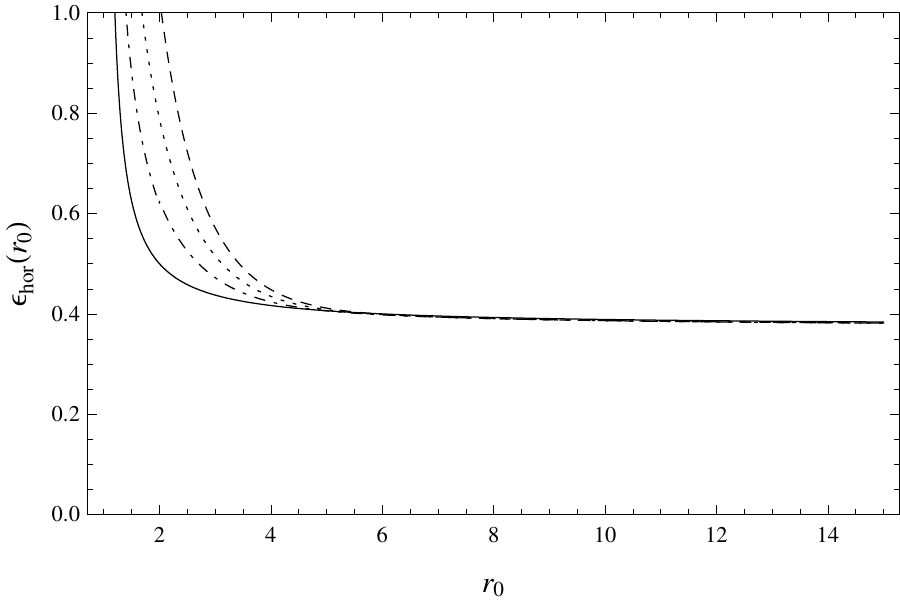}
  \caption{\footnotesize{Función de control adiabática al cruzar el horizonte como función del radio inicial~$r_0$ para diferentes observadores con retrasos~$\dtw = (0,\ 2M,\ 5M,\ \infty)$ (línea discontinua, punteada, punto-raya y continua, respectivamente). Utilizamos unidades~$2M = 1$.}}
  \label{fig_12}
\end{figure}

Repitiendo lo que hemos venido haciendo con los observadores estudiados hasta ahora, en este caso también podemos encontrar una fórmula para la función de control adiabática~$\epsilon(u)$ como función explícita de~$U(u)$ y de~$r(u)$. Y también podemos introducir la expresión explícita~$U(r)$ válida cerca del horizonte de sucesos, y encontrar una función~$\epsilon(r)$ válida allí. Finalmente, podríamos tomar el límite~$r \to 2M$ y obtener el valor~$\epsilon_{\rm hor}$ al cruzar el horizonte. Sin embargo, en este caso tanto la expresión de~$U(r)$ válida cerca del horizonte como la expresión exacta~$\epsilon(U,r)$ resultan extremadamente complejas en sus dependencias funcionales. Por tanto, no parece ni posible ni interesante encontrar una expresión explícita para~$\epsilon_{\rm hor}$ en función de los parámetros~$r_0$ y~$\dtw$. Sin embargo, sí que podemos encontrar una expresión explícita para~$\epsilon(r,r_0)$ (en cualquier punto de la trayectoria) en el caso particular del límite~$\dtw \to \infty$:
\begin{align}
\evat{\epsilon (r, r_0)}{\dtw \to \infty} = & \left[ 3 - \frac{4r}{r_0} + \left(\frac{r}{2M}\right)^2 - 4 \left(1-\frac{2M}{r_0}\right)^{1/2} \left(1-\frac{r}{r_0}\right)^{1/2} \sqrt{\frac{r}{2M}} \right]
\nonumber\\
& \times \left[ \left(\frac{r}{2M}\right)^2 -1 \right]^{-2}.
\label{epsilon.radius}
\end{align}
En esta expresión, podemos tomar~$r \to 2M$ y obtener el valor al cruzar el horizonte~$\epsilon_{\rm hor} (r_0)$ para este caso particular:
\begin{equation}
\evat{\epsilon_{\rm hor} (r_0)}{\dtw \to \infty} = \fr{8} \left[ 3 + \left( \frac{r_0}{2M} - 1 \right)^{-1} \right].
\label{epsilon.HC.radius}
\end{equation}
Este resultado toma valores desde~$3/8$ para~$r_0 \gg 2M$ hasta infinito para~$r_0 \to 2M$. Con el primer límite~$3/8$, reproducimos de nuevo el resultado obtenido en la sección~\ref{caida_libre_infinito} para tiempos largos~$\Delta t_0$ [el límite correspondiente de~$\epsilon_{\rm hor} (\Delta t_0)$ en~(\ref{epsilon.falling.EH})]. El segundo límite refleja, una vez más, el hecho de que los observadores que inician su caída desde posiciones arbitrariamente cercanas al agujero negro prácticamente no llegan a detectar radiación durante la caída. En la figura~\ref{fig_12}, se muestra la función~$\epsilon_{\rm hor} (r_0)$ en~(\ref{epsilon.HC.radius}), junto a algunas curvas obtenidas numéricamente de~$\epsilon_{\rm hor}$ para tiempos de espera~$\dtw$ nulos o finitos. Debe notarse que el límite~$3/8$ cuando~$r_0 \to \infty$ es común a todas las curvas [en consonancia con la coincidencia, ya mostrada, del valor de~$\kappa_{\rm hor}$ en tal límite, ecuación~(\ref{kappa.HC.infinity.radius})].

También podemos encontrar una expresión para~$\epsilon_{\text{caída}} (r_0)$ al comienzo de la trayectoria de caída libre, de nuevo solo para~$\dtw \to \infty$. Obtenemos
\begin{equation}
\evat{\epsilon_{\text{caída}} (r_0)}{\dtw \to \infty} = \left[ \left(\frac{r_0}{2M}\right)^2 - 1 \right]^{-1}.
\label{epsilon.init.radius}
\end{equation}
Esta expresión devuelve cero para~$r_0 \gg 2M$, e infinito para~$r_0 \to 2M$. El valor nulo aparece debido a que, para radios grandes y tiempos de espera arbitrariamente largos, lo que el observador percibe no es más que radiación de Hawking con la temperatura de Hawking constante. El valor divergente para~$r_0$ cercano al horizonte tiene la misma explicación que en el caso de~$\epsilon_{\rm hor} (r_0)$ en~(\ref{epsilon.HC.radius}): para trayectorias de caída que comienzan demasiado cerca del horizonte, la aproximación adiabática no es válida en ningún caso.

\begin{figure}[ht]
	\centering
    \includegraphics{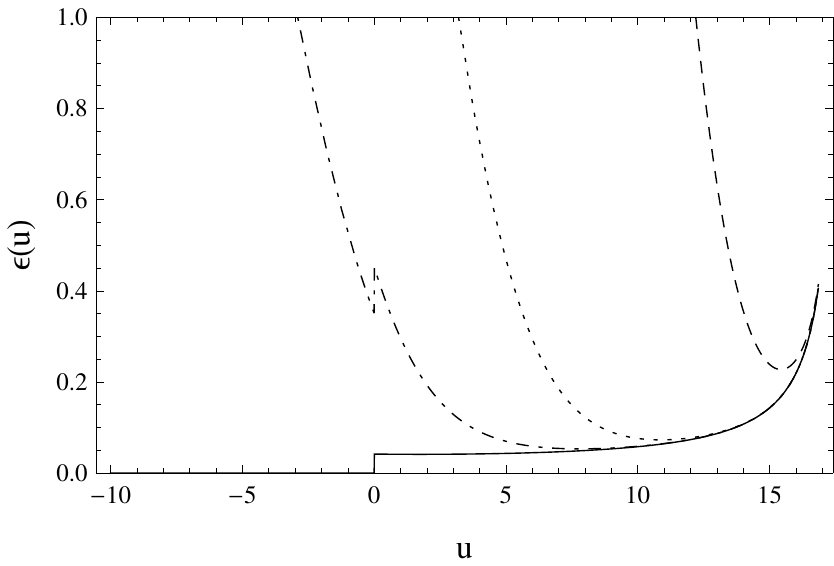}
  \caption{\footnotesize{Función de control adiabática como función de~$u$ para diferentes observadores con radio inicial~$r_0 = 10M$ y retrasos~$\dtw = (0,\ 24M,\ 36M,\ \infty)$ (línea discontinua, punteada, punto-raya y continua, respectivamente). Utilizamos unidades~$2M = 1$ y~$u_0 = 0$.}}
  \label{fig_13}
\end{figure}

\begin{figure}[ht]
	\centering
    \includegraphics{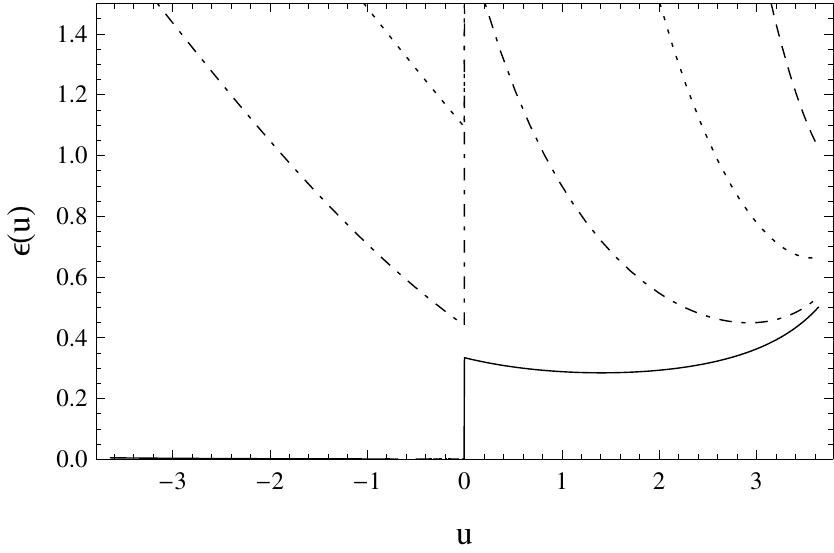}
  \caption{\footnotesize{Función de control adiabática como función de~$u$ para diferentes observadores con radio inicial~$r_0 = 4M$ y retrasos~$\dtw = (0,\ 4M,\ 10M,\ \infty)$ (línea discontinua, punteada, punto-raya y continua, respectivamente). Utilizamos unidades~$2M = 1$ y~$u_0 = 0$.}}
  \label{fig_14}
\end{figure}

\begin{figure}[ht]
	\centering
    \includegraphics{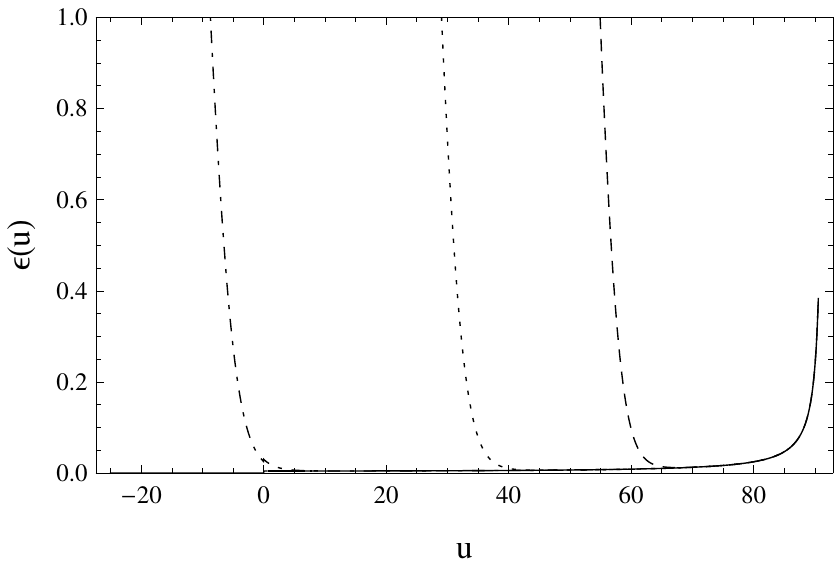}
  \caption{\footnotesize{Función de control adiabática como función de~$u$ para diferentes observadores con radio inicial~$r_0 = 30M$ y retrasos~$\dtw = (0,\ 60M,\ 140M,\ \infty)$ (línea discontinua, punteada, punto-raya y continua, respectivamente). Utilizamos unidades~$2M = 1$ y~$u_0 = 0$.}}
  \label{fig_15}
\end{figure}

Finalmente, en las figuras~\ref{fig_13}--\ref{fig_15} mostramos el valor numérico de~$\epsilon (u)$ para los mismos casos para los cuales mostramos~$\kappa(u)$. Como consecuencia del cambio brusco de~$\kappa(u)$ en el instante de iniciar la caída, las gráficas deberían exhibir un pico de tipo \emph{delta de Dirac} en ese punto, lo cual se ha omitido por simplicidad. Aparte de esto, las figuras requieren pocos comentarios. Como no puede ser de otro modo, reproducen los valores límites ya encontrados. Al igual que sucedía en la sección~\ref{caida_libre_infinito}, cuando aparece una meseta en~$\kappa(u)$ el valor de~$\epsilon(u)$ decae hasta prácticamente anularse, por lo que la radiación percibida en ese periodo es perfectamente térmica, con la temperatura determinada por~$\kappa(u)$. Por otra parte, las divergencias en tiempos tempranos solo reflejan, una vez más, que en ese límite es la propia temperatura de la radiación la que se anula.

\section{Incremento final de la temperatura efectiva}\label{sec_incremento_final}

Uno de los resultados más importantes obtenidos en el análisis de la temperatura efectiva percibida por distintos observadores es sin duda el incremento final de esta para los observadores en caída libre al cruzar el horizonte de sucesos. Tanto este resultado, como otros que hemos obtenido a lo largo de este capítulo, resultarán mucho más claros a la luz del análisis general de la función~$\kappa(u)$ que desarrollaremos en el capítulo siguiente. No obstante, ya en este punto tenemos todos los ingredientes necesarios para dar una primera explicación a este inesperado resultado.

Como ya hemos indicado en anteriores ocasiones, para tiempos de espera suficientemente largos tras el colapso, el vacío de colapso que hemos utilizado en este capítulo es indistinguible del vacío de Unruh. Es habitual encontrar en la literatura que este estado es vacío para los observadores en caída libre en el horizonte. Sin embargo, en este capítulo hemos estudiado varios observadores en caída libre, y el valor de la temperatura efectiva al cruzar el horizonte~$\kappa_{\rm hor}$ ha resultado ser no nulo para casi todos: véase el valor de~$\kappa_{\rm hor}$ para observadores en caída libre desde el infinito~(\ref{kappa.HC.falling}) y para observadores en caída libre desde un radio finito~(\ref{kappa.HC.radius}) y~(\ref{kappa.HC.Unruh.radius}). Las únicas excepciones son el observador que sigue exactamente la trayectoria del observador que define el vacío [$\Delta t_0 = 0$ en~(\ref{kappa.HC.falling})] y el observador con velocidad radial nula en el horizonte~[$r_0 = 2M$ en~(\ref{kappa.HC.radius})].

\begin{figure}[ht]
	\centering
    \includegraphics{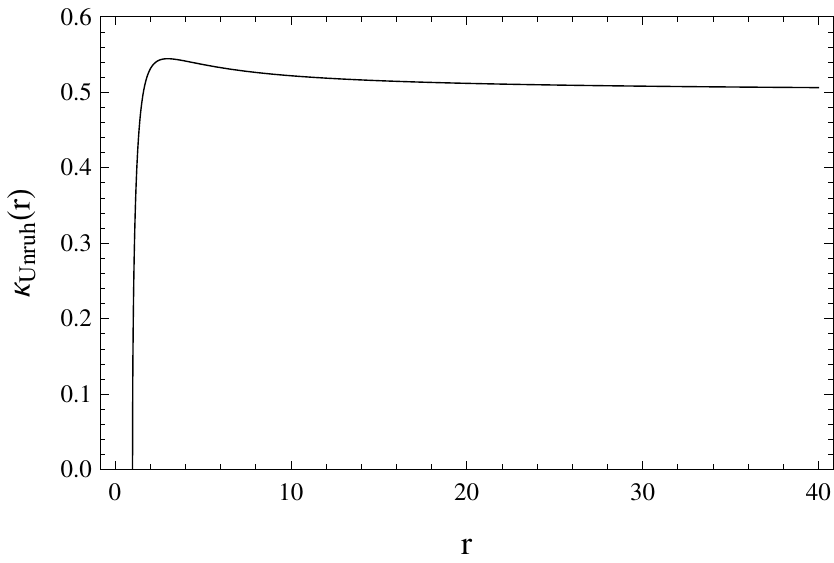}
  \caption{\footnotesize{Temperatura efectiva para el observador de Unruh en función de~$r$. Utilizamos unidades~$2M = 1$.}}
  \label{fig_16}
\end{figure}

En realidad, una vez que la radiación de Hawking está completamente encendida, solo hay una temperatura efectiva que se anula en el horizonte: la que denominamos temperatura de Unruh~(\ref{kappa.Unruh.radius}). Esta es la~$\kappa(u)$ asociada con un observador en caída libre \emph{e instantáneamente en reposo} en una posición radial~$r$:
\begin{equation}
\kappa_{\rm Unruh} (r) = \fr{4M} \left(1 +\frac{2M}{r}\right) \left(1 -\frac{2M}{r}\right)^{1/2}.
\label{kappa.Unruh.radius.r}
\end{equation}
A este observador lo denominaremos \emph{observador de Unruh.} Su temperatura efectiva~$\kappa_{\rm Unruh} (r)$ en función de la posición radial la mostramos en la figura~\ref{fig_16}. Como puede comprobarse, esta función pasa por una región en la que es creciente según nos acercamos al horizonte, hasta~$r=6M$, momento en el que decae rápidamente, anulándose en~$r=2M$.

Por otra parte, tomemos como hipótesis razonable que todo el comportamiento de la función~$\kappa(u)$ puede comprenderse en términos de las propiedades locales de la trayectoria del observador hasta segundo orden (además de la definición del estado de vacío); es decir, en términos de su posición, velocidad y aceleración.\footnote{Esta hipótesis se verá confirmada en el siguiente capítulo, y es fácil de intuir teniendo en cuenta que la definición de~$\kappa(u)$ [ecuación~(\ref{funcion_temperatura})] contiene derivadas entre las dos coordenadas nulas solo hasta segundo orden.} Bajo esta hipótesis, la única característica que distingue al observador de Unruh de cualquier otro de los observadores en caída libre es que estos últimos tienen una velocidad radial no nula. La diferencia en las velocidades radiales entre un observador en caída libre cualquiera y el observador de Unruh nos indica que sus respectivas percepciones de la radiación deben estar relacionadas mediante un factor de corrimiento Doppler. Por tanto, para obtener~$\kappa(u)$ para un observador en caída libre cualquiera a partir de~$\kappa_{\rm Unruh} (r)$, debe multiplicarse esta cantidad por un factor de corrimiento Doppler. Además, el observador de Unruh es tal que, en el límite del horizonte de sucesos, se encuentra instantáneamente viajando con la geodésica nula saliente en ese punto (que genera el horizonte). Por tanto, el factor de corrimiento Doppler será, necesariamente, divergente al azul en el horizonte. Comprobaremos que esta divergencia compensa el valor nulo de~$\kappa_{\rm Unruh} (r)$ en~$r=2M$, dando lugar a un resultado final finito (en general) para~$\kappa_{\rm hor}$ visto por otro observador en caída libre.

Veámoslo con más detalle. La forma precisa de este factor Doppler viene dada por la expresión
\begin{equation}
D_{r_0}(r):=\left(\deriv{r_{\rm l}}{t} - \deriv{r_{r_0}}{t}\right)^{1/2}  \left( \deriv{r_{\rm l}}{t} + \deriv{r_{r_0}}{t} \right)^{-1/2},
\label{doppler.kappa}
\end{equation}
donde~$r_{\rm l}(t)$ es la trayectoria de una geodésica nula saliente y~$r_{r_0}(t)$ la trayectoria en caída libre que comienza con velocidad cero en~$r=r_0$. Hemos indicado también con el subíndice~$r_0$ los distintos factores Doppler asociados a observadores en caída libre lanzados desde distintas distancias radiales~$r_0$. La forma concreta de este factor tiene una clara justificación, ya que un factor Doppler físico debe comparar la velocidad de objetos masivos con la velocidad de la luz:
\begin{equation}
D= \sqrt{\frac{c - v}{c + v}}.
\label{doppler.generic}
\end{equation}
En el caso que nos ocupa, debe notarse que en las coordenadas de Schwarzschild~$(t,r)$ la velocidad de la luz es distinta de la unidad. Este es el motivo de incluir~$\rmd r_{\rm l} / \rmd t$ en~(\ref{doppler.kappa}). Otra forma de escribir este factor es
\begin{equation}
D_{r_0}(r)=\left(1 - \deriv{r_{r_0}}{r_{\rm l}}\right)^{1/2} \left( 1 + 
\deriv{r_{r_0}}{r_{\rm l}} \right)^{-1/2},
\label{doppler.rel}
\end{equation}
donde se observa claramente que lo que se compara es la velocidad local de los observadores con la velocidad de la luz. La velocidad~$\rmd r_{r_0} / \rmd t$ la obtenemos de la ecuación de la trayectoria~(\ref{dr.of.t.radius}), en tanto que~${\rmd r_{\rm l}} / {\rmd t}$ puede calcularse de la ecuación para los rayos salientes~$t-r^* (r_{\rm l}) = {\rm const}$:
\begin{equation}
\evat{\deriv{r_{\rm l}}{t}}{r_{\rm l}=r} = \deriv{r}{r^*}= 1-\frac{2M}{r}.
\label{doppler.loc}
\end{equation}
Tras algunas manipulaciones algebraicas sencillas, se obtiene
\begin{align}
D_{r_0}(r) = & \left[ \left(1 -\frac{2M}{r_0}\right)^{1/2} + \sqrt{\frac{2M}{r}}\left(1 -{\frac{r}{r_0}}\right)^{1/2} \right]^{1/2}
\nonumber\\
& \times \left[ \left(1 -\frac{2M}{r_0}\right)^{1/2} - \sqrt{\frac{2M}{r}}\left(1 -{\frac{r}{r_0}}\right)^{1/2}
\right]^{-1/2}.
\label{doppler.expression}
\end{align}

Si a continuación se calcula~$\kappa_{\rm Unruh}(r) D_{r_0}(r)$ y se toma el límite~$r \to 2M$, puede comprobarse que se obtiene exactamente el resultado~$\kappa_{\rm hor} (r_0)$ en~(\ref{kappa.HC.Unruh.radius}). Por tanto, podemos concluir que el incremento final de la temperatura efectiva se debe a un factor Doppler divergente en el límite del horizonte, el cual ``compite'' con la tendencia a cero de la temperatura de Unruh en ese mismo límite.\footnote{Con un análisis muy similar se reproducen también los valores de~$\kappa_{\rm hor}$ para otros observadores obtenidos en~(\ref{kappa.HC.falling}) y~(\ref{kappa.HC.radius}). En tales casos, el factor Doppler es el mismo (con~$r_0 \to \infty$ en el primer caso), pero ya no compite con~$\kappa_{\rm Unruh}(r)$, sino con una cantidad más rápidamente decreciente en el límite~$r \to 2M$, lo cual se debe a que estos observadores no han esperado a que la radiación de Hawking se encuentre completamente encendida (en el vacío de Unruh).}

\section{¿Algo más que una aproximación?}

En la sección~\ref{sec_kappa}, introdujimos la función de temperatura efectiva~(\ref{funcion_temperatura}) como herramienta especialmente sencilla para estudiar la percepción de radiación en Teoría Cuántica de Campos en espacios curvos, y demostramos que bajo cumplimiento de la condición adiabática su valor se corresponde con el de la temperatura del espectro de Planck que percibe un determinado observador en un estado de vacío dado. La aplicación de esta función al estudio de los diversos ejemplos de este capítulo es ya una muestra de su potencial para tal fin, lo cual supone un valor por sí mismo. En~\cite{Smerlak:2013sga,Smerlak:2013cha} se han realizado con posterioridad estudios de algunos de estos ejemplos y de otros diferentes con un procedimiento análogo, cuyos resultados confirman los aquí expuestos.

No obstante, en principio podríamos haber propuesto otras formas de estimar la temperatura de la radiación percibida por un observador (cuando pudiera hablarse de tal). Por ejemplo, podríamos haber considerado un procedimiento distinto para aproximar la relación~$U(u)$ localmente mediante una exponencial, de manera que los parámetros de dicha exponencial hubieran sido, en general, ligeramente diferentes; o podríamos haber desarrollado formalmente el espectro realmente percibido para una relación~$U(u)$ genérica, y haber buscado el ajuste óptimo a una planckiana mediante algún algoritmo de ajuste (uno de los muchos existentes). Es de esperar que los resultados no difiriesen cualitativamente de los encontrados con la función de temperatura efectiva (que, desde esta perspectiva, no sería sino un procedimiento de aproximación más), y sin duda serían prácticamente idénticos en los casos en los que se cumple la condición adiabática.

Sin embargo, en este capítulo hemos visto claramente que algunos de los resultados concretos para~$\kappa(u)$ hemos podido explicarlos \emph{de forma analítica exacta} a partir de fenómenos físicos conocidos, como el efecto Doppler o el corrimiento al azul gravitacional, \emph{incluso en casos en los que no se cumple estrictamente la condición adiabática} que permitiría entender~$\kappa(u)$ como la temperatura aproximada del espectro térmico percibido. Esto nos lleva a pensar que esta función es mucho más que ``un número'' que nos da la temperatura del espectro térmico que ``mejor se aproxima'', en algún sentido numérico de aproximación, al espectro real, y en general no térmico, que percibe el observador considerado en el vacío considerado.

Otra idea que también apunta hacia una interpretación más profunda de~$\kappa(u)$ tiene que ver con la expansión adiabática realizada en el capítulo~\ref{adiabatica}. Es evidente que también puede definirse una función de temperatura efectiva para el espacio-tiempo de Minkowski, con el campo de radiación en el estado de vacío de Minkowski (cosa que haremos en el siguiente capítulo). En tal caso, no es difícil intuir, ni tampoco comprobar, que se obtiene~$\kappa(u) = g(u)$, es decir, que la función de temperatura efectiva no es más que la aceleración propia del observador. Y, como vimos, la contribución de orden cero a la expansión adiabática~(\ref{funcion_respuesta_expansion}) es precisamente el espectro térmico con temperatura proporcional a~$|g(u)|$, siendo el resto de contribuciones \emph{agregadas} a esta. Por tanto, es razonable pensar que la función~$\kappa(u)$ nos indica la temperatura de una radiación que, de alguna forma, \emph{siempre está ahí,} y que tiene su explicación física propia, aun cuando su papel en el espectro total percibido queda eclipsado por contribuciones con espectros diferentes, debidas a la no adiabaticidad.

Estas cuestiones nos han llevado a buscar una explicación física cerrada y general para el valor de la función de temperatura efectiva. En el siguiente capítulo mostramos que tal explicación existe, así como la forma en que la hemos hallado, un análisis detallado de la misma, y su aplicación a diversos escenarios físicos de interés.
\newpage
\thispagestyle{empty}
\hbox{}

\makeatletter
\def\cleardoublepage{\clearpage\if@twoside \ifodd\c@page\else
    \hbox{}
    \thispagestyle{empty}
    \newpage
    \if@twocolumn\hbox{}\newpage\fi\fi\fi}
\makeatother \clearpage{\pagestyle{empty}\cleardoublepage}

\chapter{Interpretación física de la función de temperatura efectiva} 
\label{fisica}

Como ya hemos adelantado, en este capítulo obtendremos una expresión general de la función de temperatura efectiva como función de las propiedades locales de la trayectoria del observador y del estado de vacío considerados. Al igual que en el capítulo anterior, trabajaremos en el sector radial del espacio-tiempo de Schwarzschild (salvo en la sección~\ref{sec_minkowski}), y nos centraremos en la radiación saliente (salvo en la sección~\ref{sec_entrante}). Veremos que una parte de esta expresión tiene una interpretación clara en términos de fenómenos físicos conocidos, en tanto que otra parte, menos evidente en un primer análisis, nos lleva a reflexionar acerca de la relación e interacción entre los dos fenómenos involucrados en la percepción de radiación de tipo térmico por parte del observador: la radiación de Hawking y el efecto Unruh. Finalmente, haremos uso de la expresión general obtenida para analizar algunos escenarios físicos concretos, y en particular propondremos un escenario realista de flotación de un objeto en las cercanías de un agujero negro debido a la radiación de Hawking.

\section{La función de temperatura efectiva del estado}

A lo largo de los capítulos anteriores se ha podido comprobar cuáles son los ingredientes para construir~$\kappa(u)$: el estado de vacío en el que se trabaja, fijado por la coordenada nula~$U$ asociada a los modos normales que definen tal vacío; el observador cuya percepción se desea conocer, descrito por la coordenada nula~$u$ asociada a los modos normales naturales para él; y la relación~$U(u)$.

Por supuesto, el estado de vacío puede fijarse sin necesidad de conocer el observador. Para ello, basta definir la coordenada nula~$U$ como función de otra coordenada nula de referencia con significado geométrico unívoco en el espacio-tiempo considerado (salvo simetrías). Dada su estaticidad, en el espacio-tiempo de Schwarzschild es evidente que resulta especialmente conveniente escoger la coordenada nula de Eddington-Finkelstein~$\bar{u}:= t- r^*$ como referencia. Tal cosa es lo que se hizo en la sección~\ref{fijado_vacio} para definir el que denominamos vacío de colapso. Por supuesto, no es la única coordenada de referencia posible, puesto que podríamos haber utilizado cualquier otra coordenada nula definida en todo el sector exterior de Schwarzschild. Sin embargo, como comprobaremos, no se trata solo de una elección de coordenada de referencia conveniente por simplicidad de cálculo: al ser una coordenada afín en la región asintótica, su uso hará que las cantidades calculadas tengan una clara interpretación física.

En el caso de un estado de vacío genérico, este quedará fijado por la relación~$U(\bar{u})$. Una vez escogido el estado de vacío, la función de temperatura efectiva [definida en~(\ref{funcion_temperatura})] para un observador cualquiera puede reescribirse como
\begin{align}
\kappa (u) = & \ -\hfrac{\deriv[2]{U}{u}}{\deriv{U}{u}} = \left(-\hfrac{\deriv[2]{U}{\ub}}{\deriv{U}{\ub}}\right) \deriv{\ub}{u} - \hfrac{\deriv[2]{\ub}{u}}{\deriv{\ub}{u}} \nonumber \\
= & \ \deriv{\ub}{u} \ks(\ub) - \hfrac{\deriv[2]{\ub}{u}}{\deriv{\ub}{u}},
\label{kappa_rew}
\end{align}
donde
\begin{equation}
\ks(\ub) := -\hfrac{\deriv[2]{U}{\ub}}{\deriv{U}{\ub}}
\label{ks_def}
\end{equation}
es lo que denominaremos \emph{función de temperatura efectiva del estado.} Es la parte de la expresión~(\ref{kappa_rew}) que contiene la información sobre la elección del estado de vacío. El resto no depende de la relación~$U(\ub)$, sino únicamente de la trayectoria del observador, descrita por la relación~$\bar{u}(u)$. Por otra parte, la definición de~$\bar{\kappa} (\bar{u})$ coincide exactamente con la definición de~$\kappa(u)$ original si identificamos~$u = \bar{u}$, es decir, si consideramos que elegimos como observadores los que se encuentran en reposo en el infinito espacial. Por tanto, $\bar{\kappa} (\bar{u})$ nos indica la función de temperatura efectiva de la radiación saliente para tales observadores. Además, al tratarse de los observadores en reposo en la región asintótica, podemos identificar tal radiación saliente como la efectivamente \emph{emitida} por el agujero negro, la cual escapa del sistema y extrae energía de este, evaporándolo. También cabe destacar que, por depender únicamente de~$\bar{u}$, se trata de radiación que se propaga con los rayos de luz salientes (es constante para cada rayo de luz saliente), lo cual resulta relevante en el caso de estados de vacío no estacionarios (como el vacío de colapso).

\begin{figure}[ht]
	\centering
    \includegraphics{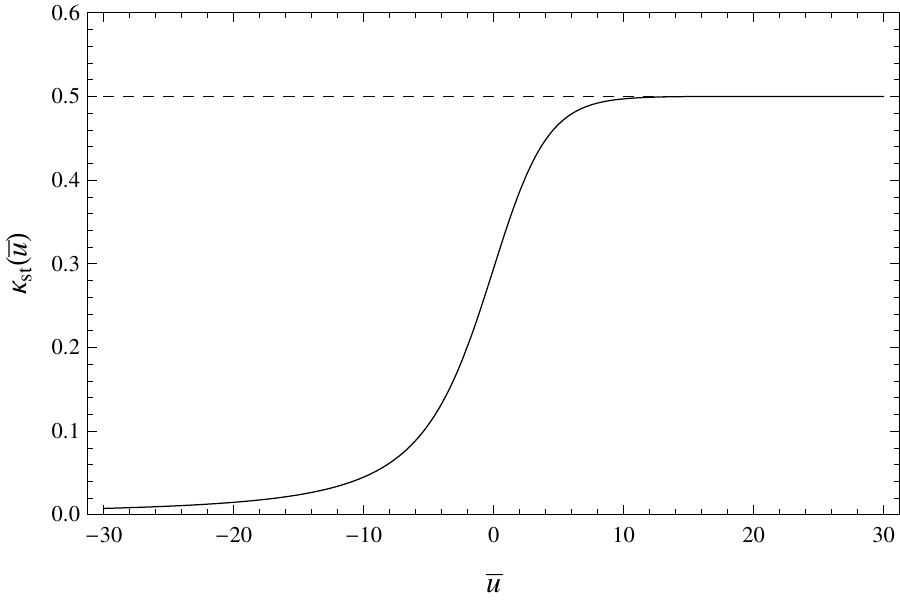}
  \caption{\footnotesize{Valor de~$\ks(\ub)$ en~(\ref{non-stat}) para el vacío de colapso (línea continua), y de~$\ks(\ub) = 1/(4M)$ para el vacío de Unruh (línea discontinua). Utilizamos unidades~$2M = 1$.}}
  \label{fig_unica}
\end{figure}

Analicemos la forma de esta función para diferentes elecciones del estado de vacío. Para el vacío de Boulware~\cite{Boulware:1974dm} tenemos que~$U(\bar{u}) = \bar{u}$, y por tanto~$\ks(\bar{u}) = 0$, lo cual significa que en este vacío no hay radiación emitida hacia el infinito espacial. Para el vacío de Unruh~\cite{Unruh:1976db}, la relación es~$U(\bar{u}) = - 4 M \exp[-\bar{u} / (4M)]$, de forma que~$\ks (\bar{u}) = 1/(4M)$, por lo que tenemos radiación de Hawking emitida hacia el infinito espacial. En el caso del vacío de colapso, de la relación implícita~$U(\bar{u})$ que puede extraerse de~(\ref{ub.of.U}) no es difícil obtener
\begin{equation}
\ks(\ub) = \fr{4M \left[\frac{3}{4M}\left(U_{\rm H} - U(\ub)\right) + 1 \right]^{4/3}},
\label{non-stat}
\end{equation}
resultado que queda a su vez en función de~$U(\bar{u})$. Una vez más, podemos comprobar que este vacío, en el pasado remoto ($U \to -\infty$), coincide con el vacío de Boulware~$\ks(\ub) \to 0$; mientras que en el futuro remoto ($U \to U_{\rm H}$) coincide con el vacío de Unruh~$\ks(\ub) \to 1/(4M)$. La interpolación entre ambos valores se muestra en la figura~\ref{fig_unica}.

\section{Expresión analítica general de la función}

Consideremos un observador cualquiera~$\mathcal{O}$ que sigue una trayectoria arbitraria~$(t(u),r(u))$ en el exterior del agujero negro, siendo~$u$ su tiempo propio. El resultado central de este capítulo es que el valor de la función de temperatura efectiva para el observador~$\mathcal{O}$ a lo largo de su trayectoria puede expresarse como
\begin{equation}
\kappa (u) = \sqrt{\frac{1-\vl}{1+\vl}} \fr{\sqrt{1-\frac{2M}{r}}} \left(\ks(\ub)-\frac{M}{r^2}\right) + \apr.
\label{local_kappa}
\end{equation}
En esta expresión, $r=r(u)$ y~$\ub = \ub (u)$; la función~$\vl = \vl(u)$ es la velocidad del observador~$\mathcal{O}$ con respecto al agujero negro \emph{medida por un observador localmente Minkowskiano,} y la función~$\apr = \apr(u)$ es la \emph{aceleración propia} del observador~$\mathcal{O}$, es decir, la aceleración medida por un observador \emph{en caída libre e instantáneamente co-móvil con el observador~$\mathcal{O}$.}

Para mostrar que la expresión~(\ref{local_kappa}) es correcta, en primer lugar describiremos la componente radial de la trayectoria~$r(u)$ localmente en torno a un instante cualquiera~$u=u_0$. Hasta segundo orden en~$u$, la trayectoria del observador~$\mathcal{O}$ puede describirse como
\begin{equation}
r(u) = r_0 + v_0 (u-u_0) + \ahalf a_0 (u-u_0)^2 + O(u-u_0)^3,
\label{local_traj}
\end{equation}
donde
\begin{equation}
r_0 := r(u_0), \quad v_0 := \derivateq{r}{u}{u_0}, \quad a_0 := \derivateq[2]{r}{u}{u_0}.
\label{local_param}
\end{equation}

Utilizando esta descripción local de la trayectoria y el elemento de línea del sector radial de la métrica de Schwarzschild~(\ref{schwarzschild_una_dimension}), pueden obtenerse las cantidades~$\rmd r / \rmd u$, $\rmd^2 r / \rmd u^2$, $\rmd t / \rmd u$ y $\rmd^2 t / \rmd u^2$ como funciones locales de~$u$ en torno a~$u=u_0$; con estas cantidades, y la definición de la coordenada nula de Eddington-Finkelstein~$\bar{u} := t - r^*$, se obtienen~$\rmd \bar{u} / \rmd u$ y~$\rmd^2 \bar{u} / \rmd u^2$; y, finalmente, de la expresión de~$\kappa(u)$ en~(\ref{kappa_rew}) se tiene
\begin{align}
\kappa (u) = & \ \fr{\sqrt{1-\frac{2M}{r_0}}} \left(\frac{\sqrt{1-\frac{2M}{r_0}+v_0^2}-v_0}{\sqrt{1-\frac{2M}{r_0}+v_0^2}+v_0}\right)^{1/2} \left(\ks(\ub(u_0))-\frac{M}{r_0^2}\right) \nonumber \\
& + \fr{\sqrt{1-\frac{2M}{r_0}+v_0^2}} \left(a_0 + \frac{M}{r_0^2} \right) + O(u-u_0).
\label{kappa_alt}
\end{align}
Es fácil comprobar que si identificamos
\begin{equation}
\vl := \frac{v_0}{\sqrt{1-\frac{2M}{r_0}+v_0^2}}
\label{vs_def}
\end{equation}
y
\begin{equation}
\apr := \fr{\sqrt{1-\frac{2M}{r_0}+v_0^2}} \left(a_0 + \frac{M}{r_0^2} \right),
\label{af_def}
\end{equation}
entonces~(\ref{kappa_alt}) es precisamente la expresión~(\ref{local_kappa}) que buscamos, evaluada en~$u=u_0$. Sin embargo, dado que~$u_0$ es un instante \emph{arbitrario} de la trayectoria, concluimos que la expresión es válida a lo largo de toda la trayectoria del observador~$\mathcal{O}$. Por tanto, para completar la demostración solo queda comprobar que las definiciones~(\ref{vs_def}) y~(\ref{af_def}) se corresponden con las cantidades físicas correctas (velocidad local y aceleración propia) en~$u=u_0$, lo que haremos a continuación.

Dada una métrica, siempre existe un sistema de coordenadas tal que en un punto concreto, pero arbitrario, la métrica resulta ser localmente Minkowski. Consideremos el punto~$(t_0 := t(u_0), r_0)$ en el sector radial de la métrica de Schwarzschild. La transformación
\begin{equation}
\left( \begin{array}{c}
	\rmd \tl \\ \rmd \rl
\end{array} \right)
=T \left( \begin{array}{c}
	\rmd t \\ \rmd r
\end{array} \right)
\label{static}
\end{equation}
entre un sistema de coordenadas localmente Minkowski~$(t_l,r_l)$ y el sistema de coordenadas de Schwarzschild~$(t,r)$ debe satisfacer
\begin{equation}
\evat{T}{u=u_0} =
\left( \begin{array}{cc}
	\gamma & -\gamma \vl \\ -\gamma \vl & \gamma
\end{array} \right)
\left( \begin{array}{cc}
	\sqrt{1-\frac{2M}{r_0}} & 0 \\ 0 & \frac{1}{\sqrt{1-\frac{2M}{r_0}}}
\end{array} \right),
\label{first_order}
\end{equation}
donde $\gamma:=1/\sqrt{1-\vl^2}$ es el factor de \emph{boost} relativista. Aquí $v_l$ representa la velocidad del sistema de referencia con respecto al agujero negro, medida por un observador localmente Minkowskiano (debe tenerse en cuenta que en coordenadas de Schwarzschild la estaticidad del espacio-tiempo es explícita).

Consideremos ahora que este sistema de referencia es co-móvil con la trayectoria~$r(u)$ del observador~$\mathcal{O}$. En tal caso, tenemos que
\begin{equation}
0=\derivat{\rl}{\tl}{\mathcal{O}, u=u_0} = \frac{ v_0 -\vl \left(1 -\frac{2M}{r_0}\right)\derivat{t}{u}{u=u_0}}{\left(1 -\frac{2M}{r_0}\right)\derivat{t}{u}{u=u_0} -\vl v_0},
\label{comoving}
\end{equation}
donde se ha indicado explícitamente que la derivada se hace a lo largo de la trayectoria del observador~$\mathcal{O}$. Directamente del elemento de línea del sector radial de Schwarzschild~(\ref{schwarzschild_una_dimension}), sabemos que
\begin{equation}
\derivat{t}{u}{u=u_0}=\frac{\sqrt{1 -\frac{2M}{r_0}+v_0^2}}{1 -\frac{2M}{r_0}}.
\end{equation}
Por tanto, de~(\ref{comoving}) podemos despejar
\begin{equation}
\vl = \frac{v_0}{\sqrt{1-\frac{2M}{r_0}+v_0^2}},
\label{vel_static}
\end{equation}
que es la expresión~(\ref{vs_def}) que buscábamos.

Una vez hemos calculado el orden cero~$T|_{u=u_0}$ de la expansión en serie de Taylor de la transformación~$T$, calcularemos el primer orden (descrito por las segundas derivadas del cambio de coordenadas evaluadas en~$u=u_0$; los términos adicionales en la expansión son irrelevantes para el cálculo), con el fin de implementar el carácter geodésico (en caída libre) de las coordenadas definidas localmente. Por conveniencia en la notación, etiquetaremos temporalmente~$(t,r) =: (x^0, x^1)$ y~$(t_l, r_l) =: (x_l^0,x_l^1)$. Si~$\{\xl^i,\,i=0,1\}$ es un sistema de coordenadas localmente en caída libre, las ecuaciones de las geodésicas \emph{radiales} escrita en estas coordenadas debe escribirse localmente
\begin{equation}
\derivat[2]{\xl^i}{s}{u=u_0} = 0.
\label{geod_xi}
\end{equation}
Pasando al sistema de coordenadas de Schwarzschild original~$\{x^i,\,i=0,1\}$, estas ecuaciones se escriben
\begin{equation}
\deriv[2]{x^i}{s} + \pderiv{x^i}{\xl^n} \frac{\partial^2 \xl^n}{\partial x^j \partial x^k} \deriv{x^j}{s} \deriv{x^k}{s} = 0.
\label{geod_mix}
\end{equation}
Pero sabemos que las ecuaciones de las geodésicas radiales en estas coordenadas tienen la forma
\begin{equation}
\deriv[2]{x^i}{s} + \Gamma^i_{j k} \deriv{x^j}{s} \deriv{x^k}{s} = 0,
\label{geod_x}
\end{equation}
donde~$\Gamma^i_{j k}$ son los \emph{símbolos de Christoffel} de la métrica de Schwarzschild~(\ref{schwarzschild}). Por tanto, podemos identificar
\begin{equation}
\pderiv{x^i}{\xl^n} \frac{\partial^2 \xl^n}{\partial x^j \partial x^k} = \Gamma^i_{j k},
\label{chris_eq}
\end{equation}
y, resolviendo para las derivadas de segundo orden de la transformación de coordenadas,
\begin{equation}
\frac{\partial^2 \xl^i}{\partial x^j \partial x^k} = \pderiv{\xl^i}{x^n} \Gamma^n_{j k}.
\label{sec_deriv_eq}
\end{equation}
Las derivadas de primer orden pueden obtenerse de la transformación a orden cero en~(\ref{first_order}), y los símbolos de Christoffel pueden calcularse utilizando la métrica de Schwarzschild~(\ref{schwarzschild}), obteniéndose
\begin{align}
\Gamma^r_{r r} = & \ -\Gamma^t_{r t} = -\Gamma^t_{t r} =   - \frac{M}{r_0^2} \fr{1-\frac{2M}{r_0}}, \nonumber  \\
\Gamma^r_{t t} = & \ \frac{M}{r_0^2} \left(1-\frac{2M}{r_0}\right),
\label{christoffel_dos}
\end{align}
siendo el resto nulos (debe recordarse que todas las expresiones tienen que considerarse evaluadas en~$u=u_0$). De esta forma, las derivadas de segundo orden de la transformación de coordenadas son localmente
\begin{align}
\pderiv[2]{\rl}{r} = & \ -\frac{M}{r_0^2} \frac{\sqrt{1-\frac{2M}{r_0}+v_0^2}}{(1-\frac{2M}{r_0})^2}, \quad & \pderiv[2]{\tl}{r} = & \ \frac{M}{r_0^2} \frac{v_0}{(1-\frac{2M}{r_0})^2}, \nonumber  \\
\frac{\partial^2 \rl}{\partial r \partial t} = & \  -\frac{M}{r_0^2} \frac{v_0}{1-\frac{2M}{r_0}}, & \frac{\partial^2 \tl}{\partial r \partial t} = & \  \frac{M}{r_0^2} \frac{\sqrt{1-\frac{2M}{r_0}+v_0^2}}{1-\frac{2M}{r_0}}, \nonumber  \\
\pderiv[2]{\rl}{t} = & \  \frac{M}{r_0^2} \sqrt{1-\frac{2M}{r_0}+v_0^2}, & \pderiv[2]{\tl}{t} = & \ -\frac{M}{r_0^2} v_0.
\label{sec_deriv}
\end{align}

Una vez descrita la transformación de coordenadas hasta primer orden, podemos calcular la aceleración del observador~$\mathcal{O}$ medida en el sistema de coordenadas~$(t_l,r_l)$. En primer lugar, usando la transformación a orden cero en~(\ref{first_order}) puede comprobarse que
\begin{equation}
\evat{\left(\deriv{\tl}{u},\deriv{\rl}{u}\right)}{u=u_0} = (1,0),
\label{com_cond}
\end{equation}
como no puede ser de otra forma, puesto que hemos impuesto que se trate de un sistema de coordenadas co-móvil con el observador~$\mathcal{O}$. Esto simplifica la expresión de la aceleración a
\begin{align}
\derivat[2]{\rl}{\tl}{\mathcal{O}, u=u_0} = \derivat[2]{\rl}{u}{u=u_0} = & \ \left[ \pderiv{\rl}{r} \deriv[2]{r}{u} + \pderiv[2]{\rl}{r} \left(\deriv{r}{u}\right)^2 + \pderiv{\rl}{t} \deriv[2]{t}{u} \right. \nonumber \\
& \left. + \pderiv[2]{\rl}{t} \left(\deriv{t}{u}\right)^2 + 2 \frac{\partial^2 \rl}{\partial r \partial t} \deriv{r}{u} \deriv{t}{u} \right]_{u=u_0}.
\label{acc_step}
\end{align}
De la evaluación de esta expresión (donde todas las cantidades son conocidas) se obtiene
\begin{equation}
\derivat[2]{\rl}{\tl}{\mathcal{O}, u=u_0} = \fr{\sqrt{1-\frac{2M}{r_0}+v_0^2}} \left(a_0 + \frac{M}{r_0^2} \right) = \apr,
\label{acc}
\end{equation}
con lo que queda demostrado que~$\apr$ dada por~(\ref{af_def}) es la aceleración del observador~$\mathcal{O}$ medida en un sistema de referencia en caída libre e instantáneamente co-móvil con él; es decir, su aceleración propia. De esta forma, hemos justificado el carácter físico de las cantidades~$v_l$ y~$a_{\rm p}$.

\section{Interpretación física. Hawking versus Unruh}\label{hawking_versus_unruh}

La interpretación física de la expresión~(\ref{local_kappa}) para la temperatura efectiva resulta en gran parte evidente. En cualquier caso, vamos a separarla en los tres términos que aparecen. En primer lugar, tenemos el término
\begin{equation}
\sqrt{\frac{1-\vl}{1+\vl}} \fr{\sqrt{1-\frac{2M}{r}}} \ks(\ub).
\label{termino_hawking}
\end{equation}
Este término corresponde a la radiación que escapa del agujero negro, dada por~$\ks (\ub)$, a lo largo del rayo saliente con el que el observador~$\mathcal{O}$ se cruza en cada momento~$\ub=\ub(u)$, y corregida por dos factores. Por una parte, encontramos el factor de corrimiento gravitacional~$1/ \sqrt{1 - 2M/r}$ en la posición del observador~$\mathcal{O}$, que aparece debido a que la radiación sufre un corrimiento al rojo según escapa del agujero negro, lo que resulta en un corrimiento al azul de la radiación percibida por observadores cercanos con respecto a la percibida por los que se encuentran en el infinito espacial. Por otra parte, tenemos el factor de corrimiento Doppler~$\sqrt{(1-v_l)/(1+v_l)}$. La velocidad que aparece en este factor es~$v_l$, que es precisamente la velocidad del observador~$\mathcal{O}$ con respecto al agujero negro, medida en un sistema de referencia localmente Minkowskiano, en el cual la velocidad de la luz es igual a~$1$, y por tanto el factor de corrimiento Doppler toma esta forma sencilla.

El segundo término con una interpretación física aparentemente inmediata es el sumando~$a_{\rm p}$. Se trata de la aceleración propia del observador~$\mathcal{O}$. Estando medida por un observador en caída libre e instantáneamente co-móvil con~$\mathcal{O}$, es la aceleración que el observador~$\mathcal{O}$ experimentará cada vez que ``encienda motores'' para modificar su trayectoria. Este término sugiere una interpretación muy extendida: se trata de la contribución del efecto Unruh, producido por la aceleración propia del observador~$\mathcal{O}$, a la percepción de radiación térmica del campo.

Sin embargo, el término que nos falta por considerar, dado por
\begin{equation}
\sqrt{\frac{1-\vl}{1+\vl}} \fr{\sqrt{1-\frac{2M}{r}}} \left(-\frac{M}{r^2}\right),
\label{termino_raro}
\end{equation}
también desempeña un papel en esta interpretación. En este término aparece la ``aceleración de la gravedad''~$-M/r^2$ multiplicada por los mismos factores que aparecen para la radiación emitida por el agujero negro en~(\ref{termino_hawking}). Si aceptásemos la interpretación, a primera vista más evidente, de~$a_{\rm p}$ como \emph{única} contribución del efecto Unruh entonces, por eliminación, este último término debería ser una corrección propia de la radiación del agujero negro, es decir, una corrección a~(\ref{termino_hawking}). Se trataría, en tal caso, del término responsable de una de las peculiaridades conocidas de la radiación de Hawking: que no es percibida por los observadores en caída libre y, según vimos en el capítulo anterior, instantáneamente en reposo cercanos al horizonte~\cite{Unruh:1976db}. En efecto, si ignoramos el factor Doppler, considerando observadores instantáneamente en reposo y en caída libre, cuando la radiación de Hawking está presente se tiene que~$\ks (\ub) = 1/(4M)$, y en el horizonte~$M/r^2 \to 1/(4M)$, por lo que los términos~(\ref{termino_hawking}) y~(\ref{termino_raro}) en principio tienden a cancelarse entre sí. Puede comprobarse que, aunque el factor de corrimiento gravitacional es divergente en el horizonte, dicha divergencia no es suficientemente rápida, y la cantidad
\begin{equation}
\kappa_{\rm Unruh} (r) = \fr{\sqrt{1-\frac{2M}{r}}} \left(\fr{4M}-\frac{M}{r^2}\right),
\label{kappa_unruh_otra}
\end{equation}
que en el capítulo anterior denominamos temperatura de Unruh [ecuación~(\ref{kappa.Unruh.radius.r})], tiende efectivamente a anularse en el límite~$r \to 2M$.

Sin embargo, la interpretación adecuada de este término es muy distinta: este término forma parte del efecto Unruh, constituyendo la parte asociada a la aceleración de la gravedad. Es decir, podemos separar la expresión de la función de temperatura efectiva total~(\ref{local_kappa}) en dos contribuciones:
\begin{equation}
\kappa(u) = \kappa_{\rm H} (u) + \kappa_{\rm U} (u),
\label{kappa_separada}
\end{equation}
donde
\begin{align}
\kappa_{\rm H} (u) := & \ \sqrt{\frac{1-\vl}{1+\vl}} \fr{\sqrt{1-\frac{2M}{r}}} \ks(\ub),
\label{kappa_rad} \\
\kappa_{\rm U} (u) := & \ \sqrt{\frac{1-\vl}{1+\vl}} \fr{\sqrt{1-\frac{2M}{r}}} \left(-\frac{M}{r^2}\right) + a_{\rm p}.
\label{kappa_efecto_unruh}
\end{align}
El término~$\kappa_{\rm H}$ es la contribución a la percepción total de la radiación emitida por el agujero negro, propia del estado y definida como aquella que escapa a la región asintótica; y el término~$\kappa_{\rm U}$ corresponde al efecto Unruh debido de la aceleración del observador, entendida esta aceleración en un sentido que explicaremos a continuación.

A esta interpretación nos llevan varias consideraciones. La primera es que la expresión de la contribución correspondiente a la radiación del agujero negro~(\ref{kappa_rad}) \emph{es totalmente consistente por sí misma.} Como ya hemos descrito, es la radiación que escapa al infinito espacial corregida por los factores adecuados al observador~$\mathcal{O}$. Es una expresión análoga a la que se obtendría para la percepción de cualquier radiación térmica emitida por un cuerpo celeste, sea esta debida a un proceso electromagnético, nuclear, etc.

La segunda consideración es que la contribución que hemos asignado al efecto Unruh~(\ref{kappa_efecto_unruh}) es toda la que no depende del estado de vacío escogido, \emph{sino únicamente del estado de movimiento del observador~$\mathcal{O}$.} No resultaría coherente asignar su procedencia a la radiación emitida por el agujero negro, sino más bien a la propia percepción del observador. Además, aunque parece razonable identificar el efecto Unruh exclusivamente con la aceleración propia, un razonamiento nos indica que tal identificación \emph{no es adecuada.} En Minkowski en $1+1$~dimensiones, entre un observador en la región asintótica y cualquier otro observador siempre media un espacio-tiempo trivial: el de Minkowski. Sin embargo, este no es el caso en general, y en particular no lo es en un agujero negro. Entre un observador en la región asintótica del agujero negro y el observador~$\mathcal{O}$ media un espacio-tiempo no trivial que acelera al segundo con respecto al primero, es decir, \emph{con respecto a la región asintótica.} A la vista de los resultados, es la aceleración con respecto a la región asintótica la que produce un efecto Unruh sobre el observador~$\mathcal{O}$, sea esta aceleración propia o aceleración debida a la curvatura del espacio-tiempo.

Puede parecer, haciendo de nuevo una comparación demasiado inmediata con Minkowski, que el hecho de que la temperatura de la radiación correspondiente al efecto Unruh no sea proporcional únicamente a la aceleración propia viola el \emph{principio de equivalencia.} Sin embargo, una correcta interpretación de nuestros resultados nos indica que esto no es así. Para comprobar que el principio de equivalencia se cumple al comparar localmente con el conocido resultado del efecto Unruh en el vacío de Minkowski, debemos considerar que el observador que define el estado de vacío es un observador \emph{en caída libre e instantáneamente en la misma posición radial~$r$ que el observador~$\mathcal{O}$.} En tal caso, es fácil ver que se requiere~$\ks(\ub) = M/r^2$ en ese instante y esa posición, por lo cual los términos~(\ref{termino_hawking}) y~(\ref{termino_raro}) se cancelan entre sí y~$\kappa(u) = a_{\rm p}$. Es decir, recuperamos localmente la expresión estándar del efecto Unruh en Minkowski. Sin embargo, en general el observador que define el vacío no tiene por qué encontrarse ni en caída libre ni en la posición del observador~$\mathcal{O}$, por lo que en general no se cumplen las condiciones para siquiera poner a prueba el principio de equivalencia.\footnote{En~\cite{Singleton:2011vh} se defiende que puede encontrarse una violación del principio de equivalencia por detectores de Unruh-DeWitt en Schwarzschild. En nuestra opinión, se comete el error aquí mencionado de no comparar con el estado de vacío adecuado, en la línea del comentario que recibe este artículo en~\cite{Crispino:2012zz}. Este comentario tiene a su vez la réplica~\cite{Singleton:2012zz}. En~\cite{Ahmadzadegan:2013iua} se hace también una discusión sobre el principio de equivalencia, pero considerando la situación en la que los detectores están aislados en cavidades.}

Pongamos un ejemplo concreto. Consideremos una estrella de neutrones, no colapsada, en cuyo exterior el campo de radiación se encuentra en el vacío de Boulware, de tal forma que~$\ks (\ub) = 0$. Para los observadores estáticos en una determinada posición radial, se tiene que~$\kappa(u)=0$ de forma constante, en cualquier posición. Basándonos en la argumentación hecha en la sección~\ref{sec_bogos_detectores}, podemos afirmar sin lugar a dudas que de una colectividad de detectores de Unruh-DeWitt siguiendo tales trayectorias estáticas \emph{ninguno se excitará.} Un resultado análogo se obtiene en~\cite{Louko:2007mu} para un campo gravitatorio newtoniano, solución de las ecuaciones de Einstein linealizadas. Sin embargo, ¿cómo es posible que no se exciten detectores de Unruh-DeWitt sometidos a aceleraciones propias arbitrariamente altas (según su posición sea más cercana al horizonte)? Una primera respuesta a tal cuestión es clara: en este caso, son los detectores en caída libre los que sí detectan una temperatura efectiva no nula, dada precisamente por el término~(\ref{termino_raro}).\footnote{Recuérdese que la temperatura del espectro térmico finalmente detectado (cuando puede hablarse de tal) es proporcional al \emph{valor absoluto} de la función de temperatura efectiva [ecuación~(\ref{temperatura_kappa})].} No hay razón para pensar que, en tal estado, los detectores con aceleración propia tengan que excitarse como si se encontrasen en el vacío de Minkowski. Y tal excitación de hecho no se produce debido a que, por una parte, la emisión de la estrella de neutrones es nula (ignorando, evidentemente, otros procesos físicos en la estrella que también dieran lugar a emisión) y, por otra, la aceleración con respecto a la región asintótica también es nula, luego ninguno de los dos fenómenos puede servir de fuente para la excitación. Algo análogo sucede con el campo electromagnético y las cargas eléctricas en electrodinámica clásica: una carga en reposo en un agujero negro no emite radiación a la región asintótica aunque esté sometida a aceleración propia, mientras que una carga en caída libre sí lo hace~\cite{Fulton1960499,kovetz}.

Un aspecto importante a resaltar es que la superposición de los dos términos en la percepción total~(\ref{kappa_separada}) puede ser \emph{destructiva}, y de hecho lo es en muchos de los casos considerados.  Es decir, el efecto Unruh puede tanto dar lugar a percepción de radiación donde no la hay, como hacer que la radiación existente pase desapercibida. Esto último sucede, por ejemplo, para los observadores en caída libre e instantáneamente estáticos en el vacío de Unruh (los que denominamos observadores de Unruh en el capítulo anterior), para los cuales la función de temperatura efectiva toma el valor~$\kappa_{\rm Unruh} (r)$ dado en~(\ref{kappa_unruh_otra}). Es decir, en las posiciones donde los observadores estáticos detectan una temperatura arbitrariamente alta, correspondiente a la radiación de Hawking que, de hecho, escapa del agujero negro, desplazada al azul por el factor de corrimiento gravitacional, los observadores de Unruh detectan esencialmente vacío.

\subsection{Temperatura efectiva y aceleración relativa}

La separación en dos términos de la temperatura efectiva hecha en~(\ref{kappa_separada}) es la misma que la que se obtuvo en la segunda línea de~(\ref{kappa_rew}), donde es evidentemente es segundo término el que se identifica con~$\kappa_{\rm U} (u)$. Dicho término puede reescribirse con un cambio trivial en una forma especialmente clarificadora:
\begin{equation}
\kappa_{\rm U} (u) = - \left( \frac{\rmd \ub}{\rmd u} \right)^{-1} \frac{\rmd}{\rmd u} \left( \frac{\rmd \ub}{\rmd u} \right).
\label{kappa_unruh_dos}
\end{equation}
Esta expresión es quizá la que mejor nos muestra la interpretación correcta de este término: la temperatura efectiva de la contribución debida al efecto Unruh es no nula cuando hay una variación en el factor de corrimiento en frecuencias, o lo que es equivalente, un reajuste en la sincronía de los relojes~$\rmd \bar{u} / \rmd u$, \emph{entre el observador$~\mathcal{O}$ y la región asintótica.}\footnote{De forma análoga sucede para la función de temperatura efectiva total: es no nula cuando hay una variación en el factor de corrimiento en frecuencias entre el observador que define el estado de vacío y el observador$~\mathcal{O}$.} Por tanto, como ya hemos dicho, el efecto Unruh no vendría dado por la aceleración propia, sino por la aceleración \emph{con respecto a la región asintótica,} entendida tal aceleración como ``variación del factor de corrimiento en frecuencias'', sea dicha variación producida por una aceleración propia, gravitatoria, o cualquier combinación de ambas. Por supuesto, en Minkowski ambos conceptos, aceleración propia y aceleración con respecto a la región asintótica, coinciden. Ello se debe a que en tal caso el factor de corrimiento en frecuencias solo responde al efecto Doppler. Por el contrario, en Schwarzschild este factor está dado por
\begin{equation}
\frac{\rmd \ub}{\rmd u} = \sqrt{\frac{1-\vl}{1+\vl}} \fr{\sqrt{1-\frac{2M}{r}}},
\label{factor_corrimiento}
\end{equation}
es decir, por el producto del efecto Doppler y el corrimiento gravitacional. La presencia de este último factor es la responsable de que la aceleración propia y la aceleración con respecto a la región asintótica no sean la misma.

Como generalización, podemos escribir dos expresiones que relacionan la función de temperatura efectiva definida para distintos observadores, y que pueden entenderse claramente como relaciones entre aceleraciones relativas. Consideremos un conjunto de coordenadas nulas salientes~$\{u_i\}$ asociadas a distintos observadores~$\{\mathcal{O}_i\}$ en el exterior de un agujero negro de Schwarzschild.\footnote{En realidad, las relaciones que escribiremos son válidas en cualquier región \emph{conforme a Minkowski,} si además se considera la teoría invariante conforme para el campo de radiación.} Definimos
\begin{equation}
D_{i,j} := \frac{\rmd u_j}{\rmd u_i}, \quad \kappa_{i,j} := - \left. \frac{\rmd^2 u_j}{\rmd u_i^2} \middle/ \frac{\rmd u_j}{\rmd u_i} \right..
\label{kappa_relativa}
\end{equation}
$D_{i,j}$ es el corrimiento en frecuencias del observador~$\mathcal{O}_i$ con respecto al observador~$\mathcal{O}_j$, y $\kappa_{i,j}$ es la función de temperatura efectiva para el observador~$\mathcal{O}_i$ en el vacío definido por el observador~$\mathcal{O}_j$. Es fácil ver que se cumplen estas dos relaciones:
\begin{equation}
\kappa_{i,j} = - D_{i,j} \kappa_{j,i}, \quad \kappa_{i,j} = D_{i,k} \kappa_{k,j} + \kappa_{i,k}.
\label{relaciones_kappa}
\end{equation}

Estas relaciones se pueden describir como propiedades de las \emph{aceleraciones relativas} entre los distintos observadores, en el sentido generalizado de aceleración que ya hemos apuntado, es decir, como \emph{variación del factor de corrimiento en frecuencias.} La primera expresa una relación de \emph{reciprocidad:} la aceleración de un observador~$\mathcal{O}_i$ con respecto al otro~$\mathcal{O}_j$ (medida por~$\mathcal{O}_i$), es igual a la aceleración de~$\mathcal{O}_j$ con respecto a~$\mathcal{O}_i$ (es decir, medida por~$\mathcal{O}_j$) con signo contrario y multiplicada por el correspondiente factor de corrimiento en frecuencias. La segunda expresa una relación de \emph{transitividad:} la aceleración de un observador~$\mathcal{O}_i$ con respecto a otro~$\mathcal{O}_j$, es igual a la aceleración de un tercer observador~$\mathcal{O}_k$ con respecto a~$\mathcal{O}_j$ (medida por~$\mathcal{O}_k$) multiplicada por el correspondiente factor de corrimiento en frecuencias (entre~$\mathcal{O}_i$ y~$\mathcal{O}_k$) más la aceleración de~$\mathcal{O}_i$ con respecto a~$\mathcal{O}_k$. Es fácil comprobar que tanto la expresión de~$\kappa(u)$ en~(\ref{kappa_rew}) como, a su vez, la expresión de~$\kappa_{\rm U} (u)$ en~(\ref{kappa_efecto_unruh}) pueden entenderse como ejemplos de la segunda de las relaciones~(\ref{relaciones_kappa}).

\section{Volviendo sobre la percepción de la radiación de Hawking}

En esta sección haremos un uso práctico de la expresión obtenida para la temperatura efectiva~(\ref{local_kappa}) volviendo de nuevo, brevemente, sobre algunos de los ejemplos de observadores estudiados en el capítulo anterior. Es evidente que, con tal expresión en la mano, los cálculos son mucho más sencillos, y el significado físico de los resultados es mucho más nítido. Además, con la expresión encontrada podemos codificar la elección del estado de vacío, sea este el que fuere, en la función~$\bar{\kappa} (\ub (u))$. Mantendremos esta función escrita de forma genérica, salvo cuando pasemos a centrarnos en el vacío de Unruh.

Para observadores estáticos en un radio fijo~$r(u) = r_{\rm s}$, se tiene que
\begin{equation}
\vl(u) = 0, \quad \apr(u) = \fr{\sqrt{1-\frac{2M}{r_{\rm s}}}}\frac{M}{r_{\rm s}^2},
\label{static_prop}
\end{equation}
y por tanto
\begin{equation}
\kappa (u) = \fr{\sqrt{1-\frac{2M}{r_{\rm s}}}} \ks(\ub(u)).
\label{static_kp}
\end{equation}
Es decir, estos observadores perciben la radiación saliente del agujero negro en ese instante, multiplicada por el factor de corrimiento gravitacional al azul correspondiente a la posición radial.

Anteriormente, hemos definido como observador de Unruh aquel que se encuentra en caída libre e instantáneamente estático en un determinado instante~$u=u_0$.\footnote{Debe notarse que la elección de la denominación ``observador de Unruh'' atiende a que estos observadores son los que ven el vacío de Unruh como vacío (cerca del horizonte), no a que no experimenten efecto Unruh.} En tal caso tenemos que~$v_l (u_0) = 0$ y~$a_{\rm p} (u_0) = 0$, y por tanto
\begin{equation}
\kappa (u_0) = \fr{\sqrt{1-\frac{2M}{r(u_0)}}} \left( \ks(\ub(u_0)) - \frac{M}{r(u_0)^2} \right).
\label{unruh_kp}
\end{equation}
Como ya hemos mencionado, estos observadores son uno de los paradigmas de ``interferencia destructiva'' entre las dos contribuciones a la función de temperatura efectiva que definimos en la sección anterior, es decir, entre la radiación que escapa del agujero negro y el efecto Unruh que ``distorsiona'' la percepción del observador, debido a la resta que aparece entre paréntesis en~(\ref{unruh_kp}). Si consideramos en particular el estado de vacío de Unruh para el campo de radiación, tenemos que~$\bar{\kappa} (u_0) = 1/(4M)$. En tal caso, cuando el observador de Unruh se encuentra arbitrariamente cercano al horizonte, se obtiene
\begin{equation}
\fr{\sqrt{1-\frac{2M}{r(u_0)}}} \left( \fr{4M} - \frac{M}{r(u_0)^2} \right) \sra[r(u_0) \to 2M] 0,
\label{unruh_limit}
\end{equation}
como ya calculamos en anteriores ocasiones. 

Consideremos a continuación la percepción de un observador en caída libre cualquiera (es decir, $a_{\rm p} (u) = 0$) al cruzar el horizonte. En este caso tenemos una velocidad radial no nula apuntando al horizonte, que llamaremos~$v := \rmd r / \rmd \tau < 0$. Teniendo en cuenta la definición de la velocidad~$v_l$ dada en~(\ref{vs_def}), esta última velocidad se comporta en el horizonte como
\begin{equation}
\vl = \frac{v}{\sqrt{1-\frac{2M}{r}+v^2}} \sra[r \to 2M] -1.
\label{free-falling_vs}
\end{equation}
Este comportamiento conlleva, por tanto, un factor Doppler divergente
\begin{equation}
\sqrt{\frac{1-\vl}{1+\vl}} \sra[r \to 2M] \infty,
\label{free-falling_Ds_infty}
\end{equation}
que competirá con la tendencia a anularse en el horizonte de la cantidad~(\ref{unruh_limit}). De esta competición es de la cual se obtiene el resultado
\begin{equation}
\kappa (u) \sra[r \to 2M] \fr{M} \sqrt{1-\frac{2M}{r_0}},
\label{kappa.HC.Unruh.radius.otra}
\end{equation}
donde aquí~$r_0$ es el radio desde el cual se inicia la trayectoria en caída libre del observador. En particular, si la trayectoria se inicia en el infinito espacial, tenemos~$\kappa = 1/M$, es decir, cuatro veces la temperatura de la radiación de Hawking asintótica.

\section{La temperatura efectiva en Minkowski}\label{sec_minkowski}

En esta sección, haremos una breve descripción del uso de la función de temperatura efectiva en el espacio-tiempo de Minkowski, considerando además un escenario físico de especial interés conceptual. En este espacio-tiempo es fácil obtener la función de temperatura efectiva para un observador cualquiera en un estado de vacío cualquiera. El resultado es
\begin{equation}
\kappa (u) = \sqrt{\frac{1-v_l}{1+v_l}} \bar{\kappa} (\ub) + a_{\rm p},
\label{kappa_mink}
\end{equation}
donde~$\ub = \ub(u) := t(u) - r(u)$ es la coordenada nula de Minkowski,
\begin{equation}
\bar{\kappa} (\bar{u}(u)) = - \left. \frac{\rmd^2 U}{\rmd \bar{u}^2} \middle/ \frac{\rmd U}{\rmd \bar{u}} \right.,
\label{kappa_estado_mink}
\end{equation}
es la función de temperatura efectiva del estado, y
\begin{equation}
v_l (u) := \left. \frac{\rmd r}{\rmd u} \middle/ \sqrt{1+\left(\frac{\rmd r}{\rmd u}\right)^2} \right., \quad a_{\rm p} (u) := \left. \frac{\rmd^2 r}{\rmd u^2} \middle/ \sqrt{1+\left(\frac{\rmd r}{\rmd u}\right)^2} \right.,
\label{vel_acc_mink}
\end{equation}
son la velocidad con respecto al sistema de referencia~$(t,r)$ y la aceleración propia, respectivamente.

Vamos a escoger en primer lugar la función de temperatura efectiva del estado. En el caso del vacío de Minkowski, es evidente que~$U \propto \ub$, y por tanto tenemos que~$\bar{\kappa} (\bar{u}) = 0$ y~$\kappa (u) = a_{\rm p} (u)$: hemos recuperado la formulación original del efecto Unruh. Imaginemos, por el contrario, que el vacío lo definimos con la coordenada
\begin{equation}
U = U_{\rm H} - A\ \rme^{-\kappa_{\rm c} \bar{u}},
\label{coordenada_U_mink}
\end{equation}
donde $\kappa_{\rm c} > 0$ es una constante. De esta forma, la temperatura efectiva del estado~$\bar{\kappa} (\ub) = \kappa_{\rm c}$ es constante. Si se calcula la trayectoria del observador que define este vacío, se comprueba que parte con velocidad igual a~$-1$ en el pasado remoto (lo cual no es un problema), pero alcanza una velocidad igual a~$1$, y agota su trayectoria, en un tiempo propio finito, lo cual es físicamente imposible. Es evidente por qué se produce esto: estamos intentando obtener un vacío equivalente al vacío de Unruh para los observadores en la región asintótica de un agujero negro de Schwarzschild, y en tal caso los observadores en caída libre que definen ese vacío abandonan la región exterior del agujero negro en un tiempo propio finito.

A pesar de ello, en lo que sigue en esta sección escogeremos el estado de vacío tal que~$\bar{\kappa} (\ub) = \kappa_{\rm c}$, asumiendo que un procedimiento físico nos ha llevado a este estado. Una forma más física de obtener~$\bar{\kappa} (\ub) = \kappa_{\rm c}$ constante es considerar que, en realidad, nos encontramos en el espacio-tiempo de Schwarzschild y en el vacío de Unruh, de forma que~$\kappa_{\rm c} = 1/(4M)$, pero a una distancia arbitrariamente grande del agujero negro, de tal manera que en el espacio que necesitamos para nuestros experimentos nunca nos acercamos significativamente a este. De esta forma, al tomar el límite~$r \to \infty$ en las expresiones asociadas a la función~$\kappa (u)$ encontradas en las secciones anteriores, estas se reducen a las expresadas aquí.

Una vez fijada la temperatura efectiva del estado, tenemos la expresión
\begin{equation}
\kappa (u) = \sqrt{\frac{1-v_l}{1+v_l}} \kappa_{\rm c} + a_{\rm p}.
\label{kappa_bar_const_mink}
\end{equation}
En la sección~\ref{hawking_versus_unruh}, resaltamos la posibilidad de que la composición entre la percepción de la radiación presente en el estado de vacío dado y la contribución del efecto Unruh fuera de carácter destructivo, de tal forma que la radiación finalmente percibida por un determinado observador fuera menor que las contribuciones por separado. Mostramos, por ejemplo, a los que denominamos observadores de Unruh como paradigma de este hecho. En el caso de Minkowski, esto puede llevarnos a plantear la siguiente situación hipotética: ¿Puede un observador en Minkowski dejar de percibir una radiación térmica propia del estado del campo, sí percibida por observadores inerciales, simplemente acelerando \emph{contra} ella (es decir, en sentido contrario a su propagación)? Es lo que parece concluirse, a simple vista, de~(\ref{kappa_bar_const_mink}): basta con escoger una aceleración en sentido de~$r$ decreciente, y por tanto negativa, tal que contrarreste el término de la radiación propia del estado, de forma que en total~$\kappa (u) = 0$.

Vamos a comprobar, sin embargo, que no es posible para un observador hacer tal cosa durante un tiempo suficientemente largo. El motivo es que, si bien la aceleración contra la radiación le permite al observador, según~(\ref{kappa_bar_const_mink}), anular la función de temperatura efectiva \emph{instantáneamente,} dicha aceleración producirá también un corrimiento al azul en la temperatura percibida de la radiación propia del estado, lo cual rápidamente compensará significativamente la anulación.

Consideremos que, en un instante inicial~$u=u_0$, el observador se encuentra en reposo en las coordenadas~$(t,r)$, es decir, $v_l (u_0) = 0$, y por tanto~$\kappa (u_0) = \kappa_{\rm c} + a_{\rm p}$. Para obtener una percepción de radiación resultante nula, dicho observador decide acelerar con~$a_{\rm p} = -\kappa_{\rm c}$. Puede comprobarse [de las definiciones de~$v_l$ y~$a_{\rm p}$ en~(\ref{vel_acc_mink})] que, para una historia de aceleración propia cualquiera, la velocidad~$v_l (u)$ se obtiene de la ecuación diferencial
\begin{equation}
\frac{\rmd v_l}{\rmd u} = a_{\rm p}(u) [1-v_l^2(u)].
\label{ecuacion_vl}
\end{equation}
En este caso, con aceleración propia constante, dicha ecuación tiene la solución~$v_l(u)=-\tanh[\kappa_{\rm c} (u-u_0)]$. De esta forma, podemos obtener el instante en el cual, debido al corrimiento Doppler, el observador que intentó ``apagar'' su percepción de la radiación vuelve a recuperarla con igual temperatura efectiva. Es decir, el instante~$u$ tal que
\begin{equation}
\kappa (u) = \left( \sqrt{\frac{1-v_l(u)}{1+v_l(u)}} -1 \right) \kappa_{\rm c} = \kappa_{\rm c}.
\label{kappa_recuperar}
\end{equation}
De esta ecuación se obtiene
\begin{equation}
v_l(u) = -3/5 \quad \Rightarrow \quad u - u_0 = \log 2 (1/\kappa_{\rm c}) \simeq 0.69(1/\kappa_{\rm c}).
\label{u_recuperar}
\end{equation}
Es decir, cuando aún no ha transcurrido el tiempo propio~$u - u_0 \sim 1/\kappa_{\rm c}$ necesario para explorar el rango de energías característico del espectro~$\omega \sim \kappa_{\rm c}$, la función de temperatura efectiva, fruto del efecto Doppler, ha recuperado su valor inicial, lo que significa que el observador volverá a detectar partículas.

En un mejor intento por apagar la radiación inherente al estado mediante una aceleración contraria a dicha radiación, el observador podría ir aumentando su aceleración progresivamente, de tal forma que en todo momento compensase el creciente factor Doppler, y en todo instante se tuviera que~$\kappa(u) = 0$. Teniendo en cuenta~$\kappa (u)$ en~(\ref{kappa_bar_const_mink}) y de nuevo la ecuación diferencial para~$v_l (u)$ en~(\ref{ecuacion_vl}), esto se traduce en la ecuación
\begin{equation}
\kappa (u) = \sqrt{\frac{1-v_l(u)}{1+v_l(u)}} \kappa_{\rm c} + \frac{\rmd v_l / \rmd u}{1-v_l(u)^2} = 0.
\label{kappa_cero}
\end{equation}
Resolviendo esta ecuación diferencial, de nuevo con la condición inicial dada por~$v_l (u_0) = 0$, obtenemos
\begin{equation}
v_l (u) = 1-\frac{1}{1-\kappa_{\rm c} (u - u_0) +[\kappa_{\rm c} (u - u_0)]^2/2};
\label{vl_bestia}
\end{equation}
y, teniendo en cuenta una vez más~(\ref{ecuacion_vl}),
\begin{equation}
a_{\rm p} (u) = \frac{1}{u - u_0 -1/\kappa_{\rm c}}.
\label{ap_bestia}
\end{equation}
Es decir, transcurrido el tiempo propio~$u - u_0 \sim 1/\kappa_{\rm c}$ necesario para explorar el rango de energías característico del espectro, el observador ha tenido que alcanzar una aceleración propia infinita y la velocidad de la luz. Por tanto, tampoco de esta forma parece físicamente posible eliminar la detección de radiación térmica del estado mediante una aceleración propia contraria al sentido de propagación de la misma.

Es importante destacar que los valores obtenidos en estos ``experimentos mentales'' con un observador acelerado en Minkowski no deben interpretarse como una descripción precisa de la percepción real de radiación de dicho observador. Las trayectorias consideradas son, en general, extremas, y los cambios bruscos, por lo que la condición adiabática que permite una interpretación directa de~$\kappa(u)$ en general se violará de forma flagrante. Lo que sí nos han permitido dichos experimentos es una comprobación de consistencia: no es sostenible un escenario en el que un observador anule su percepción de una radiación térmica que sí sea percibida por observadores inerciales, durante un tiempo suficiente, mediante un efecto Unruh que interfiera destructivamente con ella y que sea fruto \emph{únicamente} de la aplicación de una aceleración propia contra dicha radiación.

\section{La temperatura efectiva y los procesos de flotación}

Finalmente, cabe preguntarse qué implicaciones físicas tiene el considerar las distintas contribuciones a la función de temperatura efectiva como radiación emitida por el agujero negro o como efecto Unruh. Por una parte, en términos únicamente de percepción de radiación, y hasta donde podemos describir con las herramientas aquí desarrolladas, bajo cumplimiento de la condición adiabática la percepción del observador~$\mathcal{O}$ será sencillamente la de un espectro térmico con temperatura proporcional a~$|\kappa(u)|$, sin importar el peso de las distintas contribuciones a ese valor. Sin embargo, la cuestión es completamente distinta si lo que buscamos es comprender el efecto que tiene la interacción entre un dispositivo concreto (por ejemplo, un detector de Unruh-DeWitt) y el campo de radiación \emph{sobre el propio campo} o, yendo más lejos aún, la reacción que esta interacción produce \emph{sobre la propia trayectoria del dispositivo.} Una mala interpretación puede llevar a una falta de entendimiento y a esperar comportamientos opuestos a los reales.

Es conocido que la interacción de un detector acelerado con un campo en el vacío de Minkowski produce, además de la excitación del propio detector (es decir, el efecto Unruh), la excitación del campo, fenómeno al que se denomina \emph{radiación Unruh}~\cite{Parentani:1995iw,Unruh:1983ms,Unruh:1992sw}.\footnote{Sobre la existencia de tal radiación existe cierta controversia en la literatura~\cite{Grove:1986fz,Raine:1991kc,Audretsch:1994gg,Ford:2005sh}.} Dicha emisión de radiación, por conservación del momento, actúa \emph{contra} la velocidad del dispositivo, a modo de ``radiación de frenado''~\cite{Parentani:1995iw}. Sin embargo, una radiación objetiva emitida por el agujero negro tiene una consecuencia opuesta para el dispositivo con el que interacciona, pues dicho dispositivo recibe el momento de la partícula que detecta.

Esto tiene una implicación importante en el estudio de los denominados \emph{procesos de flotación} en las cercanías de un agujero negro. Se ha argumentado en algunos trabajos~\cite{Unruh:1982ic,Bekenstein:1999bh} que los fenómenos de radiación derivados de la Teoría Cuántica de Campos en espacios curvos pueden dar lugar a que determinados objetos, situados en las cercanías del horizonte de sucesos, alcancen un estado de equilibrio, en el que la fuerza neta realizada por el campo de radiación sobre ellos compense la atracción gravitatoria del agujero negro, de tal suerte que el objeto permanezca estático en el exterior.

No obstante, si la radiación percibida por los detectores estáticos muy cercanos al horizonte procediera, casi en su totalidad, del propio efecto Unruh al que da lugar la aceleración propia que necesitan tales detectores para permanecer estáticos, la argumentación para sostener el escenario de flotación entraría en un bucle lógico, y por supuesto tal escenario de flotación no resultaría plausible. La radiación producida por el efecto Unruh actuaría \emph{contra} la aceleración que la produce, y desde luego no podría ser a su vez motor de esa misma aceleración. El destino irremediable de cualquier objeto dejado a su suerte en el exterior de un agujero negro sería caer hacia el horizonte. Sin embargo, si como parece desprenderse de nuestro análisis, la radiación percibida por tales detectores estáticos no es más que la radiación que, de hecho, escapa del agujero negro, con su correspondiente corrimiento gravitacional al azul, el escenario de flotación resulta perfectamente viable.

\subsection{Un posible escenario de flotación}

A la vista de lo obtenido en secciones anteriores, la atracción gravitatoria, por su contribución de signo negativo, produce una interferencia destructiva con la percepción de la radiación procedente del agujero negro, que como decimos es la que puede dar lugar a un escenario real de flotación. Por tanto, un tipo de trayectorias de especial interés a la hora de considerar escenarios de flotación son aquellas en las que dicha interferencia destructiva no está presente, quedando como única contribución a la percepción de radiación la procedente de la emisión del agujero negro. Es decir, las trayectorias tales que~$\kappa (u) = \kappa_{\rm H} (u)$ o, de forma equivalente, $\kappa_{\rm U} (u) = 0$. Un ejemplo que ya hemos encontrado son las trayectorias estáticas (con radio constante). En general, dadas las ecuaciones para~$\kappa_{\rm U} (u)$~(\ref{kappa_unruh_dos}) y~(\ref{factor_corrimiento}), estas trayectorias son solución de la ecuación
\begin{equation}
\sqrt{\frac{1-\vl(u)}{1+\vl(u)}} \fr{\sqrt{1-\frac{2M}{r(u)}}} = C = \rm{const},
\label{corrimiento_constante}
\end{equation}
es decir, aquellas trayectorias para las que no se produce variación en el factor de corrimiento en frecuencias entre el observador~$\mathcal{O}$ y la región asintótica. Teniendo en cuenta la definición de~$v_l$ en~(\ref{vs_def}), esta ecuación tiene la solución general
\begin{equation}
r(u) = r_{\rm f} \left\{ 1 + W_0 \left[ \left( \frac{r_{\rm ref}}{r_{\rm f}} - 1 \right) \exp \left( \frac{r_{\rm ref}}{r_{\rm f}} - 1 - \frac{u - u_{\rm ref}}{2 r_{\rm f} \sqrt{r_{\rm f}/(2M)-1}} \right) \right] \right\},
\label{trayectorias_sin_unruh}
\end{equation}
donde~$r_{\rm f}$ es el \emph{radio final} al que tiende asintóticamente la solución en~$u \to \infty$, relacionado con la constante~$C$ en~(\ref{corrimiento_constante}) mediante~$C= 1 / \sqrt{1-2M/r_{\rm f}}$; $W_0(z)$ es la rama principal de la \emph{función~$W$ de Lambert;}\footnote{La función~$W$ de Lambert está definida de forma implícita como la solución a la ecuación $z=W(z) \exp [W(z)]$. La rama principal se define con las condiciones~$W_0 (z) \in \mathds{R}$ y~$W_0 (z) \geq -1$.} y~$r_{\rm ref} = r(u_{\rm ref})$ no es más que un radio de referencia arbitrario, por el cual cruza la trayectoria en el instante de referencia arbitrario~$u_{\rm ref}$.

\begin{figure}[htbp!]
\begin{minipage}[b]{0.47\linewidth}
\centering
\includegraphics[width=6.5cm]{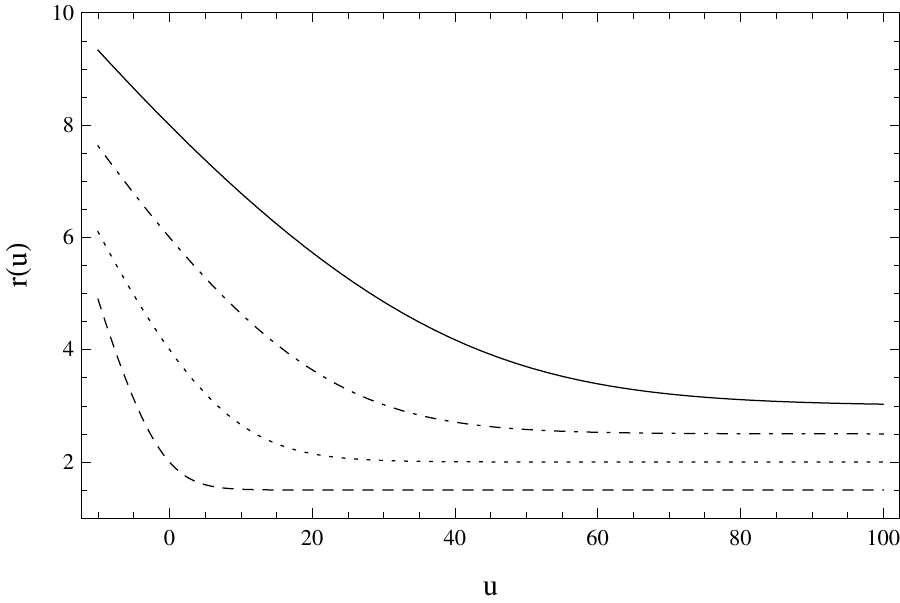}
\caption{\footnotesize{Trayectorias~$r(u)$ en~(\ref{trayectorias_sin_unruh}) para~$r_{\rm f} = (3M,\ 4M,\ 5M,\ 6M)$ y ~$r_{\rm ref} = (4M,\ 8M,\ 12M,\ 16M)$ (línea discontinua, punteada, punto-raya y continua, respectivamente). Utilizamos unidades~$2M = 1$ y~$u_{\rm ref} = 0$.}}
\label{fig_no_unruh}
\end{minipage}
\hspace{0.04\linewidth}
\begin{minipage}[b]{0.47\linewidth}
\centering
\includegraphics[width=6.5cm]{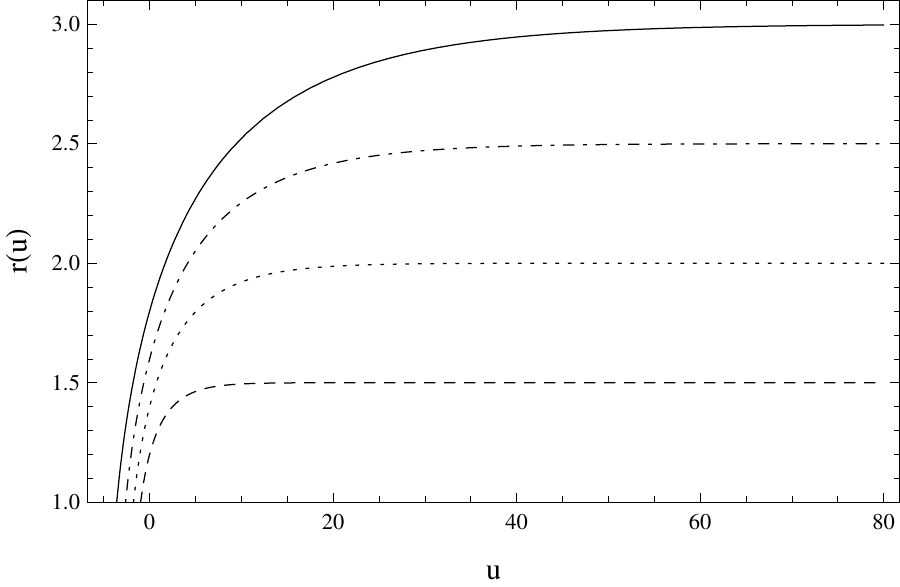}
\caption{\footnotesize{Trayectorias~$r(u)$ en~(\ref{trayectorias_sin_unruh}) para~$r_{\rm f} = (3M,\ 4M,\ 5M,\ 6M)$ y ~$r_{\rm ref} = (2.4M,\ 2.8M,\ 3.2M,\ 3.6M)$ (línea discontinua, punteada, punto-raya y continua, respectivamente). Utilizamos unidades~$2M = 1$ y~$u_{\rm ref} = 0$.}}
\label{fig_no_unruh2}
\end{minipage}
\end{figure}

Tenemos tres familias de soluciones: para~$r_{\rm ref} > r_{\rm f}$, las trayectorias parten del infinito espacial con velocidad~$v_l = (1-C^2)/(1+C^2)$; para~$2M \leq r_{\rm ref} < r_{\rm f}$, las trayectorias comienzan en el horizonte de sucesos con velocidad~$v_l=1$; y para~$r_{\rm ref} = r_{\rm f}$, se tienen las ya conocidas trayectorias estáticas~$r(u)=r_{\rm f}$. En las figuras~\ref{fig_no_unruh} y~\ref{fig_no_unruh2} se muestran ejemplos de las dos primeras familias de soluciones.

\noindent

Las trayectorias~(\ref{trayectorias_sin_unruh}) pueden entenderse como las trayectorias ``inerciales'' desde un punto de vista cuántico. Son aquellas trayectorias que no producen ninguna ``fricción'' (siempre que nos preocupemos exclusivamente del sector de radiación saliente) con el campo cuántico. Desde este punto de vista, son la generalización a espacios curvos de las trayectorias con velocidad uniforme en Minkowski. Por tanto, en un agujero negro nos encontramos con dos tipos de trayectorias especiales: las geodésicas, y las que no presentan ``fricción cuántica''.

Teniendo en cuenta la caracterización de estas trayectorias en~(\ref{corrimiento_constante}), la percepción de radiación emitida por el agujero negro [dada por~$\kappa_{\rm H} (u)$ en~(\ref{kappa_rad})] a lo largo de ellas será
\begin{equation}
\kappa (u) = \kappa_{\rm H} (u) = C\ \bar{\kappa} (\ub (u)).
\label{percepcion_sin_unruh}
\end{equation}
Es decir, a lo largo de estas trayectorias se percibe la radiación saliente del agujero negro con un factor de corrimiento en frecuencias constante.

Por otra parte, teniendo en cuenta la expresión de~$\kappa_{\rm U} (u)$ en~(\ref{kappa_efecto_unruh}) y de nuevo la caracterización las trayectorias en~(\ref{corrimiento_constante}), se puede calcular la aceleración propia que se requiere para seguir las trayectorias dadas por~(\ref{trayectorias_sin_unruh}), la cual resulta
\begin{equation}
\kappa_{\rm U} (u) = 0 \quad \Rightarrow \quad a_{\rm p} (u) = \frac{C M}{r(u)^2}.
\label{aceleracion_sin_unruh}
\end{equation}
Por tanto, la aceleración necesaria es inversamente proporcional al cuadrado de la distancia radial.

Consideremos en particular la familia de trayectorias procedentes del infinito espacial ($r_{\rm ref} > r_{\rm f}$) y el estado de vacío de Unruh~$\bar{\kappa} (\ub) = 1/(4M)$. En tal caso, de~$\kappa(u)$ en~(\ref{percepcion_sin_unruh}) se tiene que la función de temperatura efectiva es constante~$\kappa(u) = C / (4M)$ a lo largo de cada trayectoria. De este hecho se desprenden dos conclusiones: por una parte, que~$\kappa(u) = C / (4M)$ es \emph{de facto} de la temperatura del espectro térmico percibido, puesto que la condición adiabática se cumple de forma \emph{exacta;} y por otra, que la aceleración necesaria~(\ref{aceleracion_sin_unruh}), como resulta trivial, es a su vez proporcional a la radiación percibida dividida por la distancia radial al cuadrado:
\begin{equation}
a_{\rm p} (u) = 4 M^2 \frac{\kappa(u)}{r(u)^2}.
\label{aceleracion_sin_unruh_en_unruh}
\end{equation}
Esto nos lleva a cuestionarnos el siguiente escenario: En un agujero negro de Schwarzschild en $3+1$~dimensiones,\footnote{Por supuesto, ignoramos las correcciones introducidas por \emph{backscattering,} de las que no podemos dar cuenta aquí.} siendo la temperatura percibida constante, y considerando un objeto con una sección eficaz de detección constante, la \emph{intensidad total} de la radiación percibida también escala con la inversa del cuadrado de la distancia radial, debido al efecto adicional de la \textit{difusión} de la radiación en las tres dimensiones espaciales, efecto que no está presente en $1+1$~dimensiones. Es decir, si consideramos un objeto totalmente opaco y de sección eficaz radial constante, viajando hacia el agujero negro en $3+1$~dimensiones siguiendo la trayectoria dada por~(\ref{trayectorias_sin_unruh}), dicho objeto percibiría en todo momento una radiación térmica saliente de intensidad justamente proporcional a la aceleración que necesita para, de hecho, continuar en la trayectoria dada por~(\ref{trayectorias_sin_unruh}), \emph{y toda procedente de la emisión del agujero negro.} Dado que la fuerza que la radiación ejercería sobre el objeto sería proporcional a la intensidad total de radiación que este percibe, entonces tan solo ajustando una constante característica del objeto (por ejemplo, la masa, o la propia sección eficaz) quizá se pudiera preparar dicho objeto para que siguiera, sin necesidad de un mecanismo adicional, una trayectoria dada por~(\ref{trayectorias_sin_unruh}), la cual tendería asintóticamente a un radio constante. Es decir, a un estado de flotación. Esta posibilidad queda como objeto de un estudio futuro.

\subsection{Considerando la radiación entrante}\label{sec_entrante}

En el escenario de flotación que hemos planteado, jugaría también su papel un aspecto que hasta ahora, por simplificar, venimos ignorando: la radiación entrante, es decir, el sector de modos~$\bar{v} := t + r^*$ del campo. Es evidente que se puede hacer un estudio de este sector análogo al que hemos realizado para el sector de radiación saliente. En particular, también se puede separar la función de temperatura efectiva definida para este sector de una forma idéntica a~(\ref{kappa_separada}): una contribución correspondiente a la radiación propia del estado, que en este caso correspondería a radiación entrante desde el infinito espacial; y otra contribución correspondiente al efecto Unruh en este sector. Dado que para el escenario de flotación hemos considerado el vacío de Unruh, en tal escenario no existe radiación entrante desde el infinito espacial. Sin embargo, a diferencia de lo que sucede en el sector de radiación saliente, para la radiación entrante sí aparece efecto Unruh al seguir las trayectorias~(\ref{trayectorias_sin_unruh}) (salvo las de radio constante). Las trayectorias para las que no hay efecto Unruh en el sector de radiación entrante (en cuyo caso, en general sí lo hay en el de radiación saliente) son las mismas trayectorias~(\ref{trayectorias_sin_unruh}) pero recorridas \emph{en sentido contrario,} es decir, desde el radio~$r_{\rm f}$ en el pasado asintótico, hacia el infinito espacial o el horizonte. Esto se debe a que el factor de corrimiento en frecuencias para el sector entrante tiene un cambio del signo temporal:
\begin{equation}
\frac{\rmd \bar{v}}{\rmd v} = \sqrt{\frac{1+\vl}{1-\vl}} \fr{\sqrt{1-\frac{2M}{r}}},
\label{factor_corrimiento_v}
\end{equation}
donde~$v_l$ mantiene la definición dada en~(\ref{vs_def}) y~$v=u=\tau$ no es más que el tiempo propio del observador~$\mathcal{O}$. La función de temperatura efectiva para el efecto Unruh en este sector (que distinguiremos con el superíndice~$\bar{v}$) es
\begin{equation}
\kappa_{\rm U}^{\bar{v}} (v) := \sqrt{\frac{1+\vl}{1-\vl}} \fr{\sqrt{1-\frac{2M}{r}}} \frac{M}{r^2} - a_{\rm p},
\label{kappa_efecto_unruh_v}
\end{equation}
donde~$a_{\rm p}$ igualmente mantiene la definición dada en~(\ref{af_def}).

Vamos a estudiar el comportamiento de esta temperatura efectiva a lo largo de las trayectorias~(\ref{trayectorias_sin_unruh}) en las que se ha propuesto el escenario de flotación, sabiendo que dicho escenario se propone en el vacío de Unruh. Teniendo en cuenta que tales trayectorias son solución de la ecuación diferencial~(\ref{corrimiento_constante}), y que la aceleración propia a lo largo de las mismas está dada por~(\ref{aceleracion_sin_unruh_en_unruh}), puede verse fácilmente que la cantidad~(\ref{kappa_efecto_unruh_v}) para estas trayectorias como función de la posición radial toma el valor
\begin{equation}
|\kappa_{\rm U}^{\bar{v}} (r)| = \frac{C M}{r^2} \left( 1- \frac{1-2M/r_{\rm f}}{1-2M/r} \right).
\label{unruh_molesto}
\end{equation}
Considerando únicamente la familia de trayectorias que parten del infinito espacial, este valor tiende a anularse en~$r \to \infty$ y en~$r \to r_{\rm f}$ (salvo en el caso~$r_{\rm f} = 2M$), es decir, al principio y al final de la trayectoria, pero en general es no nulo. Para hacernos una idea de su importancia con respecto a la radiación que sostendría el escenario de flotación planteado, dada por~$\kappa (u) = C/(4M)$, definimos como~$R_{r_{\rm f}} (r)$ el cociente
\begin{equation}
R_{r_{\rm f}} (r) := \left| \frac{\kappa_{\rm U}^{\bar{v}}}{\kappa} \right| = \frac{4 M^2}{r^2} \left( 1 - \frac{1-2M/r_{\rm f}}{1-2M/r} \right).
\label{cociente_molesto}
\end{equation}

\begin{figure}[ht]
	\centering
    \includegraphics{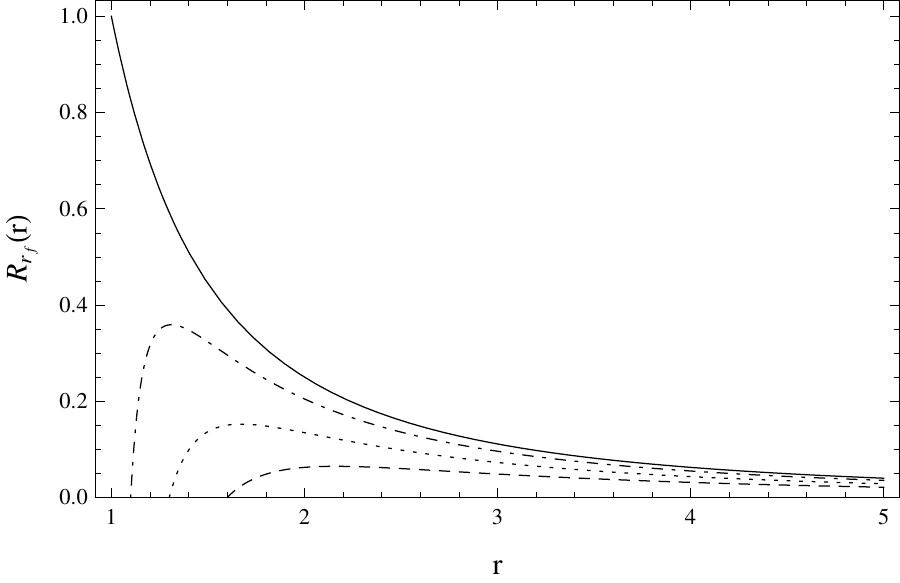}
  \caption{\footnotesize{Valor de cociente $R_{r_{\rm f}} (r)$ para~$r_{\rm f} = (3.2M,\ 2.6M,\ 2.2M,\ 2M)$ (línea discontinua, punteada, punto-raya y continua, respectivamente). Utilizamos unidades~$2M = 1$.}}
  \label{fig_ratio_unruh}
\end{figure}

En la figura~\ref{fig_ratio_unruh} se muestra un ejemplo del comportamiento de esta cantidad a lo largo de trayectorias con distintos radios finales. Para cada trayectoria, dicha cantidad presenta un máximo en el radio
\begin{equation}
r^{\text{máx}}_{r_{\rm f}} := \frac{M}{2} \left(1 + \sqrt{1-\frac{5 r_{\rm f}}{M} + \frac{9 r_{\rm f}^2}{4M^2}} \right) + \frac{3}{4} r_{\rm f},
\label{radio_molesto}
\end{equation}
cuyo valor es
\begin{equation}
R^{\text{máx}}_{r_{\rm f}} := \frac{M^2}{2 r_{\rm f}^2} \left[ \left( 1 - \frac{9 r_{\rm f}}{2M} \right) \sqrt{\left(1-\frac{r_{\rm f}}{2M} \right) \left(1-\frac{9 r_{\rm f}}{2M} \right)} - 1 - \frac{9 r_{\rm f}}{M} + \frac{27 r_{\rm f}^2}{4M^2} \right].
\label{maximo_molesto}
\end{equation}

\begin{figure}[htbp!]
\begin{minipage}[b]{0.47\linewidth}
\centering
\includegraphics[width=6.5cm]{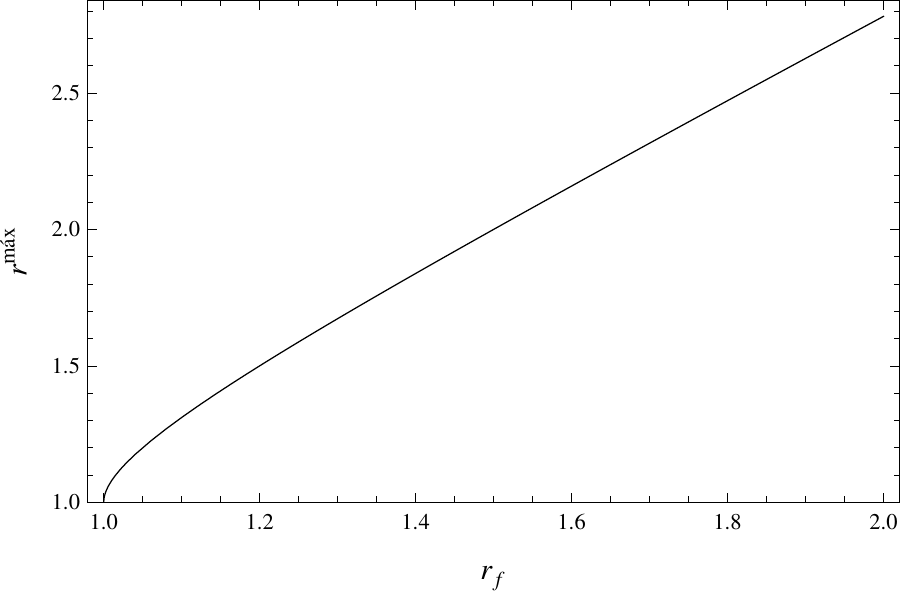}
\caption{\footnotesize{Valor de~$r^{\text{máx}}_{r_{\rm f}}$ en función de~$r_{\rm f}$. Utilizamos unidades~$2M = 1$.}}
\label{fig_r_max}
\end{minipage}
\hspace{0.04\linewidth}
\begin{minipage}[b]{0.47\linewidth}
\centering
\includegraphics[width=6.5cm]{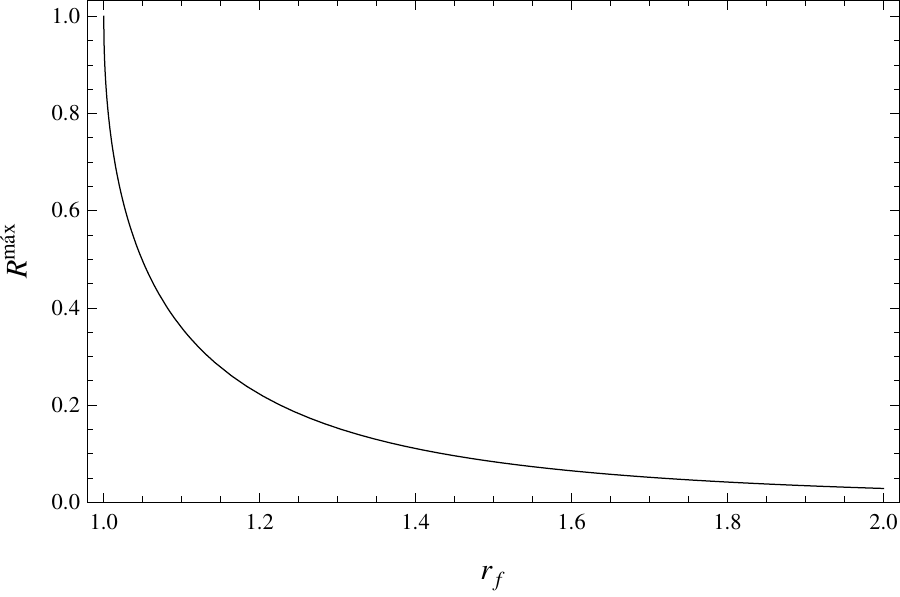}
\caption{\footnotesize{Valor de~$R^{\text{máx}}_{r_{\rm f}}$ en función de~$r_{\rm f}$. Utilizamos unidades~$2M = 1$.}}
\label{fig_ratio_max}
\end{minipage}
\end{figure}

En las figuras~\ref{fig_r_max} y~\ref{fig_ratio_max} se muestran~$r^{\text{máx}}_{r_{\rm f}}$ y~$R^{\text{máx}}_{r_{\rm f}}$ en función de~$r_{\rm f}$, respectivamente. Puede comprobarse que el máximo tiende a acercarse al final de la trayectoria según~$r_{\rm f} \to 2M$, y su valor es~$1$ en~$r_{\rm f} = 2M$, decreciendo monótona y rápidamente a cero cuando~$r_{\rm f}$ crece. De este análisis comparativo podemos intuir que el efecto Unruh en la radiación entrante solo jugaría un papel significativo sobre el proceso de flotación propuesto cuando el radio final de flotación estuviera muy cercano al horizonte de sucesos; y, en tal caso, solo lo haría en la última parte de la trayectoria de flotación.

\noindent

De cualquier forma, el resultado físico esperable de la presencia de efecto Unruh en la radiación entrante es la emisión de \emph{radiación Unruh entrante} por parte del objeto en proceso de flotación. Dicha emisión, por conservación del momento, no puede sino contribuir aun más a la propia flotación, al menos en primera aproximación. Es cierto que dicha contribución modificaría la aceleración, y por tanto sacaría al objeto de la trayectoria~(\ref{trayectorias_sin_unruh}), lo cual podría desestabilizar el proceso anteriormente planteado. El resultado final no parece que pueda conocerse salvo que se proponga un modelo concreto de interacción entre el objeto y el campo de radiación. No obstante, nada parece indicar que la presencia de efecto Unruh en la radiación entrante vaya a ser un factor decisivo en este escenario, y donde seguro que no estaría presente es en el eventual estadio final en el que se alcanzase un radio constante. En cualquier caso, el análisis en detalle del escenario de flotación planteado queda fuera de esta memoria, como propuesta de trabajo futuro.

\newpage
\thispagestyle{empty}
\hbox{}

\makeatletter
\def\cleardoublepage{\clearpage\if@twoside \ifodd\c@page\else
    \hbox{}
    \thispagestyle{empty}
    \newpage
    \if@twocolumn\hbox{}\newpage\fi\fi\fi}
\makeatother \clearpage{\pagestyle{empty}\cleardoublepage}

\chapter{El vacío pulsante}
\label{pulsante}

Una última aplicación de la función de temperatura efectiva, en particular de la \emph{función de temperatura efectiva del estado,} es la de describir un escenario en el que pueda obtenerse radiación muy similar a la radiación de Hawking emitida por un objeto estelar hacia el infinito espacial sin necesidad de formar un horizonte de ningún tipo, y por ello sin necesidad de trazar sobre modos en el interior del agujero negro ni invocar frecuencias arbitrariamente altas. Es decir, evitando los problemas conocidos como \emph{paradoja de la información}~\cite{Hawking:1976ra,Hawking:2005kf,Preskill:1992tc} y \emph{problema transplanckiano}~\cite{Unruh:1976db}, respectivamente.

En la sección~\ref{observadores_estaticos}, en la que describimos el que denominamos \emph{vacío de colapso} según es percibido por observadores estáticos en el exterior de un agujero negro, hicimos notar un hecho significativo: la radiación percibida por dichos observadores es planckiana, con la temperatura adecuada a cada posición, aun cuando el observador que define el estado de vacío se encuentra todavía relativamente lejos del horizonte. Este resultado ya se indicó anteriormente en~\cite{Barcelo:2006uw}, y está en consonancia con el hecho, mostrado en~\cite{Stephens:1993an}, de que una esfera en colapso que finalmente se detiene cerca de la formación de un horizonte, pero sin llegar a formarlo, emite una ráfaga (no necesariamente corta) de radiación antes de detenerse. En este capítulo describiremos un estado de vacío para un campo de radiación en el exterior de un cuerpo celeste que presenta una radiación muy similar a la radiación de Hawking emitida al infinito espacial, también sin necesidad de formar un horizonte ni invocar frecuencias transplanckianas, pero además de tal forma que la radiación es emitida de manera continuada.

Por sencillez trabajaremos, como venimos haciendo en los últimos capítulos, en la geometría del exterior de un agujero negro de Schwarzschild reducida a $1+1$~dimensiones~[ecuación~(\ref{schwarzschild_una_dimension})], y considerando igualmente como campo de radiación un campo escalar real sin masa, en el que se toma la aproximación de ignorar el potencial efectivo, de tal forma que la teoría resulta invariante conforme~[ecuación~(\ref{klein-gordon_schwarzschild_nulas})]. Es decir, todas las condiciones que nos permiten concentrarnos exclusivamente en la función de temperatura efectiva. En particular, nos centraremos en la que denominamos función de temperatura efectiva del estado~$\bar{\kappa} (\bar{u})$~[ecuación~(\ref{ks_def})], puesto que estamos interesados en describir un estado con radiación muy similar a la de Hawking emitida a la región asintótica. Por tanto, en este caso solo tenemos que hablar de un observador: el que define el estado de vacío. Para este observador, consideraremos una trayectoria que oscila entre dos posiciones radiales próximas al horizonte. Las dos posiciones extremas de la vibración serán~$2M + x_{\rm in}$ y~$2M + x_{\rm ex}$, con $0 < x_{\rm in}/(2M) < x_{\rm ex}/(2M) \ll 1$. Es decir, la vibración sucede muy cerca del horizonte. Identificando el tiempo propio de la trayectoria en oscilación con la coordenada nula~$U$, construimos el estado de vacío correspondiente a los modos normales asociados a tal coordenada nula, estado que denominaremos \emph{vacío pulsante.} Utilizando la métrica de Schwarzschild~(\ref{schwarzschild_una_dimension}), es fácil ver que la relación formal entre la coordenada nula de Schwarzschild~$\bar{u} := t - r^*$ y la coordenada nula~$U$ es
\begin{align}
\bar{u}(U) =  \int \rmd U \sqrt{ \frac{1+ \dot{r}^2 / (1-2M/r(U)) }{1-2M/r(U)} } - r^*(U),
\label{relation-complete}
\end{align}
donde~$r(U)$ corresponde a la ecuación de la trayectoria en oscilación, $r^*(U)$ es la ``coordenada tortuga'' [ecuación~(\ref{tortuga})] para esa misma trayectoria, y~$\dot{r}$ denota derivación con respecto a~$U$. Por otra parte, la función de temperatura efectiva del estado que queremos obtener puede reescribirse fácilmente como
\begin{equation}
\bar{\kappa} (\bar{u}) = - \left. \frac{\rmd^2 U}{\rmd \bar{u}^2} \middle/ \frac{\rmd U}{\rmd \bar{u}} \right. = \left. \left. \frac{\rmd^2 \bar{u}}{\rmd U^2} \middle/ \right. \left( \frac{\rmd \bar{u}}{\rmd U} \right)^2 \right|_{U(\bar{u})}.
\label{kappa_estado_definicion_dos}
\end{equation}
Si introducimos, por comodidad, la coordenada~$x:= r - 2M$, y tenemos en cuenta que toda la trayectoria de oscilación se encuentra en la región tal que~$x \ll 2M$, podemos aproximar~(\ref{relation-complete}) por
\begin{equation}
\bar{u}(U) \approx 2M \int \rmd U \frac{|\dot{x}|}{x} \sqrt{ 1 + \frac{x}{2M \dot{x}^2} } - 2M \log \frac{x}{2M}.
\label{relation-approx}
\end{equation}
En la misma aproximación, las derivadas que aparecen en~(\ref{kappa_estado_definicion_dos}) se escriben
\begin{align}
\frac{\rmd \bar{u}}{\rmd U} & \approx \frac{|\dot{x}|}{x} \sqrt{ 1 + \frac{x}{2M \dot{x}^2} } - 2M \frac{\dot{x}}{x}, \nonumber \\
\frac{\rmd^2 \bar{u}}{\rmd U^2} & \approx 2M \left( \frac{\dot{x}}{x} \right)^2 \left\{1 - \frac{x \ddot{x}}{\dot{x}^2} - {\rm sign} (\dot{x}) \frac{1 + x[1/(4m) - \ddot{x}]/\dot{x}^2}{\sqrt{ 1 + x/(2M \dot{x}^2) }} \right\},
\label{derivadas_pulsantes}
\end{align}
de tal forma que
\begin{equation}
\bar{\kappa} (\bar{u}) \approx - {\rm sign} (\dot{x}) \frac{1/(4m) - \ddot{x}}{\sqrt{ 1 + x/(2M \dot{x}^2) }} + \ddot{x}.
\label{kappa_pulsante}
\end{equation}

\begin{figure}[htbp!]
\begin{minipage}[b]{0.47\linewidth}
\centering
\includegraphics[width=6.8cm]{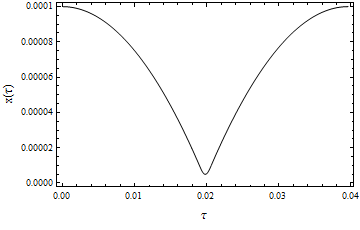}
\caption{\footnotesize{Trayectoria en caída libre desde~$x_{\rm ex} =0.0002 M$ hasta~$x_{\rm b} = 0.00002 M$, donde se produce un rebote elástico entre~$x_{\rm b}$ hasta~$x_{\rm in} = 0.00001 M$, para continuar de nuevo en caída libre de vuelta a~$x_{\rm ex}$, donde se comienza un nuevo ciclo. Utilizamos unidades~$2M = 1$.}}
\label{fig_bounce}
\end{minipage}
\hspace{0.04\linewidth}
\begin{minipage}[b]{0.47\linewidth}
\centering
\includegraphics[width=6.5cm]{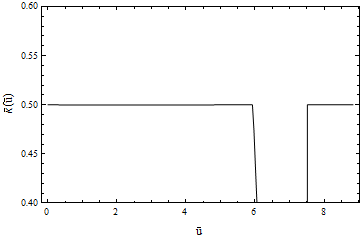}
\caption{\footnotesize{Valor exacto de~$\bar{\kappa} (\bar{u})$ resultante de la oscilación de la figura~\ref{fig_bounce}. Utilizamos unidades~$2M = 1$. El valor~$\bar{\kappa} (\bar{u}) \approx 1/(4M) = 0.5$ es prácticamente exacto en lar regiones en caída libre (la asimetría en el tiempo se debe a que la transformación de tiempos~$\bar{u} (\tau)$ no es trivial).}}
\label{fig_kappa-zoom}
\end{minipage}
\end{figure}

Consideremos a continuación una trayectoria~$x(U)$ que oscile entre~$x_{\rm in}$ y~$x_{\rm ex}$ de una forma concreta: partiendo de~$x_{\rm ex}$ con velocidad radial nula, inicia una caída libre hasta un punto~$x_{\rm b}$ muy cercano a~$x_{\rm in}$, en el cual comienza un rebote elástico de corta duración, cuyo punto más bajo es~$x_{\rm in}$, para volver de nuevo de forma simétrica hasta~$x_{\rm ex}$, punto que se alcanza con velocidad radial nula, y así sucesivamente. Excepto en el pequeño intervalo de rebote, puede comprobarse fácilmente que la trayectoria en caída libre tiene una aceleración~$\ddot{x} \approx -1/(4M)$ (la gravedad de superficie en las cercanías del horizonte). Por tanto, de~(\ref{kappa_pulsante}) se obtiene
\begin{equation}
\bar{\kappa} (\bar{u}) \approx \frac{1}{4M} =: \kappa_{\rm H}.
\label{kappa_pulsante_resultado}
\end{equation}
Es decir, salvo en el intervalo de rebote, en el infinito espacial se detecta una temperatura de radiación aproximadamente igual a la temperatura de Hawking. Como ejemplo, la figura~\ref{fig_bounce} muestra una trayectoria en caída libre con un pequeño rebote elástico regularizado, y la figura~\ref{fig_kappa-zoom} el comportamiento de la función~$\bar{\kappa} (\bar{u})$ resultante de esta oscilación, donde se comprueba el resultado~(\ref{kappa_pulsante_resultado}) con una precisión casi completa.

\noindent

Según se observa desde el infinito espacial, la duración de la mitad de un ciclo de oscilación es aproximadamente
\begin{equation}
\Delta t \approx \frac{1}{\kappa_{\rm H}} \log \frac{x_{\rm ex}}{x_{\rm in}}.
\label{vibracion}
\end{equation}
Por tanto, si imponemos~$x_{\rm ex}/x_{\rm in} \gg 1$, podemos en principio hacer que la duración de un ciclo sea significativamente mayor que el inverso de la frecuencia característica de la radiación. De esta forma, según la resolución en tiempos que se tome para la radiación percibida en el infinito, es de esperar encontrarse en uno de los dos posibles escenarios siguientes: Para una resolución temporal baja, nos encontraríamos con el espectro planckiano característico de la radiación de Hawking con un pico adicional en la frecuencia correspondiente a la oscilación~$\omega \sim 1 / (2 \Delta t)$; para una resolución alta, pero aun suficientemente baja como para resolver el espectro de radiación de Hawking, encontraríamos igualmente dicho espectro, pero acompañado de ráfagas periódicas de radiación no térmica, aproximadamente cada~$2 \Delta t$, debidas a los rebotes. En el caso en el que~$\Delta t$ no fuera significativamente mayor que~$1/|\bar{\kappa} (\bar{u})|$, es de esperar permanecer en el primer escenario. Nos queda como trabajo futuro el comprobar cómo depende la emisión acumulada en la parte del espectro asociada al rebote con las características específicas de dicho rebote.

En este punto, podemos resaltar que, como ya adelantamos, la mayor frecuencia involucrada en este cálculo para obtener la radiación de Hawking no es excesivamente alta. La máxima frecuencia involucrada en una vibración será del orden de
\begin{equation}
\omega_{\rm max} \sim \kappa_{\rm H} \rme^{\kappa_{\rm H} \Delta t}.
\label{frecuencia_maxima}
\end{equation}
($\Delta t$ es el tiempo aproximado que se tarda en realizar la mitad de la oscilación, correspondiente a la caída libre hasta el punto más bajo.) Para un agujero negro de la masa del Sol, e incluso con un semi-periodo de oscilación dado por~$\Delta t \sim 100 /\kappa_{\rm H}$, el cual corresponde a la situación extrema~$x_{\rm ex}/x_{\rm in} \sim 10^{43}$, esta frecuencia máxima sería~$\omega_{\rm max} \sim 10^4 \omega_{\rm P}$, donde~$\omega_{\rm P}$ es la frecuencia de Planck. Dicho valor está sobradamente dentro de los límites en los que la simetría Lorentz ha sido claramente verificada por las observaciones~\cite{Jacobson:2005bg,Maccione:2009ju}. A fin de tener una mejor idea del comportamiento de esta magnitud, para una oscilación entre~$x_{\rm in} \simeq 10^{-20} (2M)$ y~$x_{\rm ex} \simeq 10^{-10} (2M)$, tenemos~$\kappa_{\rm H} \Delta t \simeq 23$, de tal forma que~$\omega_{\rm max} \sim 10^{-29} \omega_{\rm P}$. La clave de que no se necesite invocar frecuencias desproporcionadas estriba en que, cuando se produce una nueva oscilación, las frecuencias involucradas no sufren un corrimiento al azul \emph{añadido,} sino que se mantienen siempre en el mismo rango, por debajo de~$\omega_{\rm max}$. Por tanto, no hay problema transplanckiano de ningún tipo en este escenario. La máxima frecuencia para la cual necesitamos asumir que la Teoría Cuántica de Campos relativista es correcta es perfectamente asumible. Por tanto, en este escenario la predicción de que en la región asintótica encontraremos radiación de Hawking (o muy similar) es completamente creíble.

Remarquemos una vez más que este vacío pulsante es apropiado para potenciales objetos estelares sin horizonte, pero cuya superficie está muy cercana al radio en el que la estructura hubiera formado un horizonte. En nuestras demostraciones, hemos hecho uso de una métrica de Schwarzschild, pero siempre a partir de un radio significativamente mayor que el horizonte. Por tanto, no ha sido necesario conjeturar la formación de un horizonte de sucesos. Además, el espectro que hemos encontrado aquí es planckiano, no necesariamente térmico en el sentido estricto de responder a una \emph{matriz de densidad} resultado de trazar sobre los modos tras el horizonte. Las partículas ``compañeras'' de las partículas de la radiación de Hawking emitidas serán igualmente emitidas a su vez, si bien con un cierto retraso temporal. Por tanto, la conocida como paradoja de la información tampoco está presente en este escenario.

En este capítulo no hemos entrado a valorar la dinámica que pudiera dar lugar a objetos estelares con una superficie oscilando muy cercana al radio de Schwarzschild del objeto. No obstante, en el marco de la \emph{gravedad semiclásica} existen varias propuestas en la literatura de objetos cuya superficie permanece muy cercana a la formación de un horizonte de sucesos, pero que, debido a la \emph{polarización cuántica del vacío,} acumulan una densidad de energía negativa suficiente en las cercanías de la superficie, por lo que no llegan a colapsar y formar un horizonte.\footnote{La polarización cuántica del vacío puede, en general, violar las denominadas \emph{condiciones de energía}~\cite{Barcelo:2002bv} en las cuales se basan los teoremas de singularidad~\cite{Penrose:1964wq,Hawking:1969sw} que predicen la formación inevitable de singularidades ``vestidas'' por horizontes de sucesos, es decir, de agujeros negros clásicos.} Algunas propuestas concretas son las \emph{gravastars}~\cite{Mazur:2001fv}, las \emph{fuzzballs}~\cite{Mathur:2005zp}, o las \emph{estrellas negras}~\cite{Barcelo:2007yk}. Oscilaciones de la superficie de estos objetos cercanas a las aquí descritas serían, muy probablemente, una fuente de radiación muy similar a la radiación de Hawking. De esta forma, es posible pensar que quizá las leyes generalizadas de la termodinámica de los agujeros negros, basadas en una temperatura \emph{exacta} proporcional a la gravedad de superficie, y en la noción geométrica del área de la superficie, no sean sino una buena aproximación a la física real de objetos mucho más complejos.

\newpage
\thispagestyle{empty}
\hbox{}

\makeatletter
\def\cleardoublepage{\clearpage\if@twoside \ifodd\c@page\else
    \hbox{}
    \thispagestyle{empty}
    \newpage
    \if@twocolumn\hbox{}\newpage\fi\fi\fi}
\makeatother \clearpage{\pagestyle{empty}\cleardoublepage}

\prefacesection{Conclusiones}
\markboth{Conclusiones}{Conclusiones}

El objetivo principal de esta tesis ha sido estudiar la percepción de los fenómenos de radiación en Teoría cuántica de Campos en sistemas de referencia no inerciales y en agujeros negros, en particular los conocidos fenómenos de la radiación de Hawking y el efecto Unruh. Se ha considerado un campo escalar de Klein-Gordon real sin masa como campo de radiación. Para llevar a cabo el estudio, se han utilizado dos herramientas: los detectores de partículas de Unruh-DeWitt macroscópicos y la función de temperatura efectiva, esta última basada en el cálculo del espectro percibido mediante transformaciones de Bogoliubov. Por otra parte, se ha demostrado que el espectro resultante del cálculo mediante transformaciones de Bogoliubov, al menos en la teoría invariante conforme en $1+1$~dimensiones, es directamente interpretable en términos de excitaciones de detectores de Unruh-DeWitt. De la aplicación de estas herramientas al estudio de la percepción de radiación por distintos observadores en diversos escenarios, extraemos las siguientes conclusiones:

\begin{itemize}
	\item La excitación de un detector de Unruh-DeWitt macroscópico siguiendo una trayectoria rectilínea con aceleración lentamente variable en el vacío de Minkowski en $3+1$~dimensiones puede calcularse mediante una expansión adiabática de su función respuesta. Dicha expansión permite descomponer el espectro total percibido en una serie asintótica de diferentes espectros de distintos órdenes. El espectro de orden cero es el espectro térmico con temperatura proporcional a la aceleración según la conocida fórmula de la temperatura de Unruh. Los siguientes son espectros con formas diversas, cuya contribución al espectro total, según aumenta su orden, está pesada por potencias cada vez mayores de derivadas cada vez mayores de la aceleración con respecto al tiempo propio. Esta expansión adiabática, además, resulta también válida como expansión en altas energías. La obtención del espectro térmico propio del efecto Unruh como contribución de orden cero nos induce a pensar en tal fenómeno como una radiación que se encuentra siempre presente cuando existe aceleración, si bien puede quedar eclipsada por contribuciones al espectro debidas a derivadas temporales de la propia aceleración.
	
	\item La función de temperatura efectiva, cuyo valor absoluto, en una teoría cuántica de campos invariante conforme para el campo de radiación, y bajo cumplimiento de la denominada condición adiabática, es proporcional a la temperatura de la radiación percibida localmente por un observador dado en un estado de vacío del campo, tiene una clara interpretación física. En general, puede descomponerse en dos contribuciones:
	
	\begin{enumerate}
		\item Una contribución a la temperatura efectiva es la que corresponde a la radiación propia del estado de vacío, la cual es aquella que escapa a la región asintótica del espacio-tiempo en el que se encuentra definido dicho vacío, o entra desde la misma, según se considere el sector de radiación saliente o entrante. La contribución de esta radiación está corregida por las características propias del observador. En el espacio tiempo de Minkowski, está corregida por el efecto Doppler dado por la velocidad del observador con respecto al sistema de coordenadas en el que se calcula la temperatura de la radiación propia del estado. En el sector radial del espacio-tiempo de Schwarzschild, está corregida por el efecto Doppler dado por la velocidad del observador con respecto al agujero negro y por el factor de corrimiento gravitacional al azul propio de la posición radial del observador.
		
		\item La otra contribución a la temperatura efectiva es la que corresponde al efecto Unruh propio del observador, el cual es debido a la aceleración del mismo con respecto a la región asintótica del espacio-tiempo, aceleración entendida como variación en el factor de corrimiento en frecuencias. En Minkowski, esta contribución está dada sencillamente por la aceleración propia del observador. En el sector radial del espacio-tiempo de Schwarzschild, está dada por la aceleración propia del observador más otra contribución debida a la aceleración gravitatoria, que igualmente acelera al observador con respecto a la región asintótica. La contribución de la aceleración gravitatoria corresponde a la aceleración dada por la ley de Newton, corregida por los factores debidos al efecto Doppler y al corrimiento al azul gravitacional ya mencionados. Esto significa que, en general, el efecto Unruh no está dado solo por la aceleración propia del observador, ni tampoco está definido localmente.
		
	\end{enumerate}
	
	\item El estado de vacío del campo puede definirse escogiendo un observador concreto, el cual no percibirá radiación en toda su trayectoria. Tal estado de vacío es el definido por los operadores destrucción correspondientes a los modos normales naturales para este observador. Dado un estado de vacío definido de esta forma, puede definirse también una función de temperatura efectiva del estado, la cual nos indica la temperatura efectiva de la radiación que escapa a la región asintótica, o entra desde la misma. Es esta temperatura, por tanto, la responsable de la primera de las contribuciones a la percepción total de un observador cualquiera enumeradas anteriormente. La temperatura efectiva del estado está dada por la aceleración del observador que define el estado de vacío con respecto a la región asintótica. Aparte de los estados de vacío manejados en la literatura (vacíos de Minkowski, Rindler, Boulware, Hartle-Hawking, Unruh, etc.), en esta tesis hemos definido y estudiado dos nuevos estados de vacío para el sector de radiación saliente de un agujero negro de Schwarzschild:
	
	\begin{enumerate}
		\item El vacío de colapso, que se define escogiendo un observador en caída libre que parte del infinito espacial con velocidad nula.
		
		\item El vacío pulsante, que se define escogiendo un observador que realiza oscilaciones en las cercanías del horizonte de sucesos, consistentes en largos periodos de caída libre (parabólica) y rebotes elásticos casi instantáneos en el punto más bajo de la oscilación.
		
	\end{enumerate}
	
	\item Las distintas contribuciones a la percepción total pueden tener carácter destructivo si presentan signos contrarios, tendiendo a anularse entre sí. Un caso paradigmático de este hecho lo constituyen los que denominamos observadores de Unruh: observadores en caída libre e instantáneamente en reposo en el exterior de un agujero negro de Schwarzschild. Para estos observadores, la contribución de la radiación procedente del agujero negro (propia del estado) y la contribución del efecto Unruh debido a su aceleración con respecto a la zona asintótica tienen signos opuestos, de manera que el efecto Unruh tiende a ``ocultar'' la radiación emitida por el agujero negro para estos observadores.
	
	\item Las expresiones analíticas obtenidas para la función de temperatura efectiva en términos de las características del estado de vacío y de la trayectoria del observador, en las cuales aparecen claramente los roles de los distintos fenómenos físicos mencionados, son válidas independientemente del cumplimiento o no de la condición adiabática que permitiría considerar el valor absoluto de tal función como proporcional a la temperatura del espectro térmico realmente percibido. Entendemos que se trata de otro indicio claro de que esta función describe un fenómeno de percepción de radiación térmica con entidad física propia, aunque en general su contribución a la percepción total no sea la única, debido a la no adiabaticidad. Este fenómeno físico, al menos en términos exclusivos de percepción de radiación, vendría a unificar y generalizar los fenómenos conocidos de la radiación de Hawking y el efecto Unruh.
	
	\item La aplicación de la función de temperatura efectiva al estudio de la percepción de radiación por distintos observadores en Minkowski nos ha permitido descartar un escenario irreal: no es posible para un observador dejar de percibir, durante un tiempo suficiente, una radiación que sí sea percibida por observadores inerciales mediante un efecto Unruh que interfiera destructivamente con tal radiación y que sea fruto exclusivamente de la aceleración propia del observador contra dicha radiación.
	
	\item La aplicación de la función de temperatura efectiva al estudio de la percepción de radiación por distintos observadores en el exterior de un agujero negro de Schwarzschild nos ha permitido encontrar los siguientes resultados físicos:
	
	\begin{enumerate}
		\item El estado que hemos denominado vacío de colapso reproduce cualitativamente el resultado esperable para la emisión de radiación durante un proceso de colapso real de un objeto estelar que acabe formando un agujero negro u otro objeto similar. A efectos de percepción de radiación emitida a la región asintótica, dicho vacío no estacionario interpola de manera monótona y suave en el tiempo entre la ausencia de radiación propia del vacío de Boulware y la radiación de Hawking propia del vacío de Unruh.
		
		\item Contrariamente a lo que se ha sostenido habitualmente en la literatura, en el vacío de Unruh los observadores en caída libre en general sí detectan radiación al cruzar el horizonte de sucesos. Salvo para el observador que al cruzar el horizonte se encuentra, además, instantáneamente en reposo, para el resto de observadores la función de temperatura efectiva es no nula al cruzar el horizonte, siendo cuatro veces mayor que la correspondiente a la radiación de Hawking para el observador que parte en caída libre desde el infinito espacial con velocidad nula. Esto se debe a un efecto Doppler divergente en el horizonte, cuya divergencia compensa el hecho de que la temperatura de los observadores de Unruh sí tiende a anularse en el horizonte, dando lugar a un valor límite finito. Este valor no nulo sugiere fuertemente que la percepción de radiación también será no nula, puesto que, aunque la condición adiabática no se cumple estrictamente durante el incremento final de la temperatura al cruzar el horizonte, el valor de la función de control adiabática resulta aun inferior a la unidad, lo cual indica que el valor de la temperatura efectiva es un razonable estimador de la cantidad de radiación percibida, aunque esta no tenga un espectro térmico.
		
		\item En un agujero negro de Schwarzschild, además de las geodésicas, existen otro tipo de trayectorias especiales, que podrían describirse como carentes de ``fricción cuántica'': aquellas para las cuales no hay efecto Unruh en el sector de radiación saliente o entrante (pero en general no en ambos). Por una parte, existen trayectorias que parten del infinito espacial o del horizonte de sucesos para aproximarse asintóticamente a un radio final, con efecto Unruh nulo para la radiación saliente. Si estas mismas trayectorias se siguen en sentido contrario, es decir, partiendo de un radio fijo en el pasado asintótico y desviándose de él hacia el horizonte o hacia el infinito espacial, entonces el efecto Unruh es nulo para la radiación entrante. Por otra parte, las trayectorias estáticas que permanecen con un radio constante son las únicas para las que no hay efecto Unruh en ninguno de los dos sectores.
		
	\end{enumerate}
	
	\item A lo largo de las trayectorias sin efecto Unruh en la radiación saliente y que parten del infinito espacial, la radiación percibida, cuando el campo se encuentra en el vacío de Unruh, es constante y toda procedente de la emisión del agujero negro. Por otra parte, la aceleración propia necesaria para seguir tales trayectorias escala con el inverso de la distancia radial al cuadrado a lo largo de las mismas. Por tanto, en un agujero negro en $3+1$~dimensiones, debido a la difusión de la radiación emitida por el agujero negro, un objeto con sección eficaz constante siguiendo una de estas trayectorias detectaría una intensidad total de radiación proporcional a la aceleración propia que necesita para continuar por esa misma trayectoria, a lo largo de la cual alcanzaría finalmente un radio constante. Esto plantea un posible escenario auto-consistente de flotación debida a la radiación del agujero negro. Si bien para estas trayectorias sí existe efecto Unruh en el sector de radiación saliente, un primer análisis parece indicar que tiene, en general, una importancia secundaria. Además, lo esperable es que, en una primera aproximación, dicho efecto Unruh sea favorable al propio proceso de flotación, y que en cualquier caso desaparezca en el estadio final del mismo. Todo este escenario será objeto de estudio futuro.
	
	\item El estado que hemos denominado vacío pulsante es una propuesta de estado de vacío para la que se obtiene una radiación muy similar a la radiación de Hawking emitida por un agujero negro, salvo por ráfagas periódicas de radiación añadidas, o por picos de emisión superpuestos en determinadas frecuencias. El uso de este vacío permite evitar el tener que invocar frecuencias extremadamente altas en el proceso de creación de partículas en el infinito, así como la necesidad de que el objeto forme un horizonte de sucesos. Sin entrar en los procesos físicos que pudieran dar lugar a un objeto estabilizado y oscilante en las cercanías de su radio de Schwarzschild (los cuales serán objeto de estudio futuro), cuya radiación creemos que podría describirse adecuadamente por este vacío, el mecanismo de pulsación aquí descrito demuestra que tales objetos podrían radiar de una manera muy similar a como lo haría un agujero negro, sin que el problema transplanckiano ni la paradoja de la información estuvieran presentes.
	
\end{itemize}
\newpage
\thispagestyle{empty}
\hbox{}

\makeatletter
\def\cleardoublepage{\clearpage\if@twoside \ifodd\c@page\else
    \hbox{}
    \thispagestyle{empty}
    \newpage
    \if@twocolumn\hbox{}\newpage\fi\fi\fi}
\makeatother \clearpage{\pagestyle{empty}\cleardoublepage}

\prefacesection{Publicaciones}
\markboth{Publicaciones}{Publicaciones}

El trabajo realizado para esta tesis doctoral ha dado lugar a cuatro artículos y dos contribuciones en actas de conferencias publicados:

\begin{itemize}
	\item Luis C.\ Barbado, Carlos Barcelo, and Luis J.\ Garay. Hawking radiation as perceived by different observers. \textit{Class.Quant.Grav.,} 28:125021, 2011.
	
	\item Luis C.\ Barbado, Carlos Barcelo, Luis J.\ Garay, and Gil Jannes. The Trans-Planckian problem as a guiding principle. \textit{JHEP,} 1111:112, 2011.
	
	\item Luis C.\ Barbado, Carlos Barcelo, and Luis J.\ Garay. Hawking radiation as perceived by different observers: An analytic expression for the effective temperature function. \textit{Class.Quant.Grav.,} 29:075013, 2012.
	
	\item Luis C.\ Barbado, Carlos Barcelo, and Luis J.\ Garay. Hawking radiation as perceived by different observers. \textit{AIP Conf.Proc.,} 1458:363--366, 2011.
	
	\item Luis C.\ Barbado and Matt Visser.\ Unruh-DeWitt detector event rate for trajectories with time-dependent acceleration. \textit{Phys.Rev.,} D86:084011, 2012.
	
	\item Carlos Barceló, Luis C Barbado, Luis J Garay, and Gil Jannes. Avoiding the trans-planckian problem in black hole physics.\ In \textit{Progress in Mathematical Relativity, Gravitation and Cosmology,} pages 129--133. Springer Berlin Heidelberg, 2014.
	
\end{itemize}

\newpage
\thispagestyle{empty}
\hbox{}

\makeatletter
\def\cleardoublepage{\clearpage\if@twoside \ifodd\c@page\else
    \hbox{}
    \thispagestyle{empty}
    \newpage
    \if@twocolumn\hbox{}\newpage\fi\fi\fi}
\makeatother \clearpage{\pagestyle{empty}\cleardoublepage}

\bibliographystyle{unsrt}
\bibliography{biblio}

\makeatletter
\def\cleardoublepage{\clearpage\if@twoside \ifodd\c@page\else
    \hbox{}
    \thispagestyle{empty}
    \newpage
    \if@twocolumn\hbox{}\newpage\fi\fi\fi}
\makeatother \clearpage{\pagestyle{empty}\cleardoublepage}

\newpage
\thispagestyle{empty}
\hbox{}

\newpage
\thispagestyle{empty}
\hbox{}

\end{document}